\documentclass{aastex62}

\graphicspath{{./}{figures/}}

\shorttitle{Large-scale filaments in inner Galactic plane}
\shortauthors{Y.-F. Ge \& K. Wang}

\usepackage[figuresright]{rotating}
\usepackage{verbatim}
\usepackage{textcomp}
\usepackage{subfigure}
\usepackage{amsmath}
\usepackage{ulem}
\usepackage{color}
\usepackage[pass]{geometry}
\usepackage{makecell}

\begin{document}

\title{A census of 163 large-scale ($\ge 10$\,pc), velocity-coherent filaments in inner Galactic plane: \\
physical properties, dense gas fraction, and association with spiral arms}

\correspondingauthor{Ke Wang}
\email{kwang.astro@pku.edu.cn}

\author[0000-0002-8727-0868]{Yifei Ge}
\affiliation{Kavli Institute for Astronomy and Astrophysics, Peking University, 5 Yiheyuan Road, Haidian District, Beijing 100871, China}
\affiliation{Department of Astronomy, School of Physics, Peking University, 5 Yiheyuan Road, Haidian District, Beijing 100871, China}

\author[0000-0002-7237-3856]{Ke Wang}
\affiliation{Kavli Institute for Astronomy and Astrophysics, Peking University, 5 Yiheyuan Road, Haidian District, Beijing 100871, China}

\begin{abstract}
The interstellar medium has a highly filamentary and hierarchical structure, which may play a significant role in star formation. A systematical study on the large-scale filaments towards their physical parameters, distribution, structures and kinematics will inform us of what kind of filaments have potential to form stars, how the material feed protostars through filaments, and the connection between star formation and Galactic spiral arms. Unlike the traditional ``by eyes'' searches, we use a customized minimum spanning tree algorithm to identify filaments by linking Galactic clumps from the APEX Telescope Large Area Survey of the Galaxy catalogue. In the inner Galactic plane ($|l| < 60^\circ$), we identify 163 large-scale filaments with physical properties derived, including dense gas mass fraction, and compare them with an updated spiral arm model in position-position-velocity space. Dense gas mass fraction is found not to differ significantly in various Galactic position, neither does it in different spiral arms. We also find that most filaments are inter-arm filaments after adding a distance constraint, and filaments in arm differ a little with those not in. One surprising result is that clumps on and off filaments have no significant distinction in their mass at the same size.
\end{abstract}


\section{Introduction} \label{sec:intro}
 Conspicuous filamentary structures are ubiquitous in the interstellar medium (ISM). But how the filaments form is not clearly understood. Numerical simulations of the multiphase atomic ISM in Galactic disk show that cold, dense gas prefer to organize itself naturally in the filamentary network \citep{Tasker2009}. Widths derived from observed filaments support the argument that the filaments may form as a result of the dissipation of large-scale turbulence \citep{Arzoumanian2011}. Filaments are constituent of molecular clouds and may play a pivotal role in the process of star formation \citep[e.g.][]{Schneider1979,Arzoumanian2011,ZhangM2019}. According to simulation, massive star forming clumps are filamentary structures in their early stages \citep{Smith2009}. And \citet{Andre2010} observed filaments in both quiescent and star-forming clouds, where a large part of prestellar cores reside. \\
 
 Filamentary structures have been observed by different tracers. For the reason that they are infrared dark, filaments can be found from extinction maps at optical \citep{Schneider1979} and infrared wavelengths \citep{Jackson2010,Kainulainen2013}. Far-infrared or submillimeter dust emission maps are also employed \citep{Andre2010,Hacar2013,Li2016}. Researches on filaments have been focused on nearby star-forming regions such as Taurus and Orion \citep{Palmeirim2013,Takahashi2013,Kainulainen2017}, or substructures such as fibers \citep{Hacar2018}. \\
 
 A variety of filaments have been found with diverse lengths and morphology. For instance, infrared dark cloud ``Nessie'' \citep{Jackson2010}, is found to be as long as 430 pc in the CO map and runs along the Scutum-Centaurus Arm in position-position-velocity space \citep{Goodman2014}. Careful studies of the individual filaments provide important hints on their properties, the role they play in the star formation and so on. However, on one hand, individuality is inevitable. On the other hand, more distant filaments which also contain high-mass star formation have not been studied comprehensively. So an unbiased sample of filaments in the Milky Way is needed.\\

Large-scale ($\ge 10$\,pc) filaments may be connected to the large-scale, Galactic spiral structures. From a position-velocity analysis, \citet{Goodman2014} suggest that Nessie with length of hundreds of parsecs forms a bone-like feature that closely follows the center of the Scutum-Centaurus Arm of the Milky Way. The inspection of this linkage between large-scale filaments and spiral arms is extended to other Galactic positions in the inner Galactic plane \citep[e.g.][]{Ragan2014,Zucker2015,Wang2015,Abreu2016}. Unlike small-scale filaments \citep[e.g.][]{Schisano2014,Schisano2020,Koch2015,Li2016,Mattern2018} that nearby ones could be found, large-scale filaments in the Milky Way are typically far from us, so researches towards them have not come up until recent years, when modern multiwavelength survey that covers Galactic plane at high resolution and sensitivity. Several powerful algorithms have been developed to identify filaments through intensity or column density map such as DisPerSE, FilFinder, local Hessian matrix, getFilaments and related algorithms based on wavelet transforms \citep{Sousbie2011a,Sousbie2011b,Koch2015,Menshchikov2013,Molinari2010b,OssenkopfOkada2018,Schisano2014}. The algorithms above perform well in finding parsec-scale filaments. However, large-scale filaments are rather different. One important feature is that, they may not be continuous in mm or submm continuum images. Therefore, although several catalogues of large-scale filaments have been produced in the past few years \citep{Ragan2014, Wang2015,Wang2016, Abreu2016,Zucker2018} and different searching methods have different selection criteria to identify filaments, of which most are by inspection of dust characteristic artificially, except \citet{Wang2016}. \\

\citet{Wang2016} adopted minimum spanning tree (MST) algorithm to automatically identify filaments by connecting velocity-coherent clumps. The minimum spanning tree, a construct from graph theory, is the unique set of straight lines (``edges'') connecting a given set of points (``vortices'') without closed loops, such that the sum of the edge lengths is a minimum. Two algorithms have been developed independently by \citet{Kruskal1956,Prim1957}. MSTs have been associated for the fist time to cluster analysis by \citet{GowerRoss1969}. In astrophysics, minimum spanning trees have so far mainly been used to analyse the large-scale distribution of galaxy or galaxy clusters, and filamentary feathers have been found \citep[e.g.][]{Barrow1985,AdamiMazure1999,Doroshkevich2004,Colberg2007,ParkLee2009,Alpaslan2014,Naidoo2020,Pereyra2020}. MSTs have also been employed to identify star clusters \citep[e.g.][]{Cartwright2004,Schmeja2006,Gutermuth2009,Wu2017}. In high energy astrophysics, MSTs have been used for $\gamma$-ray source detection \citep[e.g.][]{DiGesuSacco1983,Campana2008,Campana2013}. Recently, MSTs have also been introduced to quantify core separations and mass segregation \citep[e.g.][]{Sanhueza2019,Dib2019}. However, MSTs have also been criticized for giving unreliable cluster catalogues due to ``chaining'' of unrelated structure when noise is present \citep{Getman2018}. Therefore, apart from kinematic coherence examination carryied by \citet{Wang2016}, we also did checks on the 2D (Galactic longitude and latitude) MSTs. The clumps \citet{Wang2016} employ are from Bolocam Galactric Plan Survey (BGPS) \citep{Rosolowsky2010}, which only cover half of the Galactic plane ($7.5^\circ\leq l\leq 195^\circ$). Now, ATLASGAL gives better estimation of distance, temperature and velocity towards a complete sample of Galactic clumps \citep{Urquhart2018}. High-resolution column density map derived from the Herschel Hi-GAL survey with PPMAP \citep{Marsh2017} also makes it possible to get a more accurate estimation of filament mass in the entire Galactic plane.\\

Physical properties of large-scale filaments associated with spiral structure have been inspected recently \citep[e.g.][]{Ragan2014,Abreu2016,Wang2015,Wang2016,Zucker2018}. However, most of the studies lack a large sample size, or use spiral arm models that do not have high-accuracy distance estimation. Now we are able to obtain a large sample of large-scale filaments and an updated spiral arm model fitted from trigonometric parallaxes of high-mass star-forming regions \citep{Reid2019} which gives us better assistance to study the relation between filaments and spiral arms. With the advance of the radial velocity measurements of ATLASGAL clumps \citep{Urquhart2014} and the updated spiral arm model \citep{Reid2019}, we extend the work of \citet{Wang2016} in the northern Galactic plane to the entire (except Galactic center) inner Galactic plane covered by ATLASGAL.

The paper is structured as follows. We describe the data and methods we use in Sect. \ref{sec:method}. Section \ref{sec:results} describes our results containing physical properties of our large-scale filaments and some statistics as well as their dense gas mass fraction. Then in Sect. \ref{sec:discussion}, we examine the robustness of MST method and investigate the Galactic distribution of the filaments. Fragmentation of large-scale filaments is also inspected. And finally, our conclusions are summarized in Sect. \ref{sec:conclusion}.

\section{Data and Method} \label{sec:method}

\subsection{ATLASGAL Galactic Clumps}

The APEX Telescope Large Area Survey of the Galaxy (ATLASGAL) is an unbiased 870 \textmu m submillimeter survey of the inner Galactic plane ($|l|<60^\circ$ with $|b|<1.5^\circ$). They identified 10163 compact sources with a resolution of $19^{\prime\prime}$ and generated a comprehensive and unbiased catalogue of massive, dense clumps, which are located across the inner Galaxy \citep{Urquhart2014}. Among the total 10163 sources, $\sim$ 8000 dense clumps located away from the Galactic center region ($300^\circ<l<355^\circ$ and $5^\circ<l<60^\circ$) were studied on the physical properties, such as velocity, by \citet{Urquhart2018}. The velocity information of clumps makes it possible to investigate their coherence in position-position-velocity (PPV) space.

\subsection{Herschel Column Density Map}\label{sec:Herschel}

The Herschel satellite \citep{Pilbratt2010} provides a powerful tool to study the structure of high-mass molecular clouds on different scales \citep[e.g.][]{Beuther2010, Schneider2012}. Hi-GAL, the Herschel infrared Galactic Plane Survey, makes an unbiased photometric survey of the inner Galactic plane by mapping a $2^\circ$ wide strip in the entire Galactic Plane in five wavebands between 70 \textmu m and 500 \textmu m \citep{Molinari2010}. High resolution temperature-differential column density maps with an angular resolution of 12 arcsec have been derived from Herschel Hi-GAL data using PPMAP tool, which represents improvement over those obtained with standard analysis techniques. \citep{Marsh2017}. We employ the temperature-integrated column density maps for the following mass calculation of our filaments.\\

\subsection{Filament Identification}\label{sec:ident}
\begin{figure*}
\begin{center}
\includegraphics[scale=0.3,angle=0]{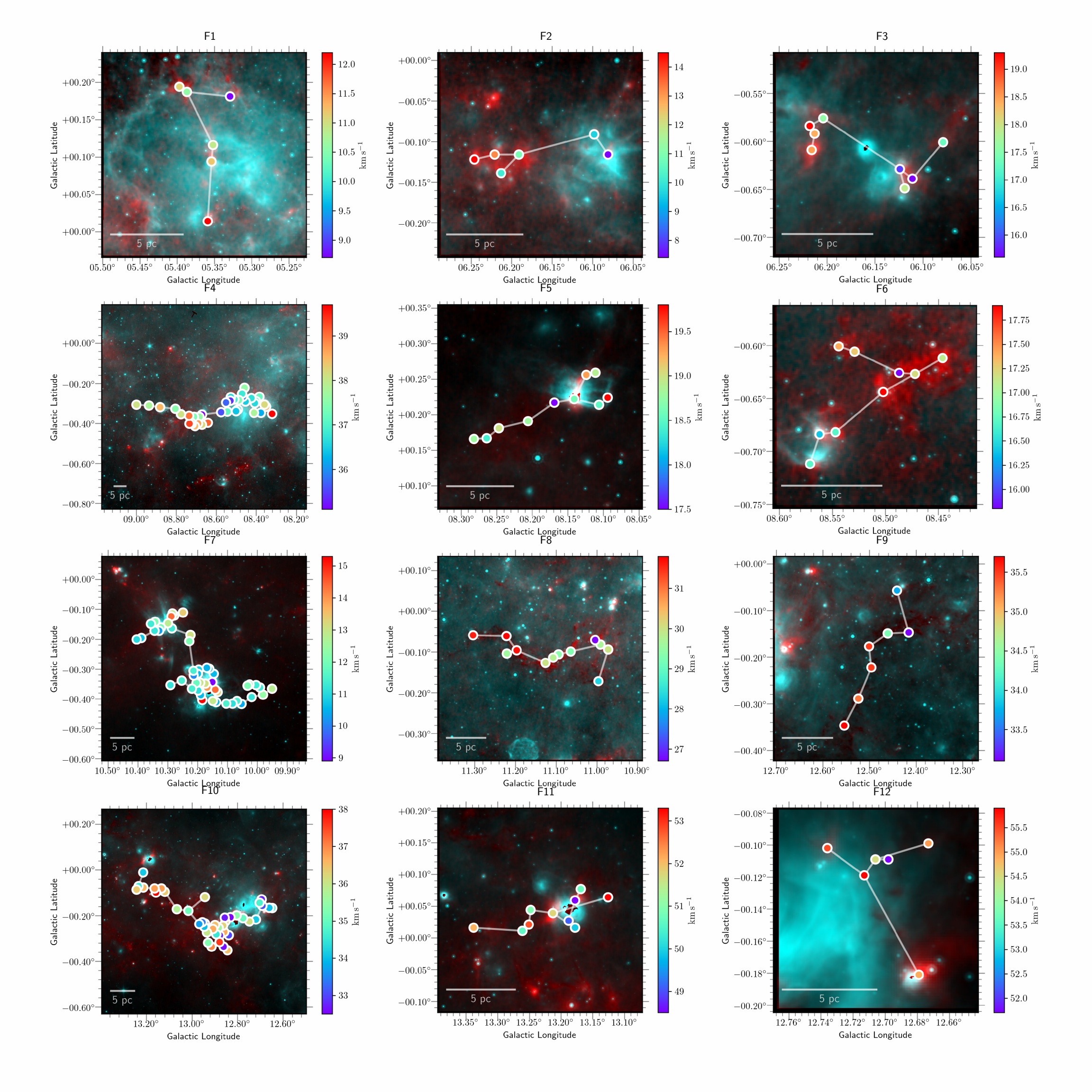}
\caption{Two-color view of filaments. The color-coded circles denote clumps in filaments with various velocities. For backgrounds, cyan represents intermediate infrared 24 \textmu m emission on logarithmic scale from MIPSGAL \citep{Carey2009} and red shows submillimeter 870 \textmu m emission on linear scale from APEX + Planck combined image \citep{Csengeri2016}. Other two-color view of filaments and their column density maps are displayed in Appendix \ref{sec:maps}}
\label{twocolor}
\end{center}
\end{figure*}

\begin{figure*}
\gridline{\fig{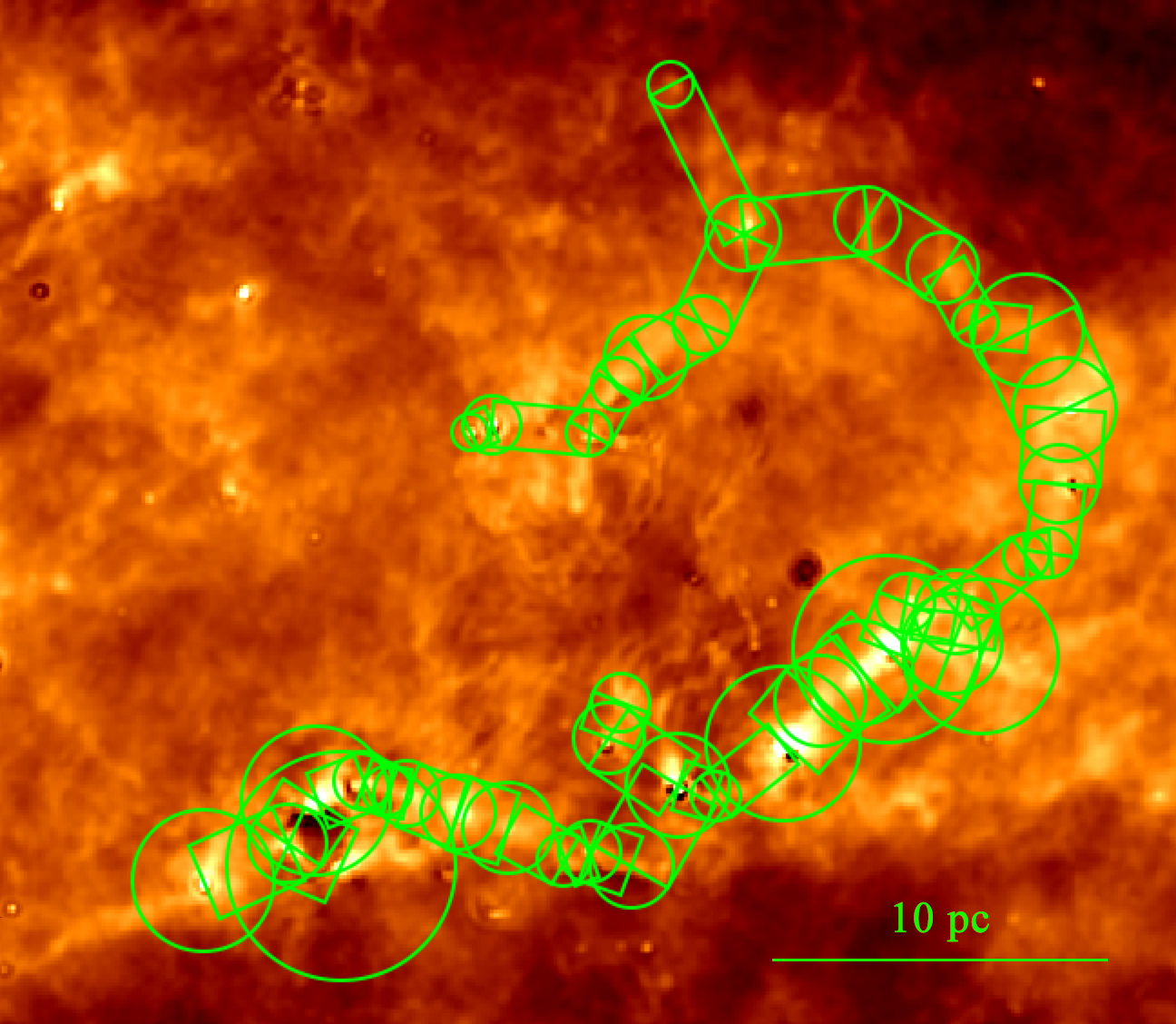}{0.485\textwidth}{(a)}
          \fig{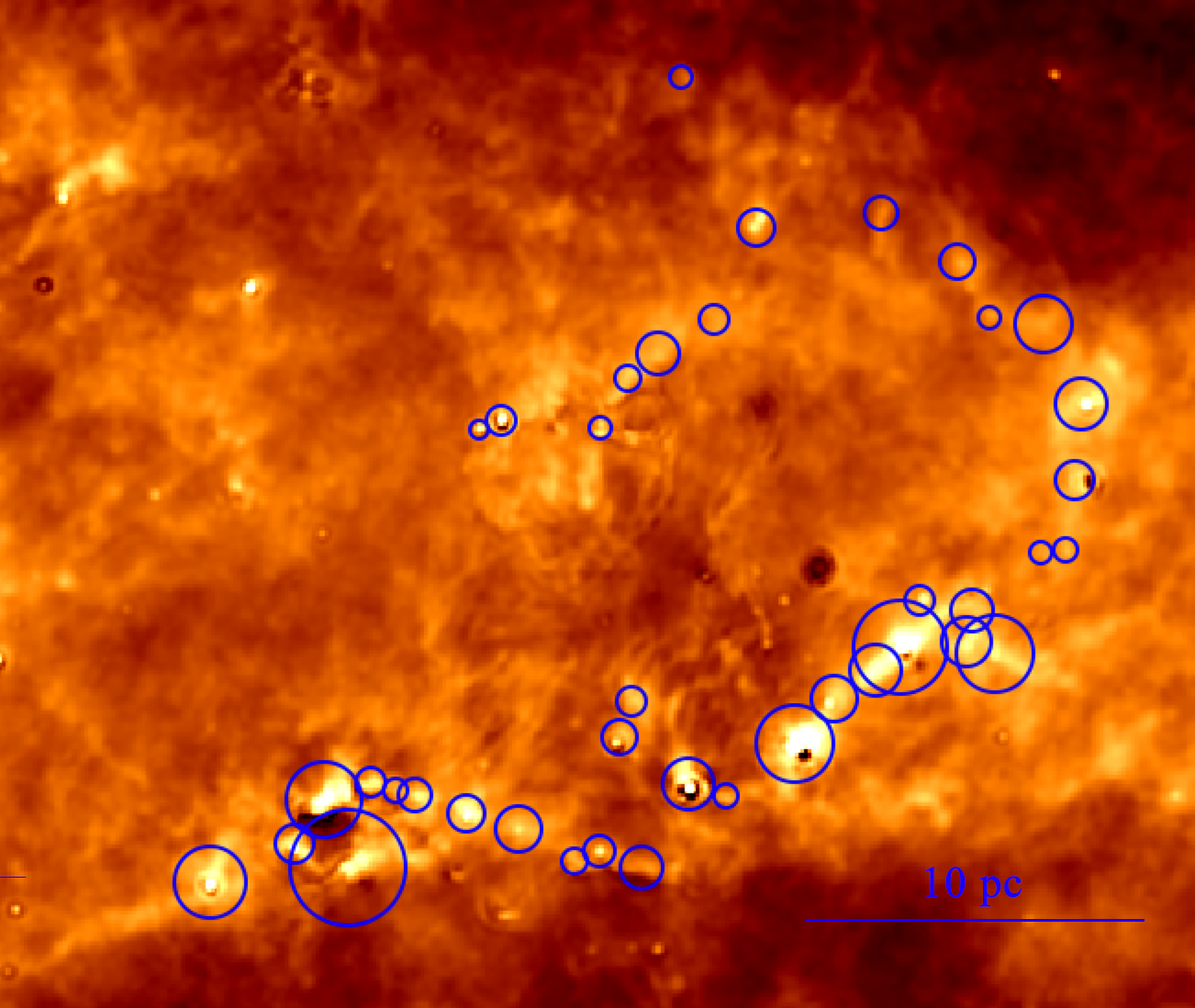}{0.5\textwidth}{(b)}
          }
\gridline{\fig{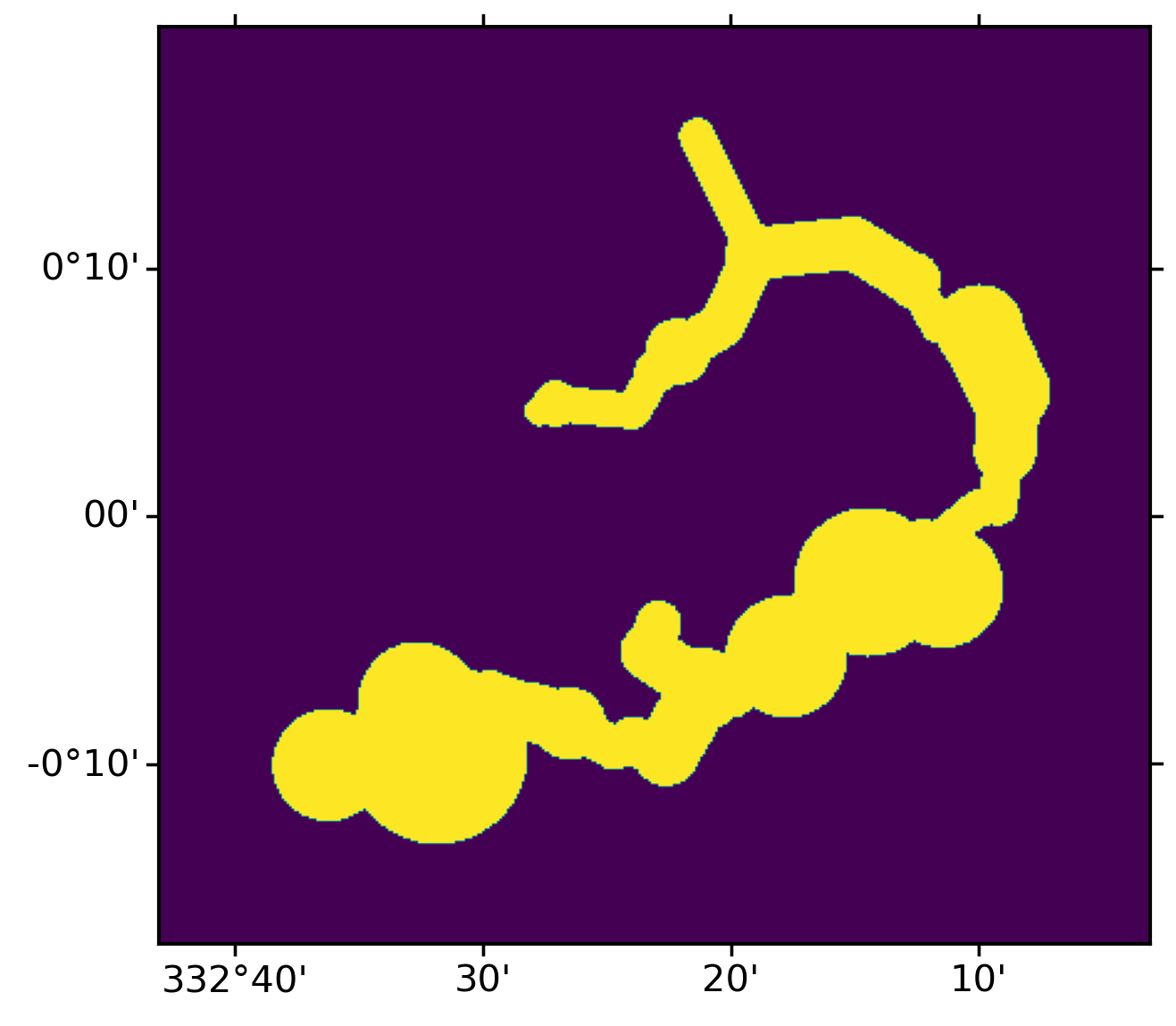}{0.5\textwidth}{(c)}
          \fig{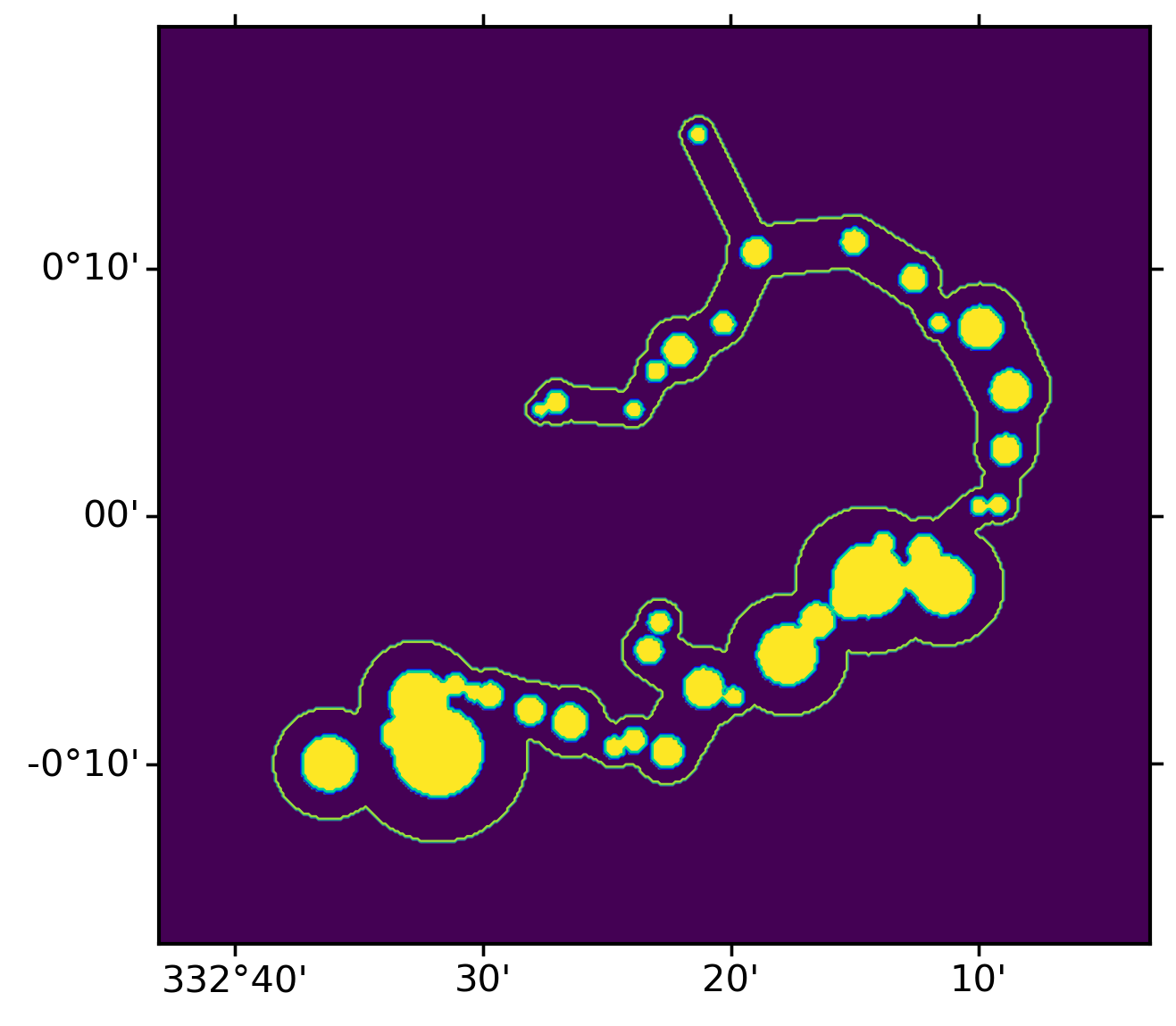}{0.5\textwidth}{(d)}
          }
\caption{Exemplification of how to define total mass and dense gas mass in a filament. In panel (a), background is Herschel column density map derived from PPMAP \citep{Marsh2017} in logarithmic scale. Circles are located in the same position with clumps in this filament and the diameters are four times the major axes of clumps. Rectangles are guided by ``edges'' of a filament. The length of a rectangle is the separation between two clumps in both ends and width is equal to the smaller diameter of the two circled in both ends. The boundary of a filament is a combination of these circles and rectangles shown in panel (c). And the mass of a filament is obtained by integrating column density within the boundary. Dense gas mass of a filament is the sum of mass within blue circles in panel (b). Blue circles are also located in the position of clumps. The diameters of circles are twice the major axes of clumps. Yellow masks in panel (d) show the boundary of dense gas and the green contour enclosing them is the sketch of the filament boundary.}
\label{Mcal}
\end{figure*}

The filaments are identified from ATLASGAL \citep{Urquhart2018} clumps usin minimum spanning tree (MST) method firstly used by \citet{Wang2016}\footnote{Code is available in \citet{Wang2021}}. The MST algorithm is adopted to isolate filaments in PPV sample, meaning clumps cluster as a filament only when they are close to each other with similar velocities. The clumps (circles in Fig. \ref{Mcal} (a) and filled circles in Fig. \ref{twocolor}) are connected with the cost of a minimum sum of edge lengths, where ``edges'' (rectangles in Fig. \ref{Mcal} (a) and white lines in Fig. \ref{twocolor}) mean straight lines linking clumps, and edge lengths are separation between each two clumps. The criteria for MST matching and filament selection follow \citet{Wang2016} :
\begin{itemize}
\item[(1)]The accepted MST must contain at least five ATLASGAL clumps: $N_{cl} \geq 5$.
\item[(2)]Only edges shorter than a maximum length (cut-off length) can be connected $(\Delta L < 0.1^\circ )$.
\item[(3)]For any two clumps to be connected, the difference in line-of-sight velocity (matching velocity $\Delta v$) must be less than 2 km s$^{-1}$.
\item[(4)]Linearity $f_L > 1.5$. Here linearity is defined to quantify the degree of similarity between the target with a linear shape.
\item[(5)]Projected length $L_{sum}\geq  10$ pc. We only focus on large-scale filaments in this work.
\end{itemize}
Initial values of cut-off length $\Delta L$ in (2), and matching velocity $\Delta v$ in (3), are chosen based on characteristics of previously known filaments. A variety of values around initial ones are tested and the best combination is chosen to identify known filaments as many as possible but not connect unrelated sources, see Appendix \ref{sec:parameter} for details. $N_{cl}$ in (1), minimum number of clumps refers to ``pruning'' level of MSTs in \citet{ParkLee2009} and \citet{Pereyra2020}. In their work, when a branch of an MST has less than 5 nodes, it is thought as a minor branch and should be removed from the tree. Linearity in (4) is the ratio between spread (standard deviation) of clumps along major axis and spread along minor axis. To define the major axis of a filament, we plot all the clumps in the projected sky (Galactic longitude as $x$ and Galactic latitude as $y$) and fitted a line as the major axis of this filament. But instead of ordinary least squares, we get this line with the help of principle component analysis (PCA), which is detailed in Appendix \ref{sec:PCA}. The minor axis is perpendicular to the major axis and pass through the mean position $(\overline{x},\overline{y})$. The larger $f_L$ means the shape is more likely to be linear. For a straight line, $f_L\rightarrow \infty $. The distances of clumps in a filament are thought to be the same. If the distances of clumps in one filament are different, a unified distance will be used (detailed in Sect. \ref{sec:properties}). The identified filaments are shown in Fig. \ref{twocolor} and Sect. \ref{sec:results}.\\

\subsection{Mass Calculation}\label{sec:mass}

Mass of the filaments is calculated by summing mass within the boundary of each filament in the Hi-GAL based column density maps. The boundary of a filament is a combination of a series of circles and rectangles as shown in Fig. \ref{Mcal}. A circle represents the position of a clump with a diameter of four times the clump's major axis. The diameter is chosen referring to the aperture radius ATLASGAL used to measure clump flux density from Herschel, which is twice the major axis \citep{Urquhart2018}. And our mass is derived from Herschel column density map. This diameter is large enough that most of the source emission is within the boundary, while still being small enough to avoid including the background. Rectangles are consist of edges linking each two clumps in a filament. The length of a rectangle is the length of the edge (that is, separation between two clumps in both ends) and the width is the same as the smaller circle diameter of two clumps in both ends. Then the mass of filament is obtained by summing contributions from the pixels of the column-density map in the boundary. Owing to gaps in coverage on column density map, 29 filaments are affected (filaments with DGMF of ``-'' in Table \ref{t1}). For a filament with incomplete column density map, we can only obtain a total clump mass, which is the sum of ATLASGAL clump mass in this filament. So for these filaments, instead of integrating column density within the boundary, we firstly calculate the mean ratio of total clump mass to filament mass obtained from column density map of other 134 filaments with complete column density map. Then we scale total clump mass for these filaments by the mean ratio.\\

For dense gas mass calculation, several approaches have been employed in previous studies. They can be classified into two categories, one is calculating mass from a denser gas tracer within the same boundary as that of the whole filament mass \citep[e.g.][]{Ragan2014}. The other is to take contour with a higher level in the extinction map as the boundary to calculate dense gas, and contour with lower level as boundary of total mass \citep[e.g.][]{ZhangM2019}. To avoid systematic bias, we adopt the latter one. That is, our dense gas mass is also acquired from Herschel Hi-GAL column density map. Dense gas mass of a filament is derived from the sum of clump mass in this filament derived by an integral of column density in circle regions (shown as blue circles in Fig. \ref{Mcal} (b)) with diameter twice the major axis of clumps. We also tried to take clump mass directly from ATLASGAL catalogue \citep{Urquhart2018}, where mass is calculated with integrated 870 \textmu m flux density as dense gas. We compared clump mass from ATLASGAL and that from Herschel column density map with the same radius. And we find some clump masses from ATLASGAL are dramatically larger than that from Herschel column density map. Considering the consistency of calculation method between the dense mass and the total mass, we did not use mass from ATLASGAL catalogue. Another reason is that, column density map obtained with PPMAP considered temperature variance pixel-by-pixel while ATLASGAL takes a single temperature from spectral grey body fitting for each clump to calculate mass.\\

Unlike employing a fixed extinction or column density contour to define boundaries of filaments or dense gas region, our boundaries vary in different Galactic longitudes. As is known to us, column density in the inner Galactic plane is not the same in various Galactic longitudes, being larger in the center and decreasing outwards. A fixed extinction or column density contour, on one hand, in Galactic longitude where column density is high, might put extra mass that should not be in a filament in that filament. On the other hand, in less dense Galactic longitudes, the contour may contain nothing. Our polygon boundaries guided by filaments avoid this problem.

\section{Results} \label{sec:results}

We identify 163 large-scale filaments in the inner Galactic plane ($|l|<60^\circ$ with $|b|<1.5^\circ$) from ATLASGAL clumps and the filaments are shown in Fig. \ref{twocolor}. The color-coded circles in Fig. \ref{twocolor} denote clumps with different velocities. For backgrounds, cyan represents intermediate infrared 24 \textmu m emission on logarithmic scale from MIPSGAL \citep{Carey2009} and red shows submillimeter 870 \textmu m emission on linear scale from APEX + Planck combined image \citep{Csengeri2016}. Other two-color view of filaments and their column density maps are displayed in Appendix \ref{sec:maps}.

\subsection{Basic Physical Properties} \label{sec:properties}
Physical properties of the 163 filaments are shown in Table \ref{t1}. The meaning of columns are listed in Table \ref{t1_explain}. The determination of ATLASGAL clump distances can be roughly divided into four steps \citep{Urquhart2018}. Firstly, if maser parallax and spectroscopic distances are available, literature solution is adopted. Secondly, kinematic distances are determined for clumps with radial velocity. Thirdly, clumps located within the Solar circle face distance ambiguities, which are resolved using archival HI data. Fourthly, they group the clumps to assign or correct their distances. Typical uncertainty for kinematic distances of ATLASGAL clumps is 0.3 kpc.  As mentioned at the end of Sect. \ref{sec:ident}, we consider clumps in a velocity-coherent filament have the same distance. This could be regarded as a correction for distance measurement of ATLASGAL clumps. Physical properties of clumps that we directly take from ATLASGAL catalogue relating with distance are radius and Galactocentric radius. Both are linear correlated with distance. So we could correct them by simply multiplying a factor $d_{fl}/d_{cl}$, where $d_{cl}$ is clump distance directly from ATLASGAL catalogue and $d_{fl}$ is the distance of the filament. In most of our filaments (95\%), clumps have the same distances. One probable reason for this is that most of velocity coherent structures found by MST are physically real structures, so clumps in the same structure naturally have the same distances. A small portion of MSTs may be a collection of clumps without physical connection. Another probable reason is that ATLASGAL applied friends-of-friends method to group the clumps \citep{Urquhart2018}. For some of clumps ($\sim10\%$) without well defined distances, distance of the cluster they belong is assigned to them. And for some well studied complexes with reliable distances, their associated clumps adopted the reliable distances.\\

We include standard deviations of Galactic longitude, Galactic latitude, velocity, velocity gradient, and height above Galactic mid-plane in Table \ref{t1}. For a fractal turbulent cloud with mass of 1000 M$_\odot$, the standard deviation of mass derived from PPMAP is 612 M$_\odot$ \citep{Marsh2015}. The uncertainties for $|\theta|$ are derived from bootstrap method. That is, we add Gaussian noise to Galactic longitude and latitude of clumps, and then get one $|\theta|$ with PCA. We repeat this process for 1000 times and take standard deviation of the 1000 values of $|\theta|$ as the uncertainty of $|\theta|$. \\

The statistics of the parameters are shown in the last few rows of Table \ref{t1}, including minimum, maximum, median, mean, standard deviation, skewness (S), and kurtosis (K). Skewness measures the asymmetry of probability distribution. For instance, normal distribution has $S=0$. Negative skewness means that the tail is on the left side of the distribution, and positive skewness indicates that the tail is on the right. Kurtosis characterizes how the distribution is compared to a normal distribution. The normal distribution has $K=0$. A distribution with $K>0$ is more centrally peaked than normal distribution while for $K<0$ is flatter.

{\setlength{\tabcolsep}{2pt}
\begin{longrotatetable}
\movetabledown=0.5in
\begin{deluxetable}{cccccccccccccccccccccccc}
\tablecaption{\label{t1}}
\tablecolumns{24}
\tabletypesize{\scriptsize}
\tablehead{
\colhead{(1)} & \colhead{(2)} & \colhead{(3)} & \colhead{(4)} & \colhead{(5)} & \colhead{(6)} & \colhead{(7)} & \colhead{(8)} & \colhead{(9)} & \colhead{(10)} & \colhead{(11)} & \colhead{(12)} & \colhead{(13)} & \colhead{(14)} & \colhead{(15)} & \colhead{(16)} & \colhead{(17)} & \colhead{(18)} & \colhead{(19)} & \colhead{(20)} & \colhead{(21)} & \colhead{(22)} & \colhead{(23)} & \colhead{(24)}\\
\colhead{ID} & \colhead{$l_{wt}$} & \colhead{$b_{wt}$} & \colhead{$v_{wt}$} & \colhead{$d$} & \colhead{N$_{cl}$} & \colhead{$L_{sum}$} & \colhead{$L_{end}$} & \colhead{$v_{grad}$} & \colhead{$T_{min}$} & \colhead{$T_{max}$} & \colhead{Mass} & \colhead{$M_{line}$} & \colhead{$N_{H_2}$} & \colhead{$n_{H_2}$} & \colhead{$\frac{L_{sum}}{w}$} & \colhead{$f_A$} & \colhead{$f_L$} & \colhead{$L_{wt}$} & \colhead{$R_{gc}$} & \colhead{$z$} & \colhead{$|\theta|$} & \colhead{Mor.} & \colhead{DGMF}
}
\startdata
1 & 5.37$\pm$0.02 & 0.13$\pm$0.06 & 11.0$\pm$1.1 & 2.9 & 6 & 12.8 & 9.3 & 0.41$\pm$0.06 & 10.8 & 18.0 & 4.6 & 364.1 & 1.8 & 5.8 & 10.0 & 7.03 & 3.02 & 11.1 & 5.4 & 25.1$\pm$3.3 & 82.9$\pm$2.7 & L,C & 0.20 \\
2 & 6.20$\pm$0.06 & -0.12$\pm$0.01 & 11.3$\pm$2.3 & 3.0 & 6 & 11.2 & 8.6 & 0.91$\pm$0.28 & 16.1 & 26.0 & 5.8 & 521.4 & 2.3 & 6.6 & 7.8 & 7.06 & 5.73 & 9.1 & 5.4 & 12.0$\pm$0.7 & 8.0$\pm$2.5 & X & 0.32 \\
3 & 6.16$\pm$0.05 & -0.61$\pm$0.02 & 17.4$\pm$1.2 & 3.0 & 8 & 11.8 & 7.2 & 1.02$\pm$0.10 & 14.6 & 25.0 & 9.7 & 824.8 & 2.2 & 3.7 & 5.0 & 4.43 & 3.11 & 8.5 & 5.4 & -13.7$\pm$1.3 & 18.8$\pm$2.7 & S & 0.38 \\
4 & 8.58$\pm$0.17 & -0.33$\pm$0.05 & 37.2$\pm$1.2 & 4.4 & 38 & 91.1 & 51.2 & 0.41$\pm$0.02 & 9.9 & 37.6 & 135.9 & 1491.3 & 3.6 & 5.6 & 34.9 & 8.07 & 3.96 & 57.3 & 4.0 & -10.2$\pm$3.7 & 7.6$\pm$0.4 & S,H & - \\
5 & 8.16$\pm$0.07 & 0.22$\pm$0.03 & 18.6$\pm$0.6 & 3.0 & 10 & 13.6 & 10.2 & 0.54$\pm$0.02 & 11.2 & 29.5 & 11.1 & 820.0 & 3.1 & 7.9 & 8.3 & 10.15 & 5.27 & 11.0 & 5.4 & 29.8$\pm$1.6 & 23.4$\pm$1.8 & L,H & 0.46 \\
6 & 8.51$\pm$0.04 & -0.64$\pm$0.04 & 17.0$\pm$0.6 & 3.0 & 9 & 12.8 & 8.4 & 0.49$\pm$0.02 & 11.9 & 26.8 & 2.6 & 200.4 & 1.1 & 4.3 & 11.7 & 5.21 & 1.86 & 12.3 & 5.4 & -15.1$\pm$2.0 & 41.2$\pm$3.6 & X & 0.26 \\
7 & 10.22$\pm$0.11 & -0.27$\pm$0.10 & 11.7$\pm$1.3 & 3.5 & 48 & 73.8 & 27.5 & 0.62$\pm$0.10 & 11.3 & 47.7 & 102.3 & 1386.5 & 3.0 & 4.2 & 25.3 & 6.61 & 2.42 & 35.7 & 4.9 & 0.9$\pm$6.1 & 41.2$\pm$0.5 & S,X & - \\
8 & 11.09$\pm$0.10 & -0.10$\pm$0.03 & 29.7$\pm$1.4 & 2.9 & 12 & 25.0 & 16.9 & 0.65$\pm$0.04 & 10.5 & 15.8 & 20.0 & 801.2 & 2.6 & 5.7 & 13.2 & 5.99 & 3.84 & 19.1 & 5.5 & 13.8$\pm$1.5 & 6.7$\pm$1.1 & S & 0.32 \\
9 & 12.49$\pm$0.04 & -0.21$\pm$0.09 & 34.8$\pm$0.9 & 2.6 & 7 & 16.8 & 14.1 & 0.24$\pm$0.01 & 10.0 & 18.5 & 8.3 & 496.1 & 2.2 & 6.7 & 12.2 & 9.75 & 5.04 & 15.2 & 5.9 & 9.9$\pm$4.1 & 65.3$\pm$1.5 & S & 0.20 \\
10 & 12.87$\pm$0.16 & -0.21$\pm$0.08 & 35.5$\pm$1.5 & 2.6 & 41 & 62.7 & 26.8 & 0.86$\pm$0.05 & 9.6 & 26.6 & 86.6 & 1382.3 & 4.5 & 9.8 & 32.8 & 6.21 & 2.36 & 35.2 & 5.9 & 9.9$\pm$3.8 & 17.8$\pm$0.5 & C,H & - \\
11 & 13.20$\pm$0.06 & 0.04$\pm$0.02 & 50.7$\pm$1.5 & 2.6 & 10 & 15.9 & 9.9 & 1.29$\pm$0.05 & 10.5 & 26.8 & 14.7 & 923.7 & 3.8 & 10.2 & 10.4 & 7.42 & 3.66 & 11.2 & 5.9 & 21.5$\pm$1.0 & 14.9$\pm$2.2 & S,H & 0.38 \\
12 & 12.69$\pm$0.02 & -0.16$\pm$0.03 & 54.9$\pm$1.4 & 4.8 & 6 & 12.9 & 8.1 & 1.21$\pm$0.04 & 13.7 & 26.9 & 36.2 & 2803.2 & 5.6 & 7.5 & 4.1 & 11.25 & 1.58 & 13.0 & 3.8 & 1.4$\pm$2.4 & 63.0$\pm$9.6 & X,H & 0.35 \\
13 & 13.21$\pm$0.10 & -0.34$\pm$0.03 & 38.9$\pm$1.2 & 2.6 & 14 & 23.9 & 12.7 & 0.91$\pm$0.04 & 9.3 & 26.0 & 15.3 & 637.5 & 3.1 & 9.9 & 18.6 & 2.81 & 3.29 & 14.9 & 5.9 & 4.0$\pm$1.6 & 8.0$\pm$1.0 & X & 0.26 \\
14 & 14.00$\pm$0.01 & -0.14$\pm$0.03 & 40.1$\pm$1.4 & 3.1 & 9 & 11.9 & 5.3 & 1.07$\pm$0.05 & 16.2 & 33.7 & 7.6 & 639.5 & 2.4 & 5.9 & 7.1 & 7.49 & 2.64 & 6.6 & 5.4 & 10.7$\pm$1.6 & 76.5$\pm$5.0 & C & 0.47 \\
15 & 14.36$\pm$0.11 & -0.14$\pm$0.05 & 38.8$\pm$1.7 & 3.1 & 30 & 51.2 & 19.4 & 0.76$\pm$0.05 & 10.1 & 29.2 & 42.5 & 830.3 & 2.8 & 6.1 & 27.3 & 4.63 & 2.69 & 24.2 & 5.4 & 10.8$\pm$2.9 & 16.7$\pm$0.7 & X & 0.43 \\
16 & 14.72$\pm$0.06 & -0.18$\pm$0.04 & 38.6$\pm$1.3 & 3.1 & 11 & 18.7 & 11.4 & 0.80$\pm$0.04 & 8.5 & 27.5 & 19.7 & 1058.4 & 3.3 & 6.7 & 9.2 & 4.11 & 2.29 & 15.1 & 5.4 & 8.5$\pm$2.0 & 26.0$\pm$2.3 & S,X & 0.33 \\
17 & 14.32$\pm$0.17 & -0.58$\pm$0.08 & 20.2$\pm$1.3 & 1.5 & 40 & 41.2 & 14.1 & 0.86$\pm$0.08 & 8.7 & 27.9 & 18.4 & 448.0 & 3.0 & 13.1 & 43.9 & 2.96 & 2.49 & 18.1 & 6.9 & 6.6$\pm$2.0 & 12.8$\pm$0.4 & L,X & - \\
18 & 15.05$\pm$0.09 & -0.66$\pm$0.07 & 19.8$\pm$1.6 & 2.0 & 38 & 40.3 & 13.5 & 1.04$\pm$0.12 & 16.5 & 53.0 & 72.4 & 1796.3 & 4.9 & 8.8 & 17.6 & 3.82 & 1.67 & 21.1 & 6.5 & -2.1$\pm$2.5 & 29.8$\pm$1.0 & H & - \\
19 & 15.65$\pm$0.03 & -0.22$\pm$0.05 & 57.0$\pm$0.2 & 11.6 & 6 & 41.2 & 39.0 & 0.07$\pm$0.01 & 16.6 & 20.9 & 44.8 & 1088.6 & 2.3 & 3.3 & 14.1 & 17.49 & 7.44 & 40.4 & 4.2 & -44.2$\pm$11.0 & 58.3$\pm$2.5 & C & 0.19 \\
20 & 13.87$\pm$0.05 & 0.25$\pm$0.03 & 48.1$\pm$1.2 & 3.9 & 6 & 19.3 & 9.8 & 0.15$\pm$0.01 & 15.6 & 33.8 & 10.9 & 565.4 & 2.1 & 4.9 & 11.2 & 2.98 & 1.51 & 16.2 & 4.6 & 34.1$\pm$2.3 & 21.2$\pm$6.5 & C,H & 0.28 \\
21 & 9.87$\pm$0.07 & -0.74$\pm$0.02 & 27.4$\pm$2.2 & 3.1 & 7 & 13.0 & 12.5 & 0.46$\pm$0.05 & 13.2 & 27.4 & 8.4 & 644.3 & 2.1 & 4.6 & 6.9 & 20.40 & 9.48 & 12.8 & 5.3 & -21.8$\pm$1.0 & 12.9$\pm$1.9 & L & - \\
22 & 16.38$\pm$0.06 & -0.60$\pm$0.06 & 40.1$\pm$2.2 & 3.3 & 17 & 26.6 & 13.1 & 0.61$\pm$0.05 & 10.7 & 16.5 & 17.3 & 651.2 & 2.8 & 7.8 & 18.1 & 4.13 & 1.97 & 18.7 & 5.2 & -16.6$\pm$3.4 & 40.5$\pm$1.6 & C,X & 0.34 \\
23 & 16.39$\pm$0.04 & -0.13$\pm$0.05 & 45.9$\pm$1.0 & 3.5 & 6 & 12.6 & 10.6 & 0.57$\pm$0.02 & 15.9 & 26.8 & 8.8 & 697.9 & 3.3 & 10.3 & 9.5 & 5.57 & 4.89 & 11.5 & 5.1 & 9.8$\pm$3.1 & 49.8$\pm$2.4 & S & 0.21 \\
24 & 18.28$\pm$0.06 & -0.26$\pm$0.02 & 68.4$\pm$1.0 & 4.9 & 6 & 15.5 & 14.9 & 0.58$\pm$0.01 & 11.1 & 19.8 & 35.2 & 2266.1 & 3.0 & 2.7 & 3.3 & 26.81 & 11.36 & 15.2 & 4.0 & -6.8$\pm$1.5 & 16.2$\pm$2.6 & L & 0.41 \\
25 & 18.61$\pm$0.05 & -0.09$\pm$0.05 & 45.3$\pm$0.6 & 3.4 & 6 & 13.7 & 12.6 & 0.27$\pm$0.01 & 12.0 & 23.7 & 13.6 & 990.5 & 3.2 & 6.8 & 7.1 & 9.72 & 5.45 & 13.5 & 5.3 & 13.0$\pm$3.1 & 44.0$\pm$2.2 & L,H & 0.29 \\
26 & 19.00$\pm$0.01 & -0.05$\pm$0.03 & 59.8$\pm$2.6 & 5.0 & 7 & 11.4 & 7.2 & 0.84$\pm$0.05 & 15.1 & 28.0 & 24.9 & 2185.5 & 4.0 & 4.9 & 3.3 & 3.39 & 3.16 & 8.6 & 4.0 & 10.9$\pm$2.6 & 76.8$\pm$5.5 & S & 0.39 \\
27 & 18.92$\pm$0.10 & -0.41$\pm$0.12 & 64.8$\pm$2.6 & 5.0 & 28 & 93.4 & 38.1 & 0.42$\pm$0.02 & 11.9 & 31.3 & 94.9 & 1015.8 & 1.9 & 2.3 & 27.8 & 8.25 & 2.83 & 46.6 & 4.0 & -20.8$\pm$10.6 & 51.2$\pm$0.6 & C & 0.45 \\
28 & 19.62$\pm$0.06 & -0.11$\pm$0.03 & 58.9$\pm$2.1 & 5.1 & 12 & 33.8 & 17.3 & 0.58$\pm$0.03 & 13.8 & 31.6 & 36.1 & 1069.5 & 2.9 & 5.2 & 14.6 & 4.49 & 2.18 & 23.4 & 3.9 & 5.0$\pm$2.9 & 20.0$\pm$2.5 & C & 0.27 \\
29 & 20.74$\pm$0.04 & -0.08$\pm$0.05 & 57.1$\pm$1.6 & 11.7 & 16 & 79.3 & 33.6 & 0.33$\pm$0.01 & 16.5 & 29.5 & 170.2 & 2146.9 & 2.0 & 1.3 & 12.0 & 2.09 & 2.04 & 47.0 & 4.9 & -14.3$\pm$10.3 & 57.0$\pm$2.2 & C & 0.43 \\
30 & 22.40$\pm$0.04 & 0.35$\pm$0.04 & 84.0$\pm$0.3 & 5.4 & 7 & 20.0 & 12.9 & 0.08$\pm$0.01 & 14.7 & 23.0 & 41.4 & 2073.4 & 3.1 & 3.0 & 4.7 & 4.57 & 1.57 & 21.0 & 3.9 & 47.2$\pm$3.4 & 37.7$\pm$5.3 & L,X & 0.39 \\
31 & 22.57$\pm$0.05 & -0.20$\pm$0.01 & 76.6$\pm$0.8 & 4.2 & 6 & 12.6 & 11.6 & 0.49$\pm$0.01 & 11.6 & 17.1 & 12.4 & 983.2 & 4.9 & 16.5 & 10.2 & 13.38 & 7.00 & 12.1 & 4.8 & 2.2$\pm$0.8 & 9.0$\pm$2.9 & S & 0.23 \\
32 & 22.94$\pm$0.01 & -0.28$\pm$0.09 & 62.2$\pm$1.8 & 4.2 & 8 & 22.7 & 17.6 & 0.62$\pm$0.02 & 13.3 & 25.0 & 19.5 & 861.5 & 2.5 & 4.7 & 10.5 & 9.78 & 10.09 & 17.9 & 4.8 & -3.5$\pm$7.0 & 87.6$\pm$1.3 & L & 0.27 \\
33 & 23.08$\pm$0.08 & -0.40$\pm$0.02 & 77.4$\pm$0.9 & 4.6 & 6 & 25.0 & 21.3 & 0.24$\pm$0.01 & 13.4 & 20.9 & 50.2 & 2009.7 & 3.0 & 2.9 & 5.9 & 13.65 & 5.54 & 22.6 & 4.5 & -15.9$\pm$2.0 & 12.8$\pm$1.8 & C & 0.31 \\
34 & 22.75$\pm$0.10 & -0.46$\pm$0.03 & 76.4$\pm$1.1 & 4.6 & 13 & 39.7 & 29.0 & 0.42$\pm$0.01 & 11.5 & 19.8 & 33.6 & 845.2 & 2.9 & 6.6 & 21.9 & 12.41 & 4.66 & 31.6 & 4.5 & -20.6$\pm$2.6 & 12.8$\pm$1.1 & S & 0.21 \\
35 & 23.40$\pm$0.05 & -0.21$\pm$0.07 & 101.5$\pm$2.7 & 5.9 & 31 & 77.7 & 30.3 & 0.47$\pm$0.02 & 12.0 & 44.1 & 322.9 & 4156.3 & 5.5 & 4.9 & 16.6 & 5.63 & 1.63 & 48.0 & 3.8 & -7.8$\pm$7.4 & 65.4$\pm$1.6 & X & 0.45 \\
36 & 23.28$\pm$0.04 & -0.27$\pm$0.04 & 61.0$\pm$1.6 & 4.6 & 6 & 16.8 & 12.3 & 0.90$\pm$0.03 & 11.9 & 26.2 & 22.4 & 1336.4 & 3.9 & 7.4 & 7.8 & 9.10 & 4.29 & 13.6 & 4.5 & -6.0$\pm$3.0 & 43.0$\pm$3.0 & C & 0.25 \\
37 & 23.22$\pm$0.03 & 0.03$\pm$0.04 & 76.0$\pm$2.1 & 5.9 & 7 & 18.2 & 11.9 & 0.77$\pm$0.03 & 13.0 & 25.6 & 29.2 & 1603.6 & 4.0 & 6.6 & 7.3 & 3.68 & 1.97 & 16.9 & 3.8 & 16.3$\pm$3.7 & 57.1$\pm$5.0 & C & 0.33 \\
38 & 23.97$\pm$0.03 & 0.15$\pm$0.02 & 79.6$\pm$1.9 & 4.8 & 10 & 18.5 & 8.7 & 0.62$\pm$0.02 & 13.5 & 38.7 & 38.4 & 2079.4 & 3.2 & 3.2 & 4.5 & 8.11 & 2.13 & 12.0 & 4.4 & 28.4$\pm$1.6 & 31.2$\pm$5.3 & L,H & 0.38 \\
39 & 24.04$\pm$0.03 & 0.18$\pm$0.07 & 109.8$\pm$4.0 & 7.8 & 11 & 45.4 & 31.7 & 0.97$\pm$0.05 & 10.9 & 27.8 & 101.1 & 2226.4 & 3.2 & 3.0 & 10.4 & 6.75 & 2.80 & 39.0 & 3.4 & 34.0$\pm$9.1 & 74.2$\pm$1.9 & L,X & 0.27 \\
40 & 23.92$\pm$0.18 & 0.51$\pm$0.06 & 95.4$\pm$1.1 & 5.8 & 22 & 84.1 & 66.7 & 0.20$\pm$0.01 & 10.3 & 24.0 & 82.4 & 979.2 & 2.1 & 2.9 & 28.3 & 15.88 & 10.22 & 68.0 & 3.8 & 65.0$\pm$5.9 & 17.2$\pm$0.4 & L & 0.27 \\
41 & 24.79$\pm$0.04 & 0.09$\pm$0.05 & 109.2$\pm$1.5 & 6.0 & 16 & 47.0 & 22.0 & 0.39$\pm$0.01 & 16.3 & 34.3 & 62.1 & 1320.5 & 2.1 & 2.2 & 11.9 & 3.27 & 2.03 & 30.9 & 3.8 & 22.7$\pm$5.3 & 53.2$\pm$2.2 & X & - \\
42 & 24.48$\pm$0.05 & -0.52$\pm$0.03 & 60.2$\pm$1.2 & 11.3 & 7 & 55.5 & 30.5 & 0.08$\pm$0.01 & 10.9 & 17.6 & 76.8 & 1382.1 & 2.3 & 2.5 & 14.7 & 4.30 & 2.31 & 40.3 & 5.1 & -99.0$\pm$5.0 & 14.6$\pm$3.7 & X & 0.19 \\
43 & 25.77$\pm$0.04 & 0.24$\pm$0.03 & 109.7$\pm$1.4 & 8.7 & 9 & 44.2 & 23.1 & 0.22$\pm$0.01 & 19.8 & 31.2 & 79.3 & 1794.8 & 1.8 & 1.2 & 7.1 & 7.54 & 3.22 & 27.2 & 3.8 & 45.0$\pm$4.3 & 34.2$\pm$3.1 & C & 0.43 \\
44 & 25.72$\pm$0.08 & -0.16$\pm$0.02 & 92.6$\pm$1.5 & 10.2 & 12 & 74.8 & 50.4 & 0.22$\pm$0.01 & 10.3 & 30.0 & 162.8 & 2176.6 & 3.2 & 3.0 & 17.3 & 8.49 & 4.47 & 55.2 & 4.5 & -21.9$\pm$4.1 & 9.0$\pm$1.4 & S & 0.29 \\
45 & 25.36$\pm$0.06 & -0.38$\pm$0.05 & 57.1$\pm$1.6 & 2.7 & 9 & 15.1 & 9.7 & 0.94$\pm$0.04 & 16.4 & 22.3 & 6.8 & 451.8 & 2.3 & 8.1 & 12.5 & 4.16 & 2.26 & 13.0 & 6.0 & 1.9$\pm$2.2 & 32.6$\pm$2.2 & L,X & 0.26 \\
46 & 28.18$\pm$0.05 & -0.05$\pm$0.03 & 97.1$\pm$1.1 & 6.1 & 7 & 26.1 & 17.1 & 0.29$\pm$0.01 & 15.3 & 29.9 & 29.7 & 1138.7 & 2.2 & 2.8 & 8.0 & 7.14 & 2.64 & 21.4 & 4.2 & 9.0$\pm$3.3 & 30.0$\pm$3.2 & S,H & - \\
47 & 28.24$\pm$0.02 & 0.04$\pm$0.02 & 107.1$\pm$1.6 & 6.1 & 9 & 18.0 & 8.5 & 0.64$\pm$0.01 & 15.5 & 29.9 & 61.3 & 3404.2 & 5.7 & 6.2 & 4.8 & 3.09 & 2.03 & 12.0 & 4.2 & 17.9$\pm$2.6 & 64.3$\pm$6.8 & S,H & 0.41 \\
48 & 28.66$\pm$0.05 & 0.03$\pm$0.03 & 98.3$\pm$2.5 & 7.4 & 6 & 30.3 & 22.4 & 0.76$\pm$0.03 & 18.3 & 27.0 & 67.9 & 2244.3 & 2.9 & 2.5 & 6.2 & 12.38 & 3.70 & 25.4 & 4.0 & 15.9$\pm$3.6 & 29.4$\pm$3.4 & C & 0.27 \\
49 & 28.79$\pm$0.02 & 0.20$\pm$0.03 & 105.3$\pm$1.4 & 7.4 & 7 & 21.7 & 13.6 & 0.72$\pm$0.01 & 18.5 & 26.5 & 41.3 & 1905.2 & 2.6 & 2.3 & 4.7 & 4.46 & 2.82 & 16.7 & 4.0 & 36.9$\pm$4.3 & 66.4$\pm$5.0 & S,H & 0.43 \\
50 & 28.57$\pm$0.03 & -0.28$\pm$0.05 & 88.0$\pm$1.8 & 4.7 & 8 & 17.3 & 12.5 & 0.43$\pm$0.01 & 10.3 & 29.7 & 40.6 & 2352.5 & 4.2 & 4.9 & 4.9 & 11.08 & 4.49 & 13.7 & 4.8 & -6.5$\pm$4.4 & 65.2$\pm$2.4 & L,C & 0.42 \\
51 & 26.55$\pm$0.03 & -0.29$\pm$0.04 & 107.7$\pm$0.8 & 7.6 & 8 & 28.5 & 16.0 & 0.31$\pm$0.01 & 16.7 & 26.8 & 25.3 & 889.1 & 2.0 & 3.1 & 10.4 & 4.35 & 1.79 & 23.9 & 3.7 & -27.3$\pm$5.1 & 54.7$\pm$4.5 & C & 0.40 \\
52 & 30.03$\pm$0.05 & 0.10$\pm$0.03 & 105.6$\pm$0.9 & 5.2 & 8 & 25.2 & 15.5 & 0.53$\pm$0.01 & 12.9 & 31.9 & 23.3 & 925.3 & 3.2 & 7.4 & 14.1 & 3.49 & 1.65 & 24.4 & 4.7 & 24.6$\pm$2.7 & 2.5$\pm$3.5 & X & 0.22 \\
53 & 30.60$\pm$0.04 & -0.11$\pm$0.02 & 113.1$\pm$1.3 & 5.2 & 9 & 15.3 & 12.2 & 0.64$\pm$0.01 & 15.0 & 23.6 & 33.8 & 2206.7 & 5.0 & 7.6 & 5.6 & 7.27 & 5.24 & 13.0 & 4.7 & 5.8$\pm$1.5 & 20.2$\pm$3.1 & S & 0.32 \\
54 & 30.34$\pm$0.10 & -0.22$\pm$0.07 & 104.0$\pm$1.5 & 5.2 & 22 & 87.6 & 33.9 & 0.33$\pm$0.01 & 12.0 & 30.0 & 96.5 & 1102.1 & 2.5 & 3.8 & 31.9 & 2.87 & 1.96 & 48.4 & 4.7 & -4.1$\pm$6.6 & 31.1$\pm$1.0 & X & 0.28 \\
55 & 30.86$\pm$0.05 & 0.10$\pm$0.03 & 38.9$\pm$1.5 & 2.7 & 6 & 13.1 & 7.8 & 0.52$\pm$0.03 & 17.1 & 31.3 & 13.7 & 1043.6 & 3.0 & 5.8 & 6.1 & 6.02 & 1.96 & 11.1 & 6.2 & 25.1$\pm$1.3 & 1.8$\pm$2.9 & C & 0.24 \\
56 & 32.03$\pm$0.05 & 0.07$\pm$0.01 & 95.6$\pm$1.0 & 5.2 & 7 & 16.6 & 15.6 & 0.86$\pm$0.01 & 10.4 & 27.2 & 30.9 & 1859.0 & 2.6 & 2.4 & 3.8 & 11.45 & 5.80 & 16.5 & 4.8 & 21.9$\pm$1.0 & 6.3$\pm$2.8 & S & 0.45 \\
57 & 33.22$\pm$0.05 & 0.01$\pm$0.03 & 100.2$\pm$1.3 & 6.5 & 6 & 32.0 & 17.0 & 0.18$\pm$0.01 & 18.8 & 27.8 & 42.0 & 1314.1 & 2.1 & 2.2 & 8.2 & 4.56 & 1.68 & 26.5 & 4.6 & 14.6$\pm$3.6 & 16.9$\pm$5.2 & S & 0.31 \\
58 & 33.63$\pm$0.10 & -0.01$\pm$0.02 & 104.0$\pm$1.0 & 6.5 & 19 & 62.5 & 41.1 & 0.32$\pm$0.01 & 10.4 & 27.3 & 92.8 & 1485.7 & 2.4 & 2.5 & 16.0 & 13.90 & 6.21 & 43.2 & 4.6 & 12.5$\pm$2.2 & 5.8$\pm$0.9 & L & 0.36 \\
59 & 34.28$\pm$0.10 & 0.16$\pm$0.07 & 57.2$\pm$1.7 & 1.6 & 25 & 24.2 & 12.3 & 1.33$\pm$0.06 & 11.4 & 33.3 & 17.9 & 739.4 & 3.8 & 12.9 & 19.9 & 8.03 & 4.01 & 13.7 & 7.1 & 26.9$\pm$1.9 & 34.3$\pm$0.7 & S,H & - \\
60 & 35.56$\pm$0.04 & 0.02$\pm$0.06 & 52.6$\pm$3.3 & 10.4 & 12 & 77.6 & 42.9 & 0.60$\pm$0.05 & 14.7 & 28.0 & 115.8 & 1491.5 & 1.3 & 0.8 & 11.1 & 4.68 & 2.48 & 55.1 & 6.1 & 11.1$\pm$11.2 & 62.6$\pm$2.0 & L,X & 0.41 \\
61 & 35.18$\pm$0.07 & -0.75$\pm$0.05 & 34.2$\pm$1.6 & 2.2 & 10 & 16.0 & 9.0 & 0.90$\pm$0.06 & 11.8 & 21.7 & 16.8 & 1050.8 & 4.2 & 10.9 & 10.2 & 4.65 & 1.91 & 13.1 & 6.7 & -7.5$\pm$1.7 & 24.2$\pm$2.4 & X,H & - \\
62 & 41.16$\pm$0.11 & -0.21$\pm$0.03 & 60.0$\pm$2.4 & 8.9 & 11 & 56.4 & 46.4 & 0.47$\pm$0.03 & 16.6 & 27.9 & 64.1 & 1138.0 & 2.0 & 2.3 & 15.9 & 9.16 & 8.32 & 47.7 & 6.1 & -20.1$\pm$4.0 & 11.8$\pm$1.0 & S & 0.29 \\
63 & 43.16$\pm$0.05 & -0.01$\pm$0.04 & 11.4$\pm$2.9 & 11.1 & 12 & 107.0 & 35.0 & 0.25$\pm$0.10 & 19.7 & 31.9 & 291.8 & 2727.4 & 1.9 & 0.9 & 12.1 & 2.70 & 1.67 & 54.6 & 7.6 & 7.3$\pm$6.9 & 16.4$\pm$3.4 & X,H & - \\
64 & 48.60$\pm$0.03 & 0.03$\pm$0.04 & 17.1$\pm$1.0 & 10.8 & 7 & 36.3 & 21.8 & 0.19$\pm$0.02 & 17.3 & 34.3 & 75.1 & 2069.3 & 1.6 & 0.8 & 4.4 & 4.12 & 1.90 & 31.7 & 8.2 & 17.6$\pm$6.8 & 62.4$\pm$5.1 & C & 0.42 \\
65 & 49.02$\pm$0.09 & -0.29$\pm$0.04 & 66.0$\pm$2.7 & 5.3 & 28 & 82.9 & 30.5 & 0.41$\pm$0.02 & 14.6 & 29.0 & 69.9 & 844.2 & 1.4 & 1.6 & 22.7 & 5.16 & 2.47 & 39.2 & 6.3 & -7.6$\pm$3.4 & 3.4$\pm$1.0 & X & 0.49 \\
66 & 49.45$\pm$0.12 & -0.38$\pm$0.05 & 60.9$\pm$5.0 & 5.3 & 34 & 102.0 & 42.9 & 0.42$\pm$0.05 & 17.3 & 35.2 & 375.2 & 3676.4 & 5.4 & 5.2 & 23.8 & 4.31 & 2.52 & 54.8 & 6.3 & -15.8$\pm$4.3 & 2.7$\pm$0.7 & L,X & - \\
67 & 49.38$\pm$0.02 & -0.27$\pm$0.05 & 50.4$\pm$1.5 & 5.3 & 12 & 25.5 & 14.6 & 0.42$\pm$0.02 & 15.8 & 30.0 & 47.0 & 1841.2 & 2.8 & 2.8 & 6.2 & 3.93 & 2.34 & 19.2 & 6.3 & -6.3$\pm$4.8 & 88.3$\pm$2.0 & S & 0.46 \\
68 & 53.18$\pm$0.07 & 0.06$\pm$0.03 & 22.8$\pm$1.4 & 4.0 & 7 & 16.5 & 15.5 & 0.36$\pm$0.03 & 12.8 & 21.8 & 6.1 & 369.2 & 1.4 & 3.5 & 10.1 & 11.06 & 6.34 & 16.3 & 6.8 & 25.5$\pm$1.7 & 18.5$\pm$2.1 & L & 0.36 \\
69 & 54.11$\pm$0.03 & -0.07$\pm$0.02 & 38.8$\pm$0.9 & 4.0 & 9 & 15.7 & 10.0 & 0.43$\pm$0.01 & 15.2 & 33.4 & 7.0 & 445.7 & 1.6 & 3.6 & 8.8 & 6.28 & 2.32 & 13.2 & 6.8 & 16.4$\pm$1.6 & 27.4$\pm$4.2 & X & 0.41 \\
70 & 59.59$\pm$0.06 & -0.21$\pm$0.02 & 27.1$\pm$0.8 & 2.2 & 6 & 10.4 & 6.4 & 0.39$\pm$0.02 & 15.3 & 20.8 & 1.7 & 162.0 & 0.9 & 3.3 & 9.2 & 3.13 & 2.67 & 7.9 & 7.5 & 15.6$\pm$0.9 & 5.5$\pm$2.9 & S & 0.37 \\
71 & 354.64$\pm$0.04 & 0.49$\pm$0.07 & -20.8$\pm$0.4 & 4.4 & 13 & 33.8 & 20.0 & 0.31$\pm$0.01 & 11.8 & 25.6 & 31.8 & 941.8 & 1.8 & 2.4 & 10.6 & 3.62 & 2.65 & 25.1 & 4.0 & 53.1$\pm$5.4 & 62.5$\pm$1.6 & S & 0.43 \\
72 & 354.18$\pm$0.04 & -0.05$\pm$0.02 & -31.3$\pm$1.9 & 11.4 & 6 & 45.2 & 25.1 & 0.19$\pm$0.02 & 13.6 & 40.2 & 63.0 & 1392.5 & 1.6 & 1.3 & 8.5 & 5.87 & 2.51 & 32.0 & 3.3 & -10.2$\pm$4.1 & 19.4$\pm$4.9 & S & 0.29 \\
73 & 353.11$\pm$0.12 & 0.72$\pm$0.20 & -3.6$\pm$2.0 & 1.4 & 124 & 98.5 & 18.8 & 1.75$\pm$1.40 & 15.4 & 49.3 & 40.1 & 407.2 & 2.1 & 7.1 & 81.1 & 2.45 & 1.71 & 29.0 & 7.0 & 39.5$\pm$4.8 & 85.8$\pm$0.3 & C,X & - \\
74 & 352.61$\pm$0.03 & -0.20$\pm$0.02 & -85.1$\pm$0.8 & 7.6 & 9 & 21.8 & 12.8 & 0.17$\pm$0.01 & 13.6 & 31.8 & 44.9 & 2063.2 & 2.4 & 1.8 & 4.0 & 2.87 & 2.43 & 16.5 & 1.3 & -18.6$\pm$3.2 & 40.1$\pm$4.5 & X,H & 0.47 \\
75 & 353.38$\pm$0.08 & -0.34$\pm$0.05 & -17.6$\pm$1.0 & 3.1 & 15 & 26.8 & 14.3 & 0.46$\pm$0.04 & 12.4 & 28.4 & 62.3 & 2326.9 & 5.4 & 8.4 & 10.0 & 4.87 & 3.92 & 16.1 & 5.3 & -0.1$\pm$2.9 & 33.4$\pm$1.2 & S,H & - \\
76 & 353.40$\pm$0.02 & -0.09$\pm$0.01 & -52.3$\pm$2.1 & 10.2 & 7 & 25.1 & 10.4 & 0.56$\pm$0.03 & 15.7 & 37.4 & 131.8 & 5257.9 & 3.8 & 1.8 & 2.9 & 4.57 & 2.27 & 13.8 & 2.2 & -12.4$\pm$1.5 & 2.5$\pm$7.1 & X & 0.51 \\
77 & 353.56$\pm$0.03 & -0.05$\pm$0.04 & -57.6$\pm$1.0 & 10.2 & 10 & 46.4 & 21.4 & 0.22$\pm$0.01 & 16.6 & 32.6 & 108.9 & 2348.5 & 2.7 & 2.0 & 8.4 & 3.11 & 2.06 & 29.8 & 2.2 & -5.4$\pm$6.9 & 61.7$\pm$3.9 & C,X & 0.35 \\
78 & 351.59$\pm$0.06 & 0.17$\pm$0.01 & -41.4$\pm$0.9 & 11.5 & 13 & 48.7 & 38.6 & 0.22$\pm$0.01 & 20.9 & 36.5 & 154.5 & 3173.0 & 2.4 & 1.2 & 5.9 & 9.20 & 4.26 & 42.6 & 3.5 & 35.1$\pm$2.7 & 3.3$\pm$1.9 & S & 0.47 \\
79 & 351.94$\pm$0.13 & 0.68$\pm$0.05 & -0.4$\pm$1.4 & 1.3 & 17 & 17.6 & 9.7 & 1.03$\pm$2.71 & 12.3 & 27.4 & 3.7 & 207.9 & 2.0 & 12.4 & 26.8 & 4.53 & 3.06 & 11.6 & 7.0 & 37.6$\pm$1.1 & 8.4$\pm$0.8 & S & 0.29 \\
80 & 351.27$\pm$0.31 & 0.69$\pm$0.11 & -4.2$\pm$2.5 & 1.3 & 106 & 78.5 & 29.1 & 1.44$\pm$1.03 & 10.2 & 46.6 & 63.3 & 806.4 & 5.0 & 20.2 & 77.5 & 12.21 & 4.89 & 31.4 & 7.0 & 37.8$\pm$2.5 & 16.3$\pm$0.1 & S,X & - \\
81 & 350.19$\pm$0.08 & 0.08$\pm$0.05 & -66.3$\pm$4.4 & 10.5 & 32 & 147.8 & 53.5 & 0.26$\pm$0.02 & 13.2 & 28.7 & 525.0 & 3552.9 & 2.6 & 1.3 & 17.4 & 5.65 & 2.57 & 67.8 & 2.7 & 16.4$\pm$8.7 & 24.2$\pm$0.9 & C,X & 0.52 \\
82 & 349.87$\pm$0.08 & 0.10$\pm$0.05 & -62.0$\pm$0.9 & 10.7 & 6 & 52.8 & 44.6 & 0.09$\pm$0.01 & 14.3 & 22.3 & 174.2 & 3300.5 & 1.8 & 0.6 & 4.6 & 5.42 & 3.11 & 53.0 & 2.9 & 21.4$\pm$9.0 & 29.8$\pm$2.1 & C & 0.29 \\
83 & 349.30$\pm$0.05 & 0.16$\pm$0.01 & -63.6$\pm$0.5 & 10.5 & 8 & 36.1 & 29.4 & 0.12$\pm$0.01 & 13.6 & 20.0 & 118.2 & 3277.0 & 3.1 & 1.9 & 5.4 & 6.30 & 4.11 & 32.7 & 2.8 & 31.5$\pm$2.6 & 8.1$\pm$2.9 & S & 0.33 \\
84 & 349.12$\pm$0.06 & 0.10$\pm$0.03 & -75.9$\pm$2.1 & 10.5 & 11 & 61.8 & 34.6 & 0.25$\pm$0.01 & 16.0 & 26.7 & 301.3 & 4878.1 & 3.0 & 1.2 & 6.0 & 3.61 & 2.68 & 43.2 & 2.8 & 20.2$\pm$4.8 & 12.6$\pm$2.3 & X & 0.38 \\
85 & 349.14$\pm$0.03 & 0.01$\pm$0.02 & 16.4$\pm$1.3 & 19.7 & 6 & 58.4 & 39.8 & 0.19$\pm$0.02 & 18.6 & 26.4 & 427.8 & 7324.3 & 2.9 & 0.7 & 3.7 & 4.56 & 2.44 & 51.4 & 11.6 & -14.4$\pm$7.3 & 27.6$\pm$5.5 & S & 0.38 \\
86 & 348.94$\pm$0.06 & 0.11$\pm$0.02 & -69.3$\pm$1.7 & 10.5 & 7 & 41.1 & 24.8 & 0.17$\pm$0.01 & 16.1 & 19.9 & 88.5 & 2152.1 & 2.7 & 2.2 & 8.2 & 3.49 & 3.08 & 29.5 & 2.8 & 23.3$\pm$3.4 & 1.8$\pm$2.0 & C & 0.26 \\
87 & 347.89$\pm$0.06 & 0.03$\pm$0.03 & -29.8$\pm$1.4 & 12.9 & 6 & 62.1 & 44.1 & 0.06$\pm$0.01 & 16.3 & 30.2 & 118.4 & 1906.5 & 1.5 & 0.8 & 8.0 & 9.12 & 4.44 & 48.4 & 5.0 & 3.3$\pm$6.0 & 22.0$\pm$2.6 & L,C & 0.31 \\
88 & 347.69$\pm$0.05 & 0.24$\pm$0.03 & -72.4$\pm$1.2 & 9.8 & 6 & 35.9 & 24.9 & 0.22$\pm$0.01 & 10.3 & 21.6 & 106.6 & 2972.1 & 3.1 & 2.1 & 5.9 & 3.41 & 2.10 & 34.4 & 2.4 & 44.8$\pm$4.7 & 16.9$\pm$4.2 & C & 0.27 \\
89 & 347.62$\pm$0.02 & 0.20$\pm$0.05 & -94.0$\pm$1.5 & 9.8 & 13 & 52.1 & 29.1 & 0.25$\pm$0.01 & 17.9 & 39.3 & 135.7 & 2604.9 & 1.9 & 0.9 & 6.1 & 4.77 & 2.90 & 35.4 & 2.4 & 38.5$\pm$8.8 & 78.4$\pm$2.4 & C & 0.54 \\
90 & 347.91$\pm$0.05 & -0.37$\pm$0.05 & -94.5$\pm$1.7 & 6.9 & 10 & 36.7 & 21.4 & 0.27$\pm$0.01 & 10.9 & 23.8 & 43.6 & 1187.3 & 1.6 & 1.4 & 7.9 & 3.92 & 1.76 & 32.4 & 2.2 & -34.4$\pm$6.3 & 47.4$\pm$2.6 & C & 0.48 \\
91 & 350.51$\pm$0.04 & -0.37$\pm$0.04 & -22.8$\pm$0.8 & 3.3 & 8 & 14.1 & 9.7 & 0.30$\pm$0.02 & 12.3 & 20.4 & 8.3 & 586.1 & 2.4 & 6.8 & 9.4 & 4.36 & 1.86 & 14.3 & 5.1 & -3.3$\pm$2.5 & 53.3$\pm$3.9 & C & 0.36 \\
92 & 349.92$\pm$0.08 & -0.53$\pm$0.02 & -24.8$\pm$2.0 & 3.3 & 13 & 25.7 & 12.4 & 0.96$\pm$0.11 & 12.9 & 31.3 & 14.5 & 563.5 & 1.3 & 1.8 & 9.2 & 5.26 & 3.25 & 14.5 & 5.1 & -12.5$\pm$1.3 & 1.9$\pm$1.3 & S & 0.43 \\
93 & 345.30$\pm$0.13 & -0.17$\pm$0.03 & -16.9$\pm$0.8 & 1.4 & 13 & 12.8 & 10.0 & 0.68$\pm$0.05 & 11.2 & 24.1 & 2.6 & 200.3 & 2.2 & 16.0 & 22.6 & 7.96 & 6.02 & 10.5 & 7.0 & 18.0$\pm$0.7 & 9.2$\pm$0.8 & L,X & 0.20 \\
94 & 345.41$\pm$0.06 & -0.94$\pm$0.04 & -20.4$\pm$1.1 & 1.4 & 19 & 14.6 & 6.0 & 1.15$\pm$0.09 & 11.7 & 38.7 & 4.9 & 333.4 & 2.0 & 7.8 & 13.9 & 4.48 & 1.77 & 9.1 & 7.0 & -1.0$\pm$1.0 & 26.9$\pm$2.0 & X,H & - \\
95 & 343.79$\pm$0.08 & -0.15$\pm$0.05 & -25.6$\pm$1.9 & 2.6 & 16 & 23.3 & 12.2 & 0.65$\pm$0.07 & 10.0 & 34.8 & 10.6 & 456.3 & 1.6 & 3.7 & 13.0 & 4.21 & 2.39 & 15.9 & 5.9 & 13.0$\pm$2.2 & 25.0$\pm$1.4 & C,S & 0.38 \\
96 & 343.49$\pm$0.03 & -0.03$\pm$0.05 & -28.9$\pm$1.1 & 2.6 & 14 & 17.8 & 9.5 & 0.81$\pm$0.04 & 17.4 & 36.9 & 8.1 & 456.6 & 1.8 & 4.8 & 11.3 & 3.84 & 2.16 & 13.0 & 5.9 & 18.2$\pm$2.3 & 67.5$\pm$2.5 & C,X & 0.49 \\
97 & 343.12$\pm$0.04 & -0.08$\pm$0.03 & -30.8$\pm$1.6 & 2.7 & 7 & 10.3 & 7.8 & 0.59$\pm$0.04 & 12.9 & 28.9 & 8.8 & 855.3 & 3.8 & 11.1 & 7.3 & 6.85 & 3.20 & 9.2 & 5.8 & 15.8$\pm$1.3 & 32.2$\pm$3.4 & L,H & - \\
98 & 344.16$\pm$0.10 & -0.61$\pm$0.05 & -24.3$\pm$1.3 & 2.2 & 34 & 35.7 & 13.7 & 1.22$\pm$0.09 & 10.8 & 27.7 & 45.3 & 1270.4 & 4.3 & 9.6 & 19.3 & 4.26 & 2.20 & 18.5 & 6.2 & -2.9$\pm$1.9 & 12.4$\pm$0.9 & X & - \\
99 & 342.54$\pm$0.05 & 0.18$\pm$0.01 & -41.7$\pm$0.5 & 12.5 & 9 & 43.0 & 36.1 & 0.10$\pm$0.01 & 11.5 & 23.2 & 122.1 & 2841.1 & 2.7 & 1.7 & 6.6 & 19.11 & 6.89 & 37.6 & 5.2 & 38.2$\pm$1.8 & 2.8$\pm$2.0 & L & 0.35 \\
100 & 342.06$\pm$0.05 & 0.42$\pm$0.01 & -70.8$\pm$0.9 & 10.9 & 7 & 29.1 & 28.1 & 0.27$\pm$0.01 & 13.1 & 22.3 & 146.6 & 5040.5 & 3.1 & 1.3 & 2.9 & 12.46 & 8.08 & 28.9 & 3.9 & 81.8$\pm$2.7 & 13.7$\pm$2.8 & L & 0.42 \\
101 & 340.93$\pm$0.04 & -0.32$\pm$0.07 & -44.8$\pm$1.0 & 3.3 & 10 & 21.5 & 11.9 & 0.37$\pm$0.01 & 10.2 & 21.5 & 19.2 & 891.4 & 2.8 & 5.8 & 10.8 & 2.85 & 1.86 & 17.5 & 5.4 & -0.1$\pm$3.9 & 86.8$\pm$2.4 & X & 0.29 \\
102 & 341.24$\pm$0.05 & -0.26$\pm$0.03 & -43.6$\pm$0.8 & 3.3 & 16 & 20.9 & 11.0 & 0.61$\pm$0.02 & 12.5 & 25.9 & 20.0 & 955.8 & 2.8 & 5.6 & 10.0 & 5.06 & 2.33 & 14.5 & 5.4 & 3.2$\pm$1.6 & 14.8$\pm$2.2 & C,X & 0.51 \\
103 & 339.95$\pm$0.03 & -0.55$\pm$0.02 & -92.3$\pm$0.9 & 10.1 & 6 & 25.9 & 18.0 & 0.25$\pm$0.01 & 12.8 & 31.0 & 40.8 & 1575.7 & 1.9 & 1.4 & 4.9 & 4.91 & 3.25 & 21.2 & 3.6 & -91.8$\pm$3.9 & 33.0$\pm$5.0 & X & 0.42 \\
104 & 338.59$\pm$0.02 & 0.03$\pm$0.05 & -22.7$\pm$2.2 & 13.6 & 11 & 80.7 & 39.6 & 0.15$\pm$0.02 & 13.1 & 27.4 & 369.9 & 4582.2 & 2.5 & 0.9 & 7.2 & 5.84 & 2.44 & 51.2 & 6.5 & 5.4$\pm$11.7 & 73.7$\pm$3.0 & C & 0.43 \\
105 & 338.41$\pm$0.05 & 0.00$\pm$0.02 & -47.6$\pm$2.0 & 2.7 & 6 & 10.5 & 6.6 & 0.95$\pm$0.06 & 22.1 & 34.3 & 8.1 & 777.7 & 2.5 & 5.2 & 5.3 & 7.01 & 3.11 & 7.8 & 5.9 & 19.7$\pm$0.7 & 1.5$\pm$2.5 & C & 0.31 \\
106 & 338.06$\pm$0.06 & -0.00$\pm$0.10 & -40.9$\pm$1.9 & 3.0 & 22 & 50.1 & 17.8 & 0.82$\pm$1.42 & 11.7 & 26.3 & 55.5 & 1109.5 & 3.6 & 7.9 & 26.3 & 3.51 & 1.70 & 27.5 & 5.7 & 18.8$\pm$5.4 & 78.9$\pm$1.3 & C,X & 0.29 \\
107 & 337.87$\pm$0.06 & -0.01$\pm$0.04 & -54.5$\pm$3.5 & 3.6 & 14 & 35.6 & 16.2 & 0.82$\pm$0.68 & 10.2 & 26.2 & 29.2 & 818.5 & 5.1 & 21.4 & 35.9 & 2.95 & 1.52 & 26.8 & 5.2 & 17.1$\pm$2.7 & 11.0$\pm$3.1 & C,X & 0.16 \\
108 & 337.67$\pm$0.04 & -0.05$\pm$0.04 & -48.1$\pm$1.7 & 12.1 & 11 & 59.4 & 33.1 & 0.29$\pm$0.11 & 15.4 & 25.5 & 308.4 & 5192.9 & 3.5 & 1.5 & 6.4 & 3.28 & 1.71 & 50.9 & 5.4 & -8.5$\pm$7.4 & 36.4$\pm$3.7 & C & 0.44 \\
109 & 337.70$\pm$0.04 & 0.10$\pm$0.04 & -75.2$\pm$1.0 & 10.8 & 6 & 31.1 & 27.7 & 0.15$\pm$0.01 & 17.0 & 22.9 & 183.4 & 5903.6 & 4.7 & 2.5 & 4.0 & 11.41 & 4.58 & 30.3 & 4.4 & 21.8$\pm$7.7 & 49.2$\pm$3.0 & C & 0.36 \\
110 & 337.41$\pm$0.03 & -0.39$\pm$0.05 & -41.5$\pm$0.6 & 3.0 & 8 & 10.7 & 7.0 & 0.43$\pm$0.01 & 15.9 & 30.4 & 11.7 & 1092.9 & 3.9 & 9.2 & 6.1 & 5.74 & 2.02 & 9.9 & 5.7 & -1.2$\pm$2.6 & 78.6$\pm$3.9 & C & - \\
111 & 336.88$\pm$0.10 & -0.02$\pm$0.15 & -75.7$\pm$3.2 & 4.7 & 64 & 212.8 & 48.9 & 0.46$\pm$0.26 & 13.3 & 47.1 & 171.1 & 804.0 & 1.6 & 2.0 & 66.2 & 3.69 & 2.39 & 63.7 & 4.4 & 14.3$\pm$12.5 & 60.0$\pm$0.3 & C,X & - \\
112 & 336.95$\pm$0.06 & -0.00$\pm$0.07 & -120.6$\pm$1.2 & 7.7 & 22 & 96.8 & 31.4 & 0.26$\pm$0.01 & 13.5 & 32.7 & 133.5 & 1379.6 & 2.1 & 2.1 & 23.3 & 3.51 & 1.56 & 51.1 & 3.3 & 9.3$\pm$9.0 & 49.8$\pm$1.7 & S,X & - \\
113 & 337.77$\pm$0.07 & -0.34$\pm$0.03 & -41.2$\pm$0.5 & 3.0 & 7 & 14.5 & 10.8 & 0.23$\pm$0.01 & 13.4 & 34.4 & 10.6 & 727.9 & 2.7 & 6.4 & 8.5 & 5.28 & 2.88 & 13.2 & 5.7 & 1.1$\pm$1.7 & 16.5$\pm$2.4 & X & 0.30 \\
114 & 337.93$\pm$0.09 & -0.48$\pm$0.03 & -39.0$\pm$1.4 & 2.9 & 15 & 25.7 & 16.0 & 0.72$\pm$0.04 & 10.8 & 34.6 & 21.8 & 846.7 & 2.2 & 4.0 & 10.9 & 7.51 & 2.90 & 19.4 & 5.8 & -4.9$\pm$1.7 & 10.1$\pm$1.3 & C & - \\
115 & 336.48$\pm$0.05 & -0.23$\pm$0.06 & -86.8$\pm$2.2 & 5.0 & 19 & 34.5 & 19.3 & 0.67$\pm$0.02 & 13.4 & 29.0 & 47.7 & 1385.2 & 2.6 & 3.3 & 10.4 & 3.37 & 2.30 & 25.5 & 4.2 & -4.7$\pm$5.1 & 49.6$\pm$1.5 & S & 0.44 \\
116 & 336.36$\pm$0.04 & -0.14$\pm$0.05 & -79.5$\pm$0.4 & 5.0 & 6 & 22.2 & 16.5 & 0.06$\pm$0.01 & 16.9 & 27.1 & 19.0 & 855.2 & 1.6 & 2.1 & 6.8 & 5.16 & 1.69 & 25.5 & 4.2 & 2.7$\pm$4.6 & 66.3$\pm$4.6 & C & 0.33 \\
117 & 335.19$\pm$0.08 & -0.34$\pm$0.07 & -41.0$\pm$2.8 & 2.9 & 25 & 45.3 & 15.5 & 0.84$\pm$0.08 & 10.3 & 29.3 & 22.1 & 487.4 & 2.0 & 5.5 & 30.1 & 3.91 & 1.61 & 24.8 & 5.8 & 2.3$\pm$3.4 & 37.7$\pm$1.4 & S,X & 0.34 \\
118 & 335.28$\pm$0.03 & -0.13$\pm$0.05 & -44.9$\pm$0.7 & 2.9 & 8 & 12.2 & 6.8 & 0.66$\pm$0.01 & 8.6 & 20.0 & 7.0 & 573.8 & 2.1 & 5.1 & 7.2 & 3.52 & 2.12 & 9.4 & 5.8 & 13.0$\pm$2.3 & 69.4$\pm$3.9 & C,X & 0.30 \\
119 & 334.04$\pm$0.02 & -0.00$\pm$0.06 & -86.1$\pm$1.3 & 5.2 & 9 & 27.3 & 17.1 & 0.47$\pm$0.01 & 11.4 & 23.1 & 20.9 & 766.0 & 2.7 & 6.2 & 15.3 & 3.39 & 2.45 & 22.0 & 4.3 & 15.2$\pm$5.3 & 87.3$\pm$2.2 & S,X & 0.22 \\
120 & 333.17$\pm$0.39 & -0.40$\pm$0.16 & -51.3$\pm$3.7 & 3.6 & 169 & 370.2 & 109.4 & 0.54$\pm$0.06 & 11.6 & 44.8 & 436.1 & 1178.1 & 2.7 & 3.9 & 133.4 & 10.46 & 4.24 & 121.0 & 5.4 & -7.0$\pm$10.0 & 18.4$\pm$0.1 & L,X & - \\
121 & 333.61$\pm$0.03 & 0.05$\pm$0.02 & -83.5$\pm$2.3 & 4.9 & 7 & 15.1 & 7.7 & 1.07$\pm$0.04 & 14.8 & 21.6 & 37.3 & 2466.5 & 3.1 & 2.6 & 3.1 & 4.77 & 1.60 & 12.3 & 4.5 & 20.2$\pm$1.7 & 3.3$\pm$6.3 & X & 0.40 \\
122 & 333.27$\pm$0.06 & 0.06$\pm$0.13 & -47.7$\pm$1.0 & 3.6 & 20 & 58.9 & 29.3 & 0.24$\pm$0.01 & 10.4 & 27.2 & 43.7 & 743.1 & 2.3 & 4.9 & 29.8 & 3.96 & 2.04 & 41.1 & 5.4 & 21.8$\pm$8.2 & 87.7$\pm$0.9 & X & 0.23 \\
123 & 333.19$\pm$0.03 & -0.07$\pm$0.05 & -87.8$\pm$1.4 & 5.2 & 12 & 41.3 & 15.3 & 0.31$\pm$0.01 & 16.1 & 27.1 & 61.0 & 1476.0 & 2.8 & 3.4 & 12.4 & 2.87 & 1.51 & 25.4 & 4.4 & 8.5$\pm$4.7 & 80.5$\pm$4.3 & C & 0.33 \\
124 & 333.06$\pm$0.04 & 0.01$\pm$0.08 & -43.9$\pm$1.6 & 3.6 & 10 & 25.3 & 13.7 & 0.66$\pm$0.03 & 17.0 & 30.8 & 28.0 & 1107.4 & 3.0 & 5.3 & 10.8 & 3.48 & 2.29 & 18.2 & 5.4 & 18.7$\pm$4.7 & 80.4$\pm$2.1 & C & 0.31 \\
125 & 332.76$\pm$0.06 & -0.00$\pm$0.02 & -95.3$\pm$1.0 & 5.6 & 9 & 24.0 & 18.5 & 0.29$\pm$0.01 & 9.6 & 31.3 & 24.6 & 1028.1 & 2.2 & 3.1 & 8.2 & 9.50 & 5.85 & 19.5 & 4.2 & 14.3$\pm$2.0 & 15.7$\pm$1.9 & S & 0.28 \\
126 & 332.36$\pm$0.13 & -0.06$\pm$0.12 & -48.4$\pm$1.6 & 3.1 & 40 & 69.4 & 28.4 & 0.54$\pm$0.02 & 8.8 & 35.5 & 57.9 & 834.2 & 2.5 & 5.0 & 33.6 & 6.40 & 1.51 & 47.0 & 5.8 & 15.9$\pm$6.3 & 39.3$\pm$0.7 & C & 0.36 \\
127 & 331.98$\pm$0.13 & -0.12$\pm$0.02 & -51.2$\pm$0.5 & 3.1 & 16 & 32.5 & 23.5 & 0.31$\pm$0.01 & 10.2 & 24.5 & 25.4 & 780.4 & 2.3 & 4.6 & 15.5 & 6.86 & 5.21 & 25.1 & 5.8 & 12.8$\pm$1.4 & 2.3$\pm$0.8 & S & 0.28 \\
128 & 331.64$\pm$0.04 & -0.07$\pm$0.04 & -85.0$\pm$0.7 & 5.3 & 8 & 22.2 & 13.8 & 0.23$\pm$0.01 & 16.8 & 26.4 & 31.8 & 1435.7 & 3.5 & 5.7 & 8.7 & 4.76 & 2.44 & 17.8 & 4.5 & 8.6$\pm$3.2 & 38.0$\pm$3.2 & S,X & 0.28 \\
129 & 331.69$\pm$0.03 & -0.22$\pm$0.05 & -47.6$\pm$0.9 & 3.1 & 7 & 13.4 & 8.7 & 0.37$\pm$0.01 & 13.6 & 21.3 & 10.5 & 778.5 & 2.1 & 3.8 & 5.9 & 4.05 & 2.13 & 11.9 & 5.8 & 7.2$\pm$2.9 & 84.1$\pm$3.5 & C & 0.25 \\
130 & 331.32$\pm$0.20 & -0.14$\pm$0.08 & -88.9$\pm$2.5 & 5.3 & 57 & 192.2 & 62.0 & 0.52$\pm$0.02 & 10.8 & 37.5 & 265.2 & 1379.7 & 2.3 & 2.6 & 51.7 & 5.22 & 3.03 & 74.3 & 4.5 & 2.6$\pm$7.0 & 9.8$\pm$0.3 & S,X & - \\
131 & 331.07$\pm$0.21 & -0.41$\pm$0.06 & -65.6$\pm$2.2 & 4.0 & 53 & 132.1 & 54.9 & 0.38$\pm$0.02 & 10.7 & 33.6 & 162.0 & 1226.2 & 2.5 & 3.5 & 43.9 & 7.73 & 3.50 & 63.2 & 5.2 & -11.1$\pm$4.3 & 1.6$\pm$0.3 & S,X & - \\
132 & 338.91$\pm$0.07 & 0.56$\pm$0.04 & -62.7$\pm$2.3 & 4.2 & 16 & 38.1 & 19.4 & 0.47$\pm$0.02 & 11.7 & 26.0 & 64.5 & 1691.0 & 3.1 & 3.8 & 11.3 & 3.93 & 2.13 & 26.7 & 4.7 & 57.4$\pm$3.2 & 24.1$\pm$1.7 & X & - \\
133 & 340.82$\pm$0.10 & -1.01$\pm$0.03 & -27.0$\pm$1.7 & 2.4 & 20 & 21.8 & 12.0 & 0.87$\pm$0.07 & 13.4 & 32.4 & 8.5 & 390.9 & 2.0 & 6.5 & 17.5 & 6.65 & 3.61 & 13.8 & 6.1 & -22.3$\pm$1.3 & 9.2$\pm$0.9 & S,X & 0.54 \\
134 & 334.64$\pm$0.03 & 0.43$\pm$0.05 & -65.2$\pm$0.4 & 4.0 & 13 & 19.3 & 13.4 & 0.30$\pm$0.01 & 10.1 & 21.1 & 8.9 & 462.2 & 1.5 & 3.1 & 9.8 & 6.92 & 3.92 & 15.1 & 5.0 & 47.7$\pm$3.7 & 66.0$\pm$1.9 & L,X & 0.45 \\
135 & 345.50$\pm$0.09 & 0.34$\pm$0.04 & -17.1$\pm$0.5 & 2.4 & 9 & 17.0 & 11.8 & 0.46$\pm$0.02 & 13.5 & 27.3 & 22.4 & 1317.6 & 4.4 & 9.8 & 9.1 & 4.75 & 2.77 & 14.6 & 6.1 & 34.1$\pm$1.7 & 18.0$\pm$1.7 & S,X & - \\
136 & 335.34$\pm$0.08 & 0.42$\pm$0.02 & -59.4$\pm$0.5 & 3.8 & 6 & 18.8 & 17.1 & 0.15$\pm$0.01 & 9.7 & 14.5 & 4.4 & 235.3 & 1.2 & 3.9 & 15.1 & 7.98 & 4.81 & 18.5 & 5.1 & 45.5$\pm$1.3 & 4.7$\pm$1.9 & S & 0.15 \\
137 & 327.99$\pm$0.04 & -0.07$\pm$0.04 & -49.7$\pm$1.5 & 3.1 & 9 & 15.3 & 10.2 & 1.44$\pm$0.06 & 12.7 & 35.4 & 9.0 & 591.0 & 3.1 & 10.7 & 12.8 & 3.56 & 1.61 & 16.3 & 6.0 & 15.7$\pm$2.3 & 42.0$\pm$3.5 & X & 0.25 \\
138 & 327.73$\pm$0.07 & -0.38$\pm$0.04 & -74.5$\pm$3.2 & 4.2 & 10 & 27.9 & 17.0 & 1.07$\pm$0.07 & 16.9 & 30.6 & 22.7 & 814.1 & 1.9 & 3.0 & 10.6 & 4.92 & 2.07 & 23.7 & 5.3 & -9.8$\pm$2.6 & 12.6$\pm$2.6 & C & 0.42 \\
139 & 327.29$\pm$0.04 & -0.56$\pm$0.03 & -46.3$\pm$1.9 & 3.0 & 11 & 14.6 & 6.5 & 0.83$\pm$0.05 & 16.1 & 38.4 & 34.8 & 2387.7 & 5.1 & 7.2 & 5.0 & 3.28 & 2.08 & 9.0 & 6.1 & -9.6$\pm$1.8 & 42.4$\pm$3.3 & C,H & - \\
140 & 327.15$\pm$0.07 & -0.29$\pm$0.07 & -61.2$\pm$1.5 & 3.5 & 19 & 41.3 & 21.4 & 0.42$\pm$0.01 & 9.6 & 27.2 & 26.6 & 643.1 & 2.4 & 5.8 & 24.5 & 4.77 & 1.93 & 30.9 & 5.8 & 1.5$\pm$4.0 & 43.0$\pm$1.4 & C,X & 0.31 \\
141 & 326.98$\pm$0.04 & -0.05$\pm$0.07 & -58.5$\pm$3.0 & 3.5 & 15 & 31.7 & 15.7 & 0.63$\pm$0.05 & 10.1 & 23.1 & 19.8 & 626.5 & 1.8 & 3.6 & 14.9 & 3.24 & 2.26 & 20.9 & 5.8 & 16.1$\pm$4.3 & 75.9$\pm$1.8 & C,X & 0.41 \\
142 & 326.80$\pm$0.04 & -0.11$\pm$0.03 & -56.3$\pm$0.7 & 3.5 & 6 & 10.8 & 6.5 & 0.74$\pm$0.01 & 12.9 & 26.1 & 7.8 & 719.2 & 2.4 & 5.3 & 5.8 & 3.20 & 1.94 & 9.4 & 5.8 & 12.5$\pm$1.9 & 37.2$\pm$5.0 & C & 0.34 \\
143 & 326.83$\pm$0.07 & -0.55$\pm$0.04 & -58.0$\pm$1.5 & 3.5 & 7 & 23.1 & 12.4 & 0.33$\pm$0.01 & 17.2 & 29.5 & 5.5 & 238.8 & 1.4 & 5.3 & 21.5 & 2.54 & 1.97 & 17.6 & 5.8 & -14.3$\pm$2.2 & 4.4$\pm$2.7 & X & 0.15 \\
144 & 326.66$\pm$0.05 & 0.60$\pm$0.11 & -39.8$\pm$1.7 & 1.8 & 40 & 37.7 & 15.4 & 0.92$\pm$0.05 & 13.7 & 34.0 & 15.4 & 409.7 & 1.8 & 5.3 & 26.7 & 5.52 & 2.31 & 20.4 & 6.9 & 40.7$\pm$3.4 & 80.5$\pm$0.7 & S,X & 0.55 \\
145 & 326.46$\pm$0.06 & 0.85$\pm$0.13 & -40.7$\pm$1.1 & 1.8 & 23 & 26.0 & 14.1 & 0.62$\pm$0.02 & 13.3 & 36.7 & 5.8 & 221.1 & 1.3 & 5.0 & 24.3 & 7.02 & 2.66 & 17.6 & 6.9 & 48.7$\pm$4.0 & 75.5$\pm$0.7 & S,X & 0.49 \\
146 & 345.24$\pm$0.06 & 1.04$\pm$0.02 & -15.2$\pm$0.9 & 2.4 & 13 & 14.3 & 10.0 & 0.88$\pm$0.07 & 16.3 & 41.5 & 7.8 & 548.7 & 1.8 & 3.8 & 7.4 & 7.16 & 4.08 & 11.2 & 6.1 & 63.4$\pm$0.7 & 5.8$\pm$1.8 & S,X & 0.52 \\
147 & 345.42$\pm$0.06 & 1.43$\pm$0.04 & -12.4$\pm$0.8 & 2.4 & 11 & 15.8 & 11.1 & 0.56$\pm$0.05 & 15.7 & 48.7 & 28.1 & 1782.2 & 5.1 & 9.5 & 7.2 & 4.72 & 1.84 & 16.4 & 6.1 & 79.9$\pm$1.6 & 17.1$\pm$2.8 & C & - \\
148 & 351.65$\pm$0.05 & -1.23$\pm$0.08 & -10.5$\pm$1.3 & 2.6 & 14 & 22.2 & 12.0 & 0.45$\pm$0.08 & 16.0 & 45.5 & 19.0 & 857.1 & 3.3 & 8.5 & 13.8 & 4.09 & 3.92 & 13.5 & 5.8 & -36.5$\pm$3.7 & 63.8$\pm$1.2 & C,X & - \\
149 & 323.88$\pm$0.06 & 0.03$\pm$0.03 & -57.8$\pm$1.1 & 10.0 & 14 & 61.6 & 42.3 & 0.17$\pm$0.01 & 12.1 & 25.9 & 62.5 & 1015.2 & 1.5 & 1.5 & 14.7 & 5.23 & 2.69 & 52.7 & 5.9 & 14.0$\pm$5.9 & 21.5$\pm$1.9 & S & 0.36 \\
150 & 322.16$\pm$0.09 & 0.63$\pm$0.04 & -54.9$\pm$1.2 & 3.3 & 13 & 25.6 & 16.3 & 0.82$\pm$0.03 & 13.3 & 30.4 & 27.4 & 1070.3 & 3.7 & 8.6 & 14.3 & 4.37 & 3.31 & 19.1 & 6.1 & 56.0$\pm$2.5 & 20.4$\pm$1.2 & L,H & - \\
151 & 321.07$\pm$0.05 & -0.52$\pm$0.02 & -60.6$\pm$0.7 & 9.3 & 12 & 44.6 & 26.2 & 0.24$\pm$0.01 & 18.5 & 30.2 & 53.8 & 1207.8 & 1.2 & 0.8 & 6.9 & 3.76 & 3.20 & 30.9 & 5.9 & -73.1$\pm$3.9 & 20.8$\pm$2.4 & C,X & 0.46 \\
152 & 320.28$\pm$0.07 & -0.30$\pm$0.02 & -65.8$\pm$1.2 & 8.6 & 11 & 46.6 & 36.4 & 0.22$\pm$0.01 & 16.6 & 31.8 & 61.0 & 1309.4 & 1.5 & 1.1 & 8.5 & 10.21 & 3.76 & 41.2 & 5.8 & -32.3$\pm$3.3 & 11.2$\pm$1.9 & S & 0.45 \\
153 & 320.40$\pm$0.02 & 0.13$\pm$0.03 & -4.9$\pm$1.5 & 12.5 & 10 & 42.0 & 20.5 & 0.27$\pm$0.11 & 12.1 & 30.8 & 81.6 & 1944.9 & 1.4 & 0.6 & 4.7 & 3.33 & 2.63 & 25.7 & 8.1 & 35.1$\pm$5.8 & 49.0$\pm$4.2 & C & 0.55 \\
154 & 320.17$\pm$0.03 & 0.83$\pm$0.04 & -39.8$\pm$2.1 & 2.4 & 14 & 13.4 & 5.4 & 0.84$\pm$0.06 & 14.1 & 33.5 & 4.7 & 350.8 & 1.4 & 3.5 & 8.3 & 2.67 & 1.75 & 8.2 & 6.7 & 56.0$\pm$1.8 & 56.6$\pm$3.3 & C,X & 0.62 \\
155 & 317.66$\pm$0.07 & 0.10$\pm$0.03 & -44.2$\pm$1.0 & 2.8 & 8 & 12.4 & 8.4 & 0.38$\pm$0.01 & 11.6 & 23.5 & 17.6 & 1416.6 & 5.7 & 15.0 & 7.9 & 3.59 & 3.08 & 10.0 & 6.6 & 26.1$\pm$1.3 & 12.9$\pm$2.2 & C & 0.40 \\
156 & 316.77$\pm$0.07 & -0.02$\pm$0.05 & -38.6$\pm$1.2 & 2.5 & 20 & 25.4 & 14.1 & 0.80$\pm$0.04 & 14.6 & 29.4 & 48.3 & 1898.4 & 6.5 & 14.7 & 13.9 & 5.50 & 2.55 & 17.9 & 6.7 & 20.6$\pm$2.0 & 32.7$\pm$1.4 & X,H & 0.51 \\
157 & 314.23$\pm$0.03 & 0.30$\pm$0.08 & -61.2$\pm$2.0 & 4.2 & 10 & 25.8 & 18.1 & 0.53$\pm$0.02 & 14.0 & 32.8 & 14.1 & 545.2 & 1.8 & 4.1 & 13.9 & 5.09 & 3.70 & 20.5 & 6.2 & 41.9$\pm$5.8 & 77.8$\pm$1.6 & L & 0.35 \\
158 & 311.60$\pm$0.04 & 0.30$\pm$0.01 & -48.1$\pm$0.9 & 4.1 & 6 & 10.2 & 7.6 & 0.33$\pm$0.01 & 12.1 & 36.5 & 11.5 & 1127.7 & 3.5 & 7.4 & 5.2 & 6.45 & 5.48 & 8.1 & 6.4 & 41.7$\pm$1.0 & 18.3$\pm$4.2 & S & - \\
159 & 309.92$\pm$0.03 & 0.39$\pm$0.08 & -57.9$\pm$1.4 & 5.5 & 6 & 34.4 & 23.6 & 0.13$\pm$0.01 & 16.7 & 35.1 & 27.3 & 793.4 & 1.4 & 1.7 & 9.8 & 4.58 & 3.04 & 28.2 & 6.4 & 56.4$\pm$7.6 & 89.0$\pm$1.5 & L,X & 0.31 \\
160 & 309.15$\pm$0.03 & -0.30$\pm$0.08 & -42.8$\pm$1.3 & 3.5 & 8 & 21.9 & 14.9 & 0.39$\pm$0.02 & 10.4 & 19.4 & 16.2 & 740.9 & 3.2 & 9.2 & 15.3 & 4.08 & 2.56 & 18.9 & 6.7 & 2.6$\pm$4.6 & 79.5$\pm$2.3 & C,X & 0.24 \\
161 & 305.55$\pm$0.06 & -0.01$\pm$0.04 & -38.2$\pm$1.4 & 3.8 & 11 & 24.3 & 17.3 & 0.52$\pm$0.03 & 17.7 & 30.0 & 15.4 & 631.3 & 1.6 & 2.7 & 9.8 & 7.14 & 2.80 & 21.3 & 6.9 & 20.8$\pm$2.5 & 30.5$\pm$2.1 & C & 0.45 \\
162 & 305.22$\pm$0.06 & -0.00$\pm$0.04 & -32.5$\pm$2.9 & 3.8 & 21 & 40.6 & 17.7 & 0.69$\pm$0.08 & 12.3 & 42.1 & 29.3 & 722.2 & 1.6 & 2.3 & 14.4 & 4.94 & 2.70 & 22.1 & 6.9 & 21.3$\pm$2.6 & 32.2$\pm$1.6 & S,X & 0.55 \\
163 & 305.30$\pm$0.09 & 0.23$\pm$0.06 & -38.8$\pm$2.1 & 3.8 & 29 & 75.5 & 22.2 & 0.66$\pm$0.05 & 16.1 & 51.2 & 100.6 & 1332.9 & 2.3 & 2.6 & 20.9 & 3.83 & 1.59 & 35.7 & 6.9 & 36.5$\pm$3.9 & 3.9$\pm$1.5 & C,X & - \\
\hline
min & 5.37 & -1.23 & -120.6 & 1.3 & 6 & 10.2 & 5.3 & 0.06 & 8.5 & 14.5 & 1.7 & 162.0 & 0.9 & 0.6 & 2.9 & 2.09 & 1.51 & 6.6 & 1.3 & -99.0 & 1.5 & - & 0.14 \\
max & 354.64 & 1.43 & 113.1 & 19.7 & 169 & 370.2 & 109.4 & 1.75 & 22.1 & 53.0 & 525.0 & 7324.3 & 6.5 & 21.4 & 133.4 & 26.81 & 11.36 & 121.0 & 11.6 & 81.8 & 89.0 & - & 0.62 \\
med & 320.40 & -0.07 & -20.4 & 4.2 & 10 & 25.9 & 16.0 & 0.47 & 13.3 & 28.0 & 31.8 & 1070.3 & 2.5 & 4.0 & 10.2 & 4.91 & 2.63 & 20.5 & 5.4 & 11.1 & 29.8 & - & 0.31 \\
mean & 202.39 & -0.07 & -2.5 & 5.3 & 16 & 39.7 & 20.6 & 0.53 & 13.5 & 29.4 & 65.0 & 1443.8 & 2.7 & 5.0 & 14.6 & 6.18 & 3.21 & 25.9 & 5.2 & 10.3 & 35.3 & - & 0.35 \\
Std & 155.33 & 0.36 & 60.9 & 3.3 & 20 & 40.8 & 14.3 & 0.32 & 2.8 & 7.2 & 89.2 & 1174.6 & 1.1 & 3.7 & 15.3 & 3.65 & 1.80 & 16.7 & 1.4 & 26.1 & 26.3 & - & 0.09 \\
S & -0.28 & 0.49 & 0.3 & 1.3 & 5 & 4.4 & 2.3 & 0.89 & 0.4 & 0.8 & 2.8 & 2.1 & 1.0 & 1.6 & 4.3 & 2.31 & 2.07 & 1.8 & 0.2 & -0.6 & 0.5 & - & 0.15 \\
K & -1.90 & 2.20 & -1.1 & 1.6 & 30 & 27.6 & 8.8 & 0.67 & -0.5 & 1.0 & 8.4 & 5.6 & 0.7 & 3.4 & 24.8 & 7.45 & 4.85 & 5.6 & 2.3 & 2.9 & -1.0 & - & -0.5
\enddata
\tablecomments {meaning of columns are detailed in Table \ref{t1_explain}. Column (1) ID of large-scale filaments; Cols. (2)-(4) flux-weighted longitude, latitude (degree), and LSR velocity (km s$^{-1}$); Column (5) distance (kpc); Column (6) Number of clumps in filaments; Column (7) sum of edge lengths (pc); Column (8) end-to-end length (pc); Column (9) velocity gradient (km s$^{-1}$ pc$^{-1}$); Cols. (10) and (11) minimum and maximum temperatures (K) of clumps; Column (12) mass of the filament ($10^3 M_\odot$); Column (13) line mass ($M_\odot$ pc$^{-1}$); Cols. (14) and (15) $H_2$ column density ($10^{22}$ cm$^{-2}$) and volume density ($10^3$ cm$^{-3}$); Column (16) sum of edge lengths over filament width; Column (17) aspect ratio; Column (18) linearity;  Column (19) linearity-weighted length; Column (20) Galactocentric radius (kpc); Column (21) height above Galactic mid-plane (pc); Column (22) orientation angle between filaments major axes and Galactic mid-plane in the projected sky (degree); Column (23) morphology class; Column (24) dense gas mass fraction, ``-'' means no DGMF measurement.}
\end{deluxetable}
\end{longrotatetable}
}

\begin{table}[b]
{\begin{tabular}{llp{40.5em}}
\toprule
\tablecolumns{3}
Column & & \multicolumn{1}{c}{Explanation} \\
\colrule
(1) ID & & Assigned filament ID \\
(2) $l_{wt}$ &(deg) & Flux-weighted Galactic longitude \\
(3) $b_{wt}$ &(deg) & Flux-weighted Galactic latitude \\
(4) $v_{wt}$ &(km s$^{-1}$) & Flux weighted local standard of rest (LSR) velocity \\
(5) $d$ &(kpc) & Distance, the median of all clump distances in this filament \\
(6) N$_{cl}$ & & The number of clumps in the filament \\
(7) $L_{sum}$ &(pc) & The sum of ``edges'' (separation of each two connected clumps) of this filament \\
(8) $L_{end}$ &(pc) & End-to-end length along the major axis \\
(9) $v_{grad}$ &(km s$^{-1}$ pc$^{-1}$) & Mean velocity gradient on edges \\
(10) $T_{min}$ &(K) & Minimum temperature of clumps \\
(11) $T_{max}$ &(K) & Maximum temperature of clumps \\
(12) Mass &($10^3 M_\odot$) & Mass of the filament, derived from Herschel Hi-GAL column density map (detailed in Sect. \ref{sec:mass}) \\
(13) $M_{line}$ &($M_\odot$ pc$^{-1}$) & Line mass, mass per unit length \\
(14) $N_{H_2}$ &($10^{22}$ cm$^{-2}$) & $H_2$ column density \\
(15) $n_{H_2}$ &($10^3$ cm$^{-3}$) & $H_2$ volume density \\
(16) $\frac{L_{sum}}{w}$ & & The ratio between sum of edge lengths and width. The width is the mean of clump diameters \\
(17) $f_{A}$ & & Aspect ratio, the ratio of area between the circle enclosing the filament and the concave hull (detailed in Appendix \ref{sec:concavehull}). For an approximately elliptical distribution, $f_{A}$ is very similar with the aspect ratio of the ellipse \\
(18) $f_{L}$ & & Linearity, the ratio between spread (standard deviation) of clumps along major axis and spread along minor axis \\
(19) $L_{wt}$ & (pc) & Linearity-weighted length, $L_{wt} = L_{end}\cdot \sqrt{1+\dfrac{4}{f_L^2}}$ \\
(20) $R_{gc}$ &(kpc) &  Galactocentric radius, median of clump Galactocentric radius \\
(21) z &(pc) & Height above Galactic mid-plane \\
(22) $|\theta|$ & (deg) & Orientation angle between filaments major axes and Galactic mid-plane in the projected sky (calculation of $|\theta|$ is from PCA, detailed in Appendix \ref{sec:PCA}) \\
(23) Mor. & & Morphology class following \citet{Wang2015}, ``L'' represents linear straight L-shape, ``C'' bent C-shape, ``S'' quasi-sinusoidal shape, ``X'' crossing of multiple filaments, ``H'' head-tail or hub-filament system \\
(24) DGMF & & Dense gas mass fraction, the ratio between dense gas mass and total mass of the filament. It is illustrated in Sect. \ref{sec:DGMF}\\
\botrule
\end{tabular}}
\caption{Explanation of physical properties listed in table \ref{t1} }
\label{t1_explain}
\end{table}

\subsection{Statistics}
\begin{figure*}
\gridline{\fig{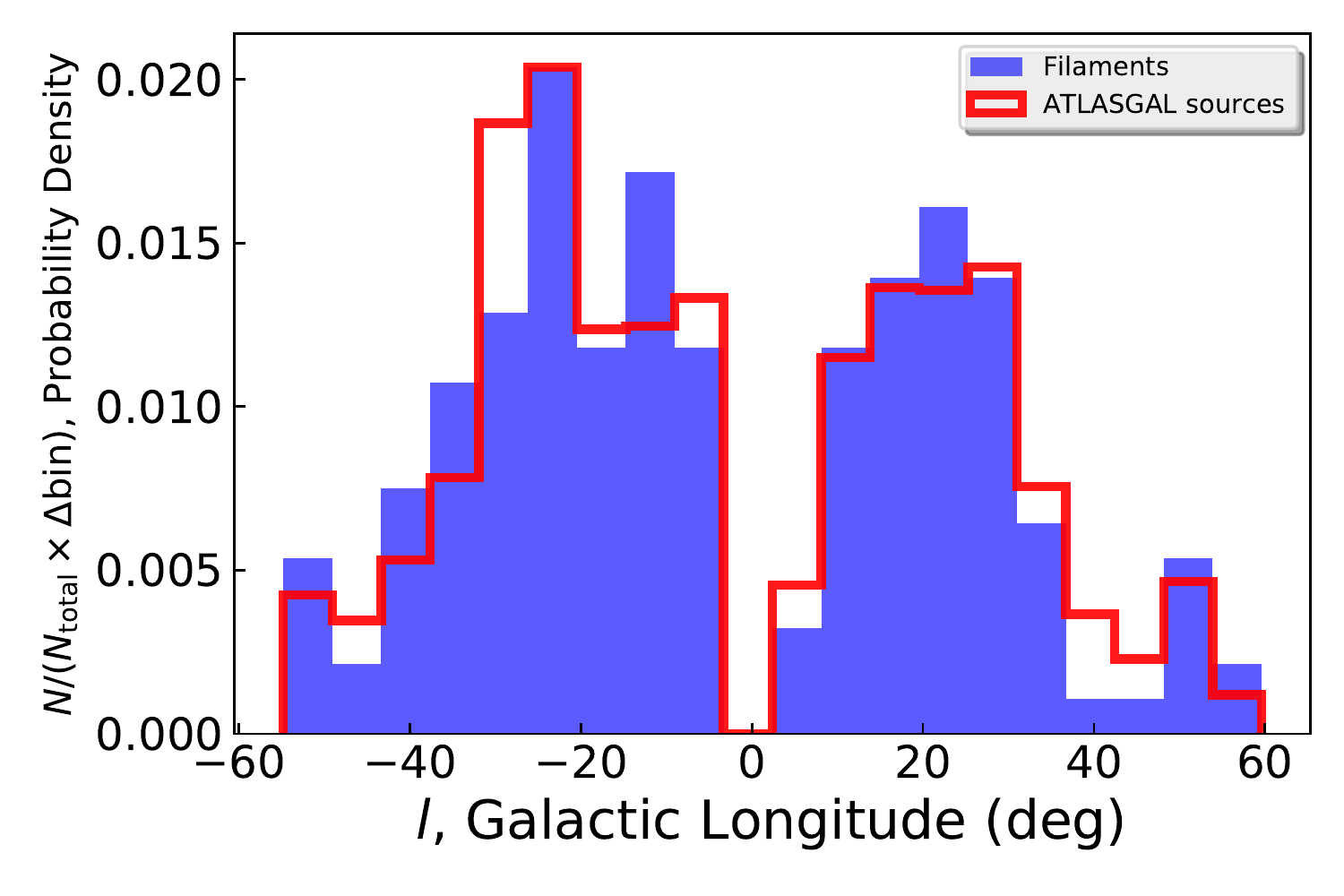}{0.3\textwidth}{(a)}
          \fig{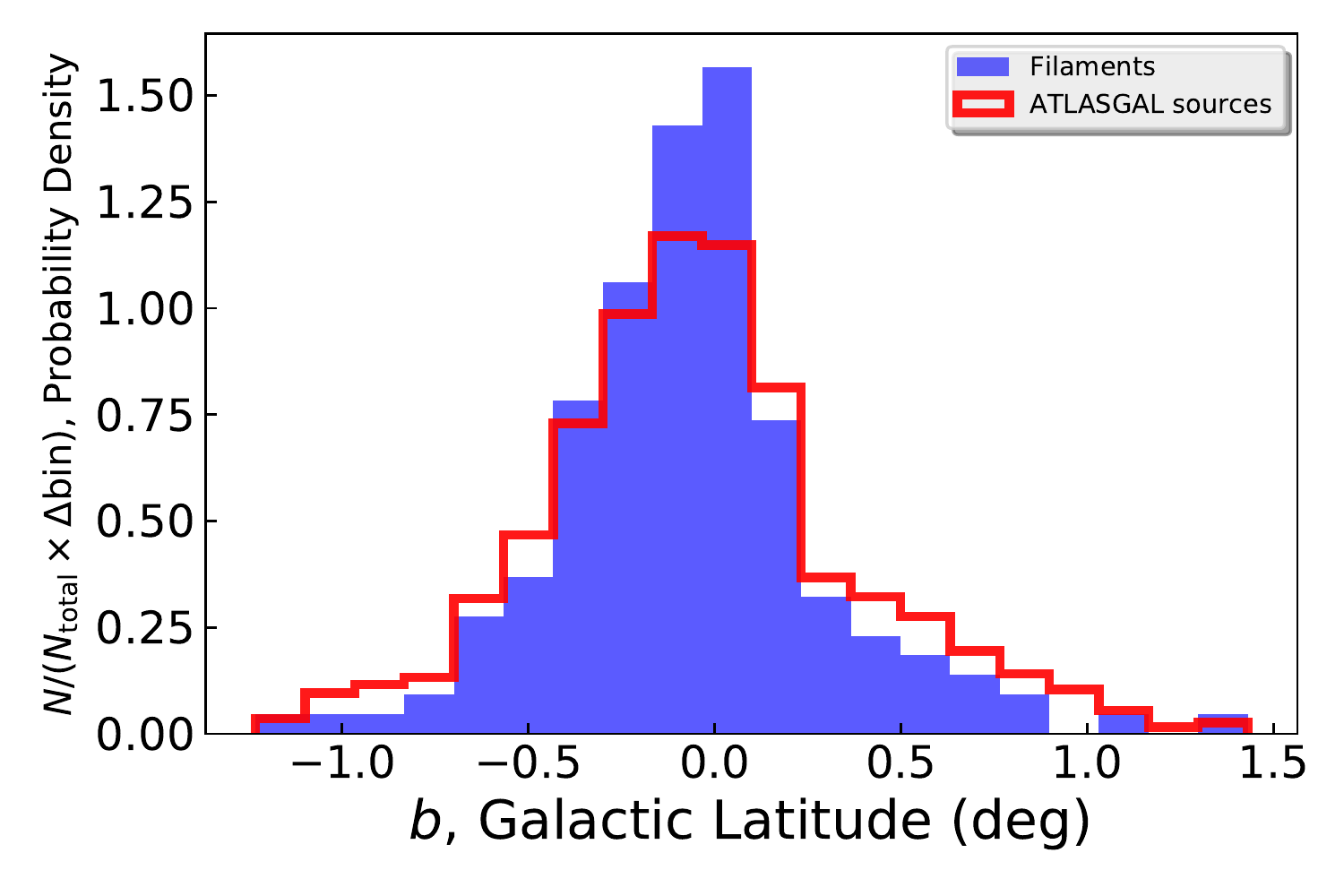}{0.3\textwidth}{(b)}
          \fig{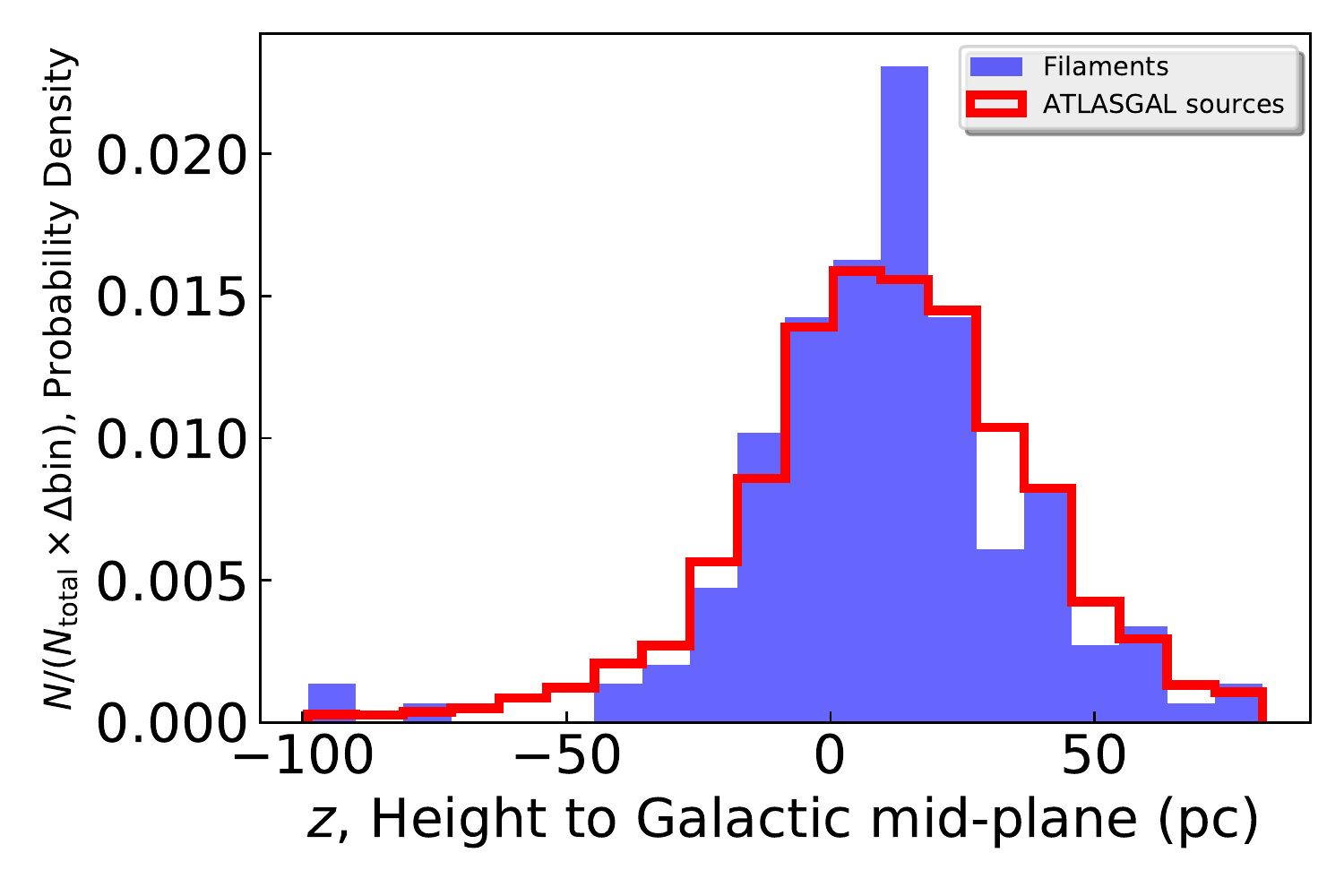}{0.3\textwidth}{(c)}
          }
\gridline{\fig{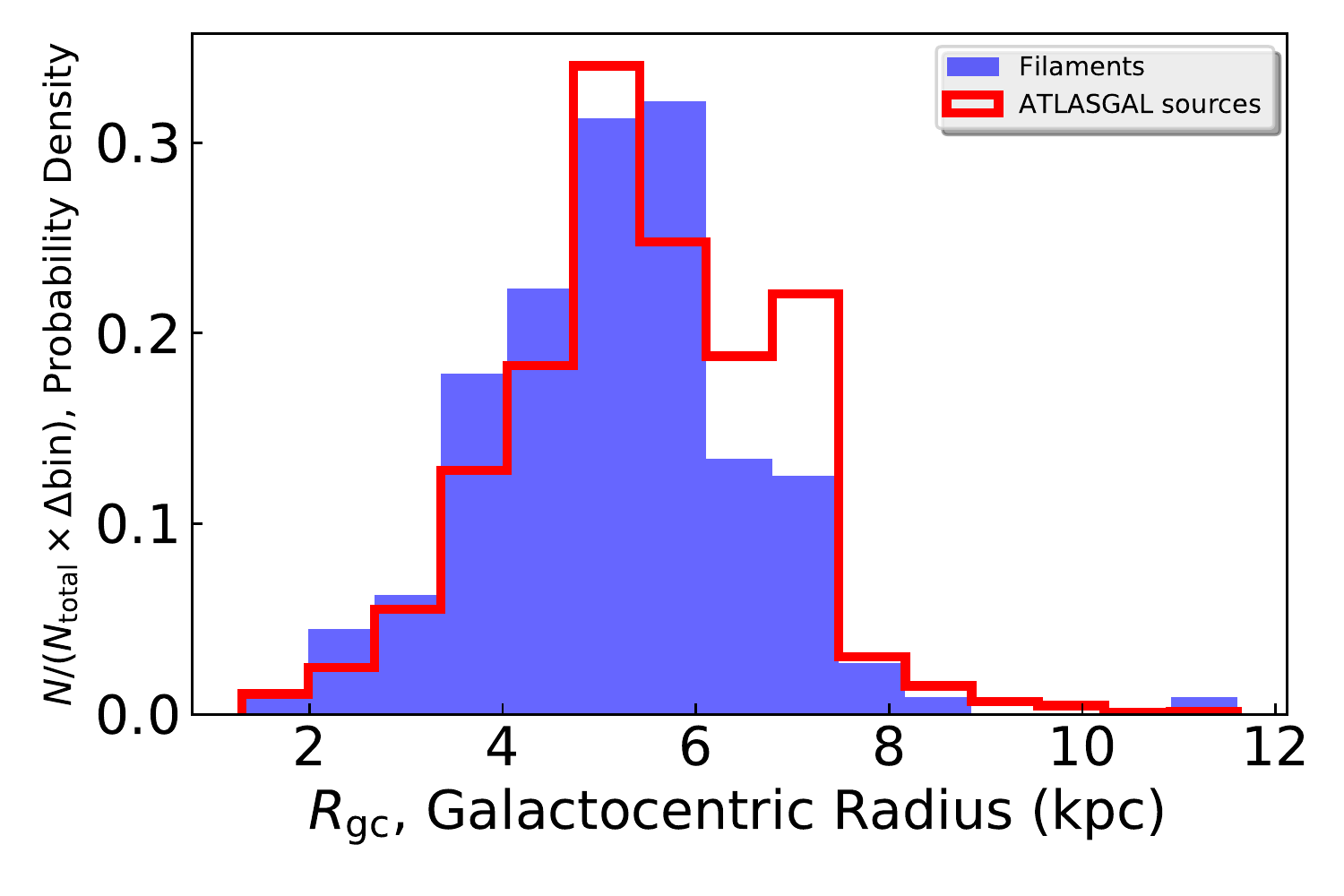}{0.3\textwidth}{(d)}
          \fig{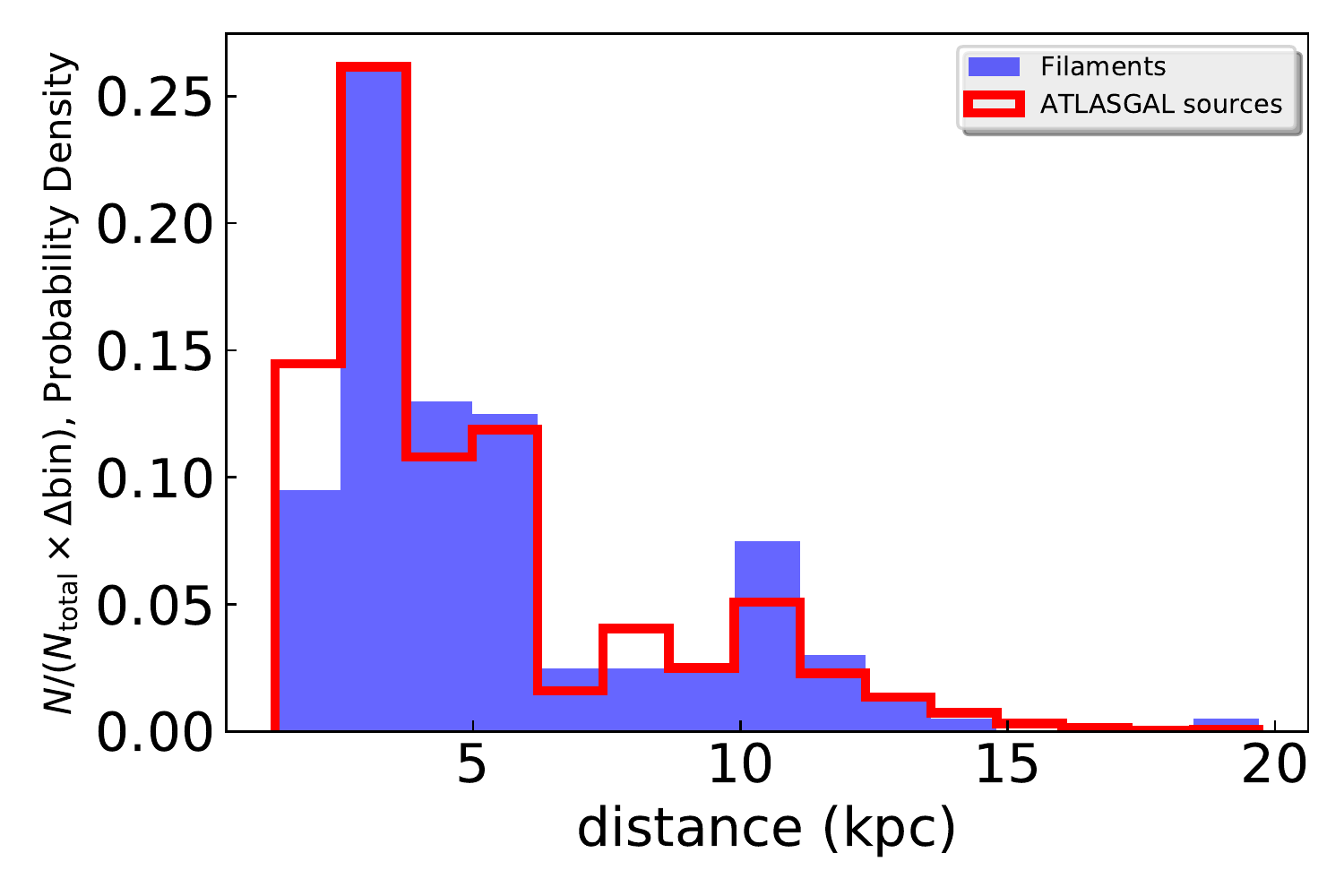}{0.3\textwidth}{(e)}
          \fig{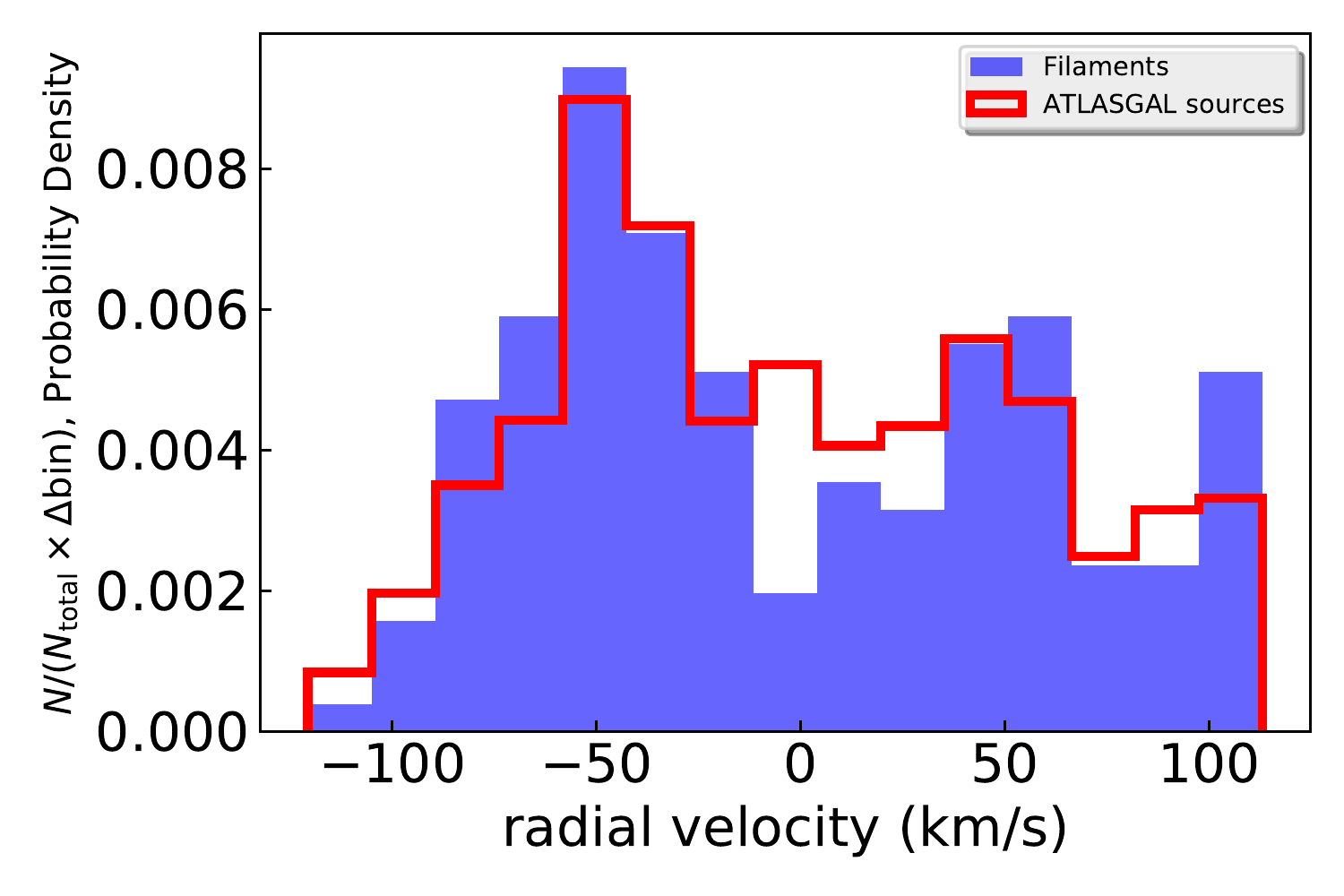}{0.3\textwidth}{(f)}
          }
\gridline{\fig{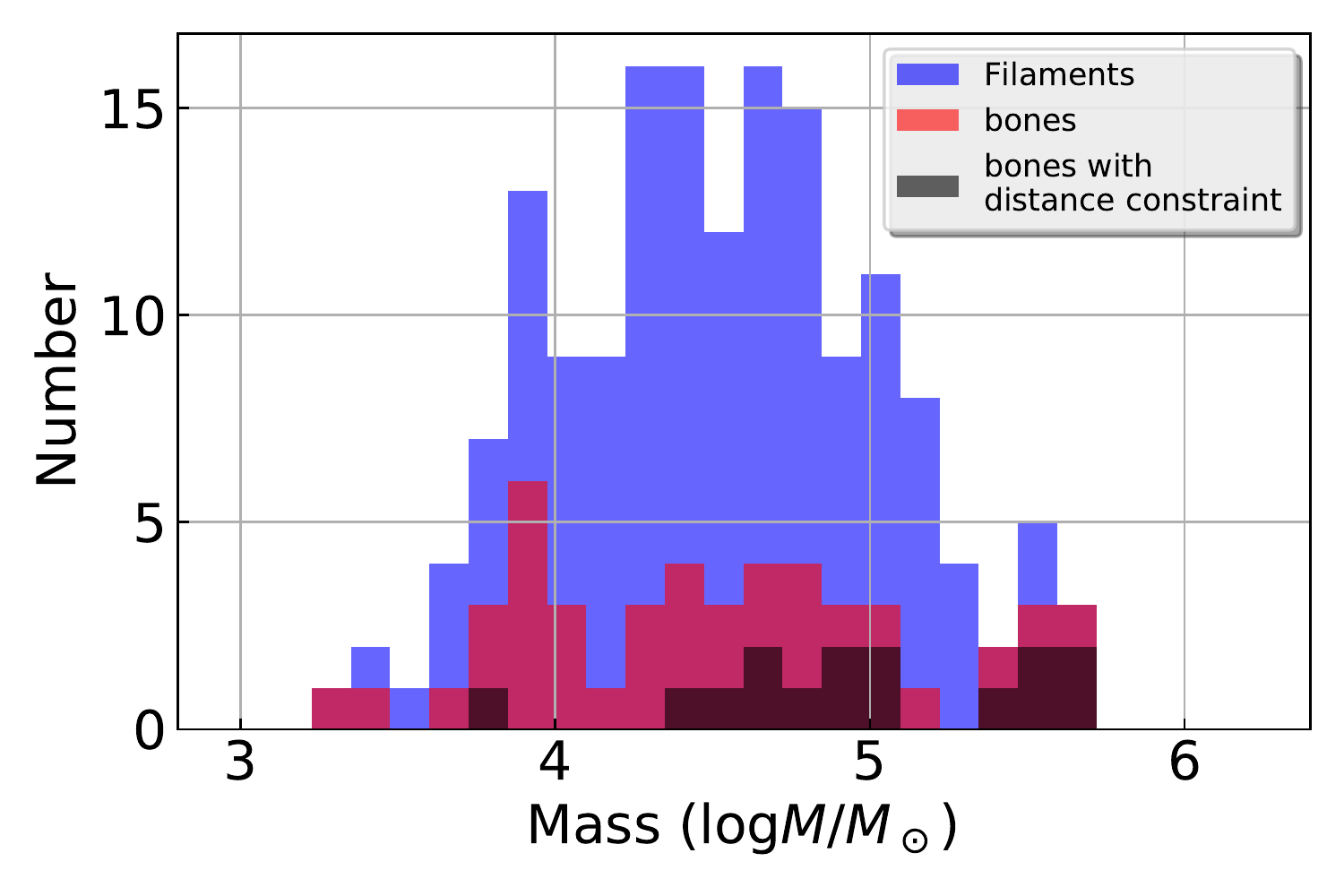}{0.3\textwidth}{(g)}
          \fig{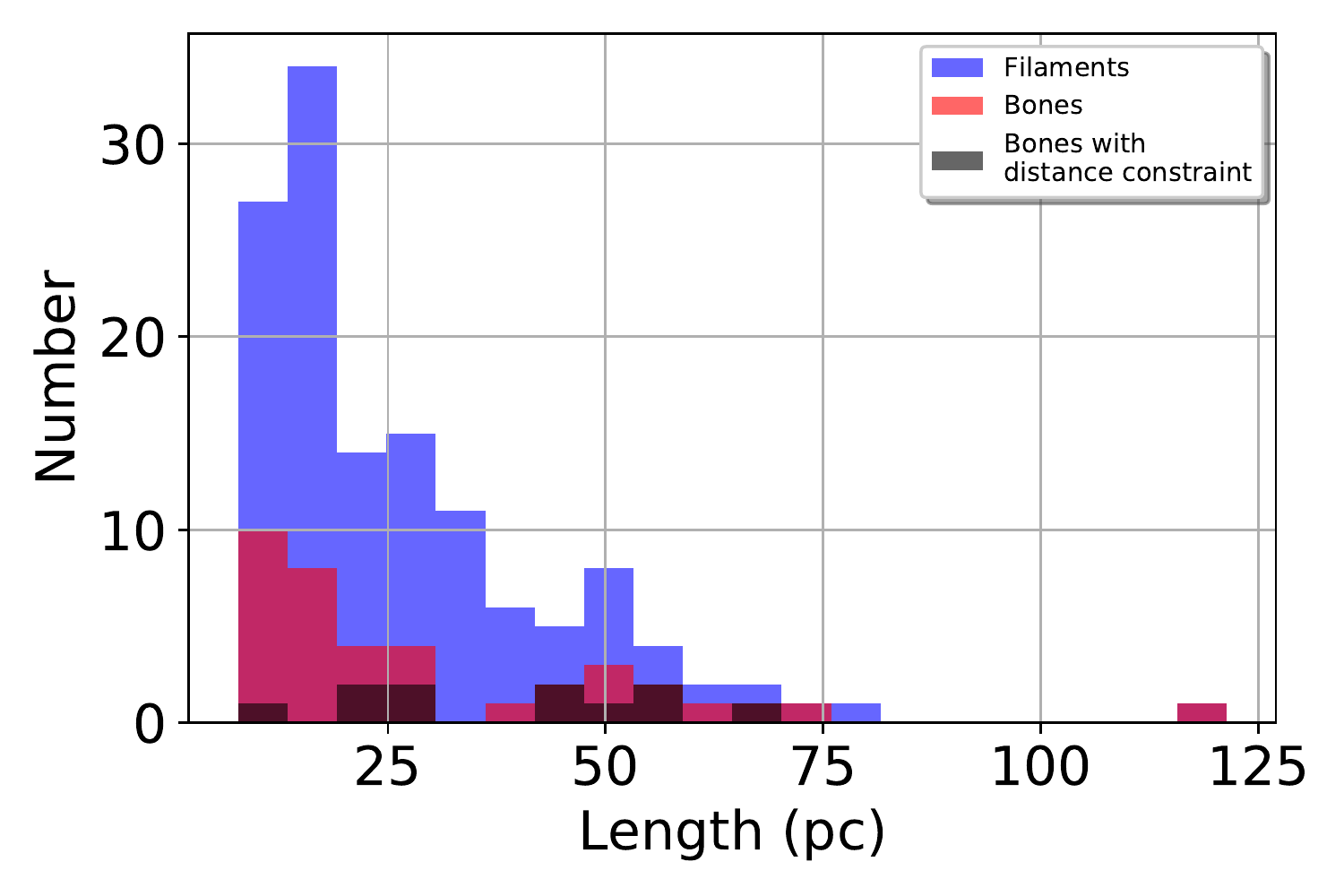}{0.3\textwidth}{(h)}
          \fig{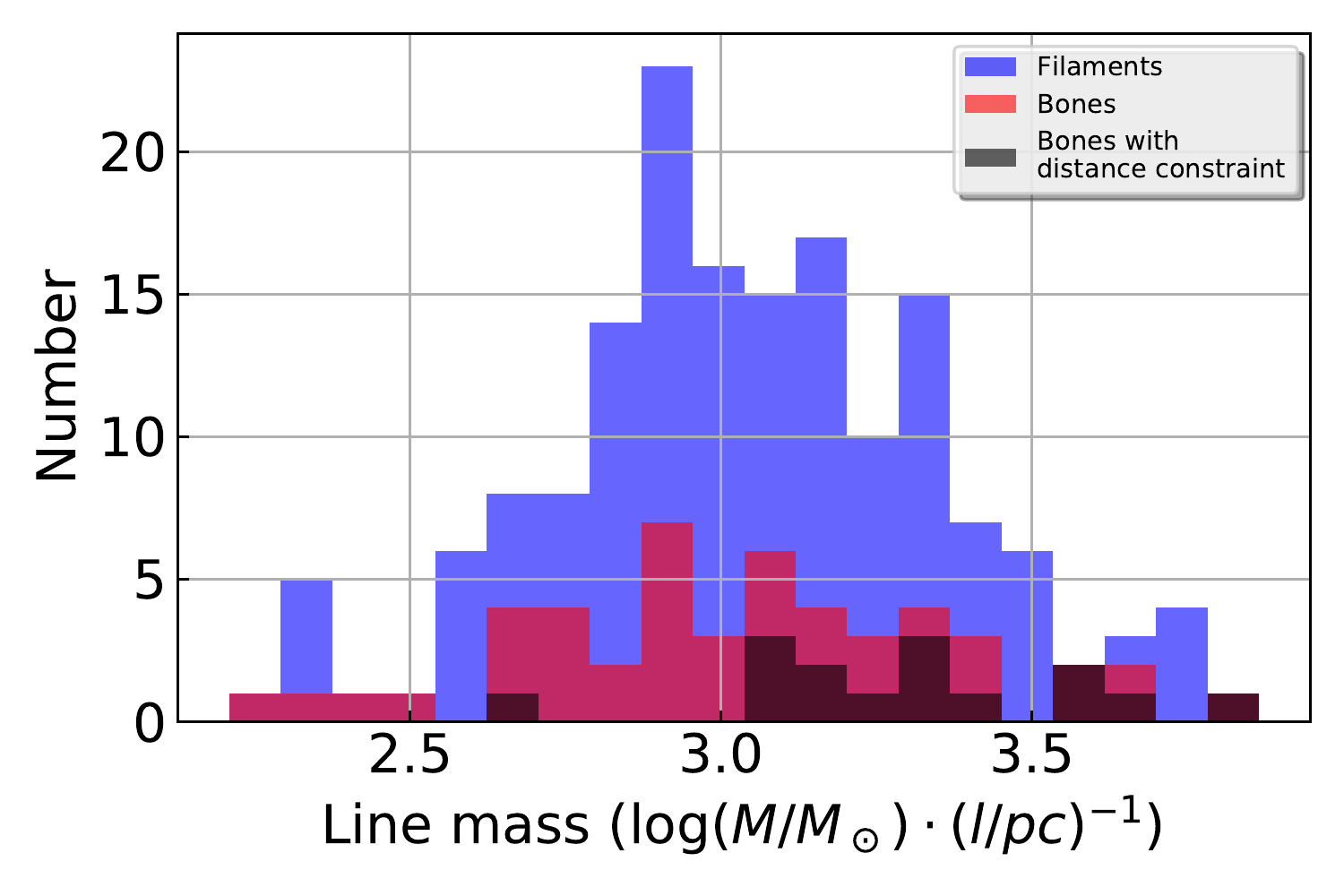}{0.3\textwidth}{(i)}
          }
\gridline{\fig{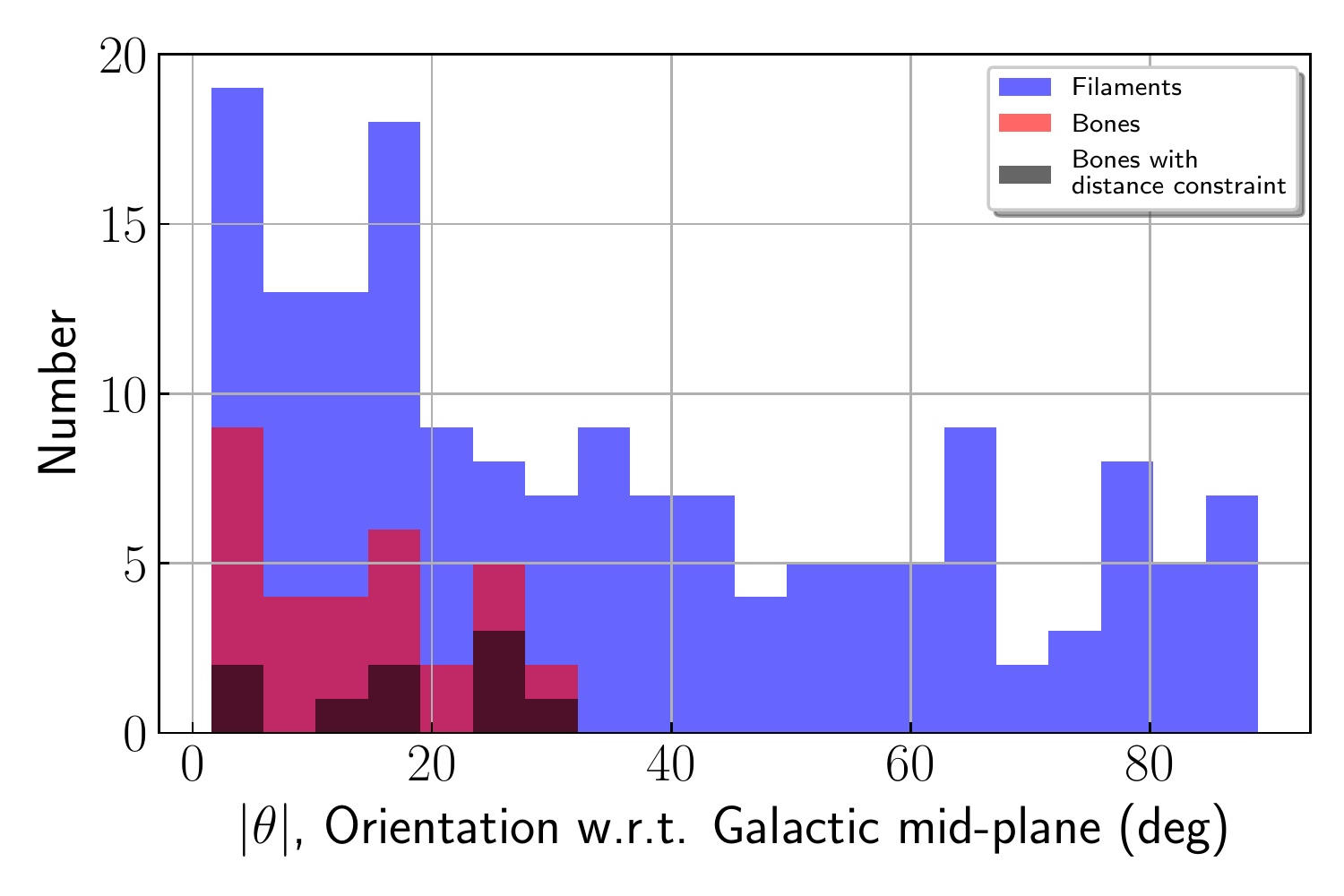}{0.3\textwidth}{(j)}
          }
\caption{Physical properties of the large-scale filaments. Blue bars show distributions of filament properties. (a)-(f) are normalized distribution of Galactic longitude, Galactic latitude, height to Galactic mid-plane, Galactocentric radius, distance, and radial velocity for filaments. Red steps show distribution of ATLASGAL clumps with velocity information from \citet{Urquhart2018}. Panels (g)-(j) are mass, linearity-weighted length, line mass, and orientation angle of filaments. Red bars denote ``bones'' with the same longitude-velocity criteria with \citet{Wang2016}, those are criteria (6)$\sim$(8) in Sect. \ref{sec:bone}. While black bars show the new definition of bones in this work. We require that a new bone not only has similar velocity with an arm in the same Galactic longitude, but also has the similar distance to us with that arm (criterion (9) in Sect. \ref{sec:bone}).}
\label{sta}
\end{figure*}

Some representative physical properties of the filaments are shown in Fig. \ref{sta}. Panels (a)-(h) are normalized distribution of Galactic longitude, Galactic latitude, height to Galactic mid-plane, Galactocentric radius, distance, and radial velocity for filaments. Panels (g)-(j) are mass, linearity-weighted length, line mass, and orientation angle of filaments.\\

No filaments are found within $\left| l\right|<5^\circ$, due to the limitation of catalogue we use. The distribution of filaments with Galactic longitude has two peaks around $l=-22^\circ$ and $l=23^\circ$ shown in Fig. \ref{sta} (a). \citet{Mattern2018} also find the former peak but not the later one in their small-scale ATLASGAL filaments, owning to their Galactic longitude ranging from $l=-60^\circ$ to $l=18^\circ$. They attribute the decrease of filaments count towards center to difficulty in identifying filaments in confused structures. They think the suppression outwards means it is unlikely to find filaments towards outer Galaxy. In our sample, the position of filaments shows a similarity with those of ATLASGAL clumps,  implying filaments tend to be found in the region where clumps are crowded, and this can be directly perceived in Fig. \ref{faceon} (c). So the peaks in distribution of filaments with Galactic longitude are merely due to peaks in distribution of ATLASGAL clumps. This result also implies that large-scale filaments are ubiquitous in the inner Galactic plane, and therefore, an unavoidable issue in the study of star formation. But we also note that this similarity in distribution of filaments and clumps does not always exist (for example, Galactocentric radius in Fig. \ref{sta} (d)). The distribution of Galactic latitude of filaments shown in Fig.  \ref{sta} (b) has a peak and mean $b=-0.01^\circ$ close to Galactic mid-plane, which is consistent with \citet{Mattern2018}.\\

Of our filaments, F120 stands out, whose linearity-weighted length is 121 pc and the mass is $4.36\times10^5$ M$_\odot$. Except this extreme filament, the linearity-weighted lengths of filaments range from 7 to 74 pc, with mass of order of 10$^3$ M$_\odot$ to 10$^5$ M$_\odot$. We also find no obvious correlation between length and distance. The vertical distances from filaments to Galactic mid-plane (Fig. \ref{sta} (c)) are not symmetrical about the Galactic mid-plane (S=-0.6), which agrees with the result of northern sky \citep{Wang2016}.\\

\begin{figure*}
\gridline{\fig{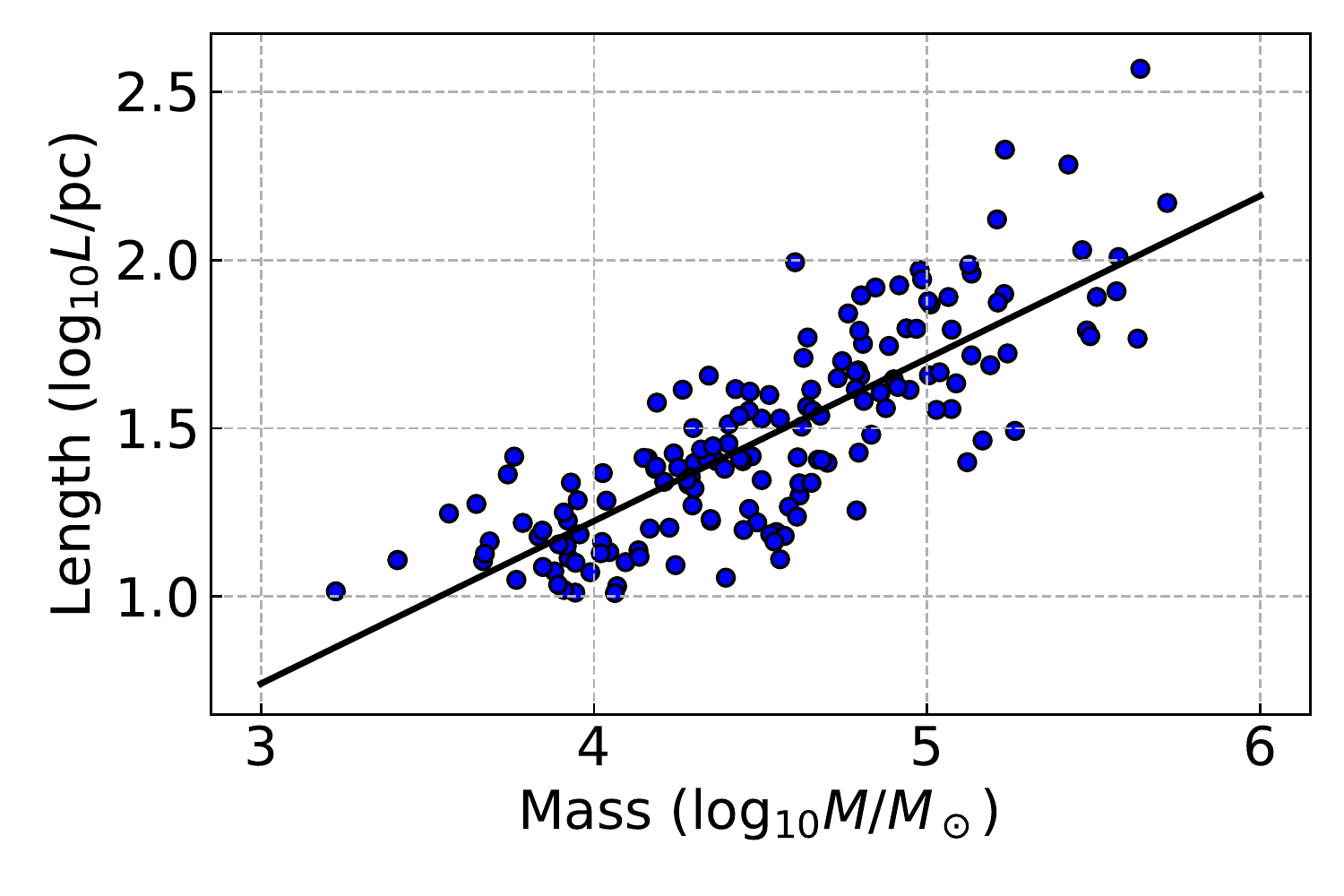}{0.5\textwidth}{(a)}
          \fig{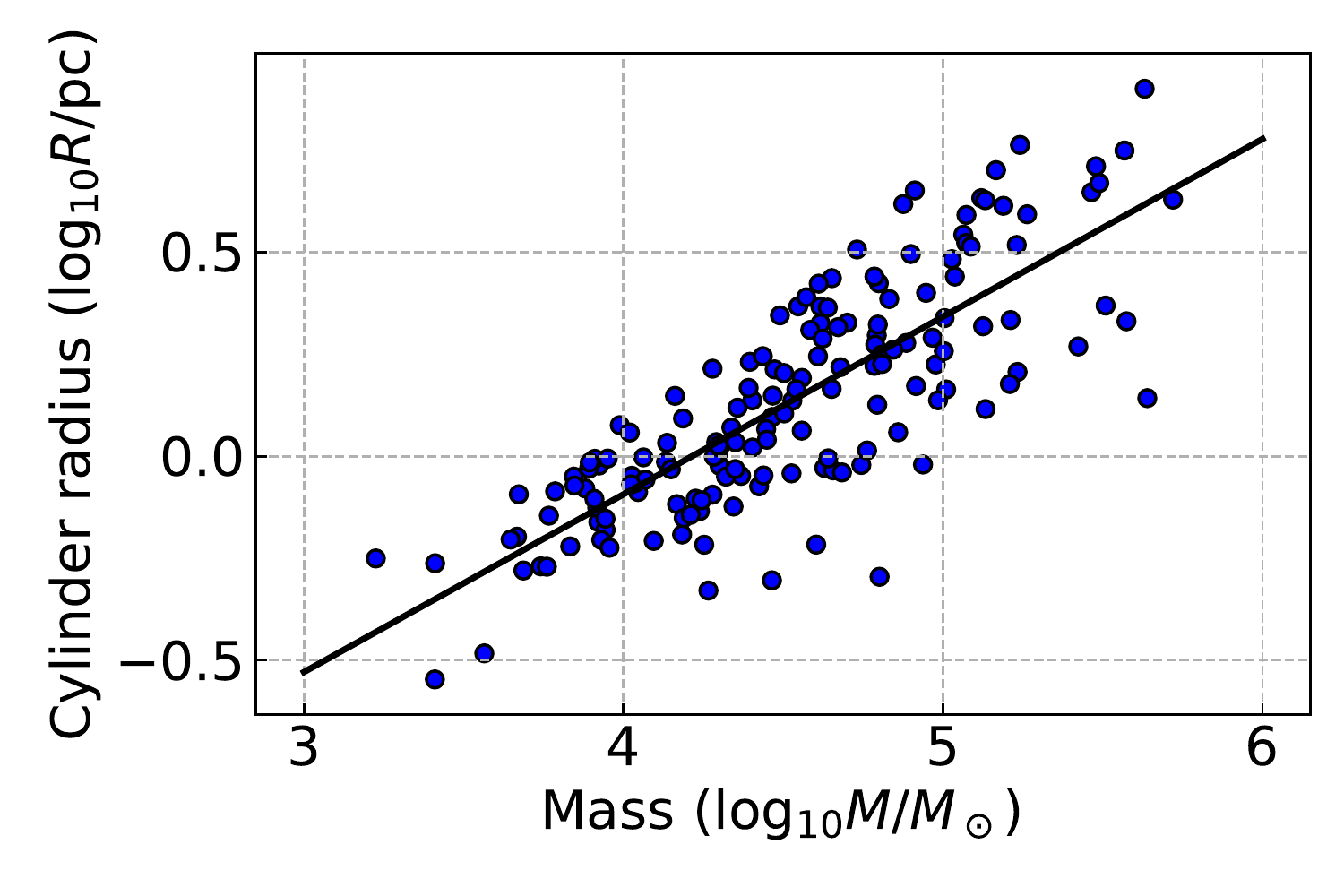}{0.5\textwidth}{(b)}
          }
\caption{Relation between the mass and lengths (a), cylinder radii (b) of large-scale filaments. Blue circles denote filaments and black lines are fitted results. }
\label{ML}
\end{figure*}

The relation between mass and lengths (sum of edge lengths) of large-scale filaments is shown in Fig. \ref{ML} (a). The black curve is the best-fitted line using the least square method, whose expression is $lgL=0.46\,lgM-0.40$. This implies a tight power-law relation between length and mass, $M\sim L^{2.17}$. The power-law index is similar with that of filaments from BGPS sources \citep{Wang2016}, indicating a fractal dimension of 2.17. For comparison, the relation between mean cylinder radii and mass of large-scale filaments is also examined (Fig. \ref{ML} (b)), which gives a relation of $lgL=0.56\,lgM-2.14$. The corresponding dimension is 1.77, out of range 2 to 3 from \citet{Roman2010}, implying the mean cylinder radii is a less essential property for large-scale filaments.\\

\begin{figure*}
\begin{center}
\includegraphics[scale=0.85,angle=0]{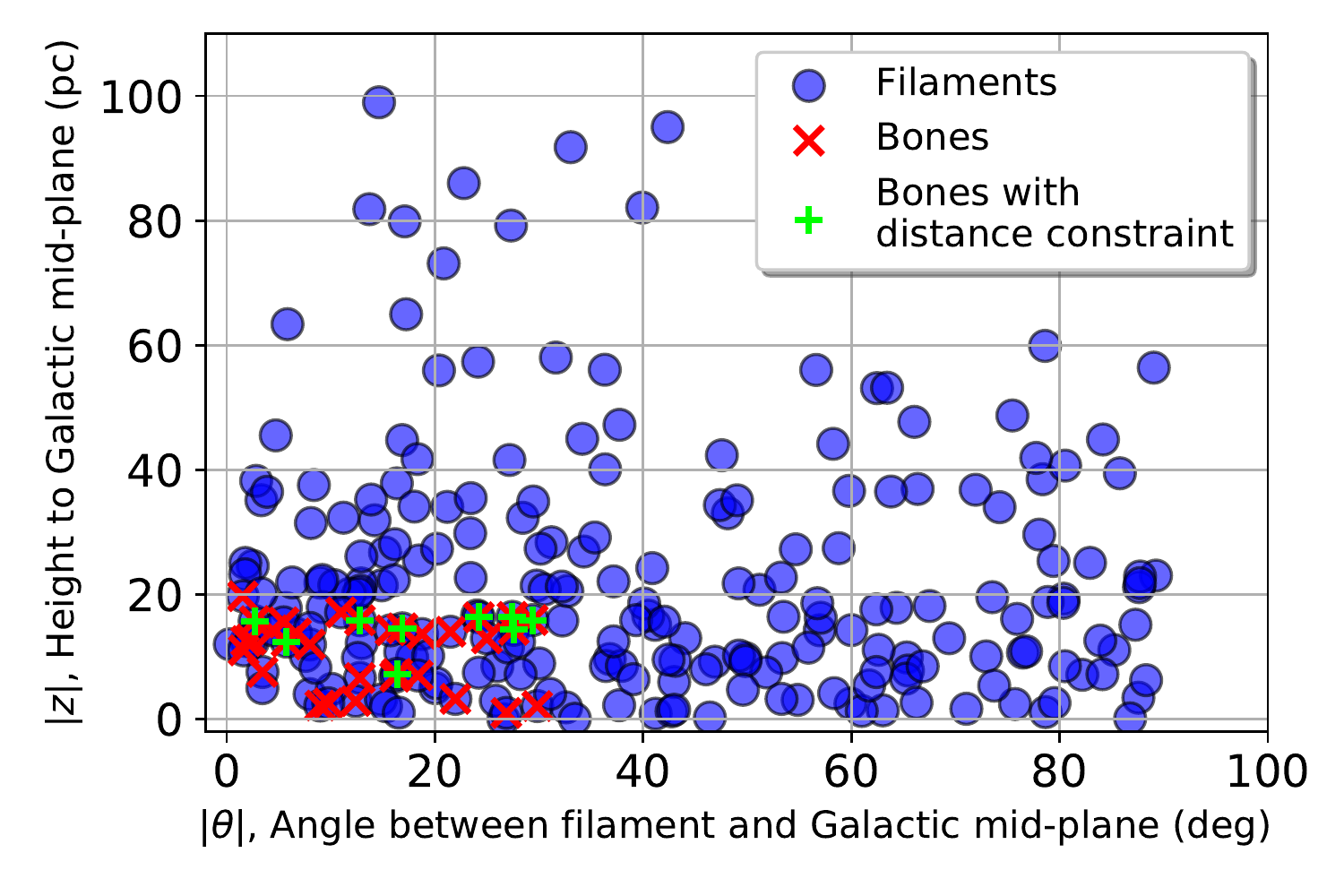}
\caption{Orientation angle vs. vertical height to the physical Galactic mid-plane of large-scale filaments. Blue filled circles show filaments while red crosses and green pluses mark bones with the same criteria with \citet{Wang2016} and bones in this work respectively.}
\label{orivsz}
\end{center}
\end{figure*}

Do filaments in the proximity of the Galactic mid-plane prefer to align with the Galactic mid-plane? The relation between orientation angle and height to the Galactic mid-plane of large-scale filaments is  plotted in Fig. \ref{orivsz}. There is no obvious correlation between the two, which is identical to the result from northern sky \citep{Wang2016}.

\subsection{Dense Gas Mass Fraction} \label{sec:DGMF}
Dense gas mass fraction (DGMF) of a filament is the ratio of dense gas mass to total mass of the whole filament, which is an important quantity related to star formation rate (SFR) and star formation efficiency (SFE) of molecular clouds \citep{Heiderman2010, Lada2010, Lada2012}, although \citet{Kainulainen2013} suggest that DGMF and SFE may not be closely linked. To avoid deviation engendered by distinct methods to get mass, both dense gas mass and filament mass are derived from Herschel Hi-GAL column density map, which is illustrated in Sect. \ref{sec:Herschel}. After discarding filaments with incomplete column density information, 130 of the 163 large-scale filaments have DGMF measurement. Their DGMFs range from 14.7\% to 62.4\%, with a mean value of 35.6\%. The result is larger than that from \citet{Ragan2014} and \citet{Abreu2016}, who take the ratio between mass from ATLASGAL 870 \textmu m dust emission and mass from $^{13}$CO emission as DGMF. But our result is consistent with a value of 50\% from ``Nessie'' \citep{Goodman2014}, where DGMF is from the ratio of mass in an envelope traced by HNC observation to mass in cylinders with a fixed diameter and above a column density threshold. We conjecture that diverse definitions of dense gas mass account for the difference. We also employ similar dense gas mass calculation with \citet{Abreu2016}, where dense gas mass is acquired from ATLASGAL flux. The DGMFs become smaller.\\

\section{Discussion}
\label{sec:discussion}
\subsection{Comparison of our MST Filaments to MST Filaments from BGPS Sources and Other Previously Known Filaments}
To see whether the MST method is robust on sources from different catalogues, we compare our filaments from ATLASGAL Galactic clumps to those identified from BGPS by \citet{Wang2016} in a common longitude range the two survey reside ($7.5^\circ<l<60^\circ$). In the common region, 42 filaments are identified from BGPS sources while 67 are from ATLASGAL. The detailed comparison is in Appendix \ref{sec:BGPS&ATLASGAL}. On the whole, in the common Galactic region of the two catalogues ($7.5^\circ<l<60^\circ$), most (70\%) of the filaments identified from BGPS sources are also found from ATLASGAL clumps despite different methods they use to extract sources, distinct lower limit of source luminosities, and various surveys referred to obtain radial velocities of sources \citep{Shirley2013,Urquhart2018}. \\

The distinction between filaments from the two catalogues is mainly caused by the number of sources used to identify filaments. BGPS sources contain over 8400 continuum sources, but only 3126 of them have velocity measurement from HCO$^+(3-2)$ and/or N$_2$H$^+(3-2)$ spectral. Since we aim to identify velocity coherent filaments, velocity is indispensable for the MST method. For ATLASGAL sources, 7809 of them have radial velocities obtained from 21 archival molecular line surveys \citep{Urquhart2018}. In the overlaid longitude we compare ($7.5^\circ<l<60^\circ$), 2201 BGPS sources have velocity measurement while that for ATLASGAL sources is 3379. Therefore, it is natural that the number of filaments identified from ATLASGAL sources is larger than that from BGPS sources. More sources with velocity information also inform us that some of filaments identified from MST may be actually a part of a larger structure (for instance, BGPS filaments F2 and F3 are eastern and western part of ATLASGAL filament F4, respectively). While missing velocity information of a portion of sources may lead to failure to identify some filaments. \\

Our MST method identified some previously known large-scale filaments in the southern sky. Among three ``bone'' candidates in southern sky from \citet{Zucker2015}, filament BC\_355.31-0.29 and BC\_332.21-0.04 are found by MST. Filament BC\_355.31-0.29 is our F117. Filament BC\_332.21-0.04 is a complex consisting of F128, F127 and the southern part of F126. Filament G350.54+0.69 found by \citet{Hong-Li2018} is a part of our F80. Filament G316.75-0.1 \citep{Watkins2019} share a portion of spatial overlap with our F156.

\subsection{Large-scale Filaments in the Milky Way}
\label{sec:bone}
\begin{figure*}
\gridline{\fig{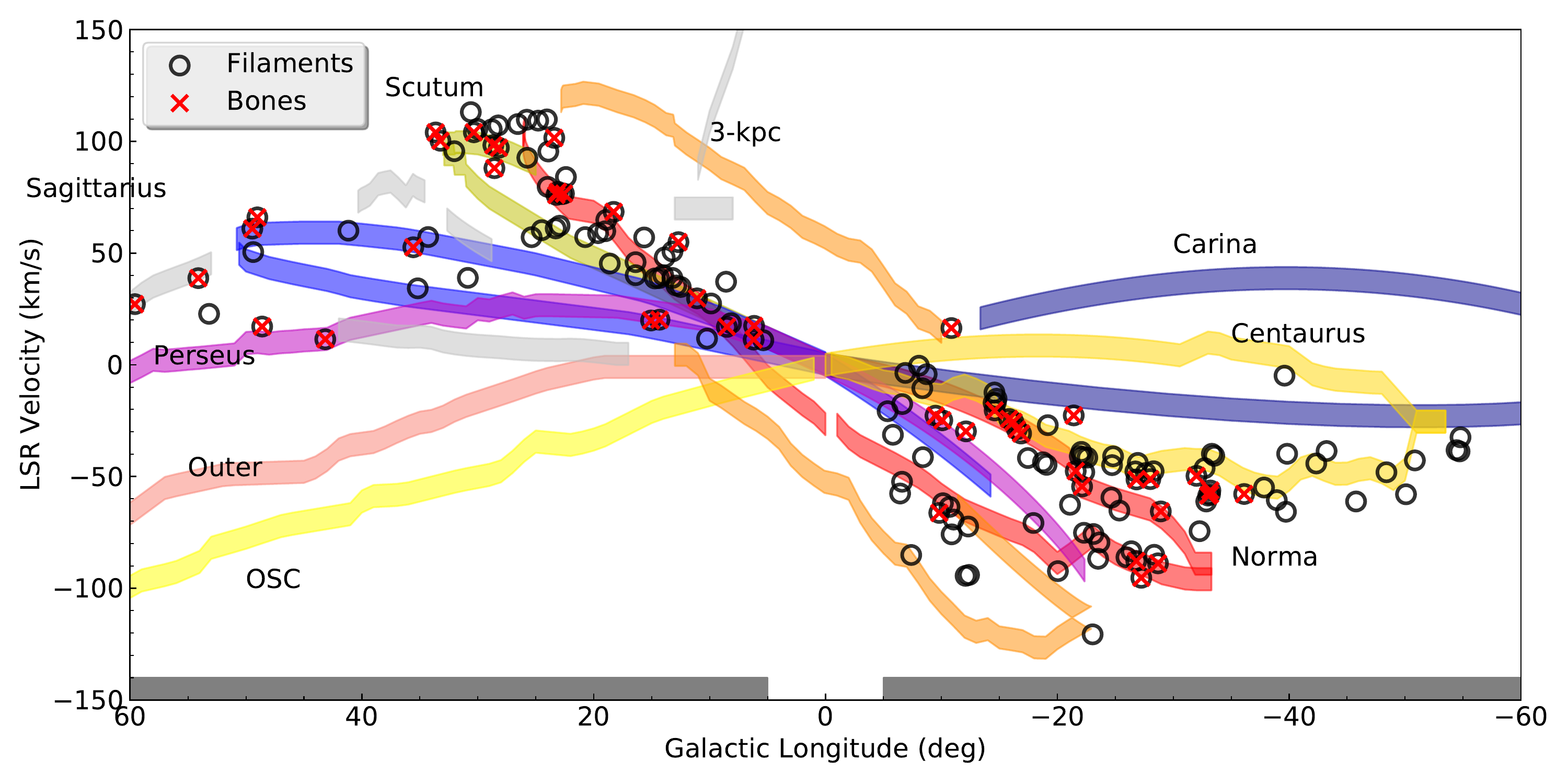}{0.8\textwidth}{(a)}
          }
\gridline{\fig{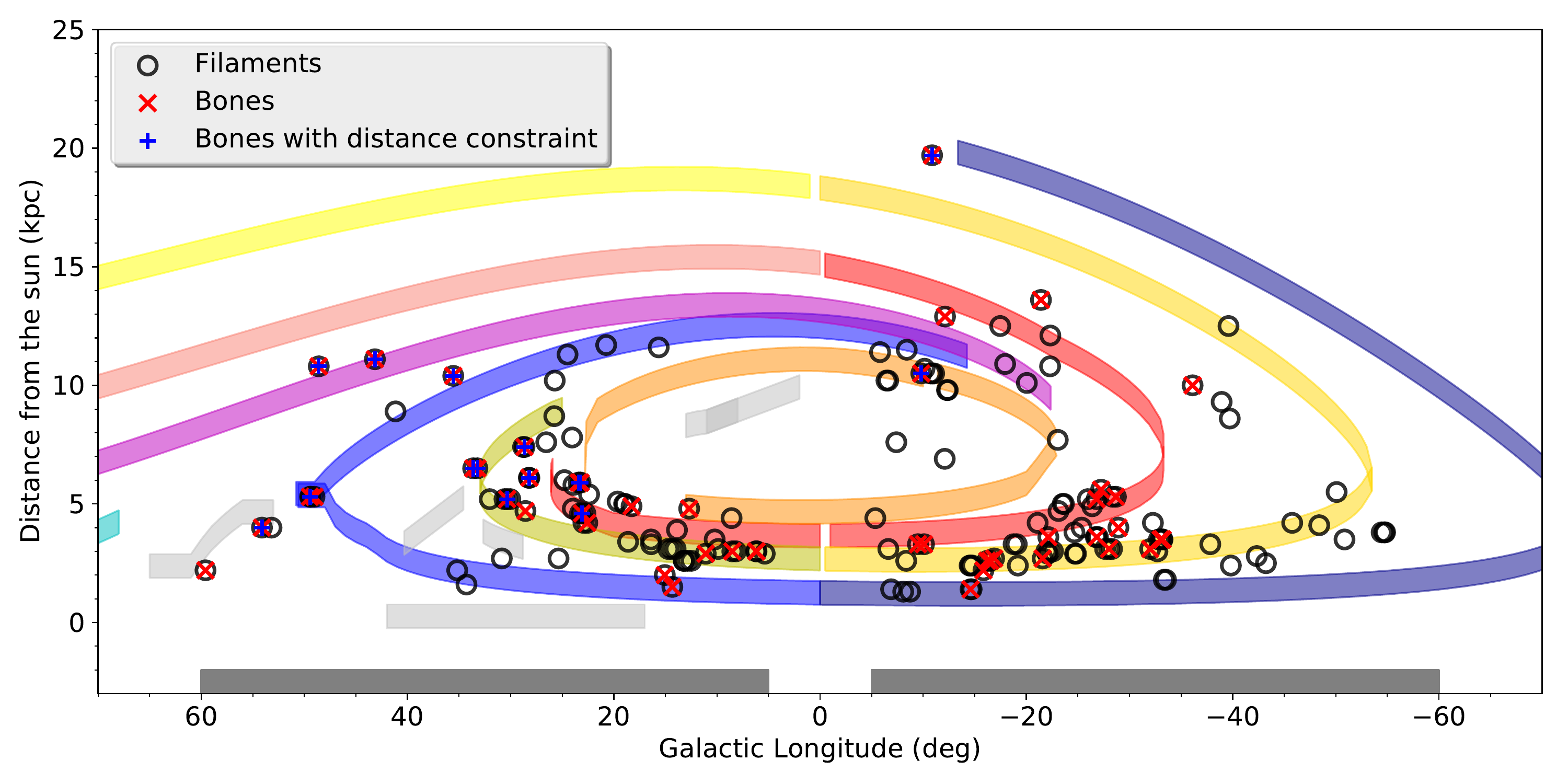}{0.8\textwidth}{(b)}
          }
\caption{(a) Longitude-velocity view of filaments and spiral arms. Spiral arms from \citet{Reid2019} are plotted as belts in a variety of colors. Our data do not support ``grand design'' spiral pattern, neither are they opposed to this pattern. We just simply show this optimistic presumption of spiral structure. Grey ``arms'' are arm segments which may join together to form arms. Filaments are shown as black circles and if it is a bone, a red cross is overlaid. Grey rectangles at the bottom mark the longitude range of our filaments. (b) Longitude-distance view of filaments and spiral arms. If a bone is also within an arm in this longitude-distance diagram, a blue plus is overlaid. (c) (In next page) Face-on view of filaments and spiral arms. Spiral arms are shown with their ``kink'' widths (taken from Table 2 of \citet{Reid2019}) which can be regarded as characteristic widths of arms. Red dots in the background mark clumps used to identify filaments from \citet{Urquhart2018}}  \label{faceon}
\end{figure*}
\setcounter{figure}{5} 
\begin{figure*}
\gridline{\fig{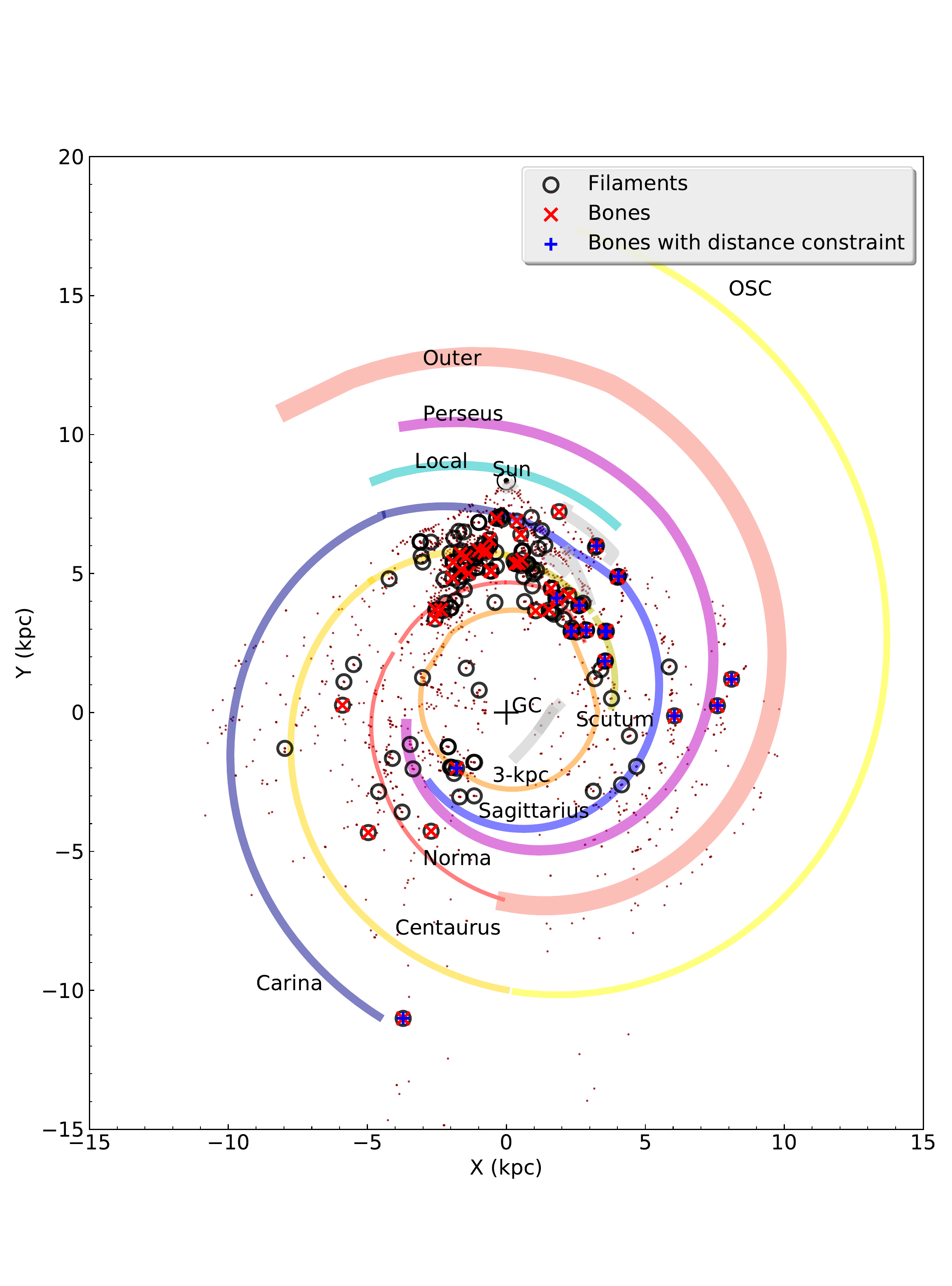}{0.8\textwidth}{(c)}
          }
\caption{(Continued) }
\end{figure*}
To examine the Galactic distribution of large-scale filaments, we exhibit locations of the filaments in the longitude-velocity (PV) space and draw Galactic spiral arms as reference (Fig. \ref{faceon} (a)). The spiral arm model is from a fitting result of Galactic high-mass star forming regions with trigonometric parallaxes \citep{Reid2019}.  According to previous criterion to judge whether a large-scale filament is associated with spiral arms \citep{Ragan2014,Abreu2016}, if a filament has LSR velocity within $\pm$5 km s$^{-1}$ \citep{Wang2016} of a spiral arm in the same Galactic longitude, it is thought to be an arm filament. This criterion is illustrated in Fig. \ref{faceon} (a). Belts show spiral arms in PV space with widths of 10 km s$^{-1}$. If a filament locates in any of the belts, it is thought as an arm filament. Of our 163 large-scale filaments, 87 (53\%) are in spiral arms or spurs under this criterion. But this value is only for comparison to previous work. We will update the criteria to judge which filaments could be thought as arm filaments later. \citet{Ragan2014} finds most of their filaments are inter-arm filaments while \citet{Abreu2016} and \citet{Wang2016} find arm filaments percentages are 67\% and 80\%, respectively.\\

There exists some filaments lying in the center of spiral arms and so sketch out bones of the Milky Way \citep{Goodman2014,Zucker2015}. ``Bones'' are also found in our large-scale filaments if we add the following three additional criteria \citep{Wang2016}: 
\begin{itemize}
\item[(6)]Lie very close to Galactic mid-plane, $|z|\leq 20$ pc;
\item[(7)] Run roughly parallel to arms in the projected sky, $\theta \leq 30^\circ $;
\item[(8)] Flux-weighted LSR velocity is within $\pm$5 km s$^{-1}$ of a spiral arm in the same Galactic longitude. 
\end{itemize}
Their Galactic distribution is shown as ``crosses'' in Fig. \ref{faceon} and physical properties are denoted as red bars in Fig. \ref{sta}. Fig. \ref{faceon} (c) shows filaments and clumps (red filled circles) overlaid to Galactic spiral arms as viewed from the Northern Galactic Pole. The exhibited widths of spiral arms are intrinsic (Gaussian $1\sigma $) arm widths at Galactic radius of ``kinks'' \citep{Reid2019}. There are 23, 12, 20, and 6 large-scale filaments in Norma-Outer, Scutum-Centaurus-OSC, Sagittarius-Carina, and Perseus arm, respectively while for bones these values are 10, 8, 12, 5, respectively (summarized in Table \ref{t_arm}). Here, to judge whether a filament is associated with an arm, we still follow the criterion from previous work. The differences between arm filaments and bones are that arm filament need only to satisfy criterion (8) while bones should fulfill not only criterion (8) but also (6) and (7). Surprisingly, most filaments in Scutum-Centaurus-OSC arm and Perseus arm are bones (8/12 and 5/6) and these two arms have been thought as two dominant spiral arms \citep{Drimmel2000, Churchwell2009}. However, we note that bone fractions in the four arms have no significant differences and we lack a statistic sample.\\

From the face-on view of filaments in spiral arms, we notice that some bones thought as arm filaments are obviously not in any spiral arms or spurs (see red crosses between arms in Fig. \ref{faceon} (c)). That is because in the previous criteria used to judge whether a filament is in a spiral arm, we only require that the filament is velocity coherent with a spiral arm in the same Galactic longitude. However, the distance of the filament to that arm was not in consideration. So this is a 2D position-velocity (PV) match. Researches on Galactic location of filaments have also been confined to two-dimension match, PV \citep{Ragan2014, Abreu2016, Wang2016} or position-position (PP) \citep{Wang2015,Mattern2018}. Now we obtain better distance estimation of both filaments and spiral arms, position-position-velocity (PPV) match becomes possible.\\

Our new definition of ``bone'' adds a distance constraint compared with the previous one when judging whether a filament is in a spiral arm. So it becomes a PPV match. This constraint is another criterion:
\begin{itemize}
\item[(9)]The difference between its distance to us and the distance of its PV-related spiral arm to us is less than 1 kpc in the same Galactic longitude.
\end{itemize}
That is, for a bone with previous definition meeting criteria (6), (7), and (8), we further examine whether it is close enough to this arm. If a previous bone also meets criterion (9), it will pass our distance constraint and be thought as a bone by new definition. Otherwise it will be excluded from bones, because in this situation the filament may just look near an arm in line of sight. The 1 kpc tolerance is from a combination of uncertainties of filament distances and $1\sigma$ widths of arms. Distances from the Sun versus Galactic longitudes of filaments and spiral arms are plotted in Fig. \ref{faceon} (b). Bones satisfying our new definition are denoted with blue pluses. As we can see, previous bones (denoted in red crosses) far from spiral arms are eliminated in our new definition. And this is also clear in the face-on map. However, some of previous bones seem to be in one arm in Fig. \ref{faceon} (b) are also excluded. This situation occurs when a filament has similar velocity with a spiral arm at a Galactic longitude, but far from this arm and near another arm in distance.\\

\setlength{\tabcolsep}{3mm}{
{
\doublerulesep=2pt
\begin{table}[b]
{\begin{tabular}{@{}cccccc}
\toprule
Arm &  Filaments & Bones & Bone fraction & Bones with distance constraint & Mean DGMF \\
\colrule
Norma-Outer & 23 (26.4$\%$) & 10 (20.4$\%$) & 43.5$\%$ & 5 (33.3$\%$) & 37.2$\%\pm2.9\%$ \\
Scutum-Centaurus-OSC & 12 (13.8$\%$) & 8 (16.3$\%$) & 66.7$\%$ & 1 (6.7$\%$) & 39.7$\%\pm2.6\%$ \\
Sagittarius-Carina & 20 (23.0$\%$) & 12 (24.5$\%$) & 60.0$\%$ & 3 (20.0$\%$) & 30.8$\%\pm2.2\%$ \\
Perseus & 6 (6.9$\%$) & 5 (10.2$\%$) & 83.3$\%$ & 1 (6.7$\%$) & 33.3$\%$ \\ 
Other Arm Segments & 26 (29.9$\%$) & 14 (33.3$\%$) & 53.8$\%$ & 5 (10.0$\%$) & 34.6$\%\pm1.7\%$\\ 
Total & 87 (100$\%$) & 49 (100$\%$) & 56.3$\%$ & 15 (100$\%$) & 35.3$\%\pm1.2\%$ \\ 
\botrule
\end{tabular}}
\caption{Number of filaments, bones with the definition of \citet{Wang2016} and our new definition of bones with distance constraint in each arm. Column ``Filaments'' means filaments meeting criteria (1)$\sim$(5) in each arm. Column ``Bones'' shows number of bones satisfying criteria (1)$\sim$(8). Column ``Bone fraction'' indicates that of filaments in each arm, how many are bones. Column ``Bones with distance constraint'' shows number of bones satisfying criteria (1)$\sim$(9). The last Column lists mean DGMF of filaments in each arm with standard error. Only one filament in Perseus has DGMF measurement, so the standard error of DGMF for Perseus is none. Apart from four main spiral arms, filaments associated with other arm segments or spurs are also listed in the 6th row. They consist of Local arm, 3 kpc and other segments and spurs, which may not be true spiral arms \citep{Reid2019}.}
\label{t_arm}
\end{table}
}
}

The amount of filaments in spiral arms decreases significantly after adding the distance constraint. Unlike bones, when a filament match a spiral arm in PPV space (at a certain Galactic longitude, they have similar distance from the Sun and similar velocity), it is thought as an arm filament. Of the 163 large-scale filaments, 138 are not in any spiral arms or spurs. While 8, 1, 4, and 1 are in Norma-Outer, Scutum-Centaurus-OSC, Sagittarius-Carina and Perseus arm (satisfy criteria (8) and (9)), respectively. The values for bones (satisfying criteria (6) $\sim$ (9)) are 5, 1, 3, and 1. The fact that 85\% of large-scale filaments are not in any spiral arms or structures seems to indicate that filament may not be associated so tightly with spiral arms as thought before. But we also note that due to the existence of distance ambiguity of filaments and uncertainty of spiral arm distance, especially in $4^{th}$ quadrant, some filaments that should be in spiral arms may be ruled out by the distance constraint, resulting in underestimation of arm filaments count. \\

If we do not care whether large-scale filaments are bones of the Milky Way and just want to know the influence of spiral arms to filaments, we could take filaments meeting criteria (1)$\sim$(5), (8), and (9) as arm filaments. And the others are inter-arm filaments. Between the two, we find no significant distinctions in some of their physical properties such as mean temperatures, non-thermal velocity dispersion, and column densities. But they also have differences in properties such as Galactocentric radius, height above Galactic mid-plane, mass, length. The detailed discussion for arm filaments and inter-arm filaments are in Appendix \ref{sec:env}. \\

\subsection{Dense Gas Mass Fraction}

Dense gas mass fraction is found to be associated with the Galactic position of filaments. From analysis of seven filaments identified from Galactic Ring Survey \citep[GRS,][]{Jackson2006}, \citet{Ragan2014} suggested that DGMF decreases as Galactocentric radius increase. They also found filaments near the Galactic mid-plane tend to have larger DGMF. In contrast, \citet{Abreu2016} find no obvious correlation between height above Galactic mid-plane and DGMF from inspection of their ten filaments identified in the fourth Galactic quadrant.\\

\begin{figure*}
\gridline{\fig{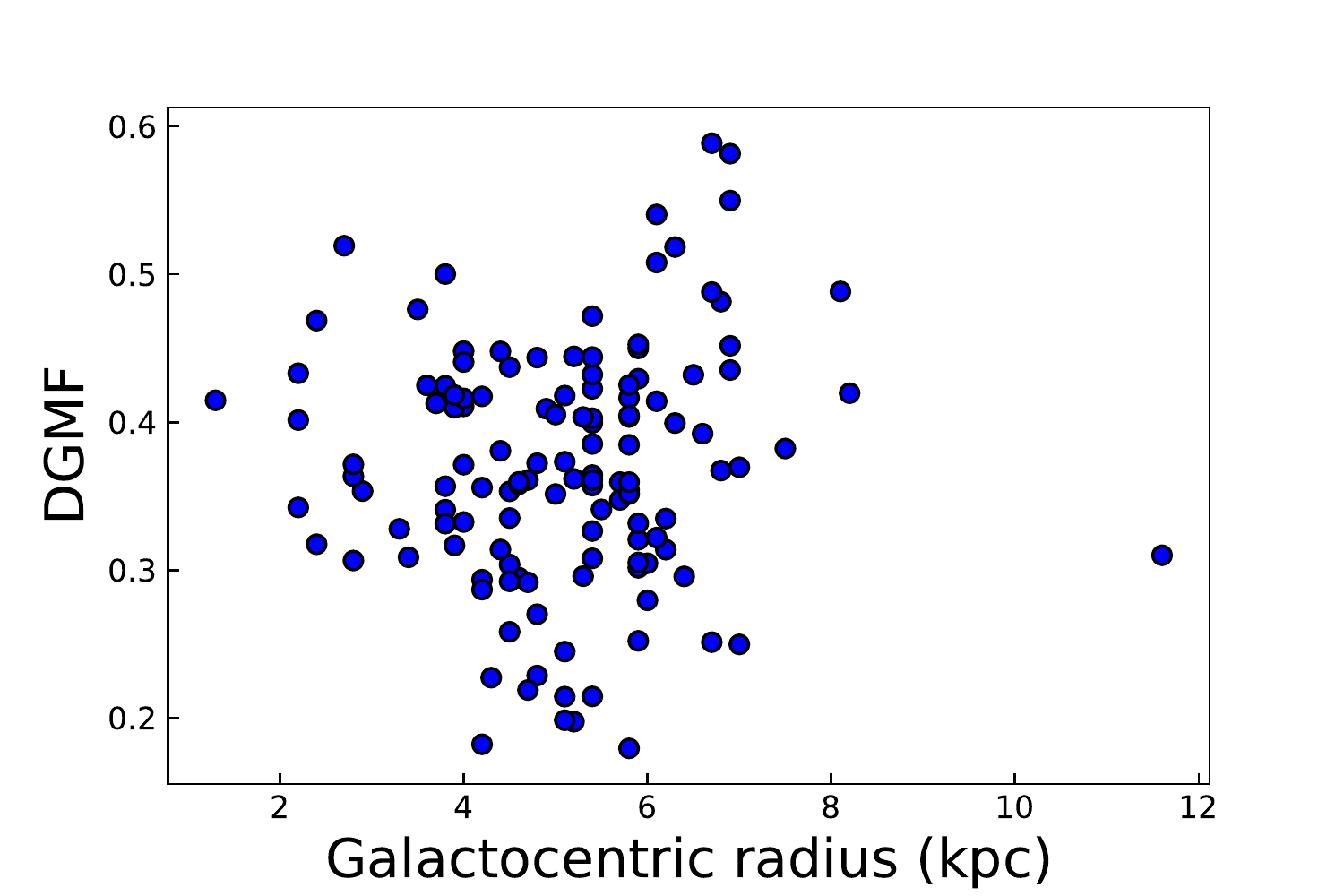}{0.45\textwidth}{(a)}
          \fig{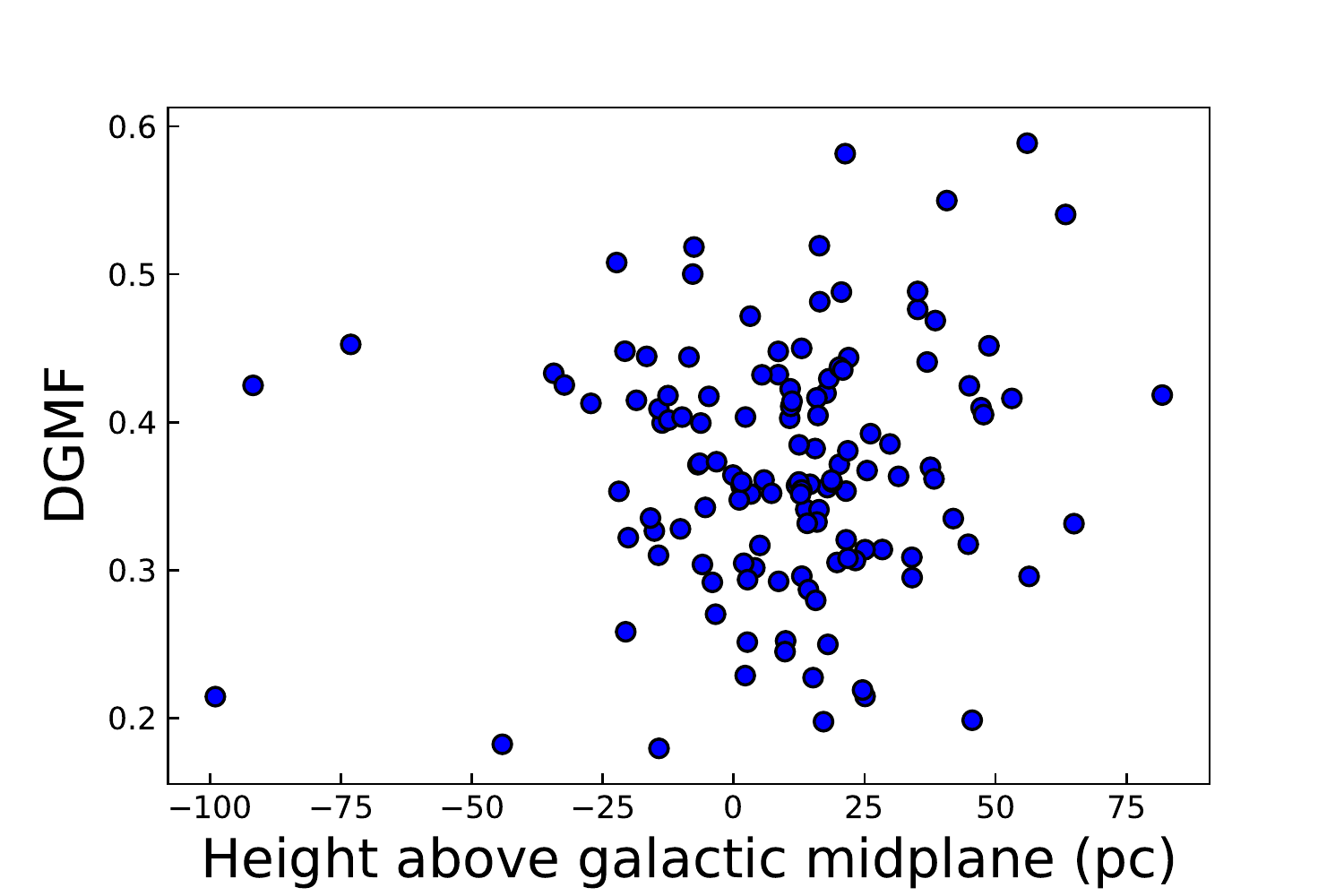}{0.45\textwidth}{(b)}
          }
\caption{Dense gas mass fraction and Galactic position of filaments}
\label{DGMF}
\end{figure*}

To investigate this issue in a larger statistical sample, we exam the DGMF and Galactic position of our 130 filaments with DGMF measurement. Results are shown in Fig. \ref{DGMF}. Surprisingly, DGMF is neither related to Galactocentric radius nor to height above Galactic mid-plane. We change the size of boundaries to get filament mass and dense gas mass, and get the similar result. We also try to use clump mass from ATLASGAL as dense gas mass, and the DGMF is still not correlated with Galactocentric radius or height above Galactic mid-plane. \\

\subsection{Dense Clumps in Filaments}

\begin{figure*}
\gridline{\fig{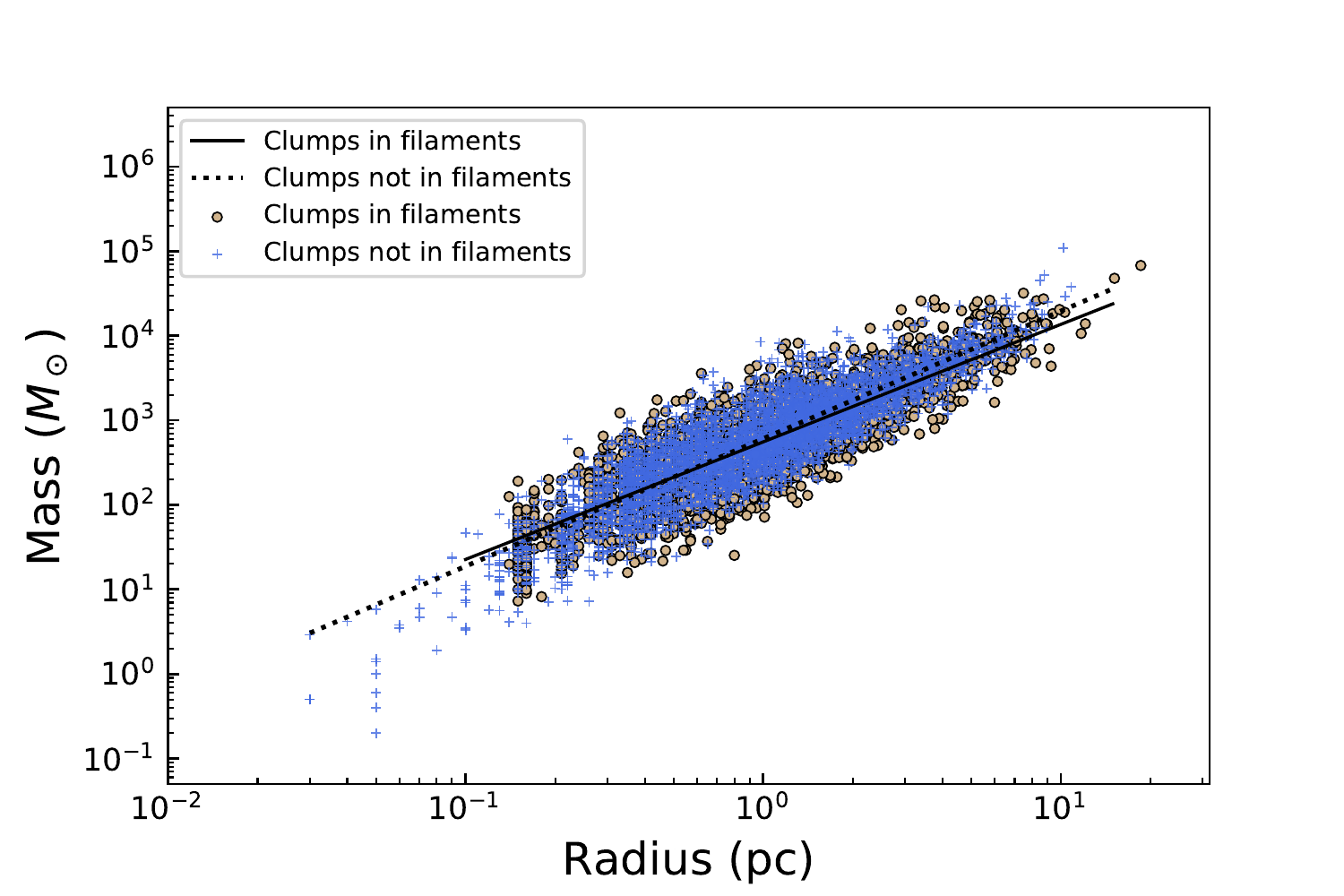}{0.5\textwidth}{(a)}
          \fig{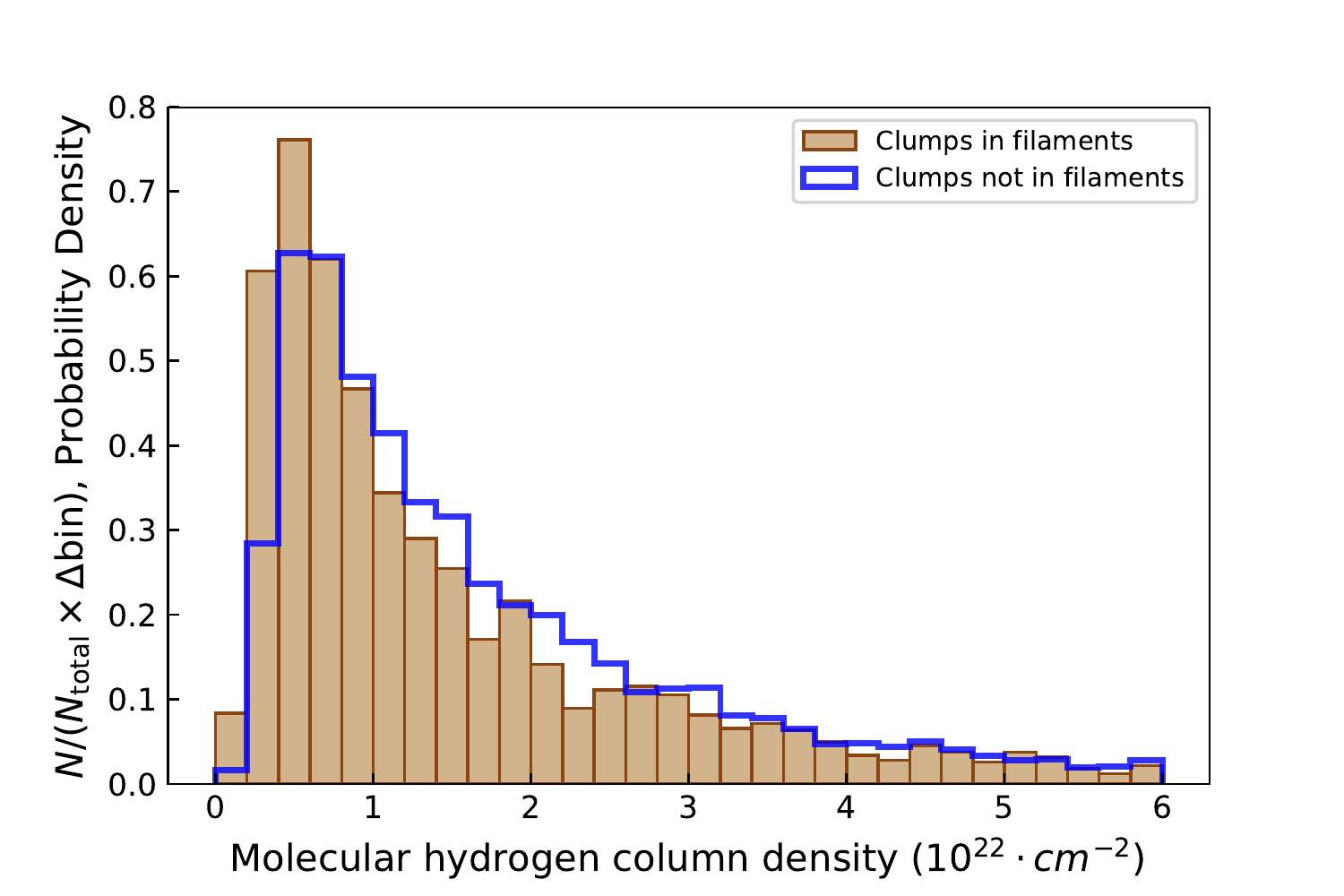}{0.5\textwidth}{(b)}
          }
\gridline{\fig{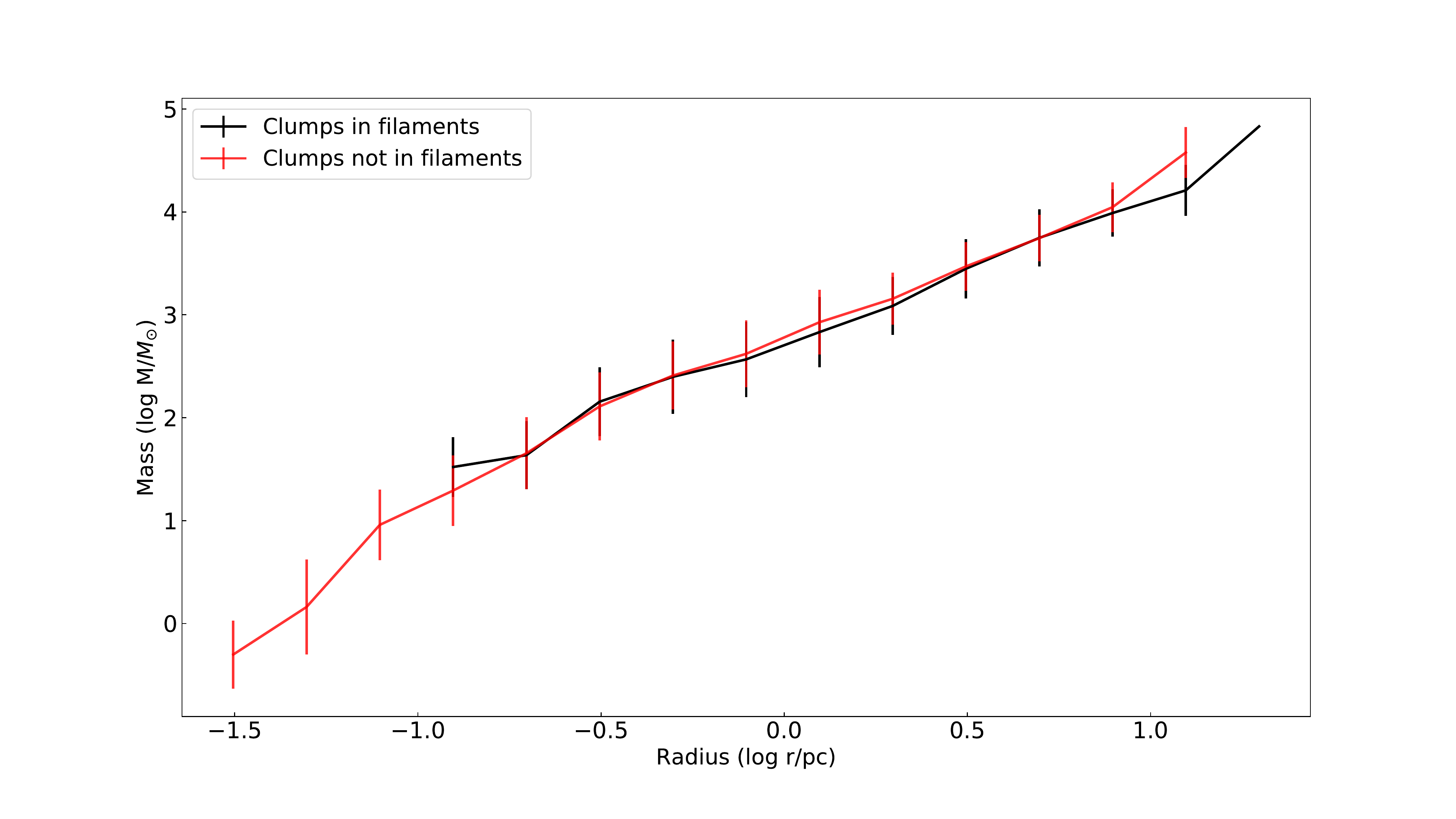}{0.5\textwidth}{(c)}
          \fig{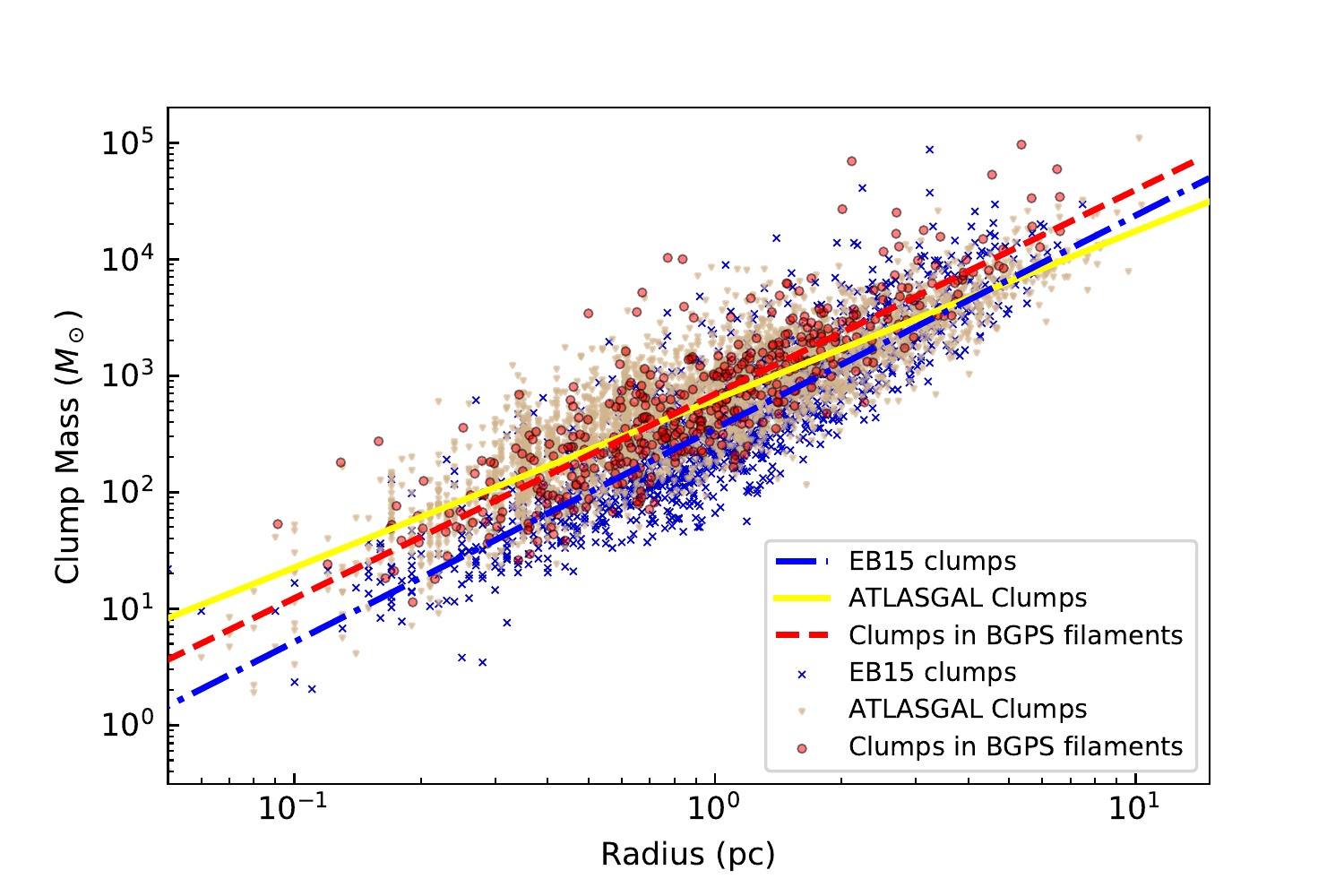}{0.5\textwidth}{(d)}
          }
\caption{Comparison between clumps on- or off-filaments. Panel (a): clump mass vs. radius for ATLASGAL clumps. Brown filled circles and blue pluses denote clumps in filaments and not in, respectively. Solid and dotted lines are fitted results. In panel (b), column density distribution of clumps in filaments or not are shown in brown bars and blue steps, respectively. We bin the logarithmic radius with bin size of 0.1 and the mean logarithmic mass in each bin is shown in panel (c). Black curve denote clumps in filaments while red curve show those not in. Error bars are standard deviations of mass in each bin. In panel (d), brown triangles are ATLASGAL clumps in the northern sky we use. Red filled circles are clumps in BGPS filaments in \citet{Wang2016}. Blue crosses are BGPS clumps from \citet{EB15}. Lines are fitted results.}
\label{infl}
\end{figure*}

Dense clumps are thought to be birthplaces of massive stars \citep[e.g.][]{Motte2018}. Their typical mass is $10^3M_\odot$, and typical molecular hydrogen column density is $10^{22}\cdot$ cm$^{-2}$ with typical radius of 1 pc. According to \citet{Wang2016}, dense clumps in large-scale filaments are slightly denser than those not in, indicating that filaments prefer to gather material and assist the formation of massive stars. To examine whether this preference can be extended to the whole inner Galactic plane, we compare mass of clumps with various effective radius between in filaments or not. Differences between clumps in filaments or not are shown in Fig. \ref{infl} (a), where lines are fitted results. Surprisingly, the two have no remarkable distinctions. We also bin the logarithmic radius to see more clear (Fig. \ref{infl} (c)). The bins range from about -1.6 to 1.2 with a step of 0.1. In each bin, we calculate the mean logarithmic mass of clumps in filaments or not, and then plot them versus median logarithmic radius of each bin. The black curve is clumps in filaments and the red curve shows clumps not in. Error bars denote standard deviations of logarithmic clump mass in each bin. As we can see, the two curves almost coincide except for both ends, where the number of clumps is small. But even at both ends, the differences are within the error bars. Therefore, mass of clumps in filaments or not have no significant distinctions on the same scale.\\

We then make a histogram to show molecular hydrogen column density of clumps in filaments or not (Fig. \ref{infl} (b)). The densest clump has column density of about $25\times 10^{22}\cdot$ cm$^{-2}$, but the number of extremely dense clumps is small. So we only show density ranging from 0 to $6\times 10^{22}\cdot$ cm$^{-2}$. The median of molecular hydrogen column density for 2628 clumps in large-scale filaments and 4213 not in are $1.03\times 10^{22}\cdot$ cm$^{-2}$ and $1.28\times 10^{22}\cdot$ cm$^{-2}$, respectively. Therefore, for our filaments, clumps not in filaments are on contrary slightly denser than those in filaments. This result may account for why star formation efficiency and star formation rate surface density in filaments are similar to that in other star-forming regions discovered by \citet{ZhangM2019}. To figure out whether this result is caused by not demanding enough for filaments, we increase the critical linearity for filaments to do the same experiment. And we also use bone to examine this issue. However, the results do not change much. In some cases, filamentary structures have been thought to fragment through cylinder fragmentation. Clumps from cylinder fragmentation will be denser than that from Jeans fragmentation \citep{Wang2014}. Therefore, our results give us a hint that the fragmentation process of filamentary structures may not be cylinder fragmentation.\\

To figure out why we find no distinction between clumps on- and off-filaments, while \citet{Wang2016} got different results, we restore Fig. 5 (b) of \citet{Wang2016} and overlay ATLASGAL clumps in northern sky in our Fig. \ref{infl} (d). We only plot ATLASGAL clumps in northern sky in this panel because BGPS clumps we compare are almost in northern sky. Brown triangles are ATLASGAL clumps in the northern sky. Red filled circles are 496 clumps in BGPS filaments in \citet{Wang2016}. Blue crosses are 1710 BGPS clumps with well-constrained distance estimates from \citet{EB15}, hereafter EB15 clumps. Lines are fitted results. On the whole, masses of clumps in BGPS filaments (red filled circles in Fig. \ref{infl} (d)) are twice as much as masses of EB15 clumps (blue crosses in Fig. \ref{infl} (d)) on the same scale. That is due to their difference in temperatures, although these two categories of clumps are both BGPS clumps and mass is calculated with the same formula. In \cite{Wang2016}, temperature of clumps in BGPS filaments are searched from three catalogues, and clumps with no reported temperature are assumed to be the average of measured temperature, 15 K (see Sect. 4.2 in \citet{Wang2016} for details). While in \citet{EB15}, temperatures of BGPS clumps are randomly assigned such that the temperature distribution is a lognormal function with a mean of 20 K (Sect. 4.1.1 in \citet{EB15}). This difference in temperature causes a systematic deviation in mass calculation, leading to an underestimation of clump mass in \citet{EB15} by a factor of about 1.5.\\

Mass of ATLASGAL clumps is also derived with the same formula as that of clumps in BGPS filaments and EB15 clumps. But their temperatures are directly from greybody fit to the submillimeter dust emission (see Sect. 4 in \citet{Urquhart2018}), which are reliable. ATLASGAL clumps are consistent better with clumps in BGPS filaments in Fig. \ref{infl} (d), while systematically above EB15 clumps. Therefore, the differences of BGPS clumps in filaments or not found by \citet{Wang2016} is caused by systematic deviation of mass calculation rather than their intrinsic differences.\\

\subsection{Fragmentation of Large-scale Filaments}
\begin{figure*}
\gridline{\fig{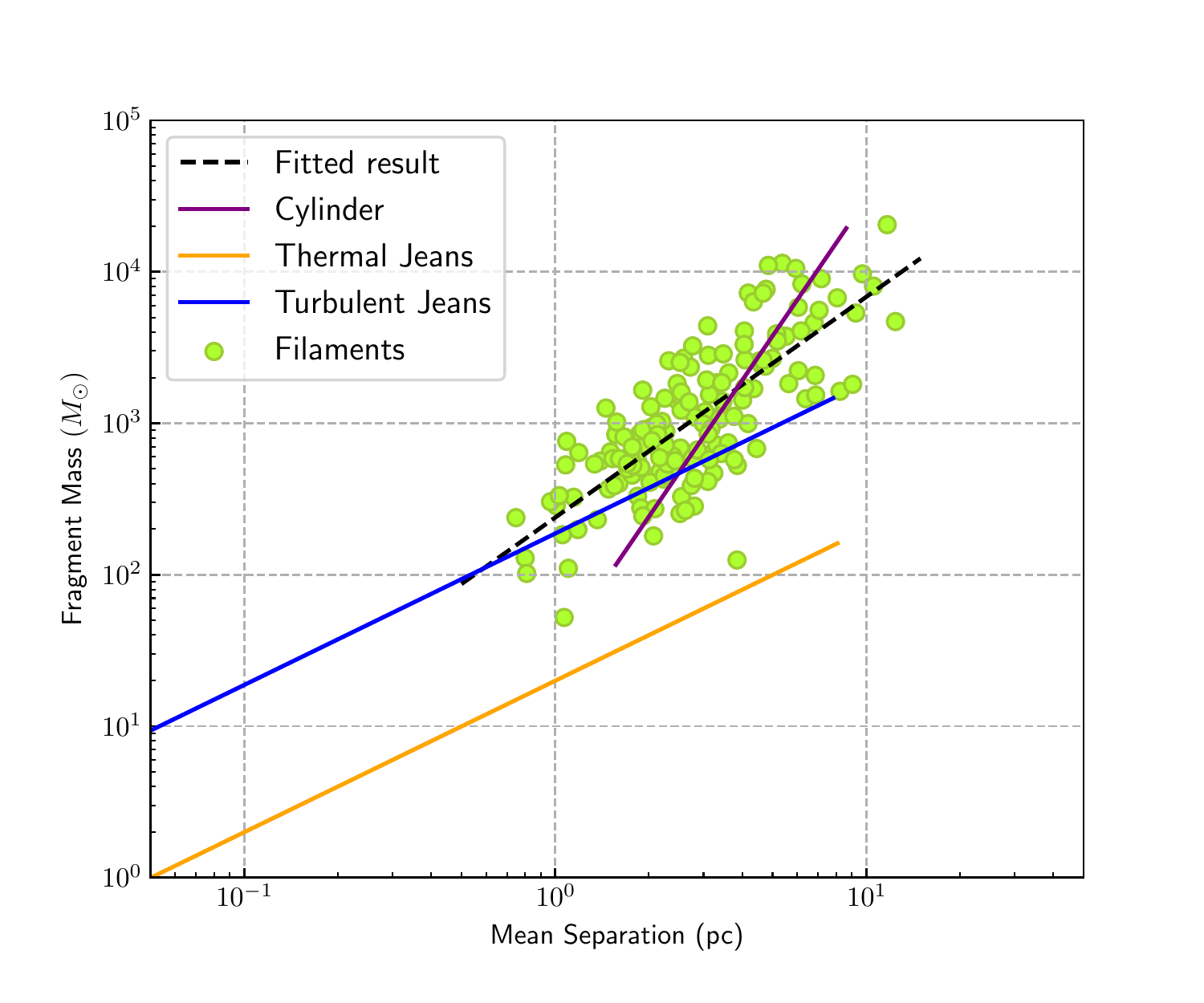}{0.5\textwidth}{(a)}
          \fig{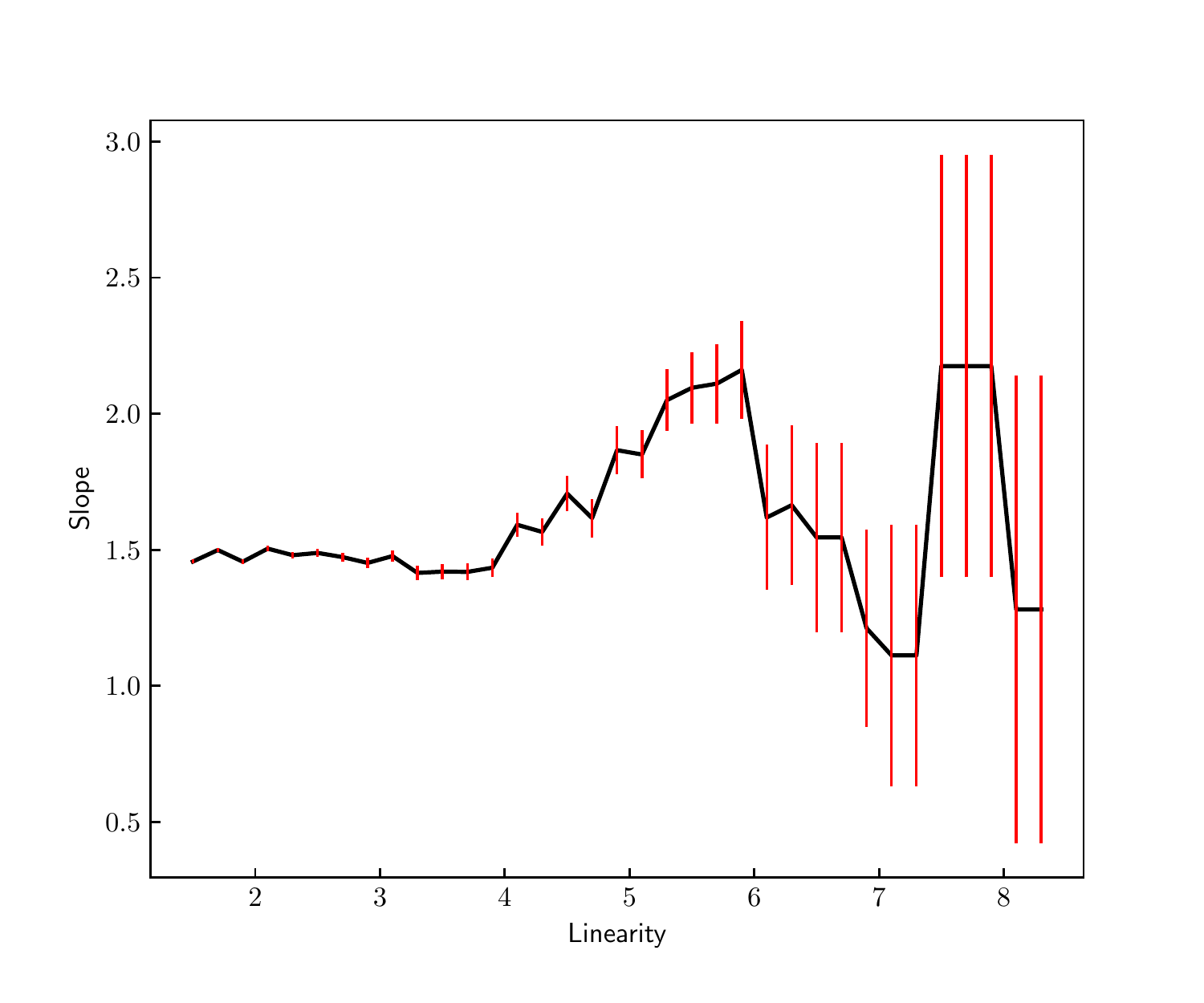}{0.5\textwidth}{(b)}
          }
\caption{(a) mean clump mass vs. mean edge length for the 163 filaments. The orange and blue lines are theoretical results for thermal Jeans and Turbulent Jeans fragmentation. The purple line is for cylinder fragmentation. The black dashed line is fitted result. (b) slope of the relation between mean fragment mass and mean separation if we only consider filaments with linearity larger than a set of values. Error bars denote statistical errors.}
\label{frag}
\end{figure*}
Linear filaments were approximately regarded as gas cylinders when investigating their fragmentation \citep[e.g.][]{Fischera2012,Wang2014,Wang2016}. According to \citet{Fermi1953} and \citet{Ostriker1964}, isothermal gas cylinder will fragment due to gravitational instability when mass per unit length exceeds a critical line mass. Under this condition, the fragments will equally space with separation $\lambda _{cl}$ and the fragment mass will be
\begin{equation}
M_{cl}=301.7M_\odot\left( \dfrac{\lambda _{cl}}{{\rm pc}}\right) ^3
\end{equation}

Alternatively, if the fragmentation is governed by Jeans instability, the relation between fragment mass and separation is different. For thermal Jeans fragmentation, the fragment mass is
\begin{equation}
M_{thJ}=13.29M_\odot\left( \dfrac{T}{{\rm 10K}}\right) \left( \dfrac{\lambda _{thJ}}{{\rm pc}}\right) 
\end{equation}

While for turbulent fragmentation, the fragment mass is
\begin{equation}
M_{turbJ}=381.7M_\odot\left( \dfrac{\sigma}{{\rm km\cdot s^{-1}}}\right)^2\left( \dfrac{\lambda _{turbJ}}{{\rm pc}}\right)
\end{equation}

where $\sigma$ is turbulent linewidth which is well approximated by the velocity dispersion measured from dense gas tracers such as $\rm NH_3$. The above equations are deduced from \citet{Wang2014}. 
As can be seen, the power law index for the relation between fragment mass and separation is 3 for cylinder fragmentation while 1 for Jeans fragmentation. So we show mean clump mass versus mean clump separation in each filament in Fig. \ref{frag} (a). The slope of the fitted line is 1.46, representing the power law index between mean clump mass and mean separation. If we take the projection effect into consideration, the slope will be even shallower. This power law index indicates that large-scale filaments are more likely to have Jeans fragmentation rather than cylinder fragmentation, which is in contrast to \citet{Wang2016} who suggest a cylinder fragmentation. We note that our sample is larger (163 filaments) compared with \citet{Wang2016} (54 filaments). However, it is too early to draw a conclusion because Eq. (1) for cylinder fragmentation is based on isothermal linear gas cylinder with infinite length. In practice, a filament could not be strictly isothermal, and it also has finite length with some bents.\\

To inspect whether straight filaments would prefer to have cylinder fragmentation, we get the power law index for filaments with linearity larger than a set of values. For instance, mean clump mass versus mean separation of filaments with linearity larger than 3 are plotted. We fit a line for mass and separation, and take the slope of this line as a power law index for this linearity. For filaments with linearity larger than 5, we repeat the procedure and get another power law index. Similarly, we choose critical linearity from 1.5 to 8.5 and get a series of slopes (power law index). Then we plot power law index versus linearity in Fig. \ref{frag} (b). As can be seen, the power law index has a rising trend in linearity between 4 and 6. This result gives us a hint that linearity might play an important role in fragmentation mechanism of filaments. And the larger the linearity is, the more likely to be cylinder fragmentation. For linearity larger than 6, the trend of the curve is uncertain due to large error resulted from small sample size.

\section{Conclusion}
\label{sec:conclusion}

We employ MST method \citep{Wang2016} to search for large-scale filaments in PPV space in the inner Galactic plane. The algorithm is applied to 7809 clumps with velocity measurement in ATLASGAL Galactic clumps catalogue in the range $|l|<60^\circ$ with $|b|<1.5^\circ$ \citep{Urquhart2018}. We produce a sample of large-scale filaments in the inner Galactic plane consisting of 163 filaments. We derive their physical properties and examine their DGMFs with the help of Herschel Hi-GAL column density map derived from PPMAP \citep{Marsh2017}. We inspect the Galactic distribution of the filaments and compare them with the latest spiral arm model \citep{Reid2019}. Fragmentation of large-scale filaments is also investigated. The main results are:
\begin{itemize}
\item[1.] Robustness of MST method is verified by a comparison between the results from ATLASGAL Galactic clumps and BGPS dust clumps. Most (70\%) of the filaments identified from BGPS dust clumps are found from ATLASGAL clumps in the common Galactic region of the two catalogues despite different methods used to extract them, distinct lower limit of clump luminosities and various surveys referred to obtain radial velocities of clumps.
\item[2.] The Galactic position of filaments are asymmetric about Galactic mid-plane, which may be an observational bias.
\item[3.] Dense gas mass fractions of filaments have no significant distinctions in different Galactic radii and vertical height above Galactic mid-plane, in contrast to previous studies.
\item[4.] Filaments are compared with updated spiral arm model and a new PPV match is employed to judge whether a filament is in a spiral arm. Under this matching method, a number of filaments that are thought to be in arm from PP or PV match are eliminated. So the number of filaments actually associated with arms decreases a lot. 
\item[5.] Bone fraction and DGMF do not vary too much in different spiral arms.
\item[6.] Dense clumps in filaments have no obvious distinction in mass compared with those not in on the same scale.

\end{itemize}

\acknowledgments
We are grateful to Chao Wang, Wenyu Jiao, K. A. Marsh, Nannan Yue, Siju Zhang, Fengwei Xu, and an anonymous referee for helpful discussion that improved the scientific content as well as clarity of the manuscript. We acknowledge support by the National Key Research and Development Program of China (2017YFA0402702, 2019YFA0405100), the National Science Foundation of China (12041305, 11973013, 11721303), and the High-performance Computing Platform of Peking University through the instrumental analysis fund of Peking University (0000057511). This research has made use of SAOImageDS9, developed by Smithsonian Astrophysical Observatory. This research made use of Montage. It is funded by the National Science Foundation under Grant Number ACI-1440620, and was previously funded by the National Aeronautics and Space Administration's Earth Science Technology Office, Computation Technologies Project, under Cooperative Agreement Number NCC5-626 between NASA and the California Institute of Technology.

\appendix

\renewcommand\thefigure{\Alph{section}\arabic{figure}} 
\section{robustness of the algorithm}
\label{sec:robust}

\subsection{Comparison between Filaments from BGPS and ATLASGAL sources}\label{sec:BGPS&ATLASGAL}
To figure out whether MST method is sensitive to dataset, we compare filaments identified from BGPS and ATLASGAL Galactic clumps. The result is summarized in table A1. The common part of the two Galactic clumps catalogues is from Galactic longitude $7.5^\circ$ to $60^\circ$, corresponding with 42 filaments (F1 to F42) found from BGPS sources by \citet{Wang2016} and 67 filaments identified from ATLASGAL sources by us. We list the 42 filaments from BGPS sources and scrutinize whether they are found from ATLASGAL sources. In table A1, if a filament from ATLASGAL is similar with a filament from BGPS, the sequence number of that ATLASGAL filament will be filled below the ID of the BGPS filament. If a BGPS filament is a part of an ATLASGAL filament, ``E'', ``W'', ``N'' or ``S'' will be added to the sequence number of the ATLASGAL filament on behalf of east, west, north or south part. On the other hand, if an ATLASGAL filament is a part of a BGPS filament, ``P'' will be added before the sequence number of the ATLASGAL filament. If both a BGPS filament and an ATLASGAL filament are at the same position but look different, a ``D'' will be added. If a BGPS filament is not found from ATLASGAL sources, an ``NF'' will be filled below. 
\newcounter{Rownumber}
\newcommand{\Rown}{\stepcounter{Rownumber}\theRownumber}
\begin{center}
\setlength{\tabcolsep}{1.2mm}{
{
\doublerulesep=5pt
\begin{tabular}{ccccccccccccccc}
 \hline
BGPS filaments & F\Rown & F\Rown & F\Rown & F\Rown & F\Rown & F\Rown & F\Rown & F\Rown & F\Rown & F\Rown & F\Rown & F\Rown & F\Rown & F\Rown  \\
 \hline
ATLASGAL filaments & NF & F4-E & F4-W & F5 & F7 & NF & F8 & NF & F9 & F10 & P-F13 & D-F11 & D-F17 & F16 \\
 \hline
 \hline
BGPS filaments & F\Rown & F\Rown & F\Rown & F\Rown & F\Rown & F\Rown & F\Rown & F\Rown & F\Rown & F\Rown & F\Rown & F\Rown & F\Rown & F\Rown  \\
 \hline
ATLASGAL filaments & D-F15 & NF & NF & F18 & D-F26 & F29 & F33 & F35-N & F35-S & NF & D-F40 & NF & F41 & NF \\
 \hline
  \hline
BGPS filaments & F\Rown & F\Rown & F\Rown & F\Rown & F\Rown & F\Rown & F\Rown & F\Rown & F\Rown & F\Rown & F\Rown & F\Rown & F\Rown & F\Rown  \\
 \hline
ATLASGAL filaments & F44 & NF & NF & P-F54 & F56 & F57 & F58 & F59 & NF & F62 & NF & NF & F65 & P-F66 \\
 \hline
\end{tabular}
}}\\
\textbf{Table A1}~~Comparison between Filaments from BGPS and ATLASGAL Sources\\
\end{center}

\subsection{Choice of Parameters in MST}\label{sec:parameter}

\setcounter{figure}{0}
\begin{center}
\setlength{\tabcolsep}{1.2mm}{
{
\doublerulesep=5pt
\begin{figure}
\includegraphics[width=1\linewidth]{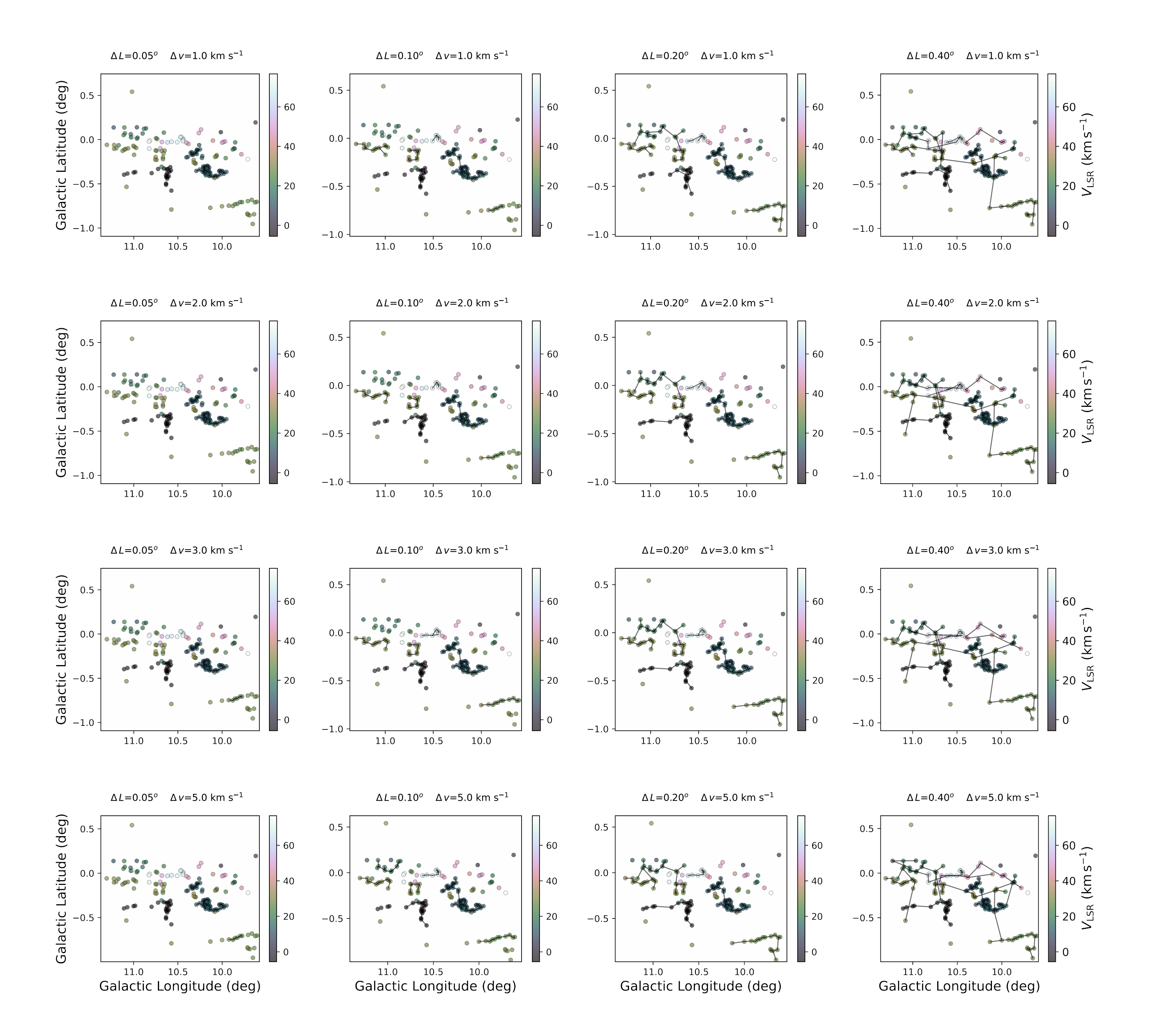}
\caption{MSTs with different parameters in the region around ``Snake''. Panels in each row show MSTs obtained with the same matching velocity while cut-off lengths ranging from 0.05$^\circ$ to 0.40$^\circ$. Different rows of panels have different matching velocities ranging from 1 km s$^{-1}$ to 5 km s$^{-1}$. Dots are ATLASGAL clumps color-coded by LSR velocities. Black line segments are ``edges'' of MSTs.}
\end{figure}
}}
\end{center}

The choice of parameters starts from physical consideration. Since the scale we are interested in is tens of parsec, and the observed angle for a filament with length of 10 pc at a distance of 5 kpc is about 0.1$^\circ$, we choose 0.1$^\circ$ as the initial cut-off length. The velocity difference between two ends of that filament is has an order of 1 km s$^{-1}$ if we treat velocity gradient as 0.1 km s$^{-1}$ pc$^{-1}$. This order of velocity gradient is estimated from velocity difference divided by total length of filaments from several large-scale filament catalogues \citep{Ragan2014,Zucker2015,Wang2015,Abreu2016}. \\

After initial cut-off length and matching velocity are chosen, we test other values around them. For cut-off length, we give 0.05$^\circ$, 0.10$^\circ$, 0.20$^\circ$, and 0.40$^\circ$. Meanwhile, 1 km s$^{-1}$, 2 km s$^{-1}$, 3 km s$^{-1}$, and 5 km s$^{-1}$ are given to matching velocity. For illustration, tests on a small part of the sky are shown in Fig. A1. Fig. A1 exhibits MSTs with different parameters in the region around ``Snake'' nebula, one of the first identified IRDCs  \citep{Carey1998} located at Galactic longitude 11.11$^\circ$ and latitude -0.12$^\circ$. Panels in each row show MSTs obtained with the same matching velocity while cut-off lengths range from 0.05$^\circ$ to 0.40$^\circ$. Different rows of panels have different matching velocities ranging from 1 km s$^{-1}$ to 5 km s$^{-1}$. Dots are ATLASGAL clumps color-coded by LSR velocities. Black line segments are ``edges'' of MSTs. As we can see, ``snake'', the MST in Galactic longitude $\sim$ 11$^\circ$, has been successfully found with cut-off lengths 0.10$^\circ$, 0.20$^\circ$, and 0.40$^\circ$. When cut-off length is 0.05$^\circ$, ``snake'' cannot be found. When cut-off length is set to 0.20$^\circ$ or 0.40$^\circ$, too many unrelated structures are linked. So the best cut-off length from the test of ``snake'' is 0.10$^\circ$. Since MSTs do not vary much when we alter matching velocity in this region, tests in other parts of the sky show that very strict matching velocity will cause loss of real structures. Results obtained with matching velocity of 2 km s$^{-1}$, 3 km s$^{-1}$, and 5 km s$^{-1}$ do not vary too much. That is reasonable considering the possibility that a physically isolated source is observed exactly within a structure. Then we choose the relatively strict matching velocity, 2 km s$^{-1}$. \\
\begin{center}
\setlength{\tabcolsep}{1.2mm}{
\doublerulesep=5pt
\begin{figure}
\includegraphics[width=1\linewidth]{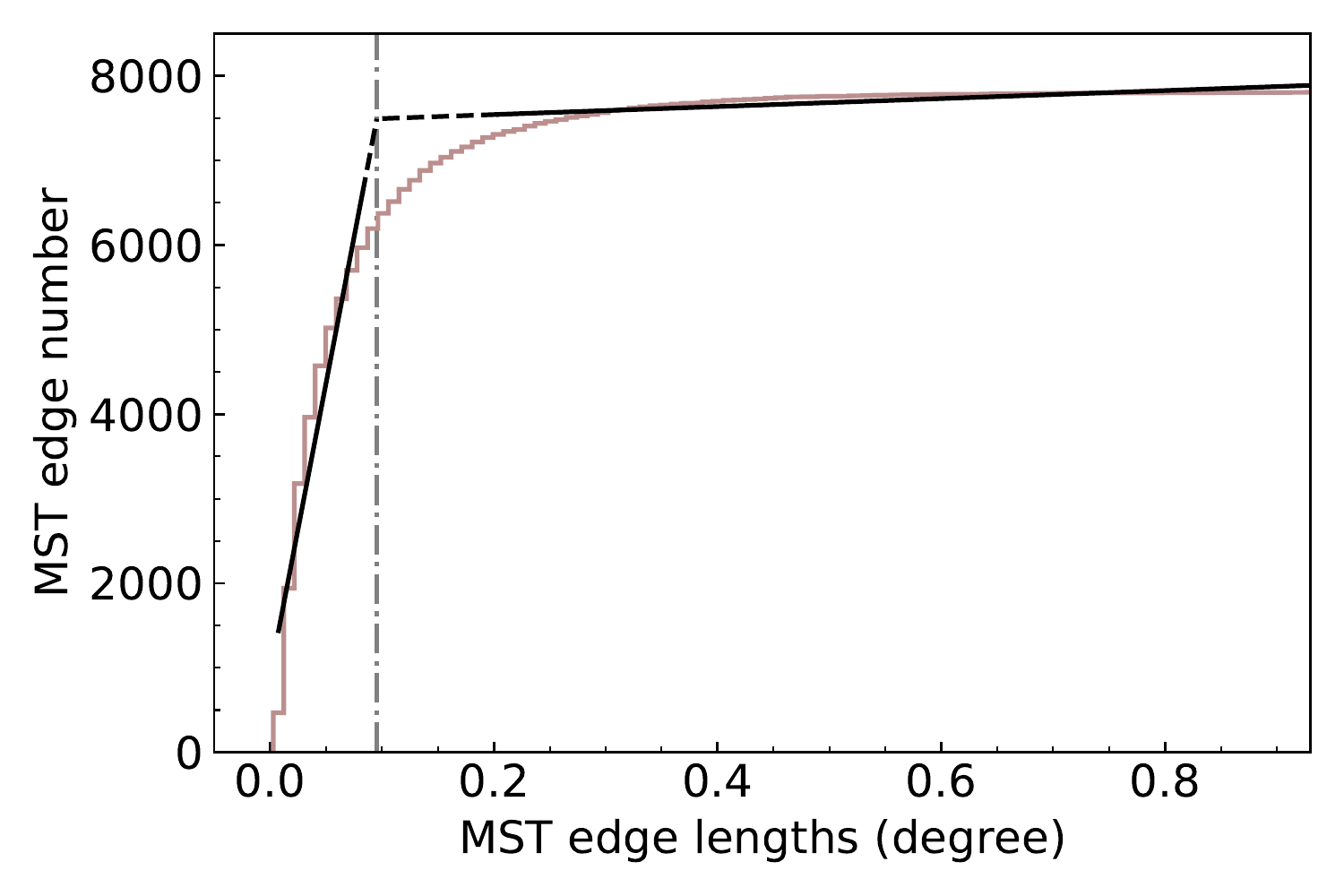}
\caption{An alternative approach to get cut-off length. Brown steps show cumulative distribution function (CDF) of MST edge lengths. Two black lines are fitted from teep-sloped segment and shallow-sloped segment of CDF. The grey dotted line denote the edge length at intersection point.}
\end{figure}}
\end{center}

Since results of MST are relatively sensitive to cut-off length, we apply a widely used approach in star cluster \citet[e.g.][]{Gutermuth2009} to examine the chosen cut-off length. Firstly, we set cut-off length to maximum and cluster all clumps in a single MST. Next we plot the cumulative distribution function (CDF) of MST edge lengths in Fig. A2 as brown steps. As we can see, the CDF is steep on the short edge part while shallow on the long edge part. So the CDF can be approximated by three parts: a steep-sloped segment, a shallow-sloped segment, and a transition segment. We fit a line for the steep-sloped segment and shallow-sloped segment, respectively. Then the intersection of the two lines gives us the cut-off length. The cut-off length acquired from this approach is 0.95$^\circ$, which is very close to our criteria. If we alter the fitting range with reason, the result changes less than 10 percent.

\subsection{Comparison on Methodological Approach}

\begin{figure*}
\gridline{\fig{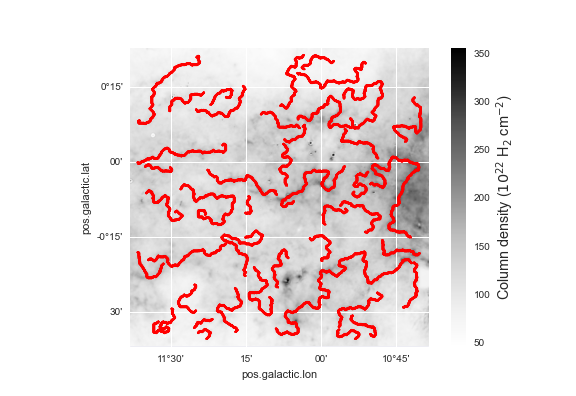}{0.55\textwidth}{(a)}
          \fig{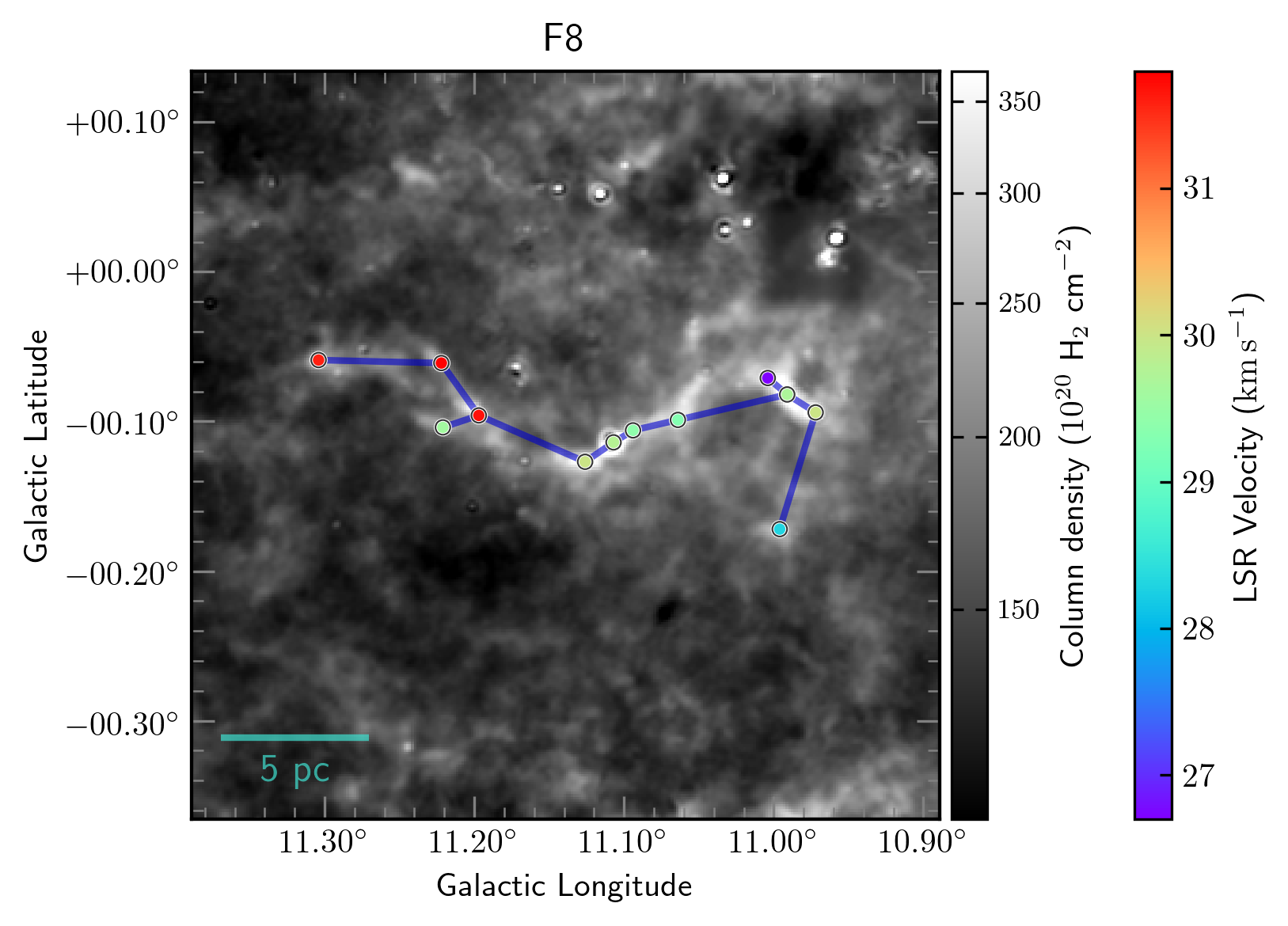}{0.5\textwidth}{(b)}
          }
\caption{Comparison of filament finding methods applied to ``snake'' nebula \citep{Wang2014}. The result of FilFinder \citep{Koch2015} is shown in panel (a), while that from MST \citep{Wang2016} is in panel (b). Background is column density map from \citet{Marsh2017}.}
\end{figure*}
Take the well studied filament ``snake'' nebula as an example. We employ FiFinder \citep{Koch2015} to find filamentary structures in that region through column density. In Fig. A2, red curves in panel (a) are ``skeleton'' found by FilFinder. Background is the ``flattened'' (a preprocess to the map of FilFinder which suppresses objects that are significantly brighter) column density map. Panel (b) is the filament found by MST as a comparison. As we can see, FilFinder is powerful in finding minor structures. But if we would like to extract large-scale structure, by eye inspection has to be acted on the result from FilFinder.\\

\renewcommand\thefigure{\Alph{section}\arabic{figure}} 
\section{additional remarks on methods to get physical properties}
\subsection{Orientation Angles and Major Axes of Filaments}
\label{sec:PCA}
\setcounter{figure}{0} 
\begin{figure*}
\gridline{\fig{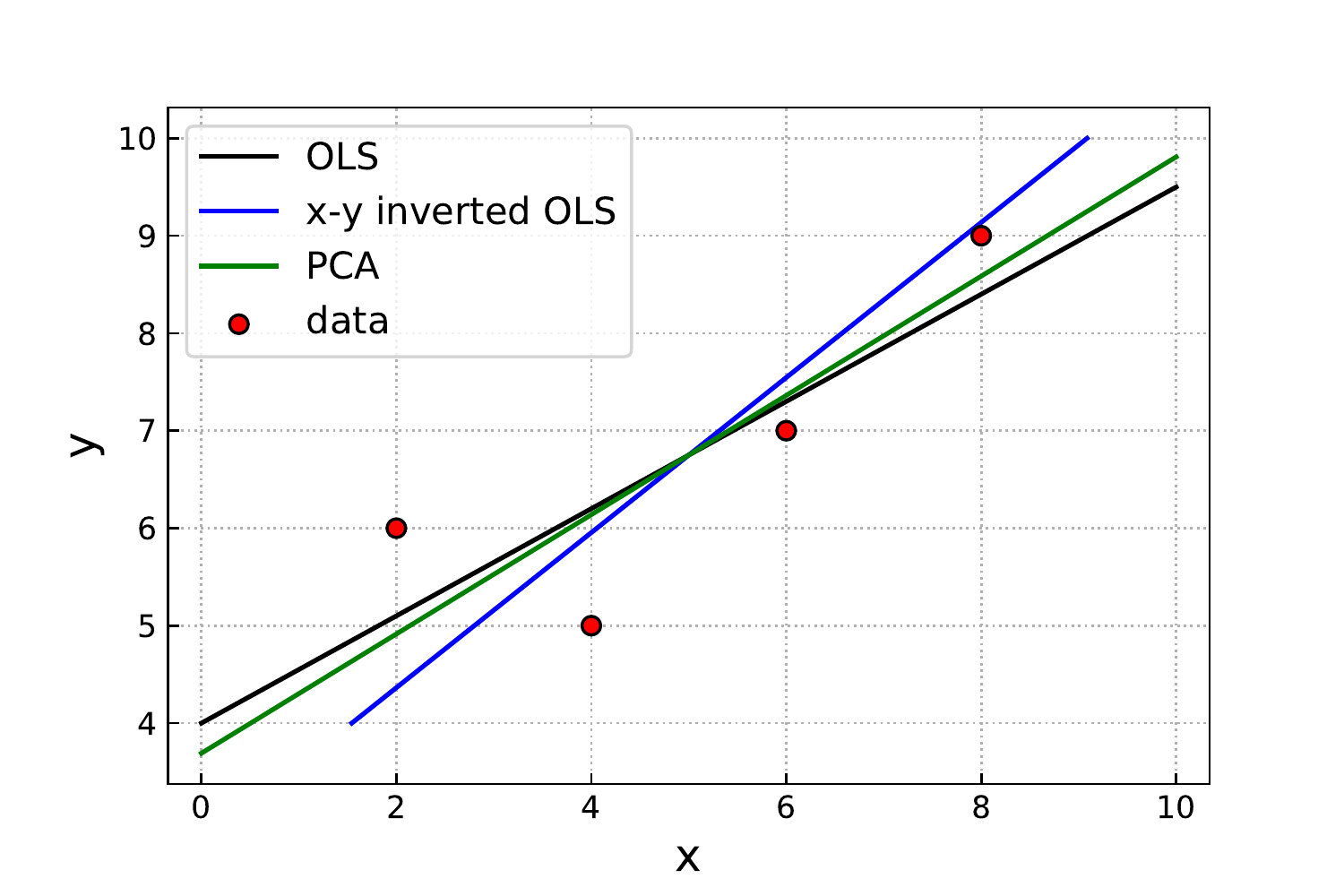}{0.5\textwidth}{(a)}
          \fig{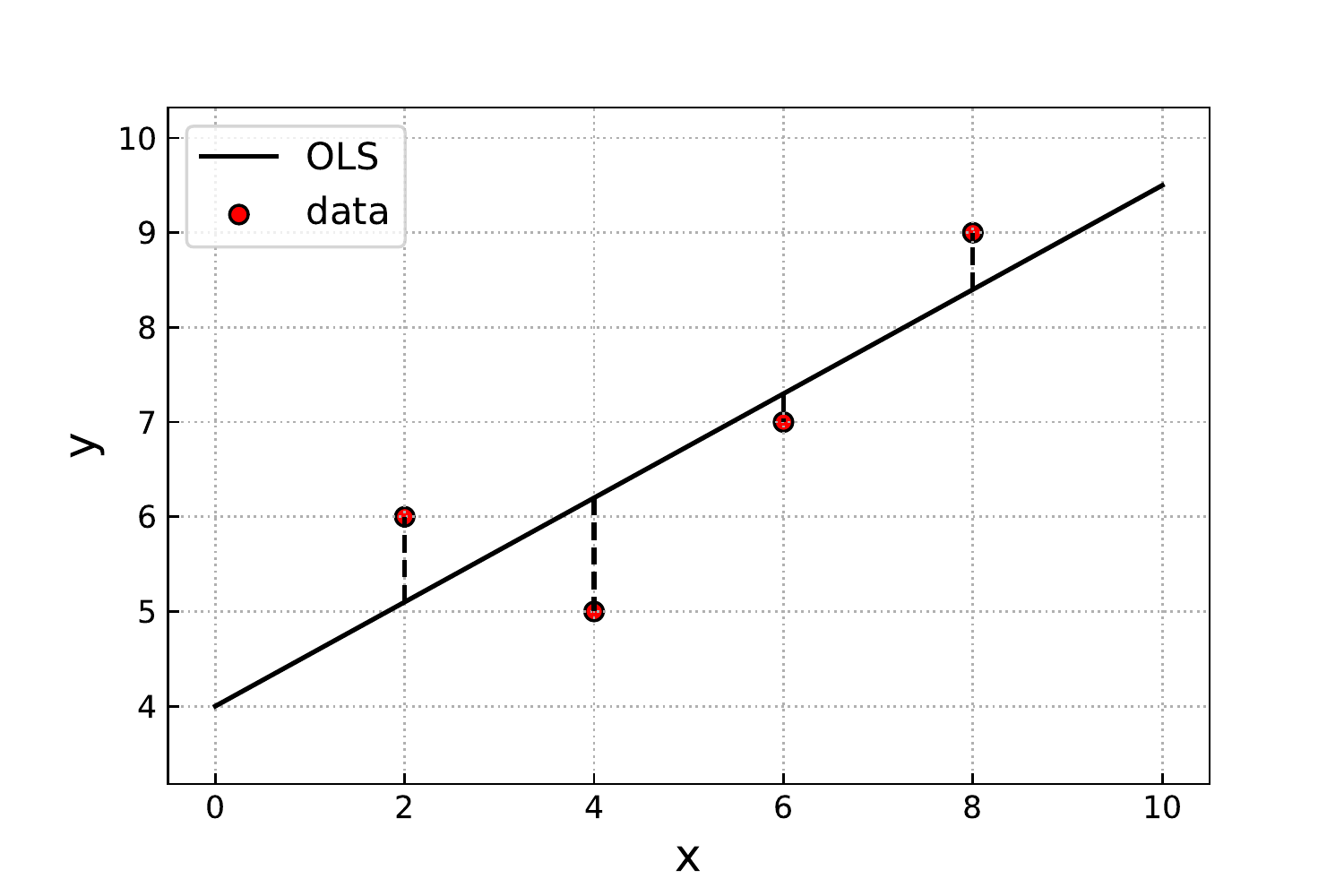}{0.5\textwidth}{(b)}
          }
\gridline{\fig{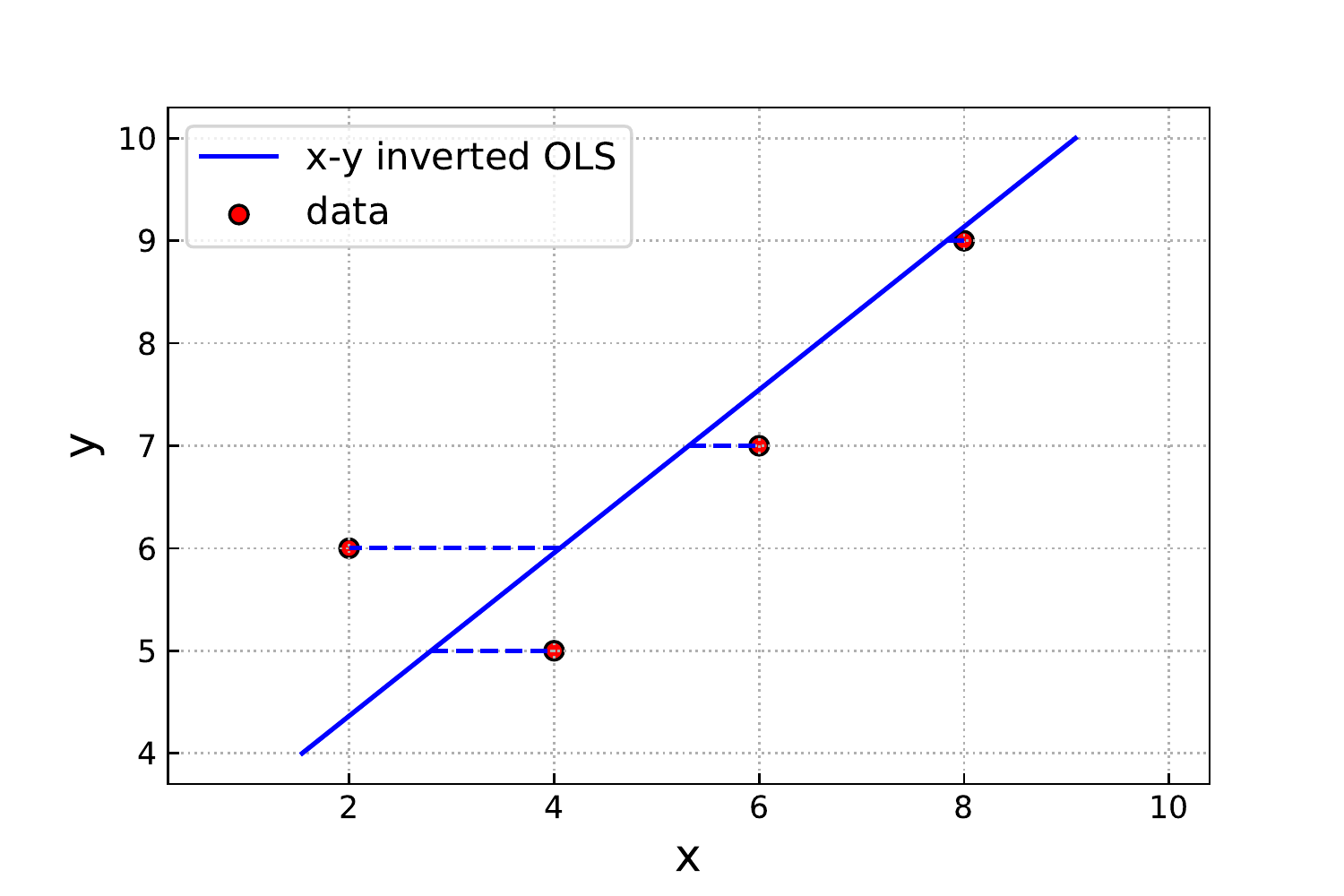}{0.5\textwidth}{(c)}
          \fig{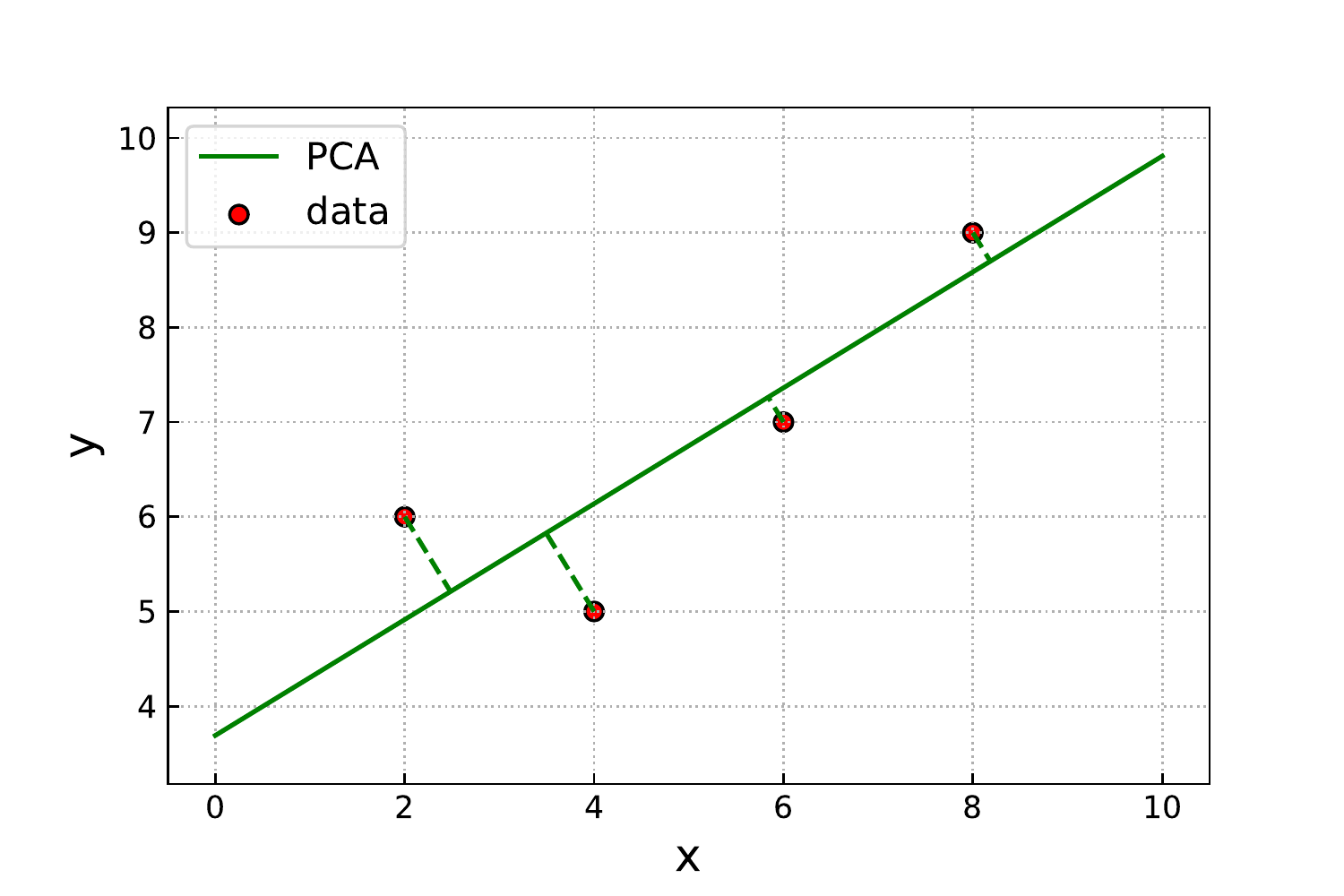}{0.5\textwidth}{(d)}
          }
\caption{Different methods to get orientation angle. Panel (a) A comparison of fitted results between three fitting methods. Red dots are hypothetical data. The black line is the fitted line with ordinary least squares to the data while the blue one is also the fitted line with OLS but the x and y are inverted. The green line is the fitted line with principle component analysis. (b) OLS fitting with vertical offsets. (c) x and y inverted OLS fitting with horizontal offsets. (d) PCA fitting  with perpendicular offsets.}
\end{figure*}
The orientation angle is the angle between filament major axis and the Galactic mid-plane. That is, it is the arctan of the slope of the fitted line. But in the fitting process, we employ principle component analysis (PCA) to get the slope rather than using ordinary least squares (OLS). The reason to do this is that OLS is not rotation invariant. As can be seen in Fig. B1 (b), OLS acts by minimising the sum of the squared vertical distances, between the line that best represents the data, and the data points. If we invert x and y, we expect that the fitted line does not change. However, the fitted lines from OLS and OLS after inverting x and y are not the same, which can be seen from black and blue lines in Fig. B1 (a). That is because after x and y are inverted, the offsets used to minimize in OLS become horizontal, as seen in Fig. B1 (c). As a result, x and y are not equivalent, which is accessible in data fitting but in feature extraction of image.\\

Therefore, it suggested to use another variant of least square method, where the sum of the squared offsets (or distances) perpendicular to a line is minimized. PCA \citep{Pearson1901} uses the principle component of data on behalf of the whole data. Specifically, it reduce data from n-dimensional to m-dimensional. Take the 2D data of red points in Fig. B1 as an example. We aim to reduce them into a line (1D data) and at the same time the loss of information should be as low as possible after the reduction. To realize this, points are projected to the line and the spread of the projected data should be as large as possible. Since the mean position of the points is determined, maximizing the sum of variance along the line is equivalent to minimizing the sum of squared distances between points and the line. So the procedure to find this line (strictly, the direction of the line as the first base vector) with PCA in our 2D case is the same as least square with perpendicular offsets shown in Fig. B1 (d). Then the orientation angle is obtained from the slope of the line (or the direction of the first base vector from PCA). It is the property of least square method that the fitted line passes through the mean position ($\overline{x},\;\overline{y})$), so the major axis is also determined.

\subsection{Concave Hulls and Aspect Ratios}
\label{sec:concavehull}
\begin{figure*}
\gridline{\fig{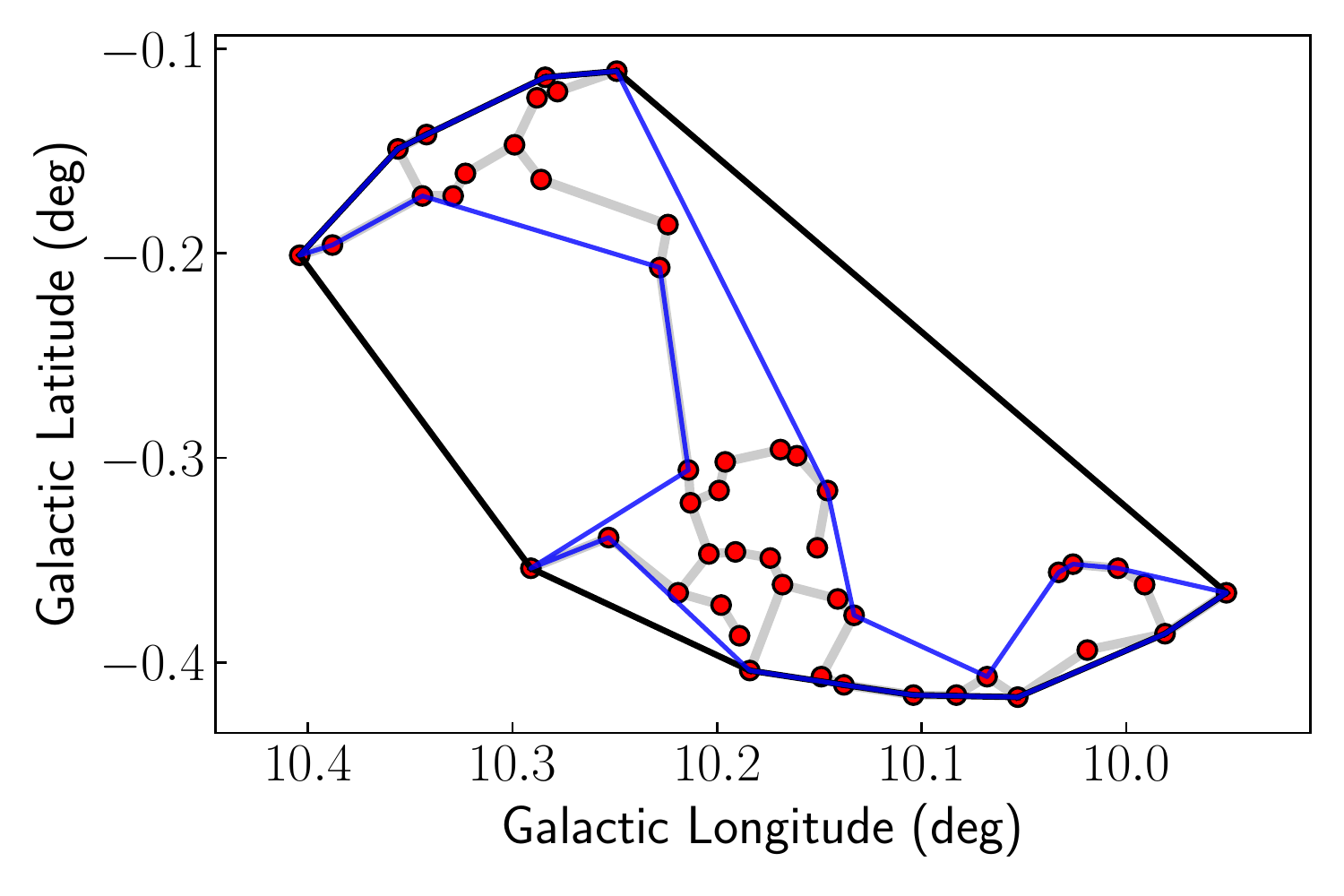}{0.5\textwidth}{(a)}
          \fig{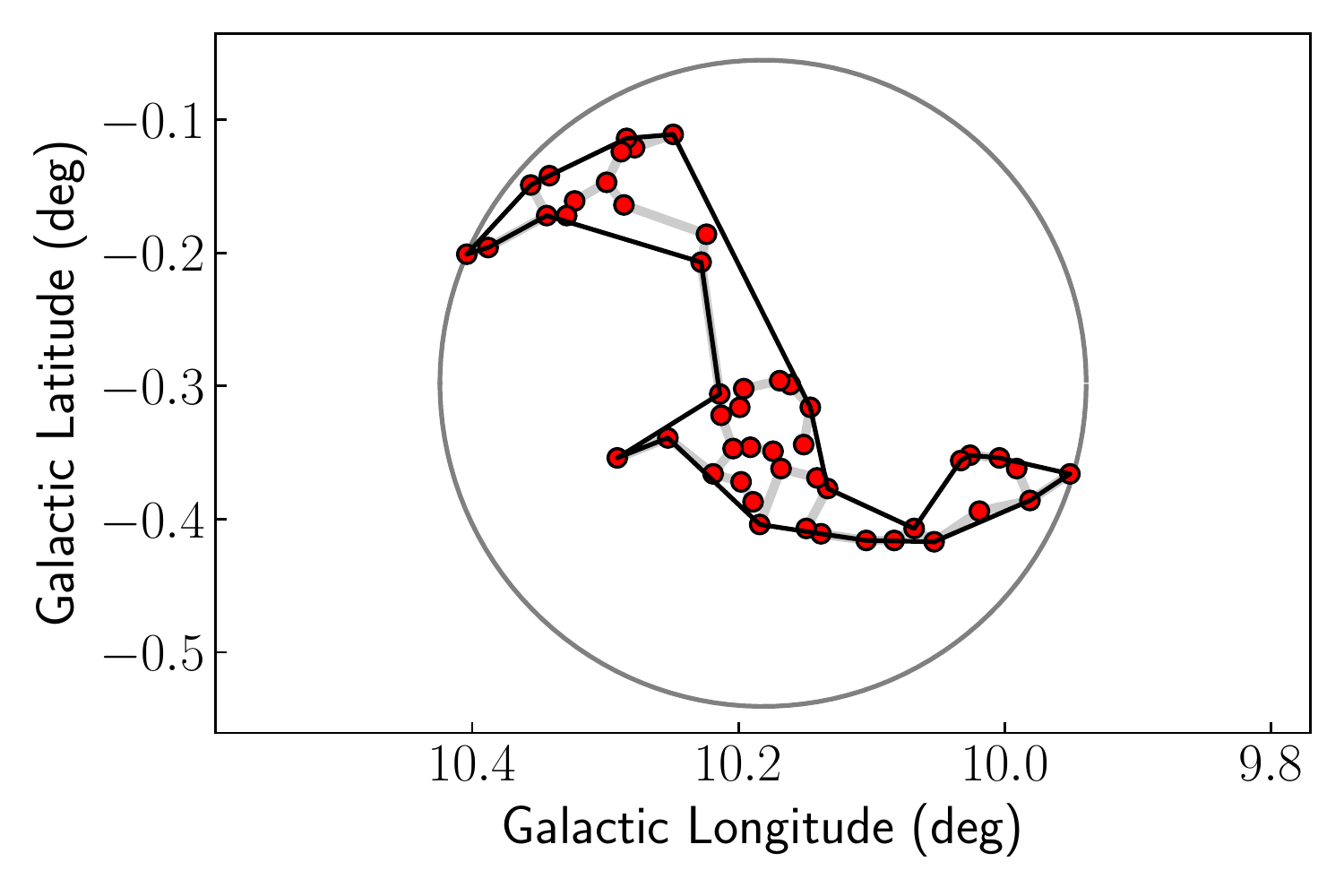}{0.5\textwidth}{(b)}
          }
\caption{Convex hull, concave hull, and aspect ratio. In panel (a), red dots and gray segments show one of our filaments F7. The black polygon enclosing it is the convex hull, and the blue one is the concave hull. Panel (b) illustrates how we get aspect ratio. The aspect ratio of a filament is the ratio of area between the circle enclosing vertices and concave hull.}
\end{figure*}
The convex hull of a set of points in two dimensions is the minimum area polygon that contains those points such that all internal angles between adjacent edges are less than 180$^\circ$. For example, the black polygon in Fig. B2 (a) is the convex hull of our filament F7. Convex hull has been used to characterize size and aspect ratio of MSTs since \citet{Schmeja2006}. However, in some situations such as F7 in Fig. B2 (a), the convex hull does not represent well the boundaries of a given set of points. So we adopt concave hull instead. A concave hull has at least one reflex interior angle. That is, an angle with a measure that is between 180$^\circ$ and 360$^\circ$. Blue polygon in Fig. B2 (a) is the concave hull for F7 based on works of \citet{Moreira2007}. As can be seen, the concave hull better describes the boundary of the MST.\\

We then take the ratio of area between the circle enclosing vertices and concave hull as the aspect ratio of a filament, similar with \citet{Gutermuth2009} for convex hull. As shown in Fig. B2 (b), black polygon is the concave hull of the filament. We take the mean position of hull vertices as the center to get a circle, that is gray circle in Fig. B2 (b). The radius of the circle is the distance of the farthest vertice to the center. Then the aspect ratio is defined as the ratio of circle area to concave hull area, $f_A=\frac{S_{circ}}{S_{hull}}$. For an approximately elliptical distribution, the $f_A$ is very similar with aspect ratio of the ellipse \citep{Schmeja2006}.

\renewcommand\thefigure{\Alph{section}\arabic{figure}} 
\section{Influence of Galactic environments on large-scale filaments}
\label{sec:env}

\setcounter{figure}{0}    
\begin{figure*}
\vspace{10mm}
\gridline{\fig{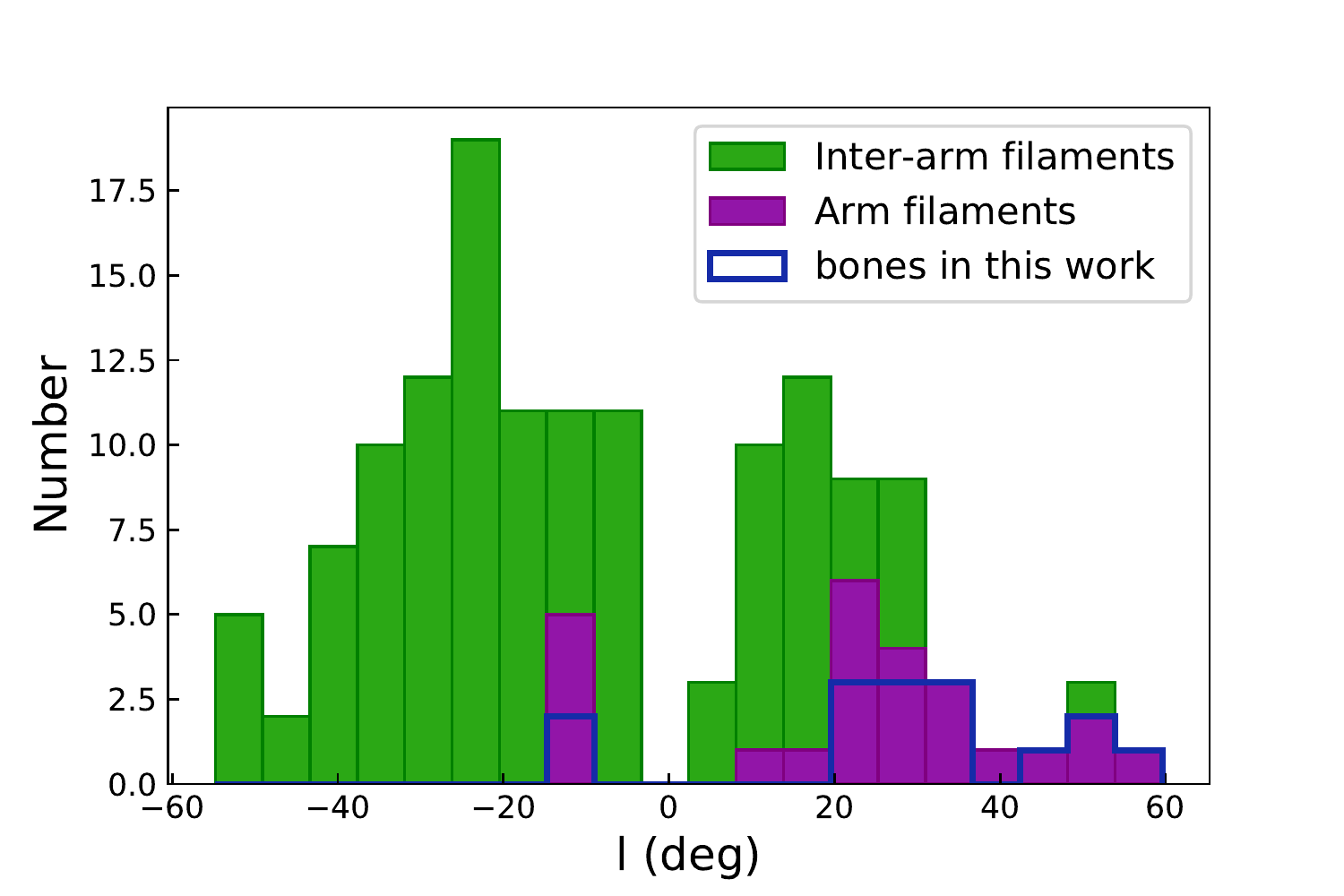}{0.3\textwidth}{(a)}
          \fig{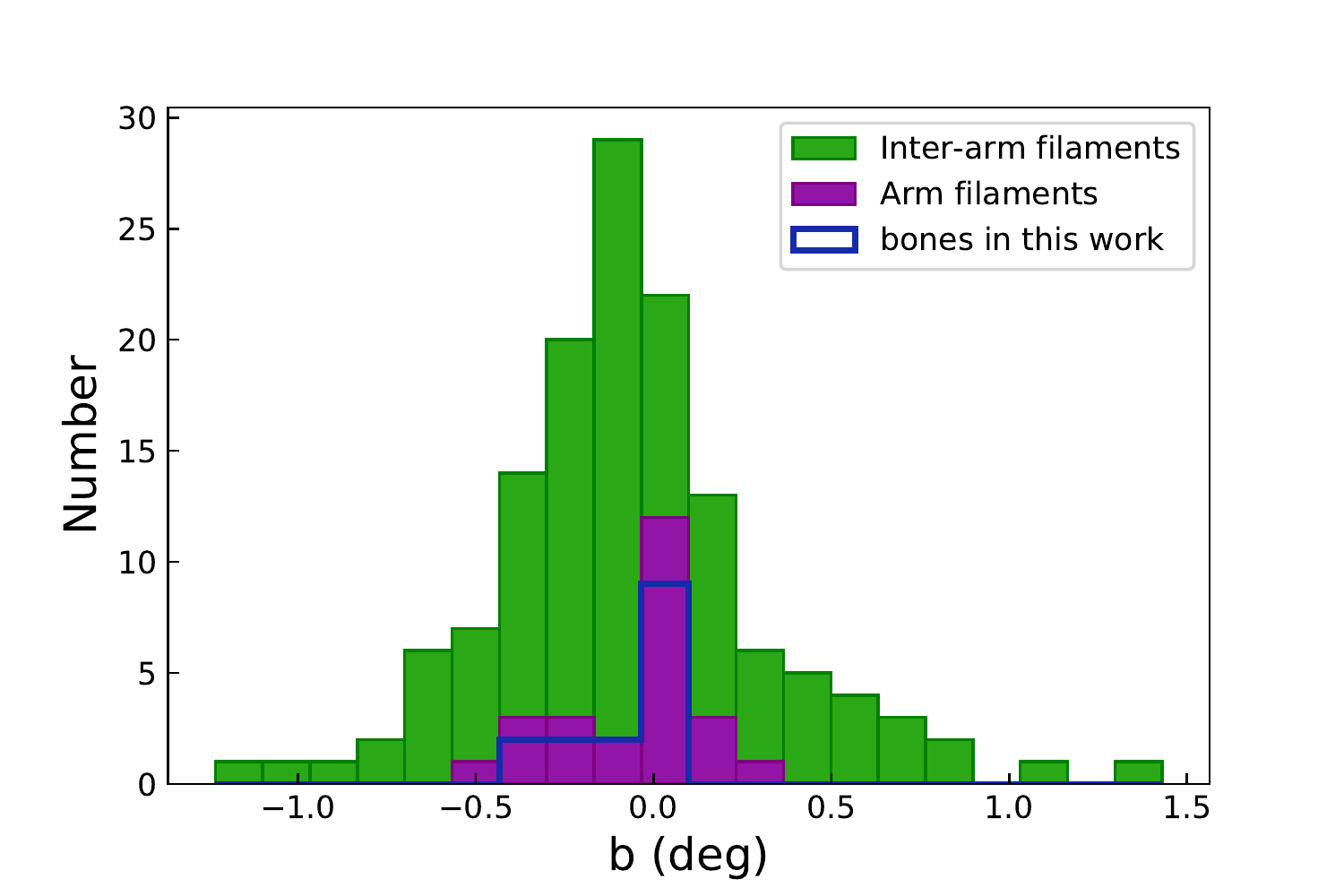}{0.3\textwidth}{(b)}
          \fig{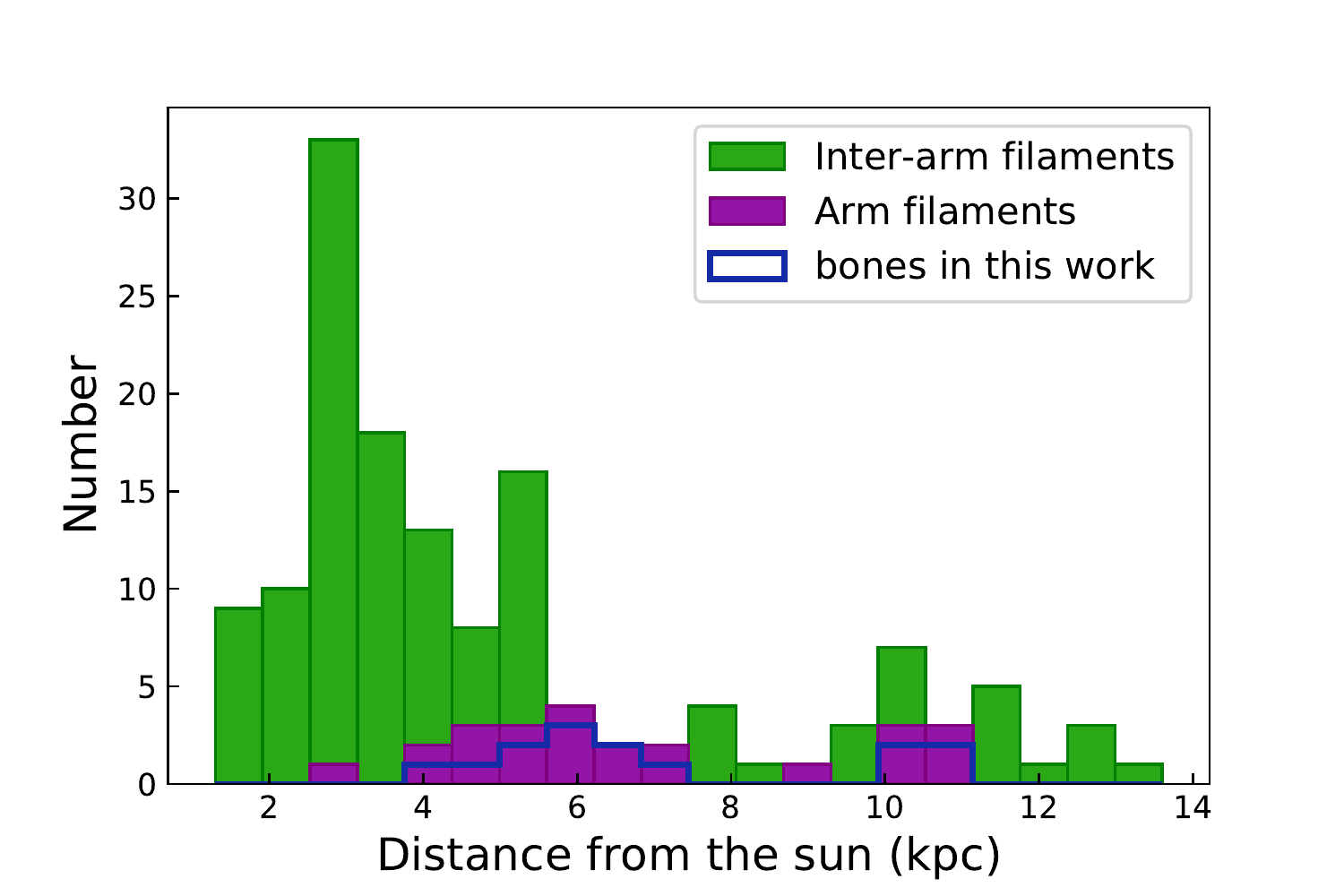}{0.3\textwidth}{(c)}
          }
\gridline{\fig{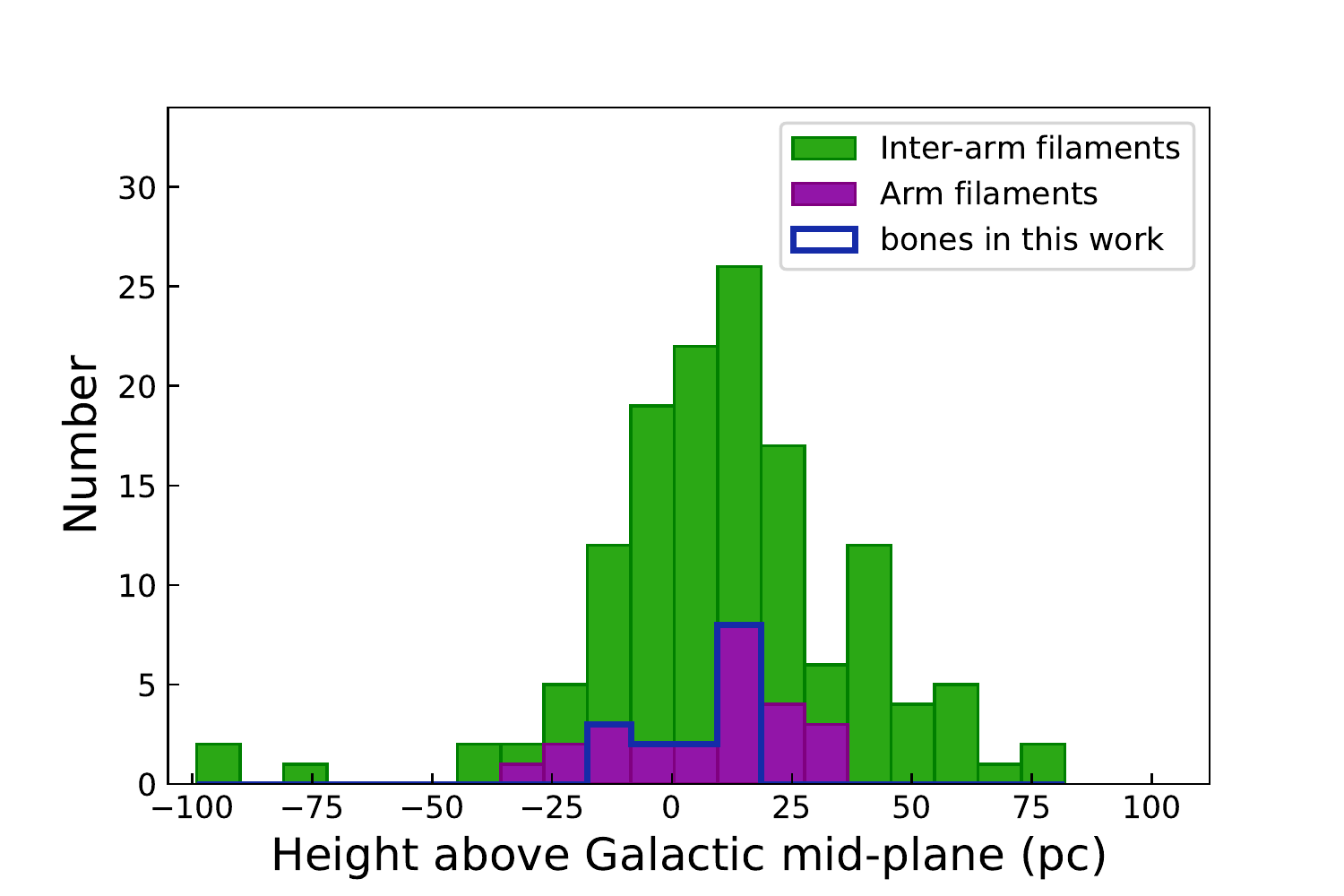}{0.3\textwidth}{(d)}
          \fig{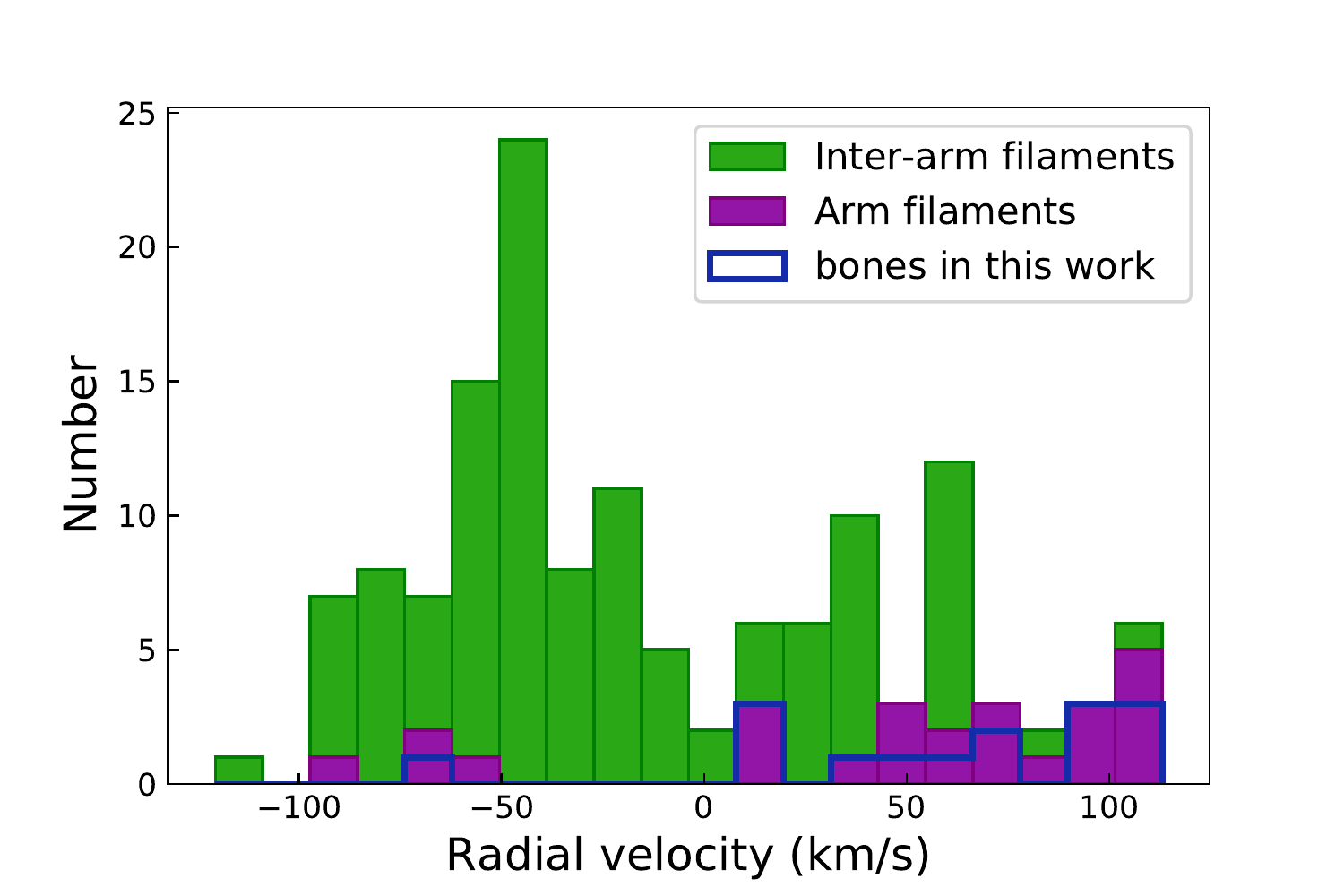}{0.3\textwidth}{(e)}
          \fig{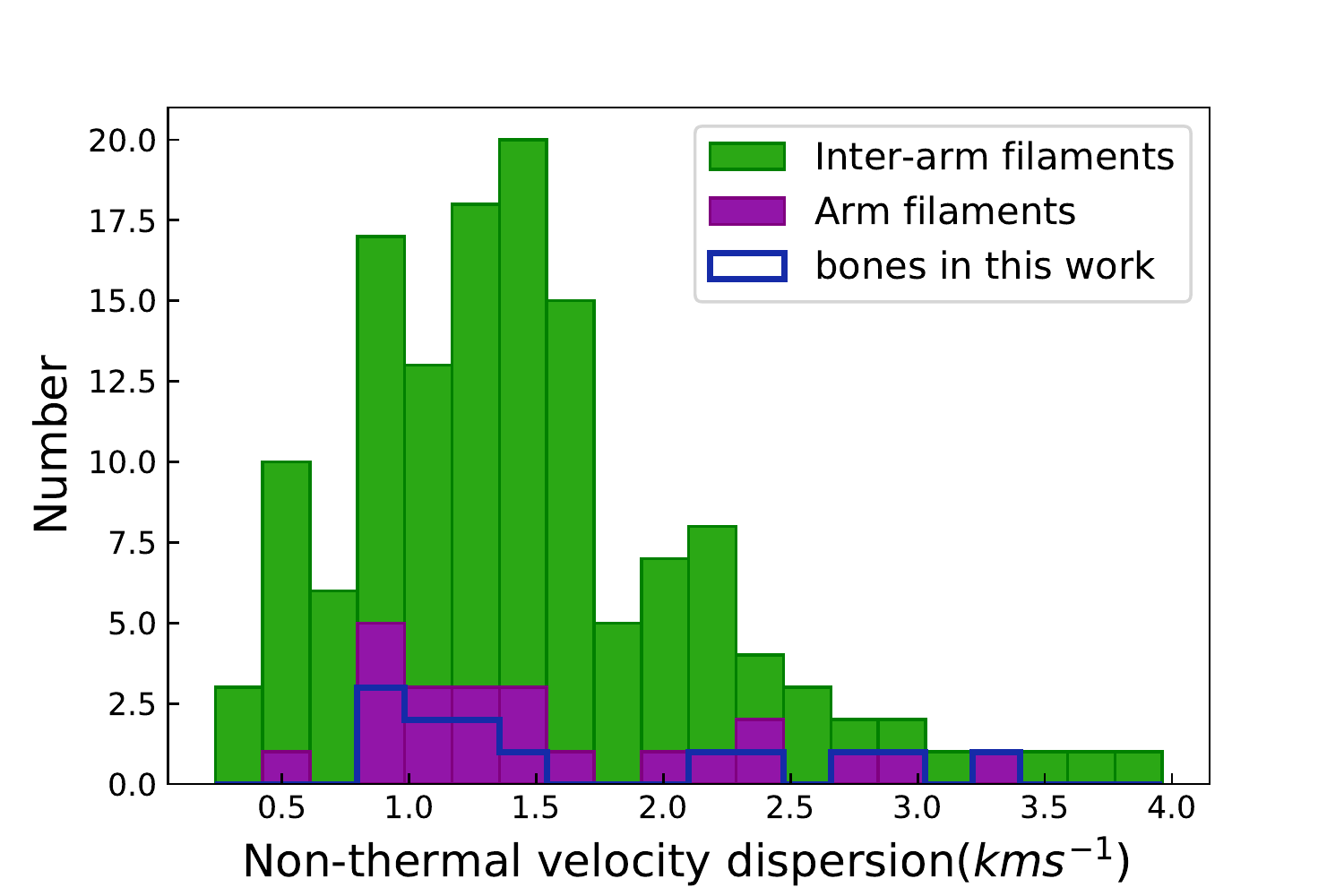}{0.3\textwidth}{(f)}
          }
\gridline{\fig{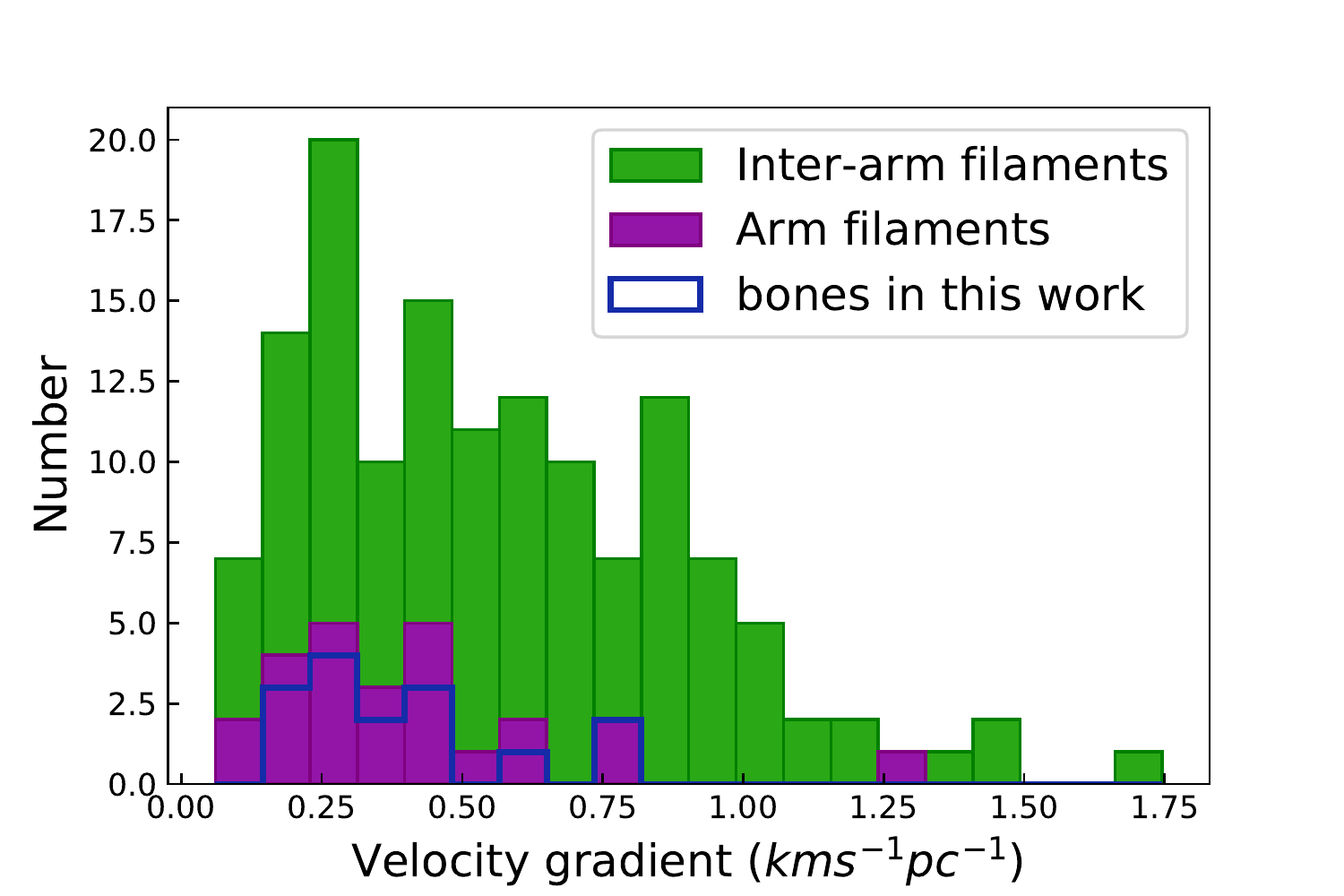}{0.3\textwidth}{(g)}
          \fig{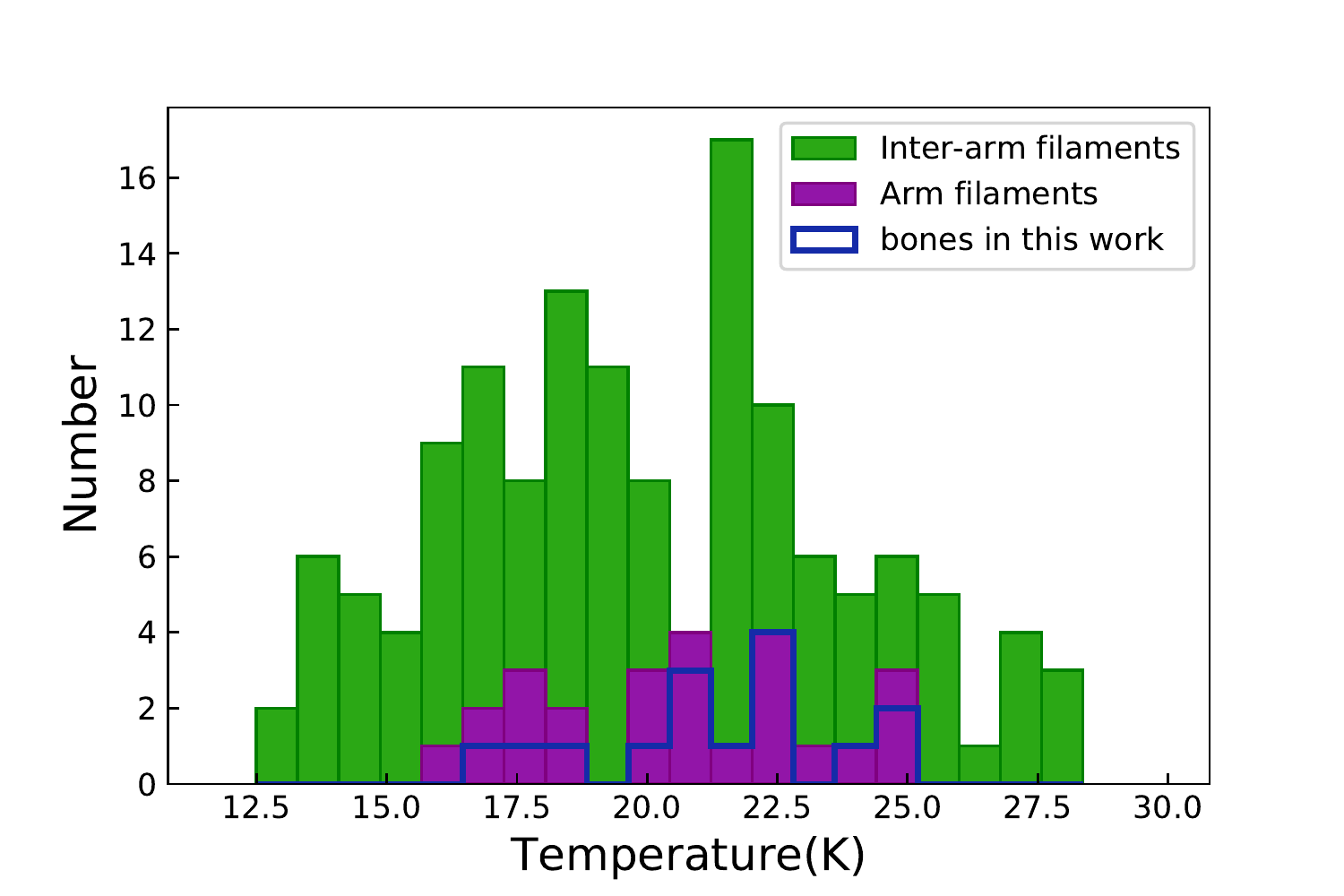}{0.3\textwidth}{(h)}
          \fig{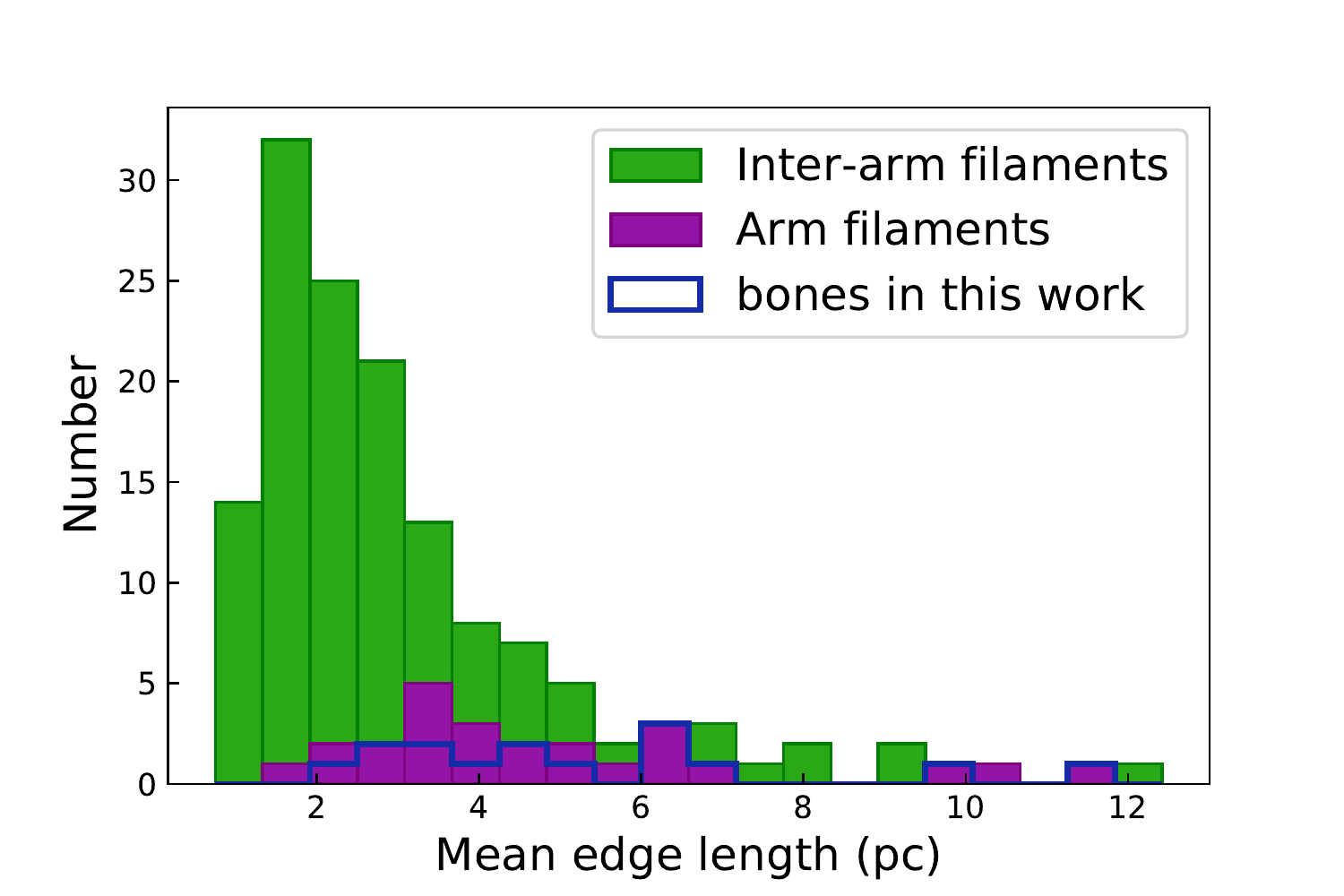}{0.3\textwidth}{(i)}
          }
\gridline{\fig{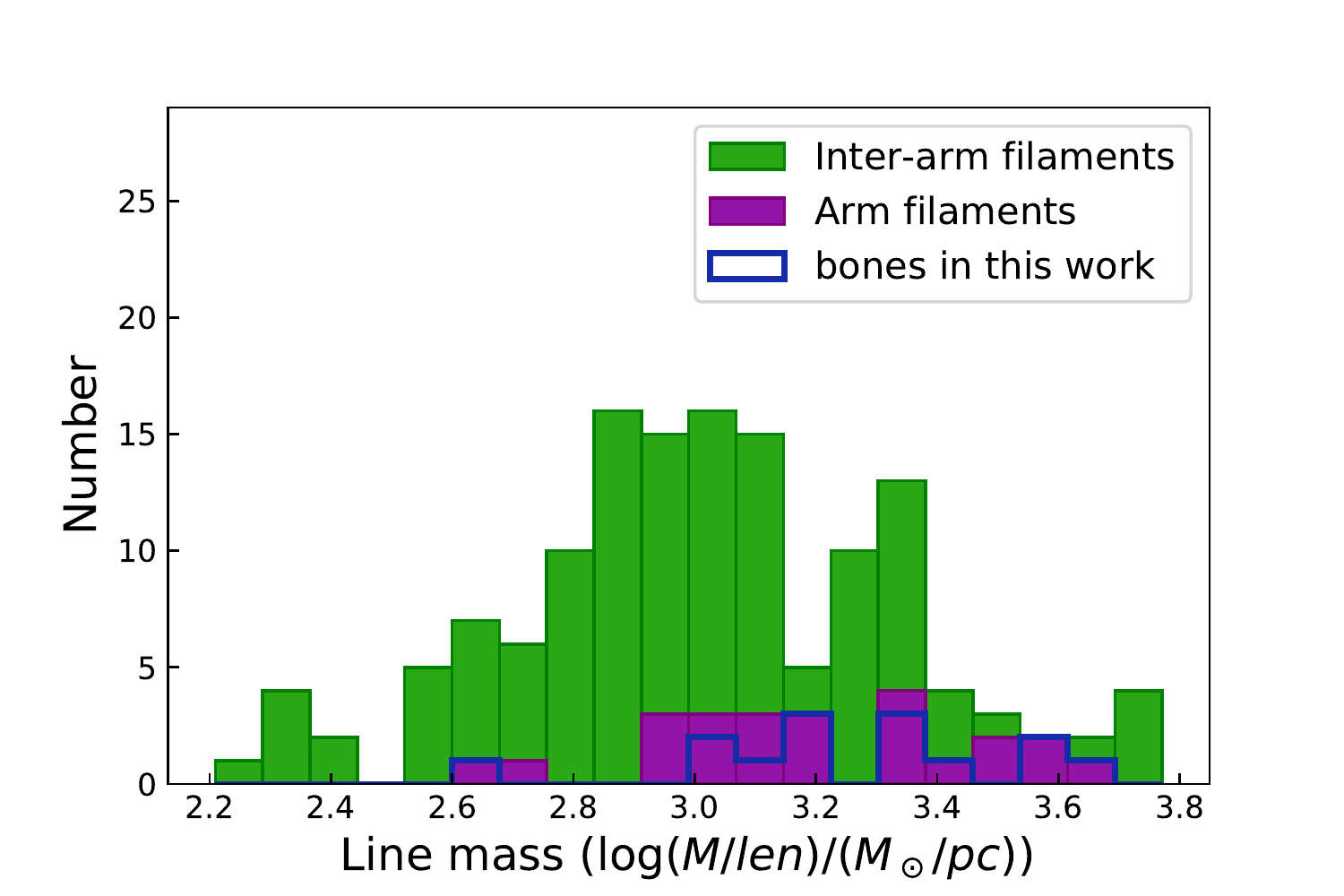}{0.3\textwidth}{(j)}
          \fig{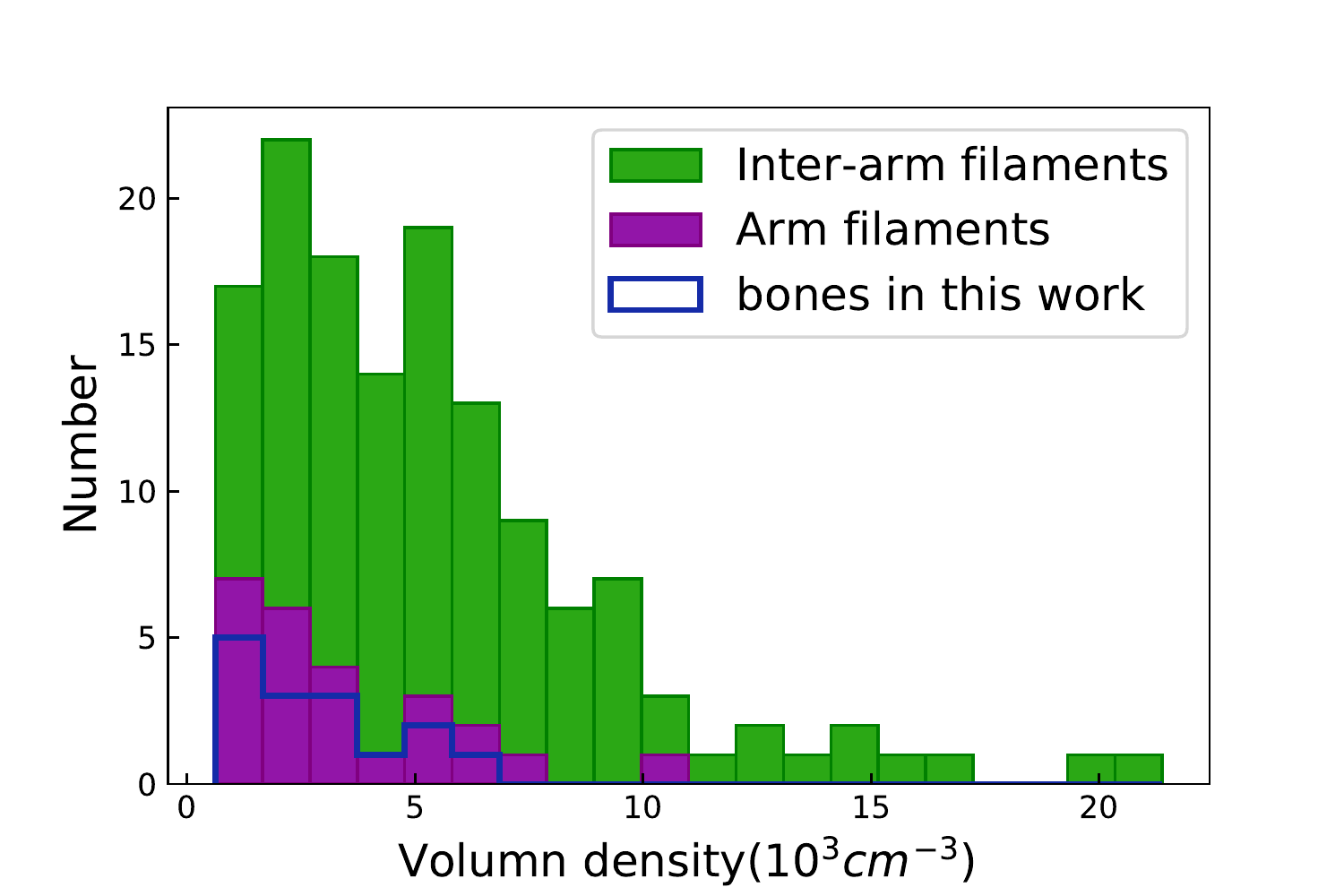}{0.3\textwidth}{(k)}
          \fig{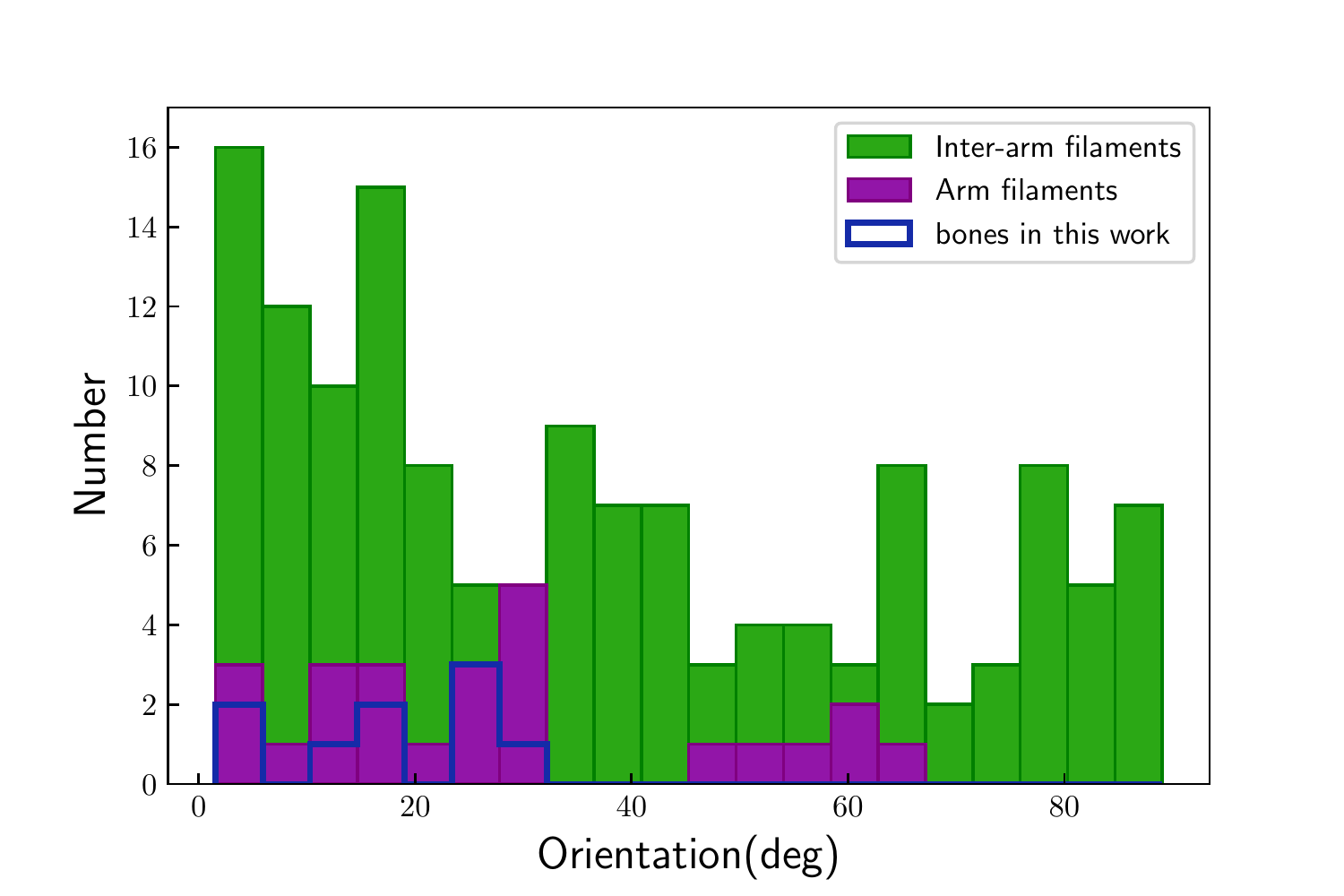}{0.3\textwidth}{(l)}
          }
\gridline{\fig{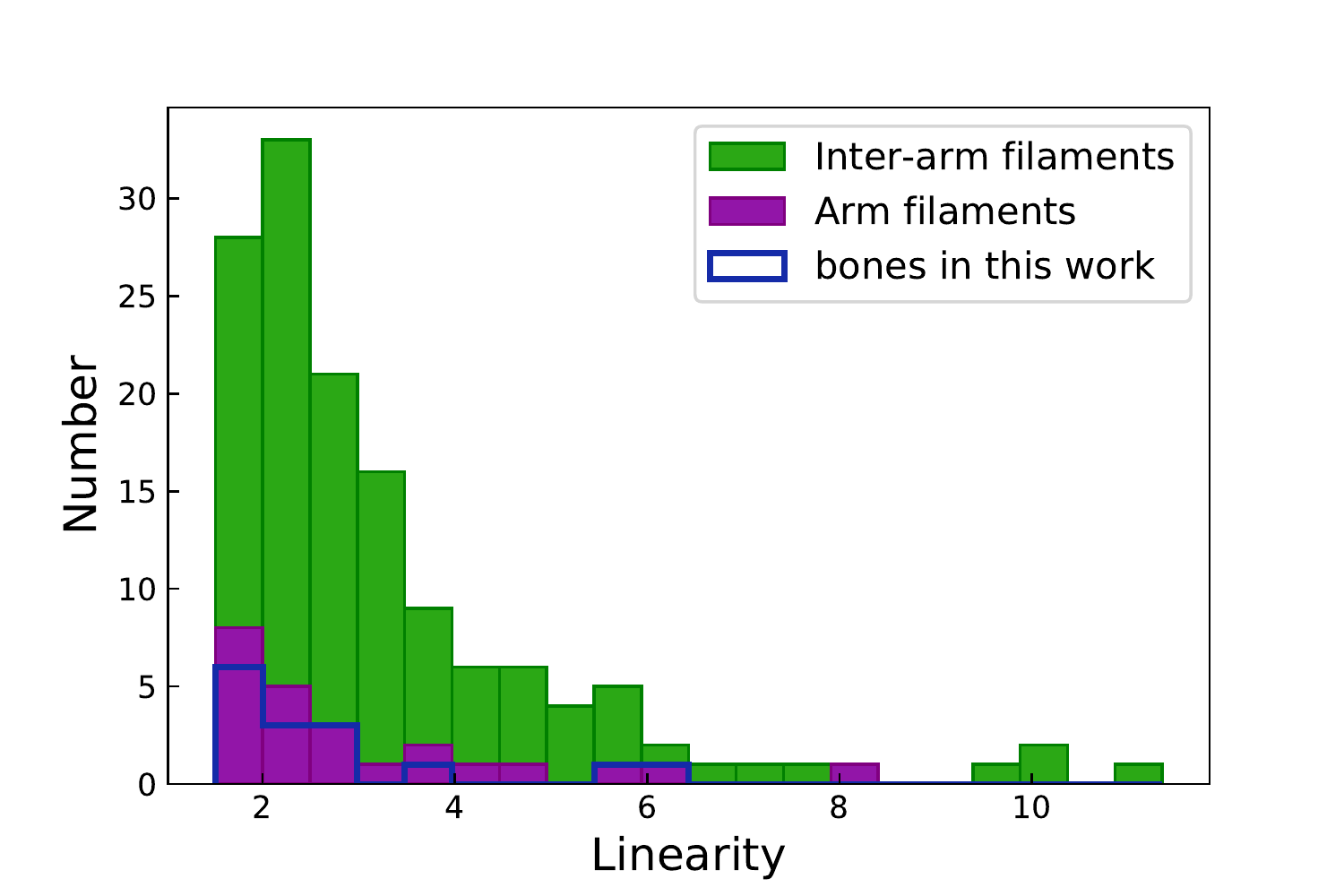}{0.3\textwidth}{(m)}
          \fig{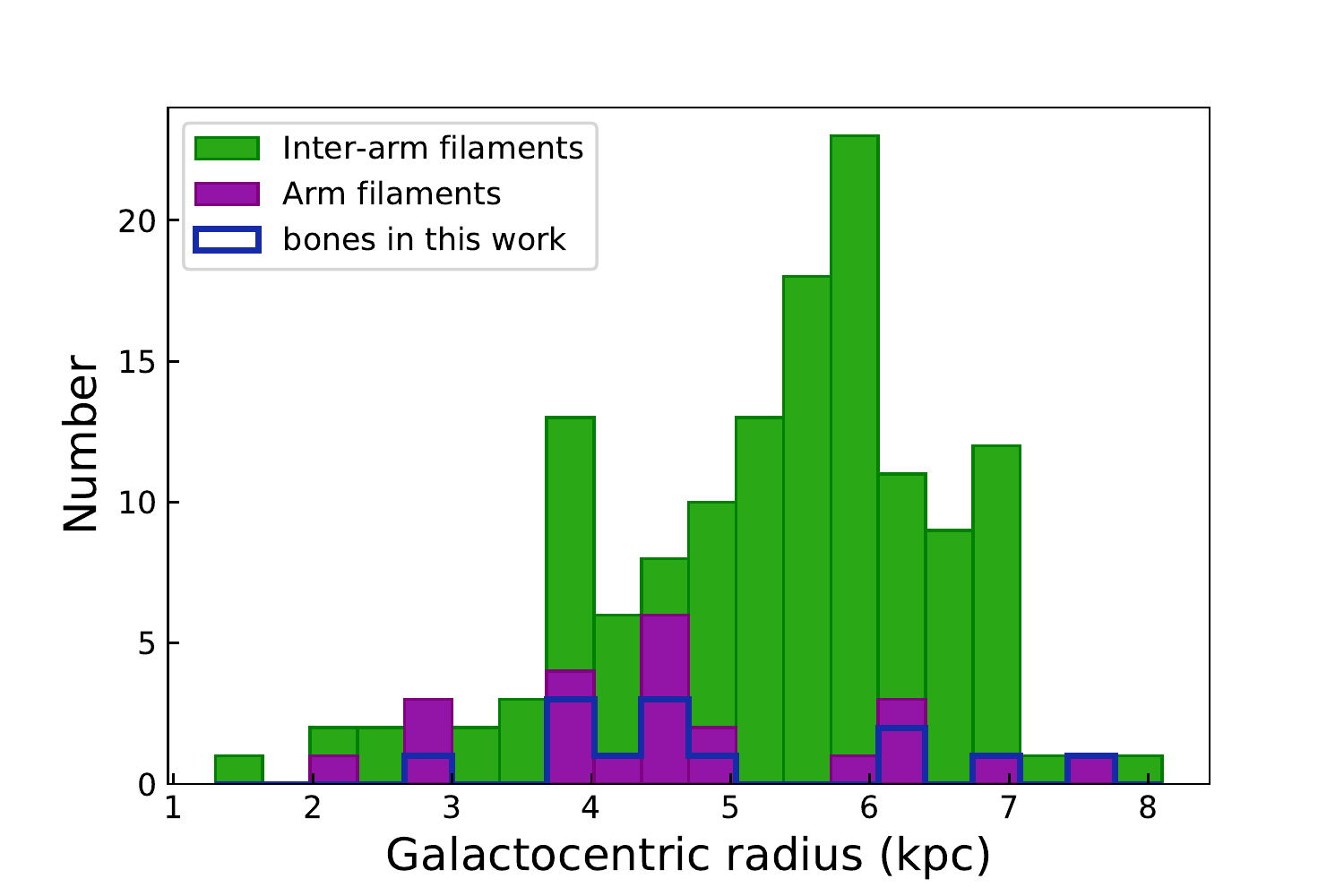}{0.3\textwidth}{(n)}
          \fig{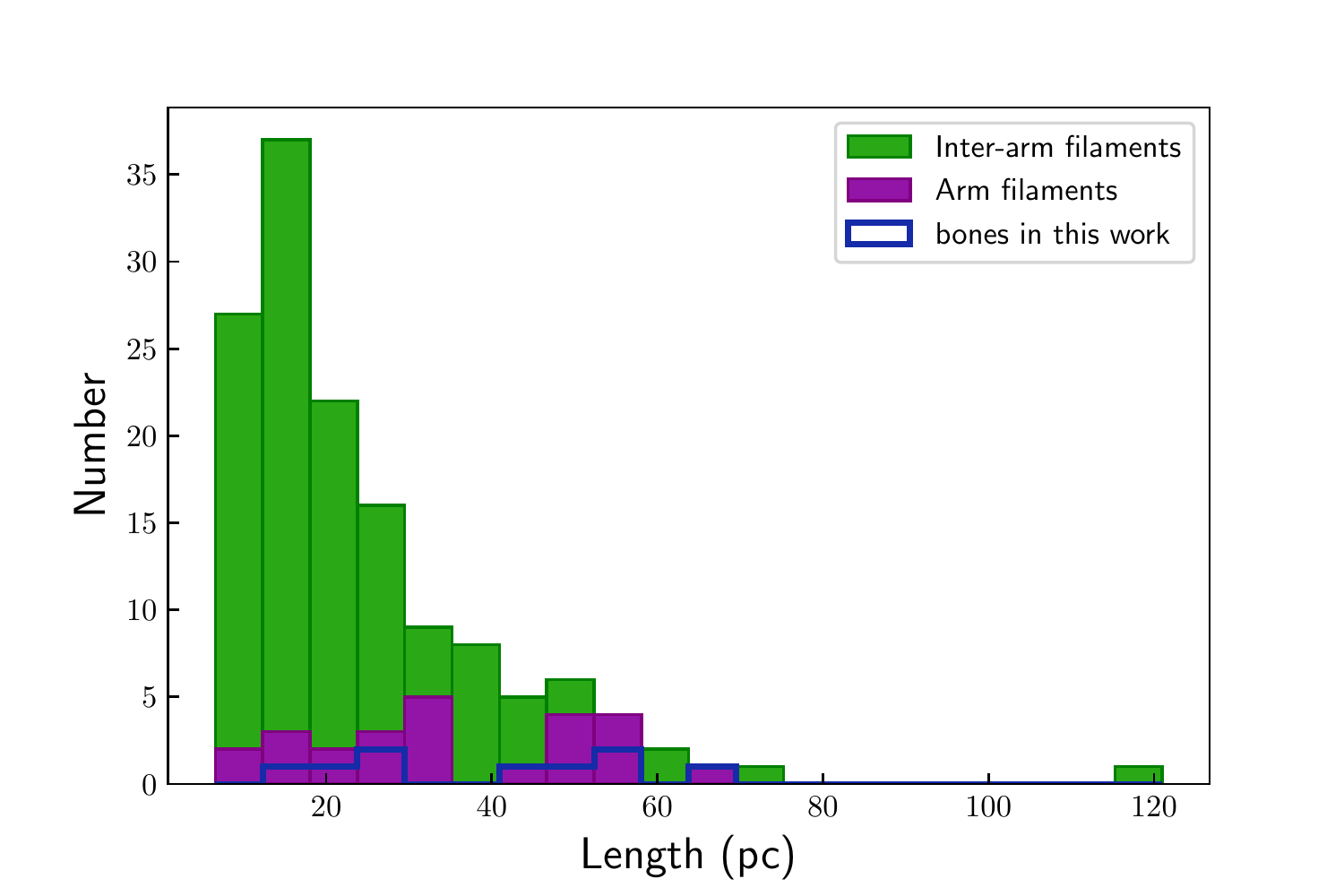}{0.3\textwidth}{(o)}
          }
\end{figure*}
\begin{figure*}
\vspace{10mm}
\gridline{\fig{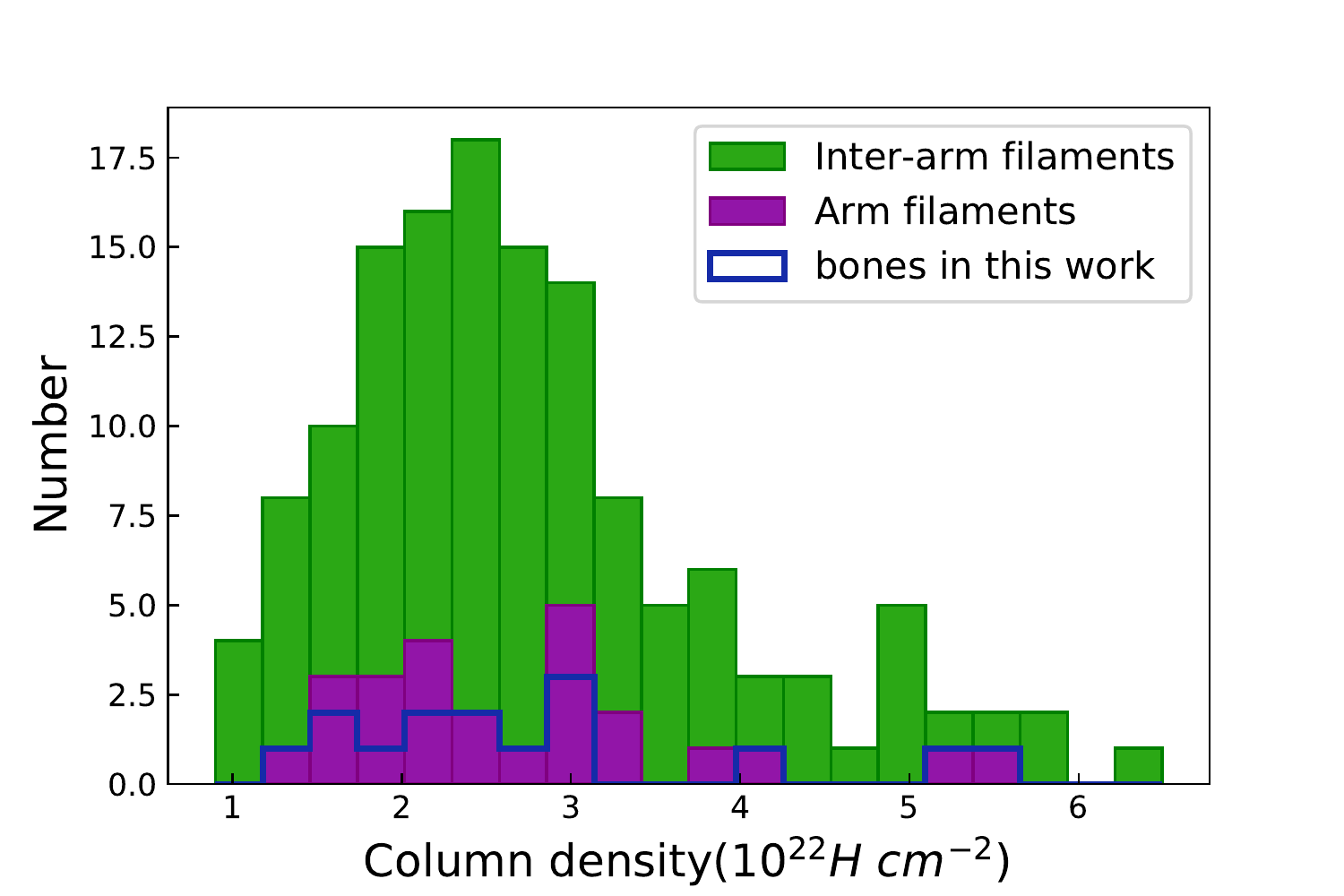}{0.3\textwidth}{(p)}
          \fig{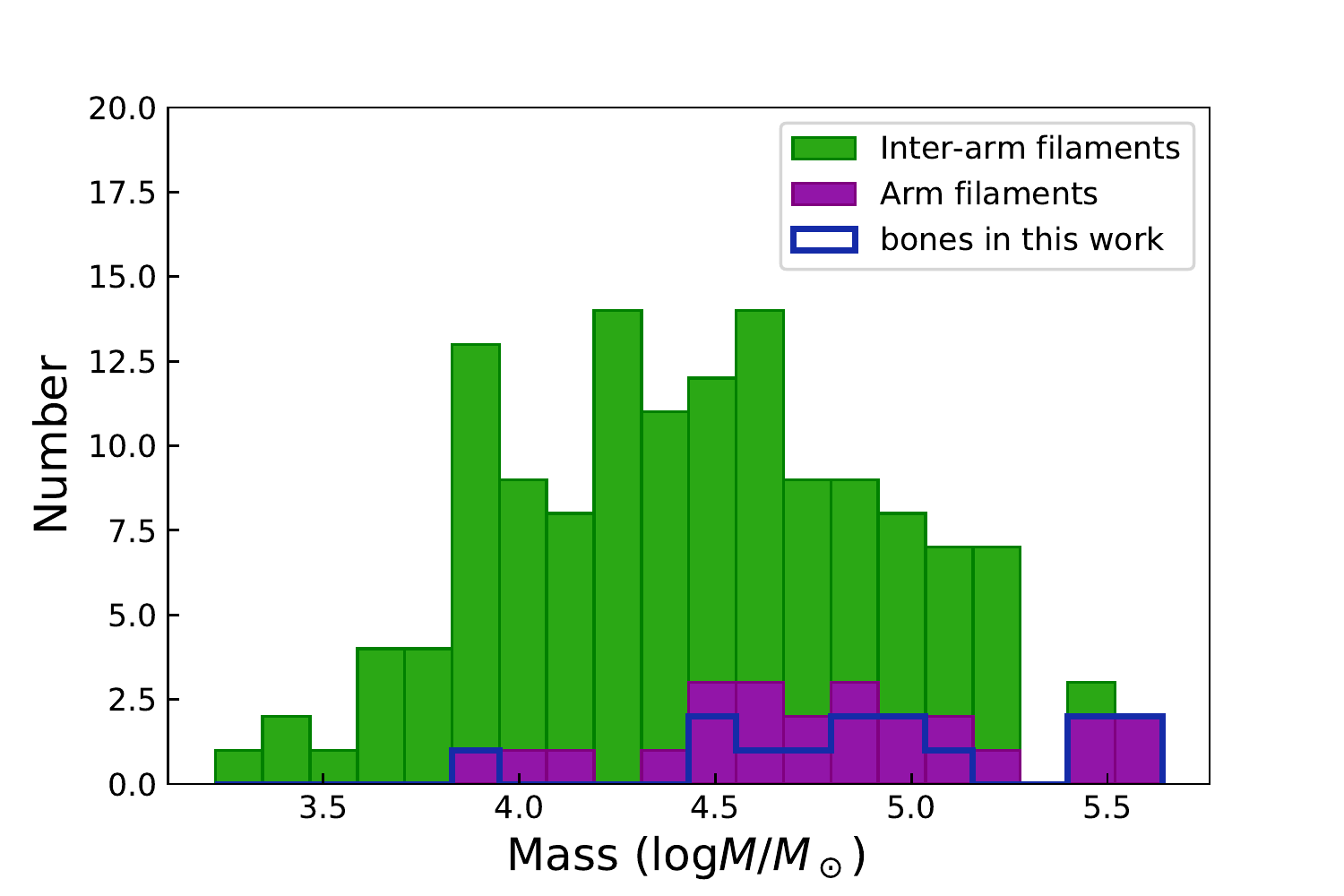}{0.3\textwidth}{(q)}
          }
\caption{Differcences of filaments in arms or not in. (a) Galactic longitude, (b) Galctic latitude, (c) distance from the Sun, (d) height above Galactic mid-plane, (e) mean radial velocity of each filament, (f) non-thermal velocity dispersion of clumps in each filament, (g) mean velocity gradient of edges in each filaments, (h) mean temperature, (i) mean separation of clumps in each filament, (j) line mass, (k) volume density, (l) angle between major axes and Galactic mid-plane in the projected sky, (m) linearity, (n) Galactocentric radius, (o) linearity-weighted length, (p) column density, (q) mass.}
\end{figure*}

To investigate the influence of Galactic environments on large-scale filaments, we compare properties of our filaments in spiral arms (25 of 163 filaments) or not (the rest 138 filaments). We find no significant distinctions in some of their physical properties such as mean temperatures, non-thermal velocity dispersion, and column densities (Fig. C1 (p)). \citet{Zucker2019} also find length of filaments in arm or not in invariant. They find arm filaments tend to have larger column density than inter-arm ones, but they remind us that their results need more advanced numerical simulations to confirm. Filament catalogue from \citet{Schisano2020} containing both large-scale and small-scale filaments also shows that column densities are similar in arm filaments and inter-arm filaments .\\

Apart from similitude, large-scale filaments also have distinctions on such as Galactocentric radius, height above Galactic mid-plane, mass. The distribution of Galactocentric radius of arm filaments shows a peak in 4.6 kpc (Fig. C1 (n)), meaning most population of filaments are found from Scutum arm, which is consistent with \citet{Zucker2019}. Arm filaments, on the whole, have larger mass and linearity-weighted length than inter-arm filaments (Fig. C1 (q) and (o)). But this might be caused by the fact that arm filaments are mostly far from us (Fig. C1 (c)). Distributions of bones in this work with distance constraint have also been overlaid in Fig. C1 with blue steps. The differences between arm filaments and bones in this work are that bones in this work have two additional requests: (6) height above Galactic mid-plane $|z| \leq  20$ pc and (7) orientation angle between the filament’s major axis and the physical Galactic mid-plane, $|\theta| \leq  30^\circ$. We do not find obvious distinctions between the distribution of the two samples.\\

Other panel in Fig. C1 are: (a) Galactic longitude, (b) Galctic latitude, (c) distance from the Sun, (d) height above Galactic mid-plane, (e) mean radial velocity of each filament, (f) non-thermal velocity dispersion of clumps in each filament, (g) mean velocity gradient, (h) mean temperature, (i) mean separation of clumps in each filament, (j) line mass, (k) volume density, (l) angle between major axes and Galactic mid-plane in the projected sky, (m) linearity. Arm filaments in northern sky are obviously more than in southern sky (Fig. C1 (a)), this might be caused by inaccuracy of spiral arm model where Quadrant 4 is a fitted result \citep[Fig. 1 in][]{Reid2019}. Distribution of arm filaments' height above Galactic mid-plane (Fig. C1 (d)) has a doublet peaked at about $z = \pm 20$ pc, and this might be observation effect considering our Sun resides 25 pc above Galactic mid-plane. Distribution of radial velocity of arm filaments (Fig. C1 (e)) is due to combination of Galactic latitude and rotation of the Milky Way.\\

We also find that DGMF of filaments in spiral arms have no significant distinction with inter-arm filaments, neither are bones. For arm filaments, DGMFs vary in different spiral arms according to \citet{Abreu2016}, but lack a statistical sample. Our filament catalogue provides a larger sample to conduct a statistical analysis. We find about ten filaments with DGMF measurement in each spiral arm, except Perseus arm with only one. Mean DGMFs of filaments in each spiral arm are $35.8\%\pm 9.8\%$, $41.0\%\pm 5.2\%$, $32.5\%\pm 7.0\%$, and $44.8\%$ for Norma-Outer, Scutum-Centaurus-OSC, Sagittarius-Carina, and Perseus arm, respectively. The larger DGMF of Scutum-Centaurus-OSC and Perseus arm along with the larger bones fraction give us a hint of the peculiarity of these two arms.

\section{Filaments on emission maps and column density maps}
\label{sec:maps}

We have shown two-color view of some filaments in Fig. \ref{twocolor}. Two-color view of other filaments are shown in Fig. D1. The color-coded circles denote clumps in filaments with various velocities. For backgrounds, cyan represents intermediate infrared 24 \textmu m emission on logarithmic scale from MIPSGAL \citep{Carey2009} and red shows submillimeter 870 \textmu m emission on linear scale from APEX + Planck combined image \citep{Csengeri2016}.\\

Filaments are also overlaid on Herschel Hi-GAL column density map from PPMAP \citep{Marsh2017} in Fig. D2. The same as in Fig. D1, the color-coded circles also denote clumps in filaments with various velocities. There are some gaps in coverage on the map such as F21 in Fig. D2. That is due to the fact that in order to get the entire Galactic Plane done in a reasonable amount of time \citet{Marsh2017} had to truncate the field of each tile, and unfortunately the truncation was excessive in some places. And column density map for F147 (with Galactic latitude 1.43$^\circ$) is absent because it is outside the coverage of Herschel Hi-GAL map.
\renewcommand\thefigure{\Alph{section}\arabic{figure}} 
\setcounter{figure}{0} 
\begin{center}
\setlength{\tabcolsep}{1.2mm}{
{
\doublerulesep=5pt
\begin{figure}
\includegraphics[width=1\linewidth]{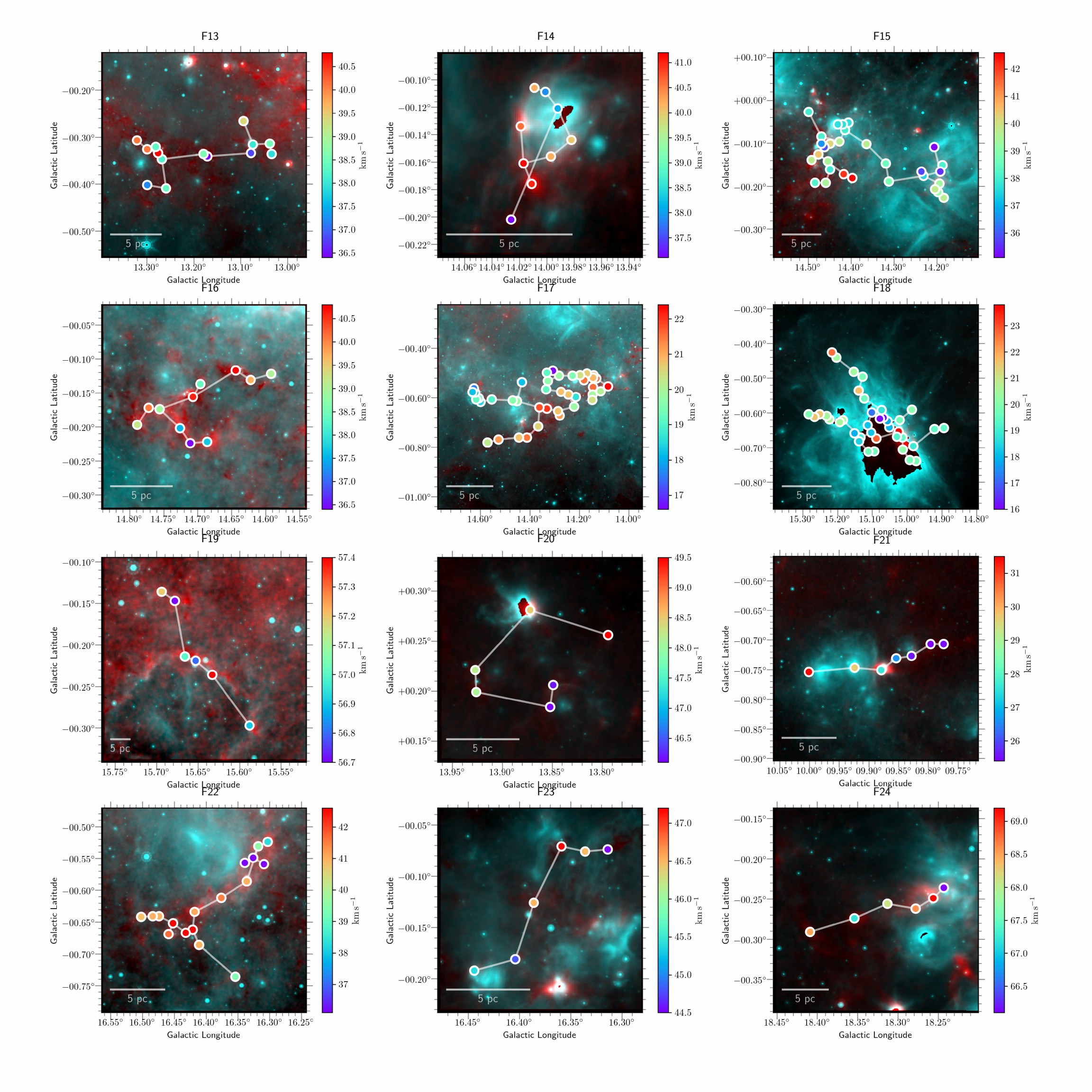}
\caption{Two-color view of filaments. The color-coded circles denote clumps in filaments with various velocities. For backgrounds, cyan represents intermediate infrared 24 \textmu m emission on logarithmic scale from MIPSGAL \citep{Carey2009} and red shows submillimeter 870 \textmu m emission on linear scale from APEX + Planck combined image \citep{Csengeri2016}.}
\end{figure}
}}
\end{center}
\setcounter{figure}{0} 
\begin{center}
\setlength{\tabcolsep}{1.2mm}{
{–––
\doublerulesep=5pt
\begin{figure}
\includegraphics[width=1\linewidth]{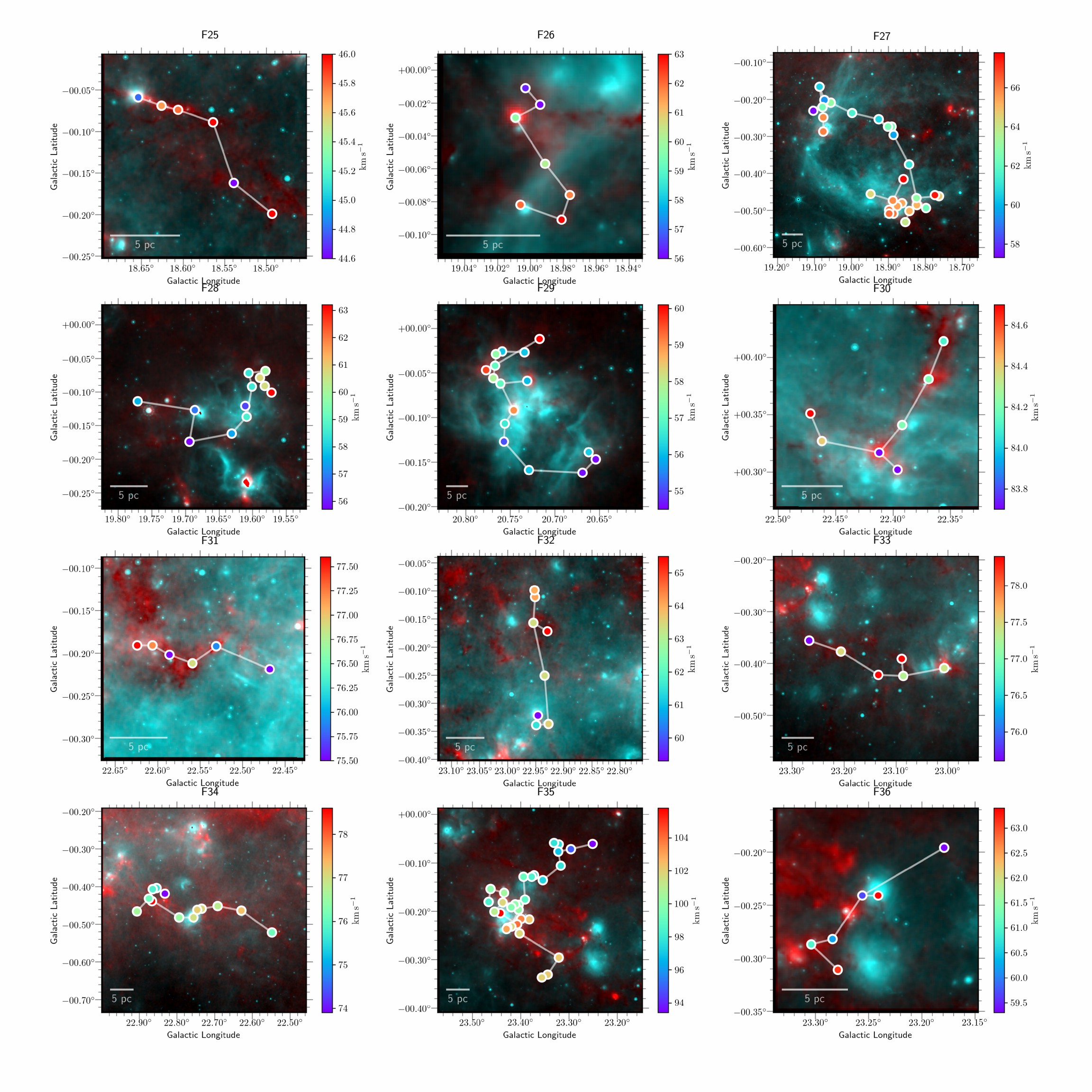}
\caption{(continued)}
\end{figure}
}}
\end{center}
\setcounter{figure}{0} 
\begin{center}
\setlength{\tabcolsep}{1.2mm}{{
\doublerulesep=5pt
\begin{figure}
\includegraphics[width=1\linewidth]{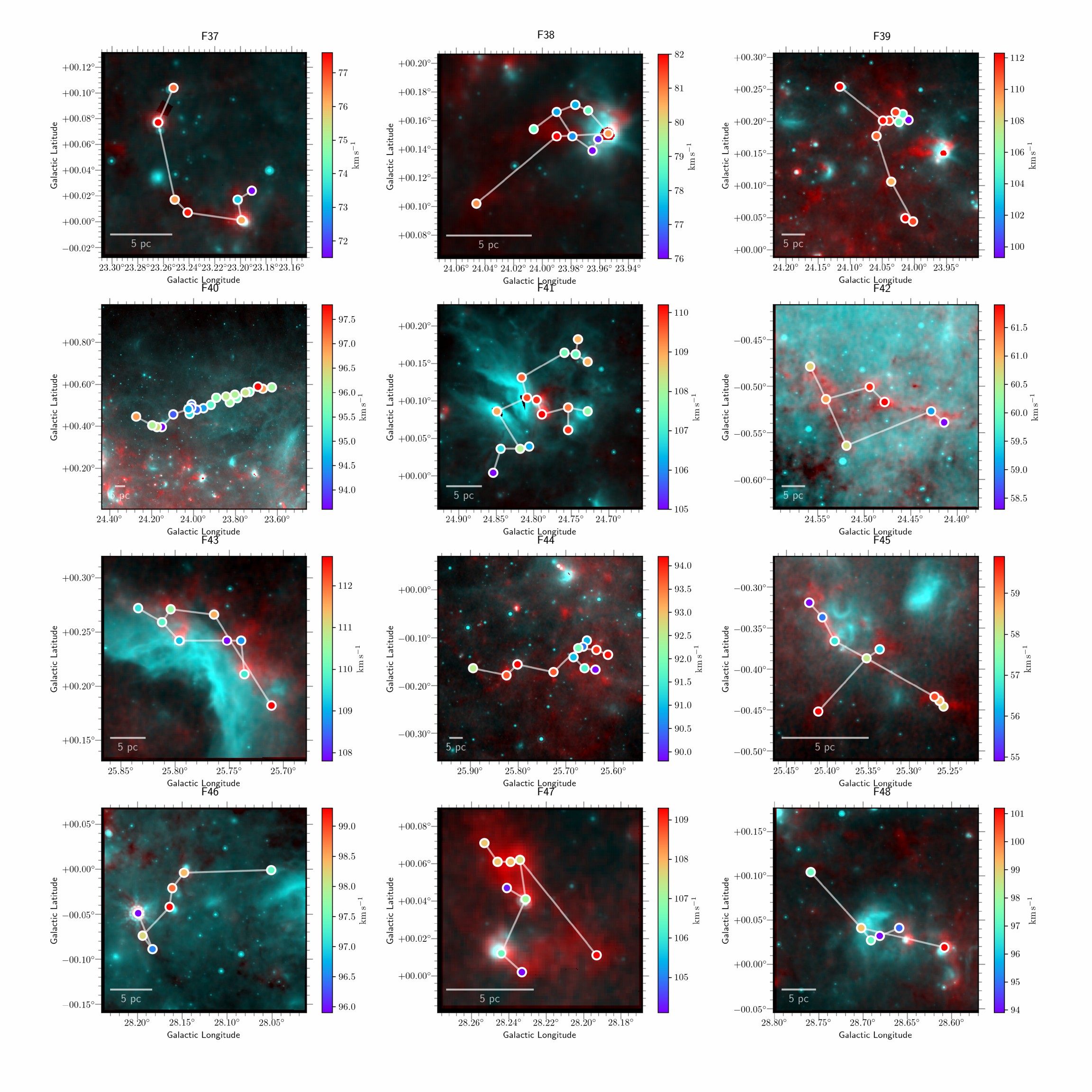}
\caption{(continued)}
\end{figure}
}}
\end{center}
\setcounter{figure}{0} 
\begin{center}
\setlength{\tabcolsep}{1.2mm}{{
\doublerulesep=5pt
\begin{figure}
\includegraphics[width=1\linewidth]{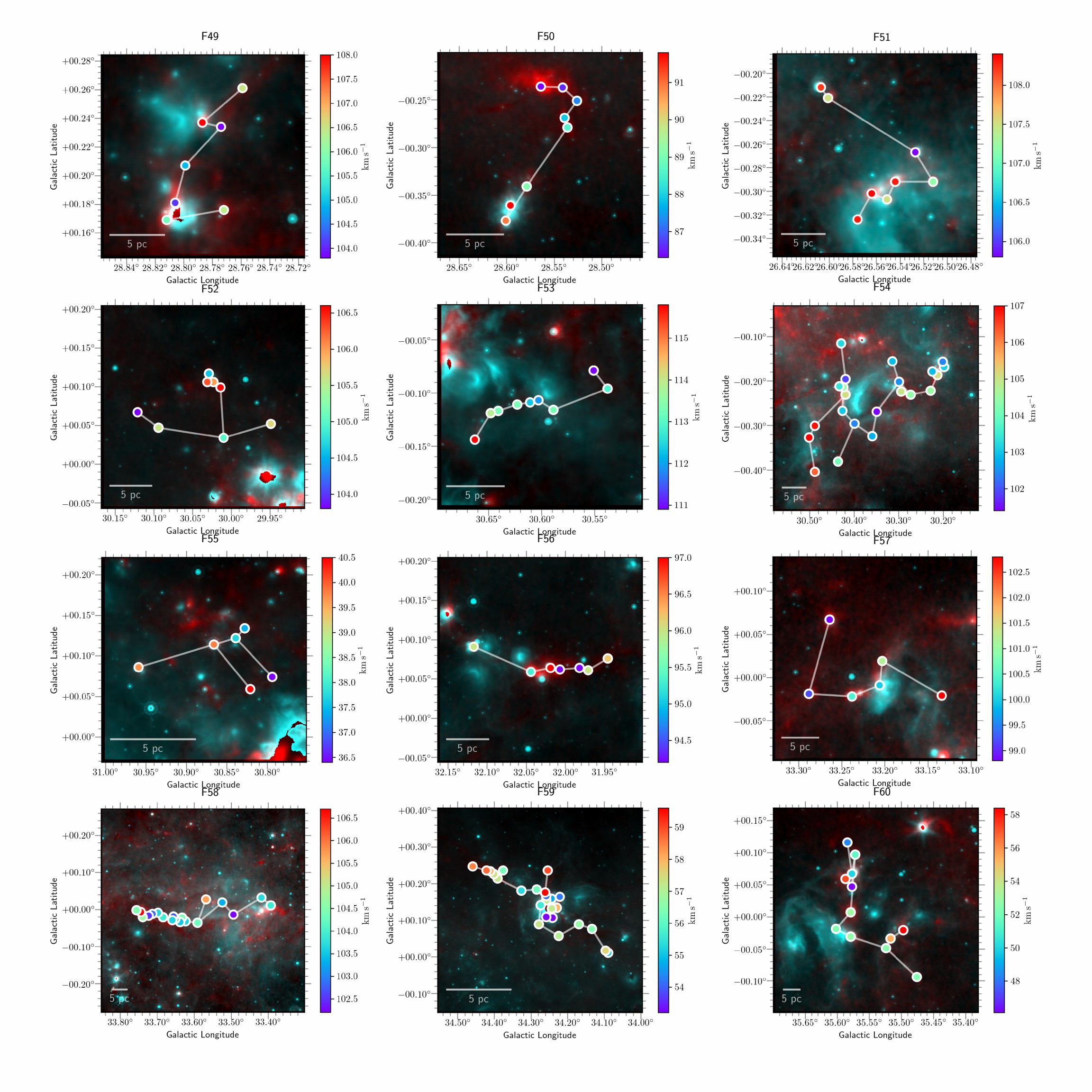}
\caption{(continued)}
\end{figure}
}}
\end{center}
\setcounter{figure}{0} 
\begin{center}
\setlength{\tabcolsep}{1.2mm}{{
\doublerulesep=5pt
\begin{figure}
\includegraphics[width=1\linewidth]{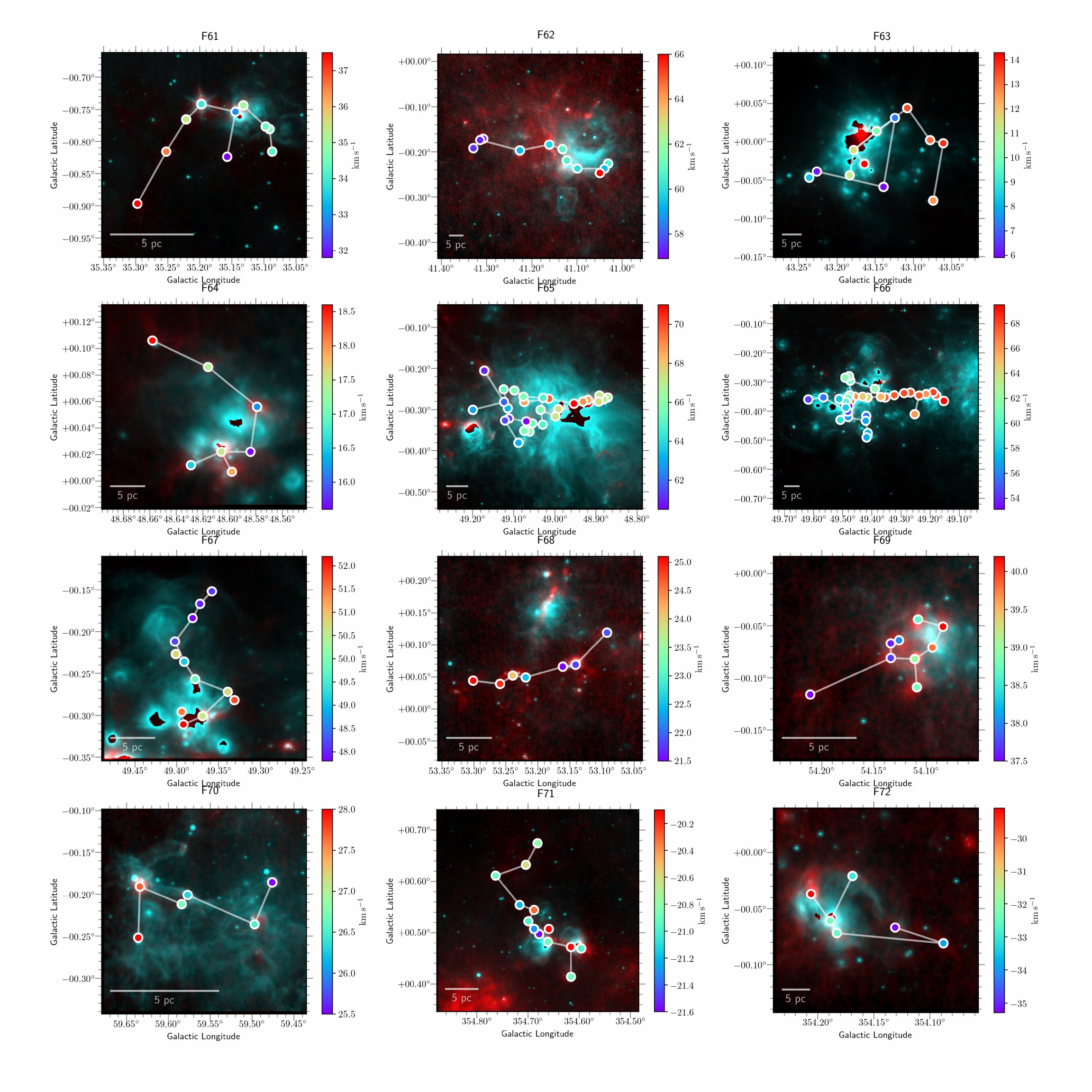}
\caption{(continued)}
\end{figure}
}}
\end{center}
\setcounter{figure}{0} 
\begin{center}
\setlength{\tabcolsep}{1.2mm}{{
\doublerulesep=5pt
\begin{figure}
\includegraphics[width=1\linewidth]{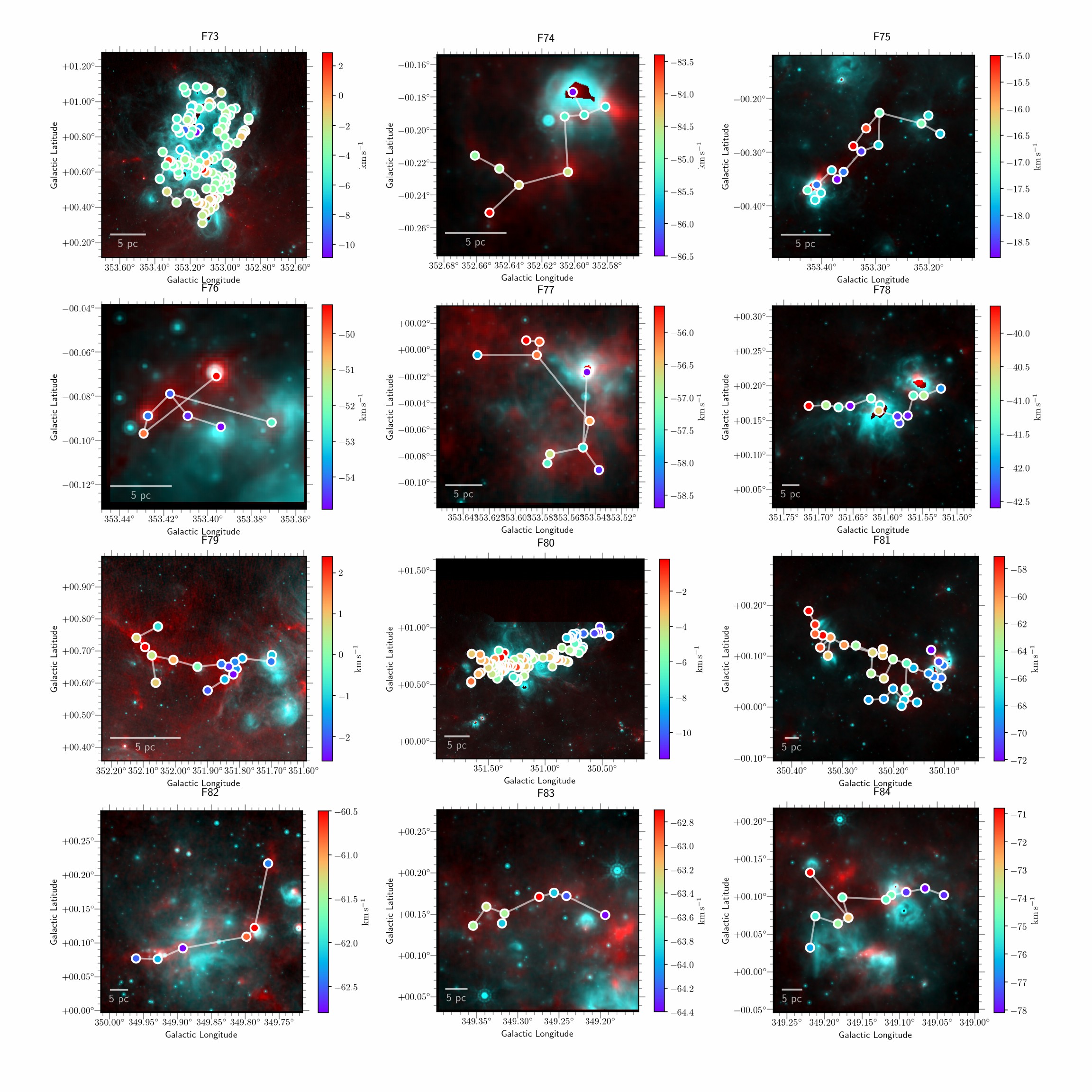}
\caption{(continued)}
\end{figure}
}}
\end{center}
\setcounter{figure}{0} 
\begin{center}
\setlength{\tabcolsep}{1.2mm}{{
\doublerulesep=5pt
\begin{figure}
\includegraphics[width=1\linewidth]{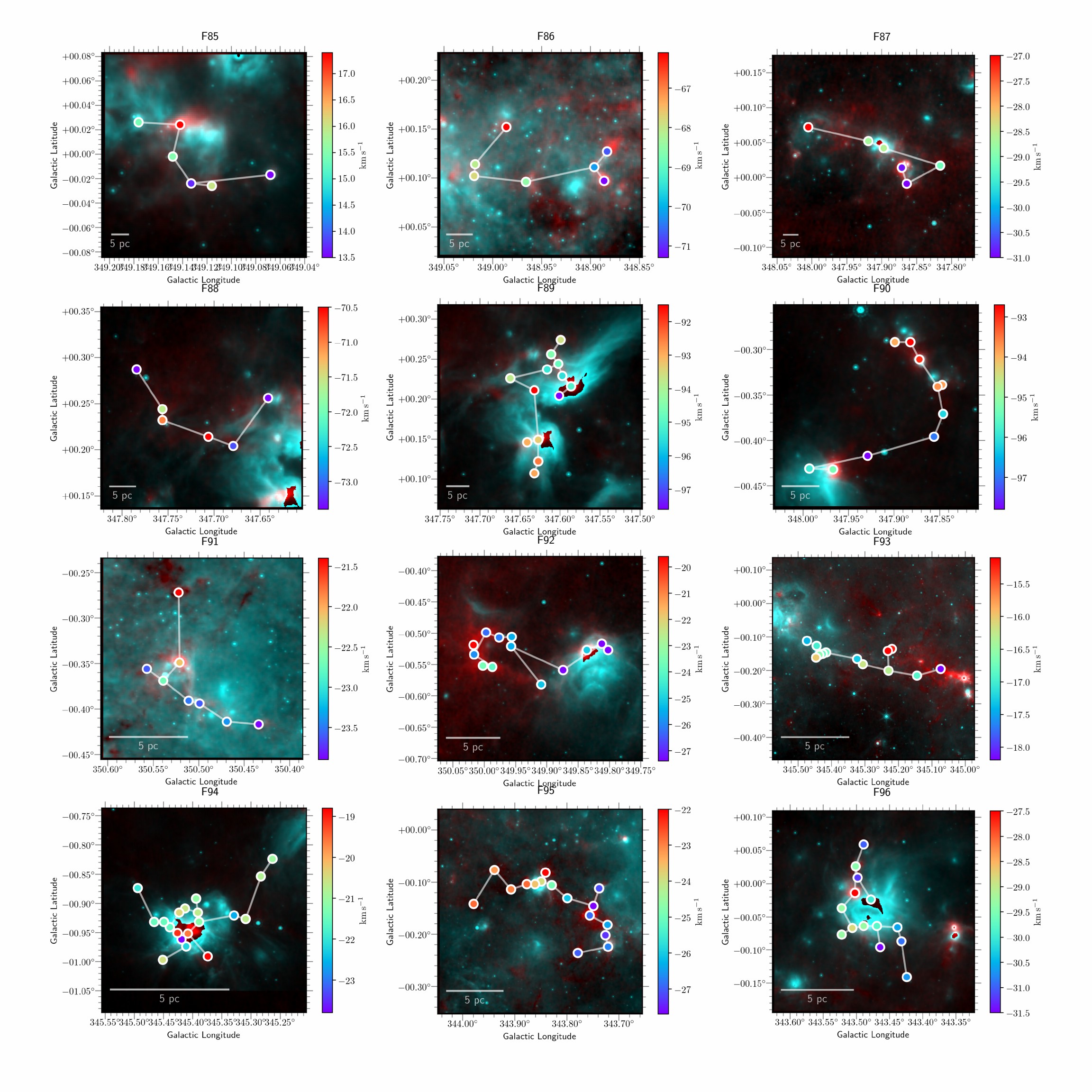}
\caption{(continued)}
\end{figure}
}}
\end{center}
\setcounter{figure}{0} 
\begin{center}
\setlength{\tabcolsep}{1.2mm}{{
\doublerulesep=5pt
\begin{figure}
\includegraphics[width=1\linewidth]{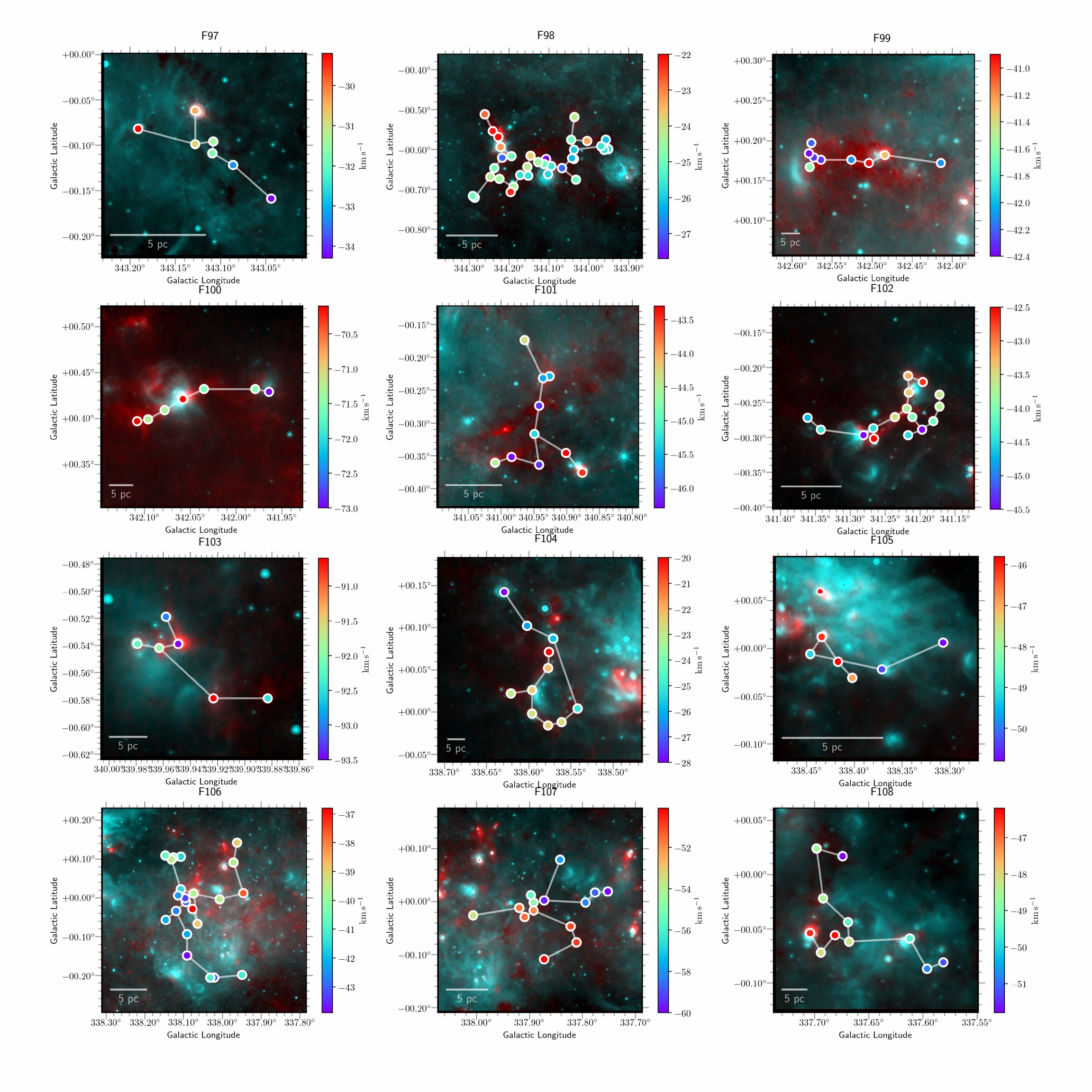}
\caption{(continued)}
\end{figure}
}}
\end{center}
\setcounter{figure}{0} 
\begin{center}
\setlength{\tabcolsep}{1.2mm}{{
\doublerulesep=5pt
\begin{figure}
\includegraphics[width=1\linewidth]{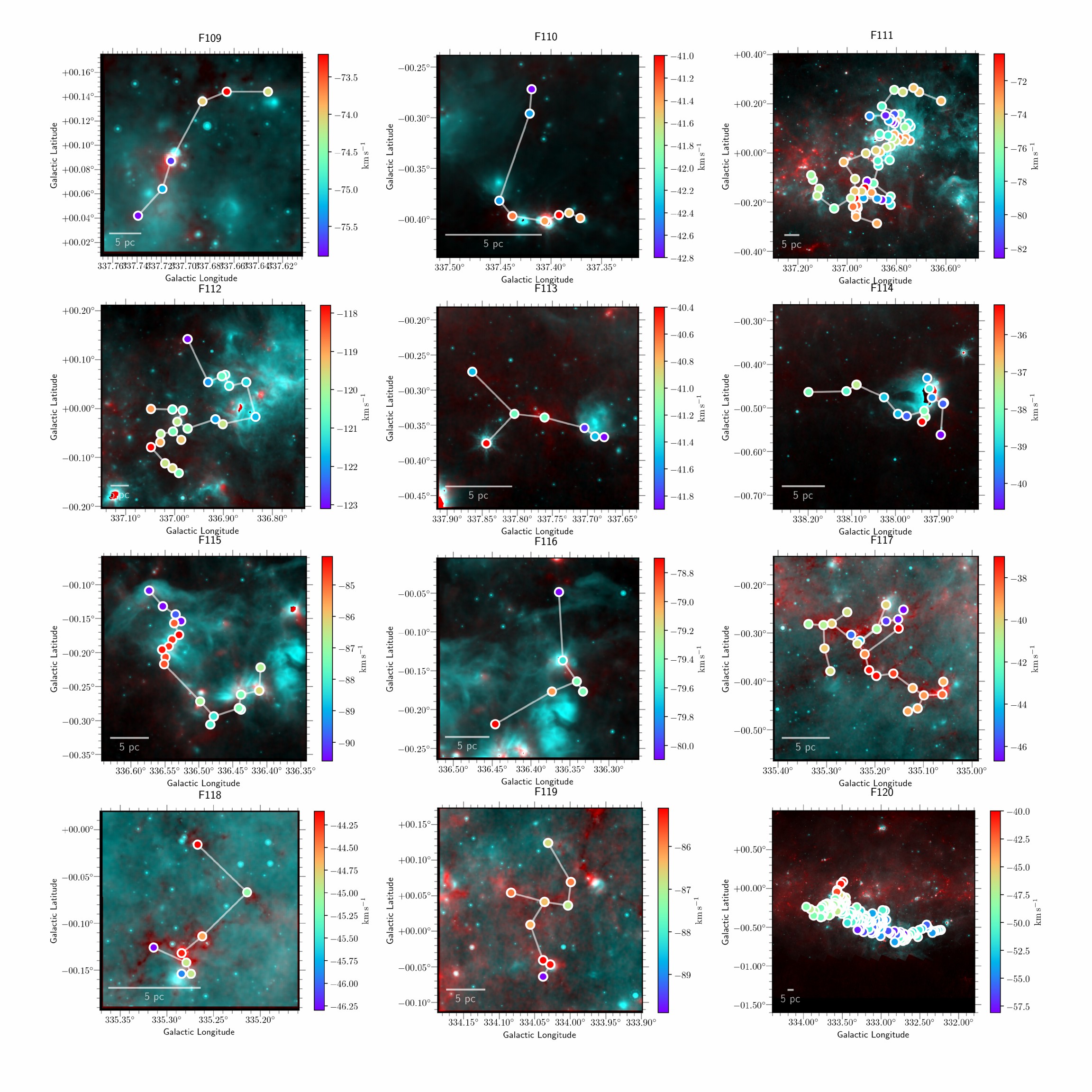}
\caption{(continued)}
\end{figure}
}}
\end{center}
\setcounter{figure}{0} 
\begin{center}
\setlength{\tabcolsep}{1.2mm}{{
\doublerulesep=5pt
\begin{figure}
\includegraphics[width=1\linewidth]{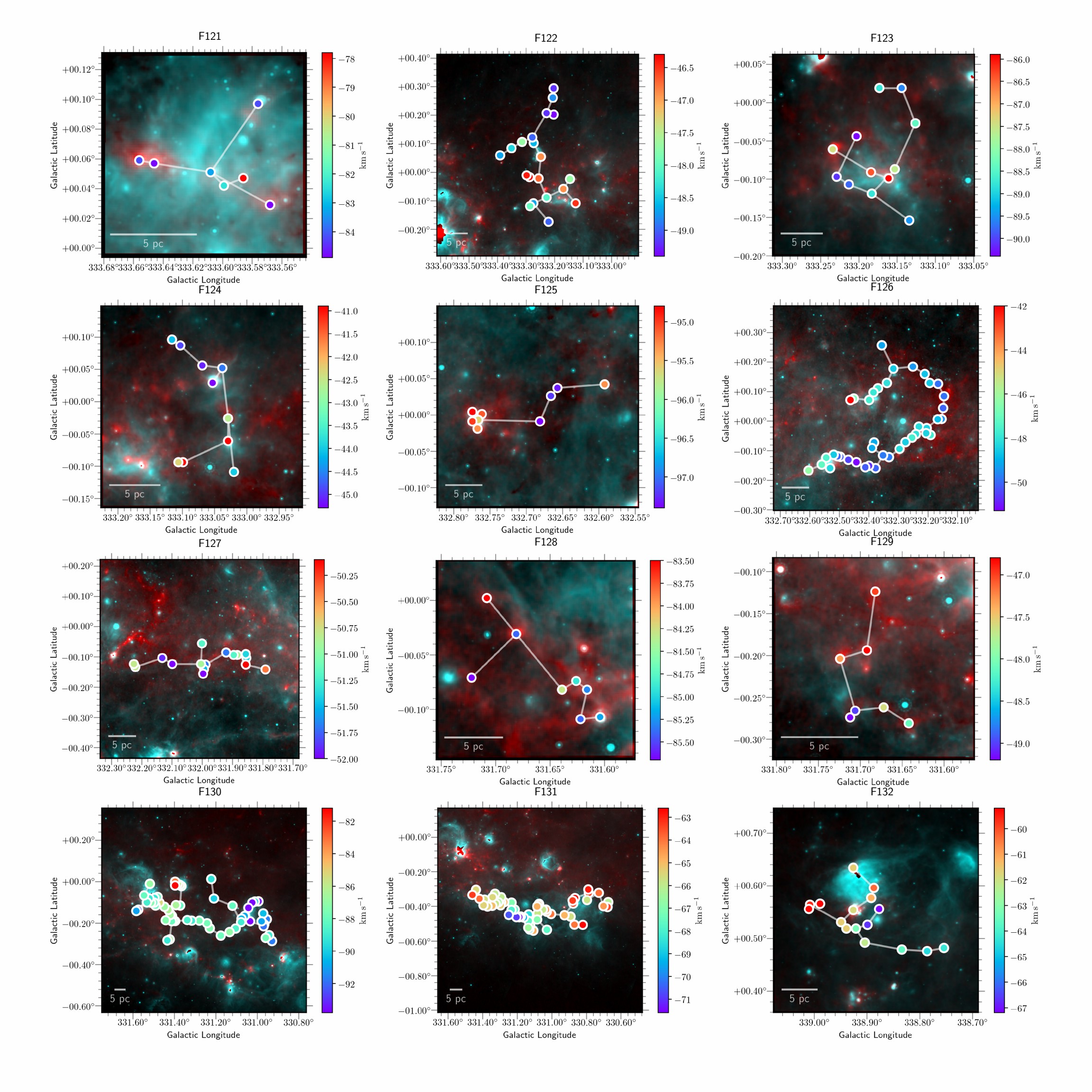}
\caption{(continued)}
\end{figure}
}}
\end{center}
\setcounter{figure}{0} 
\begin{center}
\setlength{\tabcolsep}{1.2mm}{{
\doublerulesep=5pt
\begin{figure}
\includegraphics[width=1\linewidth]{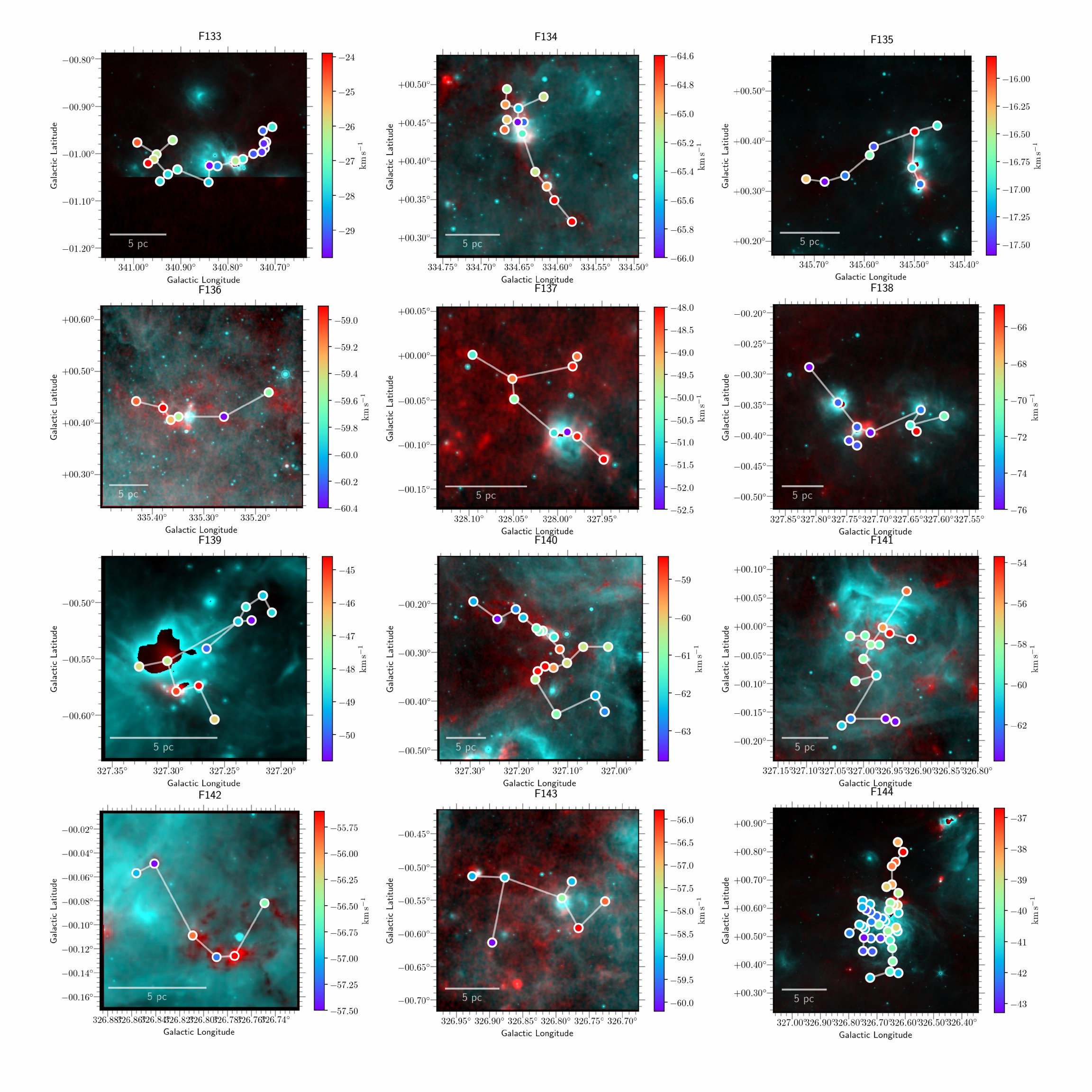}
\caption{(continued)}
\end{figure}
}}
\end{center}
\setcounter{figure}{0} 
\begin{center}
\setlength{\tabcolsep}{1.2mm}{{
\doublerulesep=5pt
\begin{figure}
\includegraphics[width=1\linewidth]{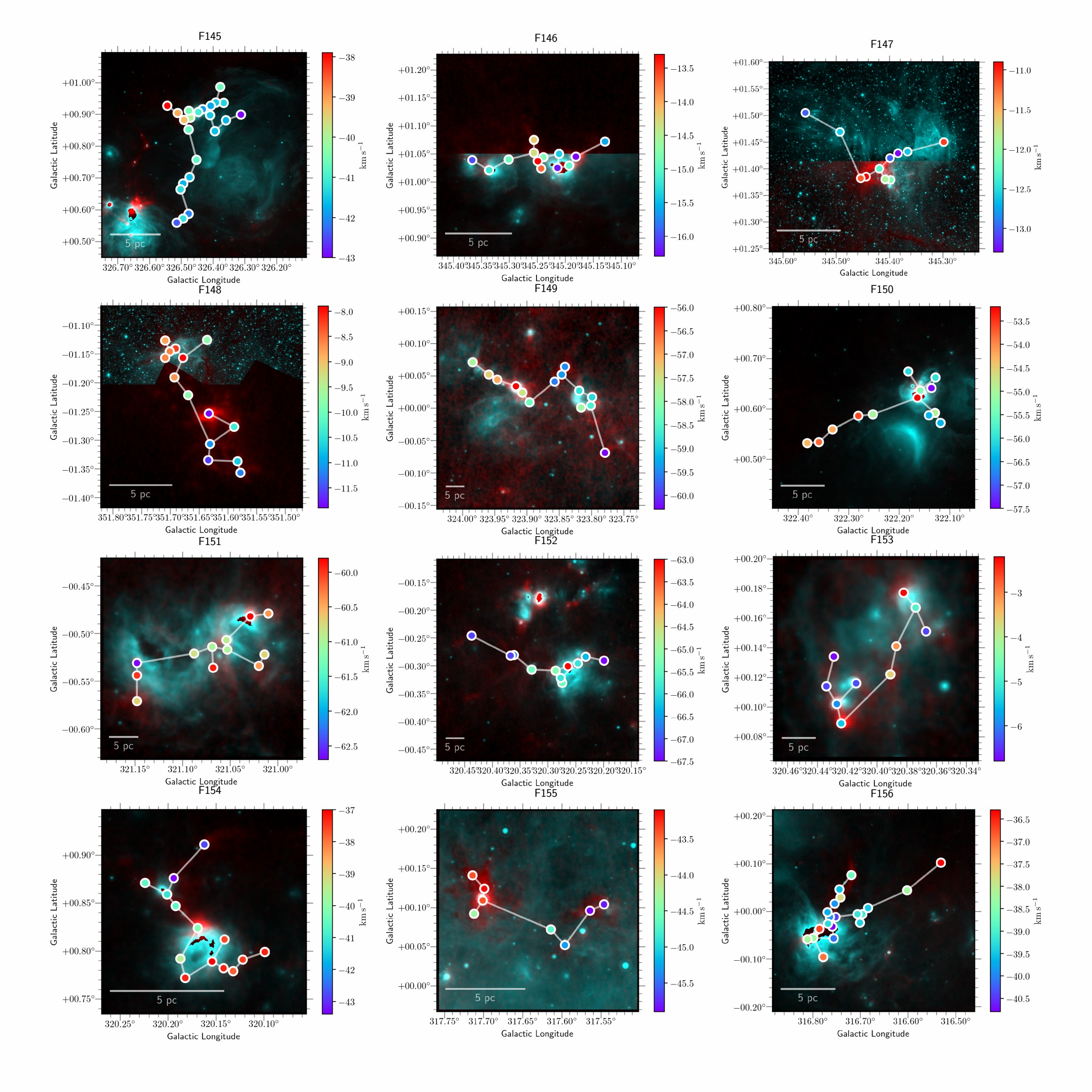}
\caption{(continued)}
\end{figure}
}}
\end{center}
\setcounter{figure}{0} 
\begin{center}
\setlength{\tabcolsep}{1.2mm}{{
\doublerulesep=5pt
\begin{figure}
\includegraphics[width=1\linewidth]{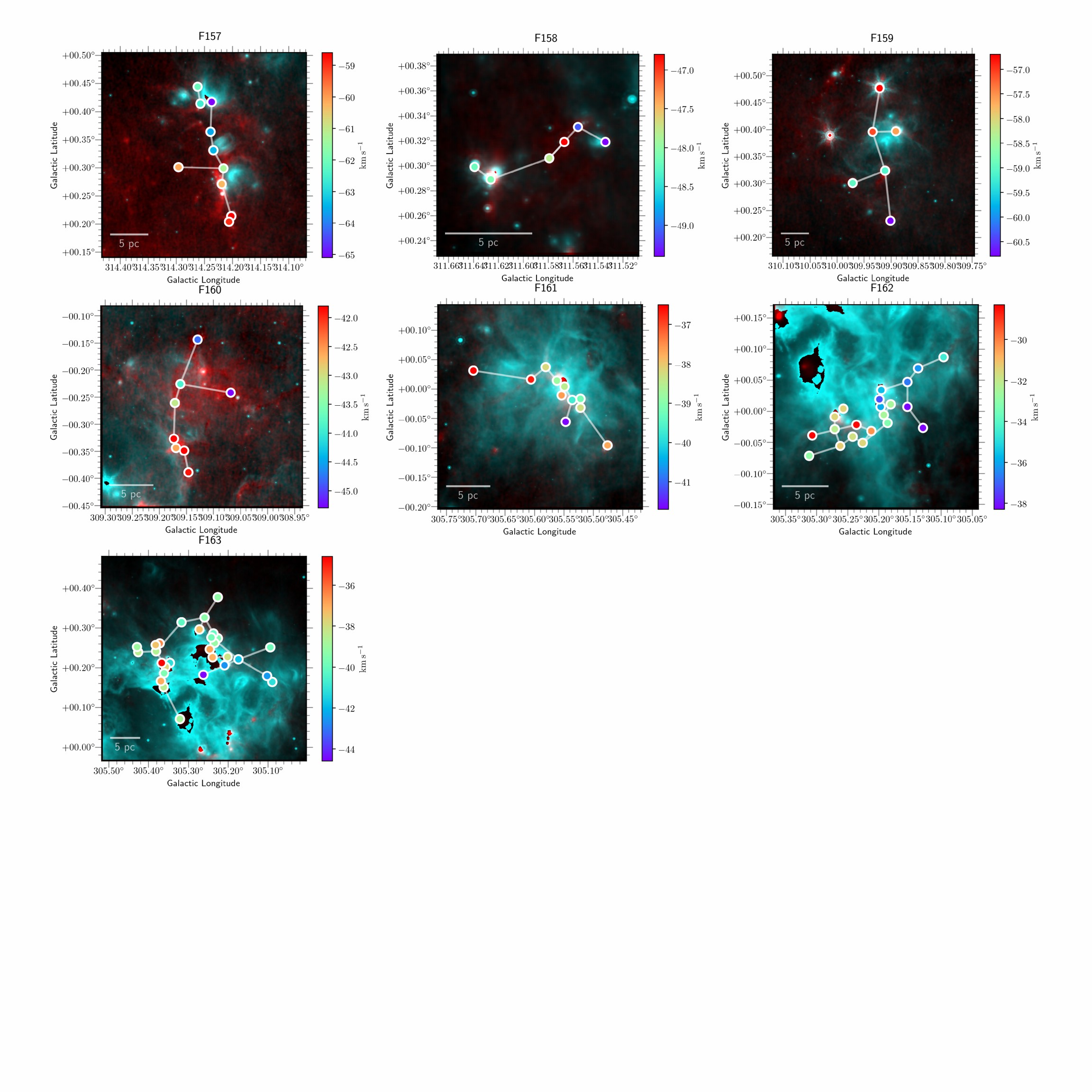}
\caption{(continued)}
\end{figure}
}}
\end{center}
\setcounter{figure}{1} 
\begin{center}
\setlength{\tabcolsep}{1.2mm}{{
\doublerulesep=5pt
\begin{figure}
\includegraphics[width=1\linewidth]{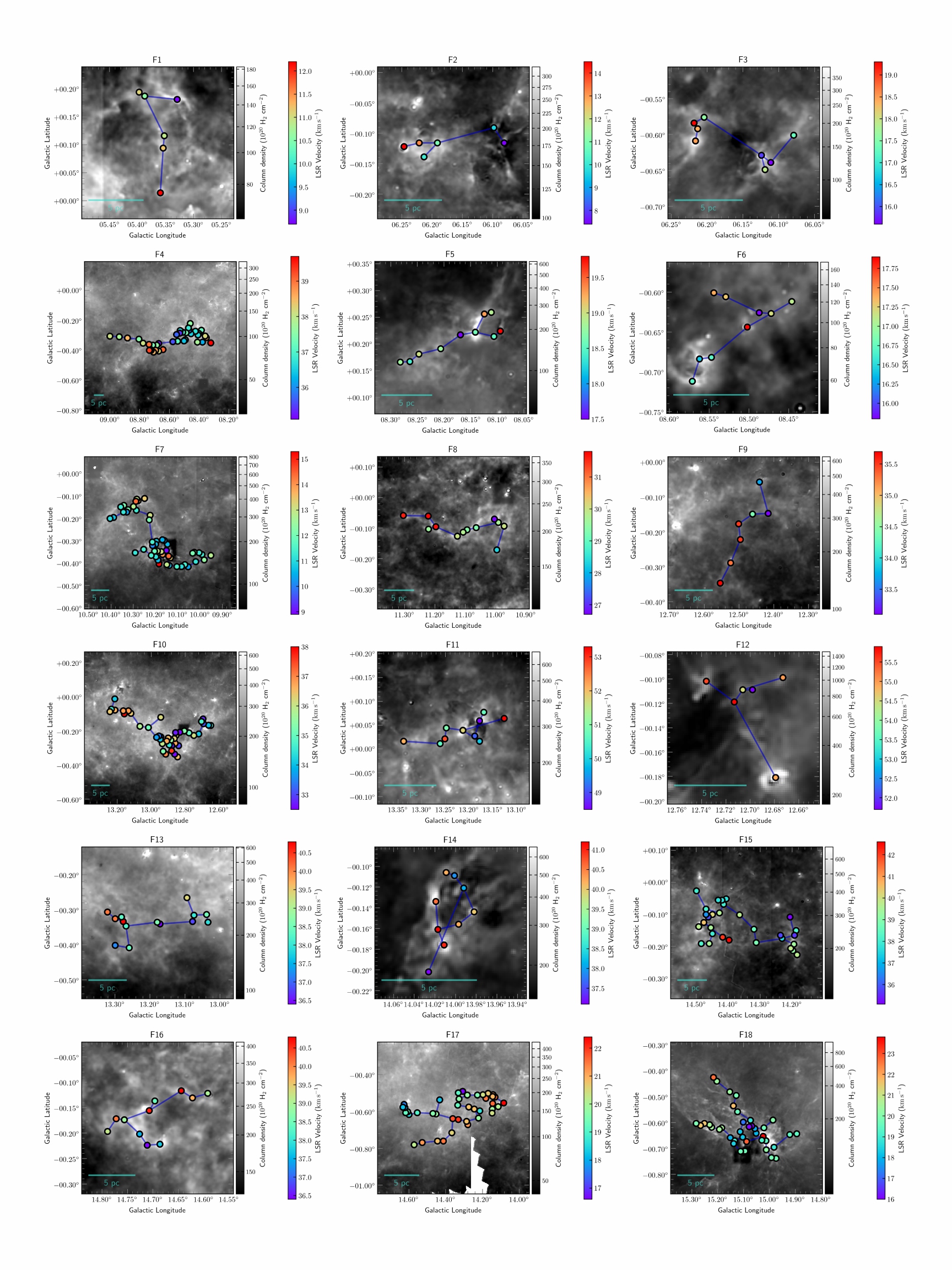}
\caption{Filaments overlaid on Column density maps from PPMAP \citep{Marsh2017}.}
\end{figure}
}}
\end{center}
\setcounter{figure}{1} 
\begin{center}
\setlength{\tabcolsep}{1.2mm}{{
\doublerulesep=5pt
\begin{figure}
\includegraphics[width=1\linewidth]{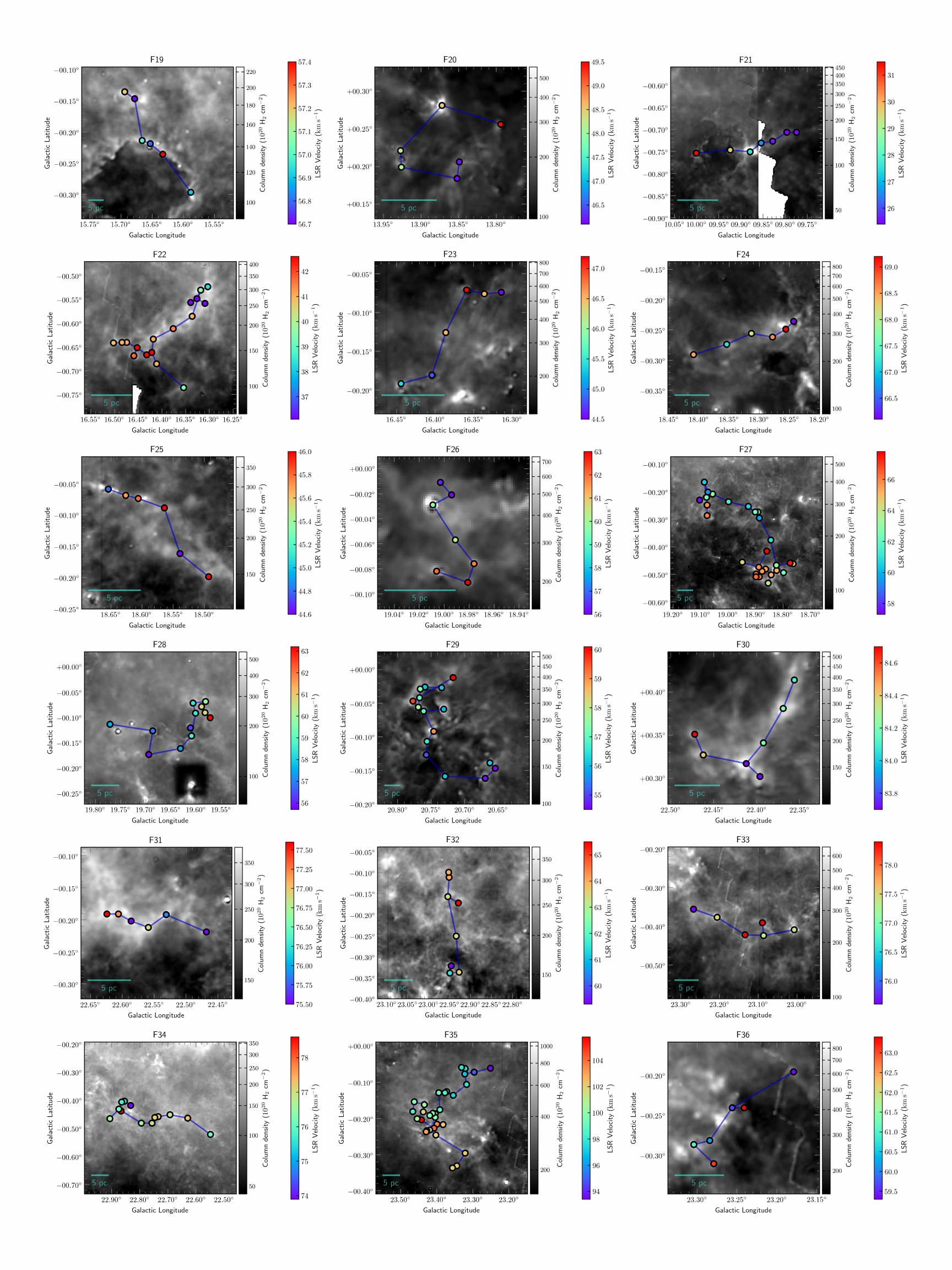}
\caption{(continued)}
\end{figure}
}}
\end{center}
\setcounter{figure}{1} 
\begin{center}
\setlength{\tabcolsep}{1.2mm}{{
\doublerulesep=5pt
\begin{figure}
\includegraphics[width=1\linewidth]{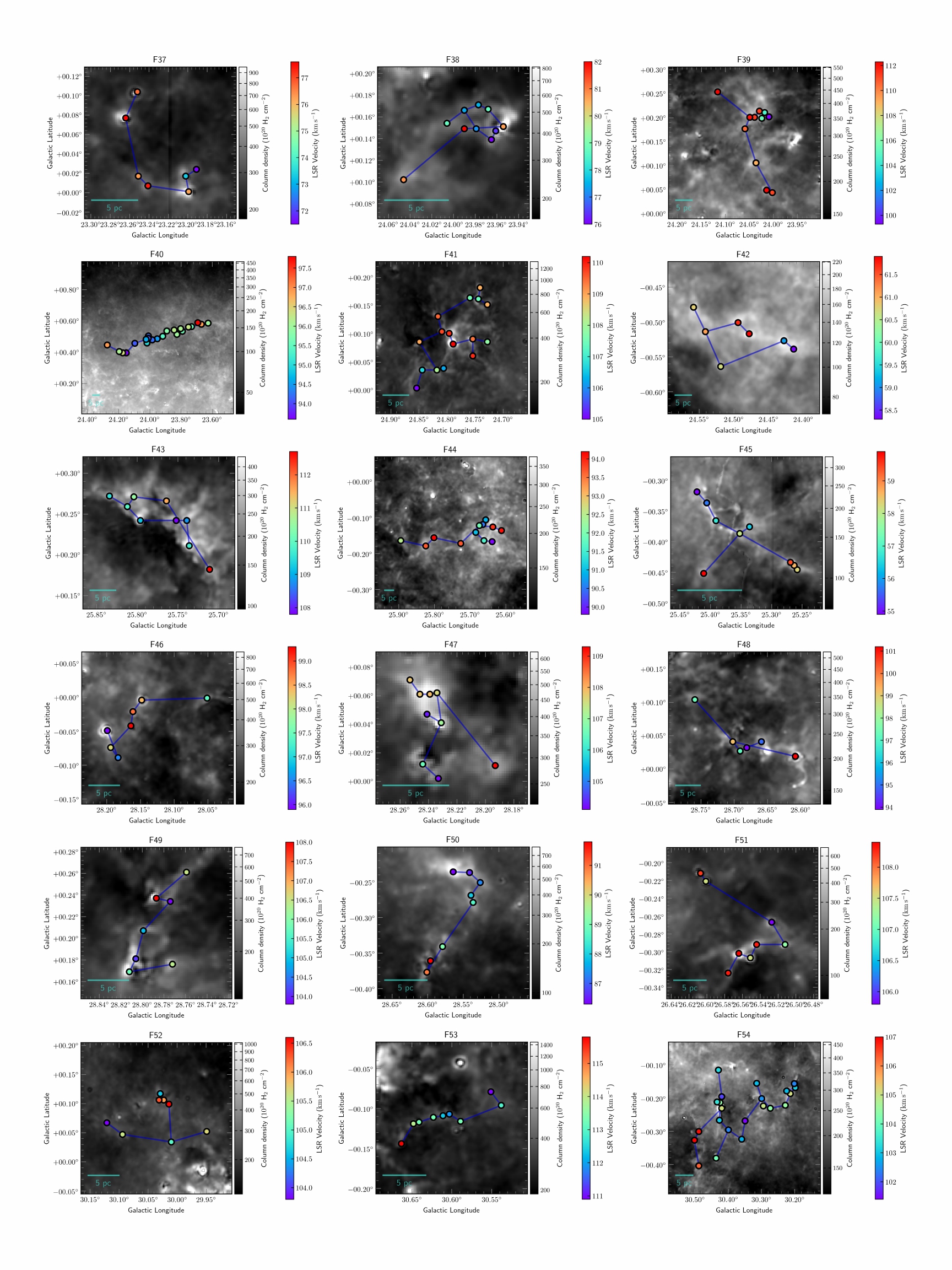}
\caption{(continued)}
\end{figure}
}}
\end{center}
\setcounter{figure}{1} 
\begin{center}
\setlength{\tabcolsep}{1.2mm}{{
\doublerulesep=5pt
\begin{figure}
\includegraphics[width=1\linewidth]{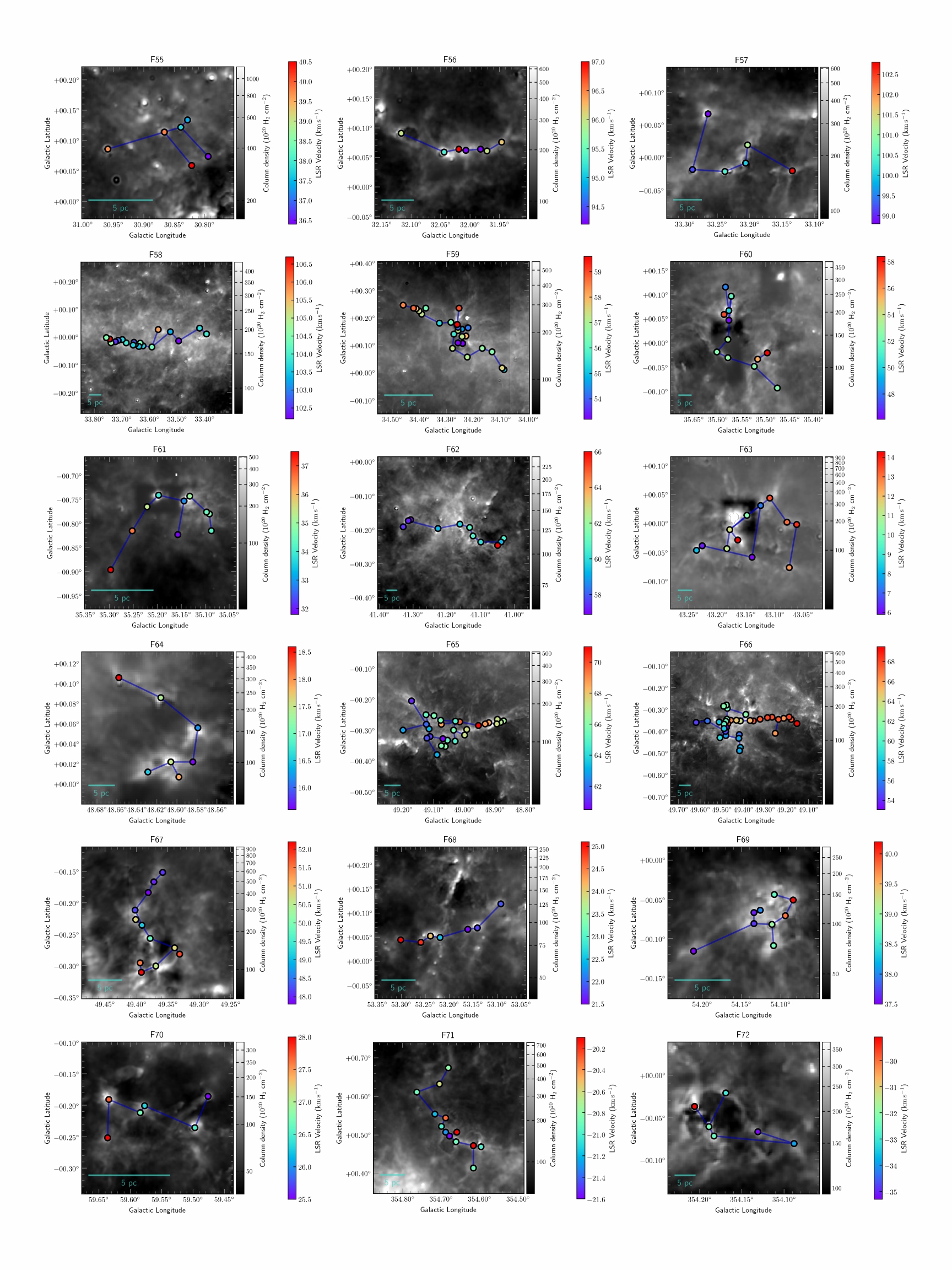}
\caption{(continued)}
\end{figure}
}}
\end{center}
\setcounter{figure}{1} 
\begin{center}
\setlength{\tabcolsep}{1.2mm}{{
\doublerulesep=5pt
\begin{figure}
\includegraphics[width=1\linewidth]{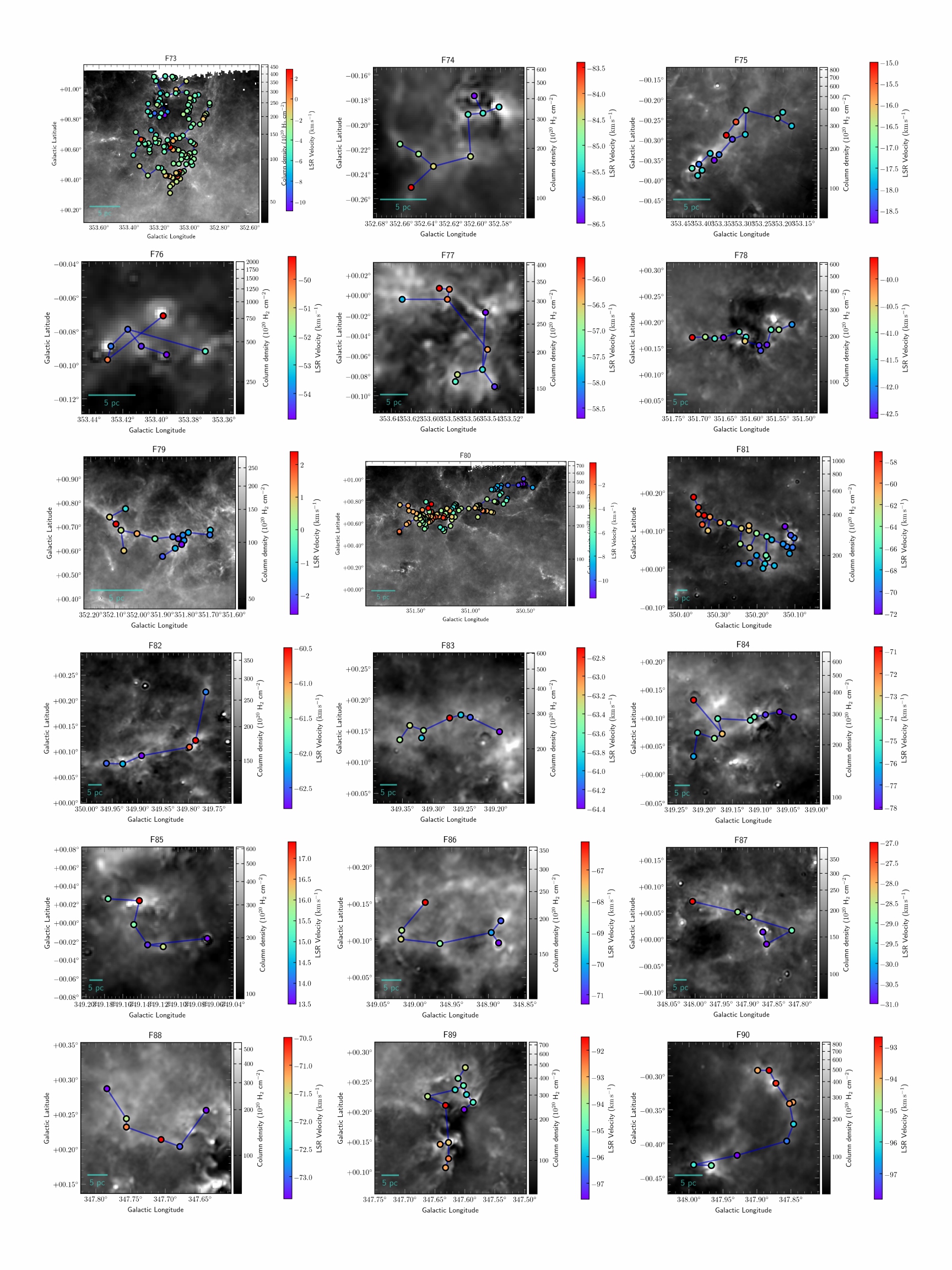}
\caption{(continued)}
\end{figure}
}}
\end{center}
\setcounter{figure}{1} 
\begin{center}
\setlength{\tabcolsep}{1.2mm}{{
\doublerulesep=5pt
\begin{figure}
\includegraphics[width=1\linewidth]{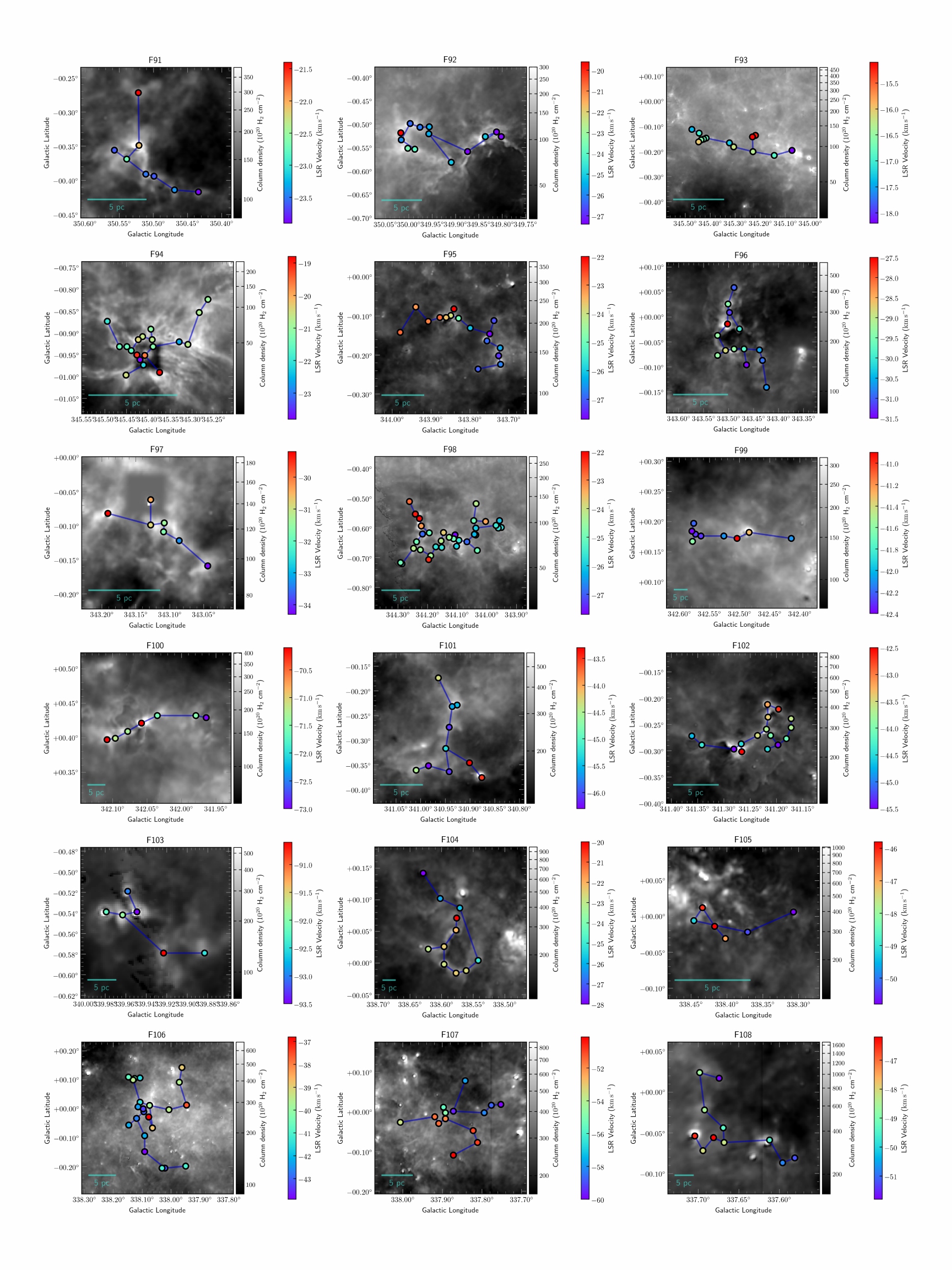}
\caption{(continued)}
\end{figure}
}}
\end{center}
\setcounter{figure}{1} 
\begin{center}
\setlength{\tabcolsep}{1.2mm}{{
\doublerulesep=5pt
\begin{figure}
\includegraphics[width=1\linewidth]{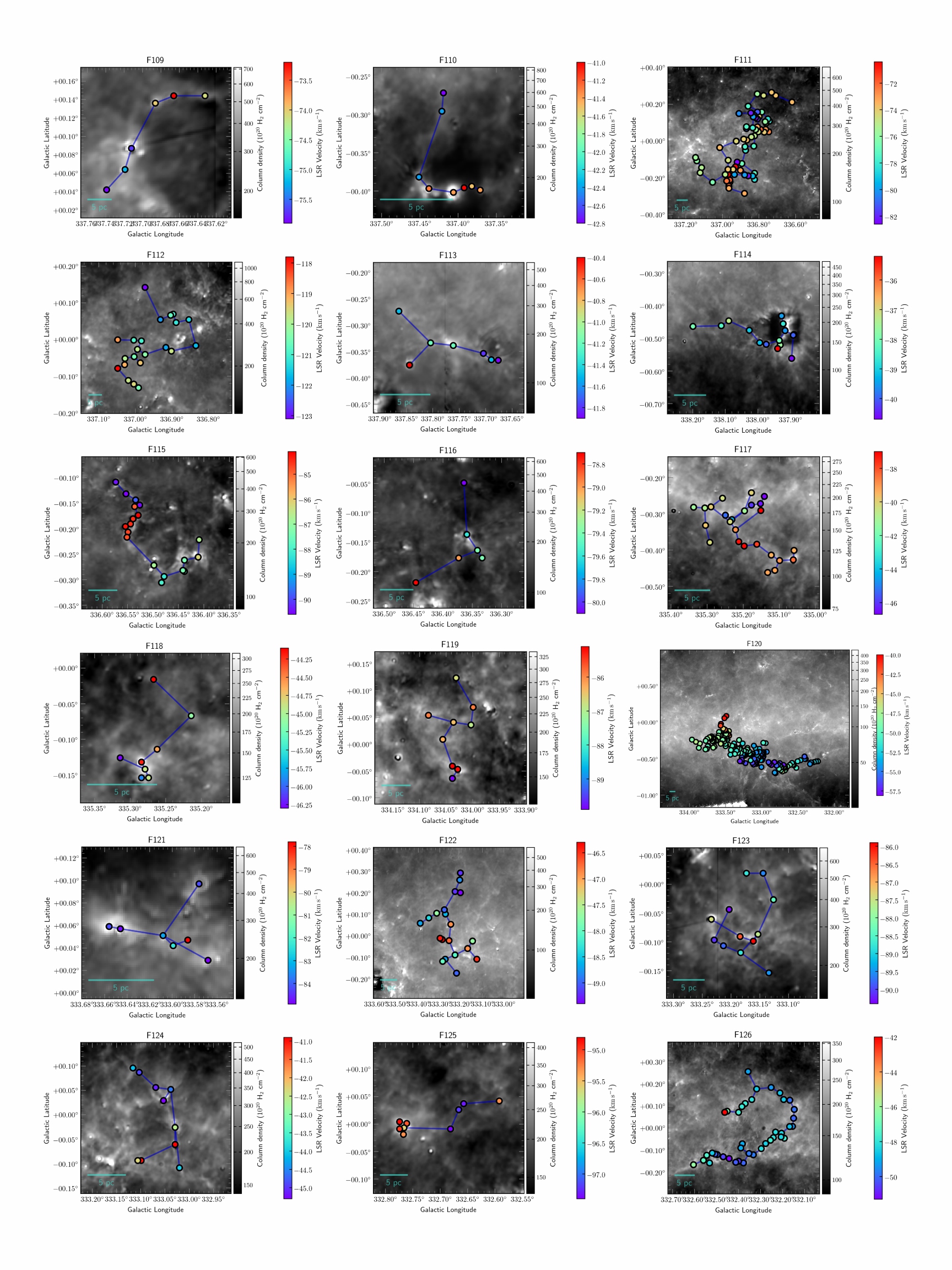}
\caption{(continued)}
\end{figure}
}}
\end{center}
\setcounter{figure}{1} 
\begin{center}
\setlength{\tabcolsep}{1.2mm}{{
\doublerulesep=5pt
\begin{figure}
\includegraphics[width=1\linewidth]{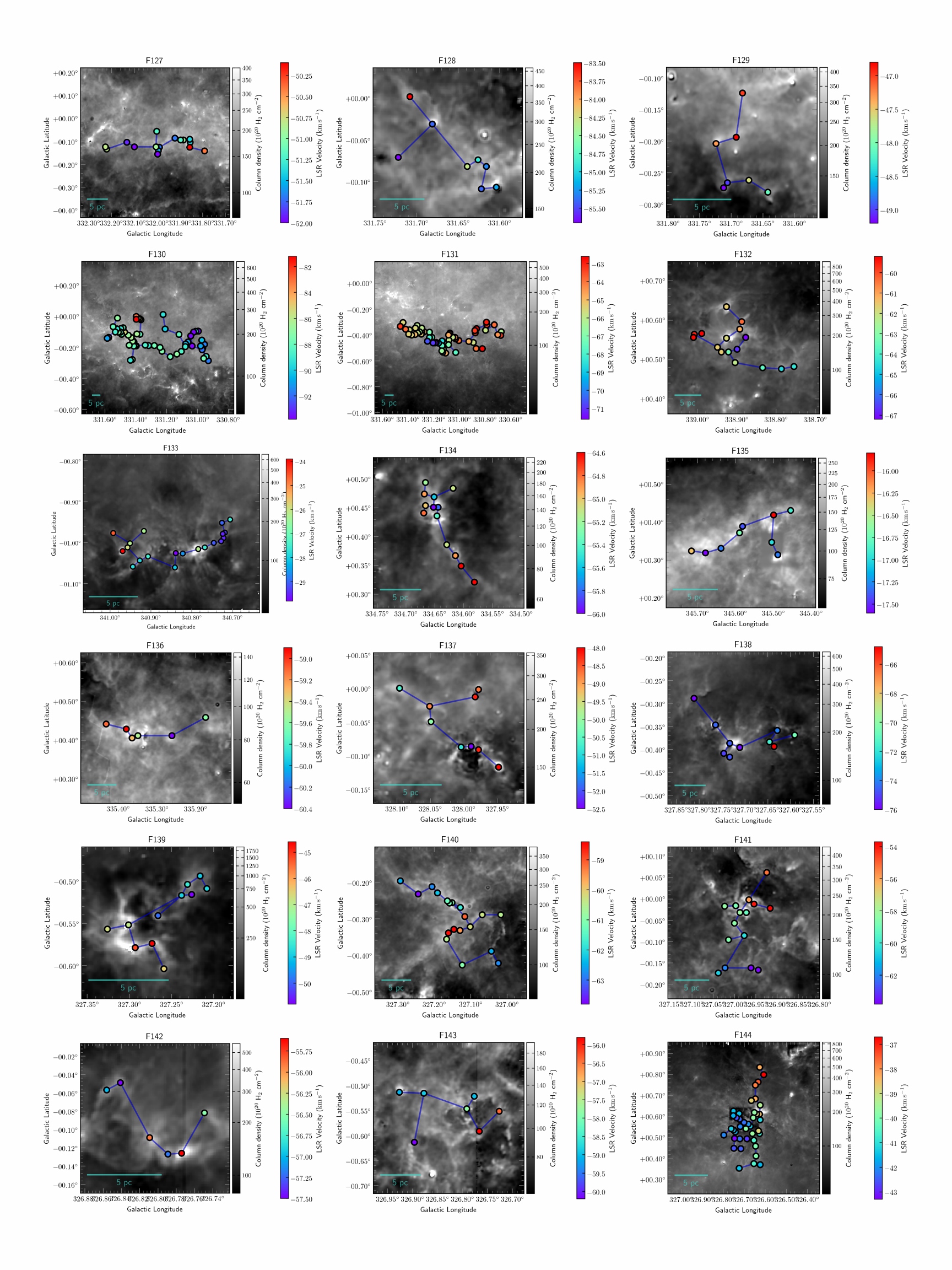}
\caption{(continued)}
\end{figure}
}}
\end{center}
\setcounter{figure}{1} 
\begin{center}
\setlength{\tabcolsep}{1.2mm}{{
\doublerulesep=5pt
\begin{figure}
\includegraphics[width=1\linewidth]{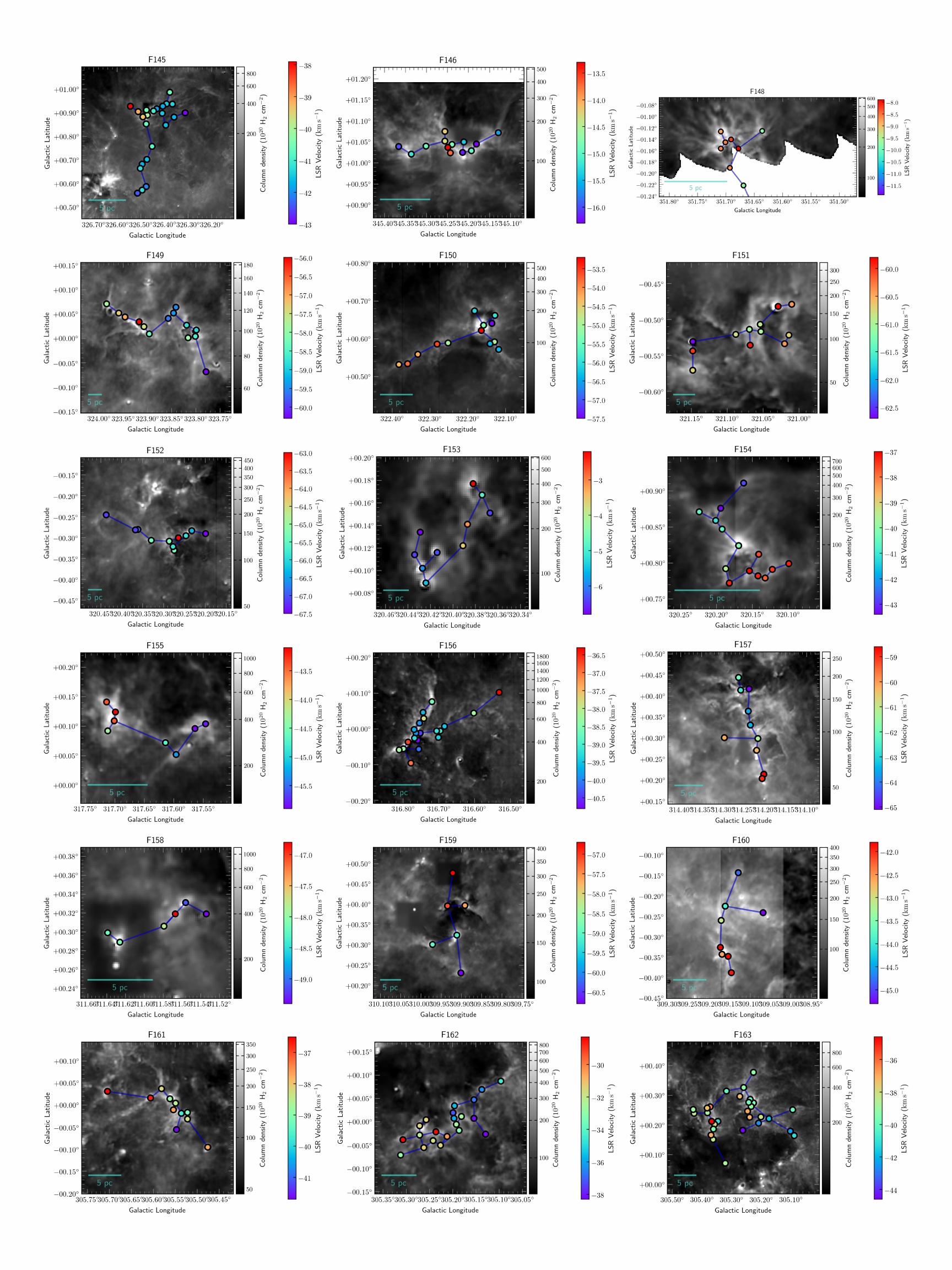}
\caption{(continued)}
\end{figure}
}}
\end{center}

\bibliographystyle{apj.bst}

%\bibliography{ref.bib}

\end{document}